\documentclass[11pt,a4paper]{article}
\pdfoutput=1
\usepackage{amssymb,amsmath,amsfonts, mathtools, mathrsfs}
\usepackage[utf8]{inputenc} 
\usepackage[dvipsnames]{xcolor}

\newif\ifnatbibsort\natbibsorttrue

\relax

\ifnatbibsort\RequirePackage[numbers,sort&compress]{natbib}\else\RequirePackage[numbers,compress]{natbib}\fi
\RequirePackage[colorlinks=true
,urlcolor=blue
,anchorcolor=blue
,citecolor=blue
,filecolor=blue
,linkcolor=blue
,menucolor=blue
,pagecolor=blue
,linktocpage=true
,pdfproducer=medialab
,pdfa=true
]{hyperref}

\usepackage{graphicx}
\usepackage{caption}
\usepackage{tensor}
\usepackage{subfigure}
\usepackage{enumerate}
\usepackage{dsfont}
\usepackage{verbatim}
\usepackage{amsfonts,euscript,amssymb,stmaryrd,braket}
\usepackage{tikz}
\usetikzlibrary{arrows,decorations.markings,patterns}
\usepackage{slashed}

\def\clock{{\count0=\time
		\divide\count0 60
		\ifnum\count0<10 0\fi\the\count0
		\multiply\count0 -60 \advance\count0 \time
		:\ifnum\count0<10 0\fi \the\count0
}}
\newcommand{\timestamp}{{\small\vbox{\hbox{\tt\jobname.tex}
			\hbox{\the\day/\the\month/\the\year, \clock}}}}

\newcommand{\bea}{\begin{eqnarray}}
\newcommand{\eea}{\end{eqnarray}}

\DeclareMathOperator{\diag}{diag}

\newcommand{\be}{\begin{equation}}
\newcommand{\ee}{\end{equation}}

\makeatletter
\let\old@startsection=\@startsection
\let\oldl@section=\l@section
\renewcommand{\@startsection}[6]{\old@startsection{#1}{#2}{#3}{#4}{#5}{#6\mathversion{bold}}}
\renewcommand{\l@section}[2]{\oldl@section{\mathversion{bold}#1}{#2}}
\makeatother

\numberwithin{equation}{section}

\usepackage{color}

\begin{document}
	\renewcommand{\thefootnote}{\arabic{footnote}}

	\overfullrule=0pt
	\parskip=2pt
	\parindent=12pt
	\headheight=0in \headsep=0in \topmargin=0in \oddsidemargin=0in

	\vspace{ -3cm} \thispagestyle{empty} \vspace{-1cm}
	\begin{flushright} 
		\footnotesize
		\textcolor{red}{\phantom{print-report}}
	\end{flushright}

\begin{center}
	\vspace{.5cm}

		{\Large\bf \mathversion{bold}
	Complexity of mixed Gaussian states
	}
	\\
	\vspace{.25cm}
	\noindent
	{\Large\bf \mathversion{bold}
	from Fisher information geometry}


	\vspace{0.8cm} {
		Giuseppe Di Giulio\footnote[1]{gdigiuli@sissa.it}
		and 
		Erik Tonni\footnote[2]{erik.tonni@sissa.it}
	}
	\vskip  0.7cm
	
	\small
	{\em
		SISSA and INFN Sezione di Trieste, via Bonomea 265, 34136, Trieste, Italy 
	}
	\normalsize
	
\end{center}

\vspace{0.3cm}
\begin{abstract} 
We study the circuit complexity for mixed bosonic  Gaussian states 
in harmonic lattices in any number of dimensions. 
By employing the Fisher information geometry for the covariance matrices,
we consider the optimal circuit 
connecting two states with vanishing first moments,
whose length is identified with the complexity to create a 
target state from a reference state through the optimal circuit. 
Explicit proposals to quantify 
the spectrum complexity and the basis complexity are discussed. 
The purification of the mixed states is also analysed.
In the special case of harmonic chains 
on the circle or on the infinite line, 
we report numerical results 
for thermal states and reduced density matrices.
\end{abstract}

\newpage

\tableofcontents

\section{Introduction}
\label{intro}

The complexity of a quantum circuit is an insightful notion of quantum information theory
\cite{Nielsen06,NielsenDowling06, DowlingNielsen08, 
 Watrous2008quantum,  Aaronson:2016vto,
 aharonov1998quantum}.
 During the last few years it has attracted increasing attention also
because it has been proposed as a new quantity to explore
within the (holographic) gauge/gravity correspondence between quantum (gauge) field theories 
and quantum gravity models from string theory.
In this context, 
different proposals have been made to evaluate the complexity of a quantum state 
by considering different geometric constructions in the gravitational dual
\cite{Susskind:2014rva, Stanford:2014jda, Susskind:2014moa,Alishahiha:2015rta, Brown:2015bva,Brown:2015lvg,
Barbon:2015ria,Barbon:2015soa, Couch:2016exn, Carmi:2016wjl}.

A quantum circuit constructs a target state 
by applying a specific sequence of gates to a reference state. 
The circuit complexity is given by the minimum number of allowed gates that is needed to construct 
the target state starting from the assigned reference state.
This quantity depends on the target state, on the reference state, 
on the set of allowed gates and, eventually, on the specified tolerance for the target state.
Notice that this definition of complexity does not require the introduction 
of ancillary degrees of freedom.

Remarkable results have been obtained over the past few years
in the attempt to evaluate complexity in quantum field theories
\cite{Jefferson:2017sdb,Guo:2018kzl,Khan:2018rzm,Hackl:2018ptj,Bhattacharyya:2018bbv,Chapman:2018hou,Caceres:2019pgf,Braccia:2019xxi,Chapman:2019clq,Doroudiani:2019llj,guo2020circuit,Jaiswal:2020yif,Agon:2018zso,Caputa:2017urj,Czech:2017ryf,Caputa:2017yrh,
Chapman:2017rqy,
Bhattacharyya:2018wym,Caputa:2018kdj,Camargo:2018eof,Chapman:2018bqj,Camargo:2019isp,Ge:2019mjt,Bueno:2019ajd,Sato:2019kik,Erdmenger:2020sup,
Flory:2020eot}. 
Despite these advances,
it remains an interesting open problem that deserves 
further investigations.

In order to understand the circuit complexity in continuum theories,
it is worth exploring the complexity of a process that constructs a quantum state
in lattice models whose continuum limit is well understood. 
The free scalar and the free fermion are the simplest models to consider.
For these models, it is worth focussing on the Gaussian states 
because they provide
an interesting arena that includes important states (e.g. the ground state and the thermal states) 
and that has been largely explored in the literature of quantum information
\cite{ferraro2005gaussian,Holevo11book,Weedbrook12b, Adesso14, Serafini17book}.
The bosonic Gaussian states are particularly interesting because,
despite the fact that the underlying Hilbert space is infinite dimensional,
they can be studied  through techniques of finite dimensional linear algebra.

Various studies have explored the complexity of quantum circuits 
made by pure Gaussian states in lattice models
\cite{Jefferson:2017sdb,Bhattacharyya:2018bbv,Chapman:2018hou,Guo:2018kzl,Hackl:2018ptj,
Khan:2018rzm,Caceres:2019pgf,Braccia:2019xxi,Chapman:2019clq,Doroudiani:2019llj,guo2020circuit}.
In these cases the gates implement only unitary transformations of the state. 
It is important to extend these analyses by considering quantum circuits
that involve also mixed states;
hence it is impossible to construct them 
by employing only unitary gates \cite{aharonov1998quantum}.
A natural way to construct mixed states consists in 
considering the system in a pure state and tracing out some degrees of freedom.
This immediately leads to consider the entanglement entropy and other entanglement quantifiers
(see \cite{Eisert:2008ur,Casini:2009sr,Calabrese:2009qy,Peschel_2009} for reviews).
The same consideration holds within the context of the holographic correspondence,
where the gravitational dual of the entanglement entropy has been found in 
 \cite{Ryu:2006bv,Ryu:2006ef,Hubeny:2007xt}
 (see \cite{Rangamani:2016dms,Headrick:2019eth,Tonni:2020bjq}
 for recent reviews).

The notions of complexity are intimately related to the geometry of quantum states \cite{GeomQuantumStates_book}. 
While for pure states a preferred geometry can be defined, 
when mixed states are involved, different metrics have been introduced in a consistent way \cite{LinkStrunz15}.
Furthermore, for quantum circuits made also by mixed states, the notions of spectrum complexity and basis complexity can be introduced \cite{Agon:2018zso}.

A method to quantify the complexity of circuits involving mixed states
has been recently investigated in \cite{Caceres:2019pgf}.
In this approach, the initial mixed state is purified by adding ancillary degrees of freedom
and the resulting pure state is obtained by minimising the circuit complexity within the set of pure states.
This procedure requires the choice of a fixed pure state 
to evaluated this circuit complexity for pure states.

In this manuscript we explore a way 
to evaluate the complexity of quantum circuits made by mixed states
within the framework of the Information Geometry
\cite{Chentsov82book,Amari16book,nielsen2018elementary}.
The method holds for bosonic Gaussian states 
and it does not require the introduction of ancillary degrees of freedom.
It relies on the fact that, whenever the states provide a Riemannian manifold
and the available gates allow to reach every point of the manifold, 
the standard tools of differential geometry can be employed to find
the optimal circuit connecting two states.
Since the pure states provide a submanifold of this manifold,
this analysis also suggests natural quantum circuits to purify a given mixed state.

We focus only on the bosonic Gaussian states 
occurring in the Hilbert space of harmonic lattices
in any number of dimensions. 
These are prototypical examples of
continuous variable quantum systems;
indeed, they can be described by the positions and the momenta,
which are continuous variables.
The bosonic Gaussian states 
are completely characterised by their covariance matrix,
whose elements can be written in terms of the two point correlators,
and by their first moments. 
The covariance matrices associated to these quantum states
are real symmetric and positive definite matrices 
constrained by the validity of the uncertainty principle 
\cite{ferraro2005gaussian,Weedbrook12b, Adesso14, Serafini17book, Holevo11book}.
We mainly explore the bosonic Gaussian mixed states with vanishing first moments. 
This set can be described by a proper subset of the 
Riemann manifold defined by the symmetric and positive definite matrices 
\cite{PETZ199681,Lawson01,Forstner2003,Bhatia07book,Bathia15}
equipped with the metric provided by the Fisher information matrix
\cite{Rao45,Atkinson81,Chentsov82book,Amari16book,Felice17}.
We remark that our analysis considers quantum circuits that are made by Gaussian states only.
Despite this important simplifying assumption, 
the resulting quantum circuits are highly non trivial because
non unitary states are involved in the circuit. 
In this setting, by exploiting the Williamson's theorem \cite{Williamson36}, 
we can consider circuits whose reference and target states have either the same spectrum or can be associated to the same basis.
This allows us to propose some ways to quantify the spectrum and the basis complexity for bosonic Gaussian states with vanishing first moments.

The manuscript is organised as follows.
In Sec.\,\ref{sec:FR-distance} we introduce the quantities 
and the main results employed throughout the manuscript:
the covariance matrix through the Gaussian Wigner function,
the Fisher-Rao distance between covariance matrices
and the corresponding geodesics, that provide the optimal circuits.
The particular cases given by pure states, thermal states
and coherent states (the latter ones need further results discussed in
Appendix\;\ref{app:first-moments}) are explicitly considered. 
In Sec.\,\ref{sec:spectrum-basis} 
we provide explicit expressions to evaluate the 
spectrum complexity and the basis complexity,
by employing also the first law of complexity 
\cite{Bernamonti:2019zyy,Bernamonti:2020bcf}.
The purification of a mixed state is explored
in Sec.\,\ref{sec:purification}, where
particular optimal circuits are mainly considered.
In Sec.\,\ref{sec:bounds} 
we discuss some lower and upper bounds on the complexity.
In Sec.\,\ref{sec:comp-eh} 
we focus on the circuits that do not contain pure states
because they can be also parameterised through
the entanglement hamiltonian matrices.
The Gaussian channels underlying the 
optimal circuits are briefly discussed in Sec.\,\ref{sec:gauss-channel}.
In Sec.\,\ref{sec:purification-approaches} 
we describe the approach to the complexity of mixed states 
based on the purification of a mixed state through ancillary degrees of freedom. 
The last analysis reported in Sec.\,\ref{sec:examples} 
focuses on the periodic harmonic chain in one spatial dimension
and on its limiting regime given by the harmonic chain on the
infinite line. 
Numerical results are reported both
for some quantities introduced in the other sections 
and for other quantities like the mutual complexity
for the thermofield double states and 
for the reduced density matrices. 
Finally, in Sec.\,\ref{sec:discussion} we summarise our results
and discuss future directions. 

Some appendices 
(\ref{app:wwm}, \ref{app:sec_first_law} and \ref{app:matrix_id})
contain the derivation of selected results reported in the main text
and related technical details.
Other appendices, instead, provide
complementary analyses that expand the discussion of the main text,
adding further results.
In particular, 
in Appendix\;\ref{app:first-moments} 
we explore Gaussian states with non vanishing first moments,
in Appendix\;\ref{app:other_dist} 
the Bures and the Hilbert-Schmidt distances are discussed,
in Appendix\;\ref{app-tfd} 
the complexity of the thermofield double states is explored
and in Appendix\;\ref{app:c1} 
we describe the two particular bases 
employed in \cite{Caceres:2019pgf}
to study the complexity of mixed states
through the $F_1$ cost function.

%
%

\newpage
\section{Complexity as Fisher-Rao distance and the optimal path}
\label{sec:FR-distance}

In Sec.\,\ref{subsec:phase-space} we introduce Gaussian Wigner functions
(defined in terms of the covariance matrix and of the first moment)
to characterise a generic Gaussian state.
The Fisher-Rao distance and other distances are defined in Sec.\,\ref{subsec:FRdist}.
In Sec.\,\ref{sec:williamson} we discuss the Williamson's decomposition of the 
covariance matrix, a crucial tool largely employed throughout the manuscript. 
The optimal circuit in the Fisher information geometry is 
analysed in Sec.\,\ref{subsec:mixed}.
The special cases given by pure states 
and thermal states are explored in Sec.\,\ref{subsec:pure}
and Sec.\,\ref{subsec:thermal} respectively. 
Finally, in Sec.\,\ref{sec-coherent-states} some results
about the complexity for the coherent states are discussed.

\subsection{Gaussian states in harmonic lattices}
\label{subsec:phase-space}

The hamiltonian of a spatially homogeneous harmonic lattice made by $N$ sites with  
nearest neighbour spring-like interaction with spring constant $\kappa$ reads
\be
\label{HC ham}
\widehat{H} 
\,=\, 
\sum_{i=1}^{N} \left(
\frac{1}{2m}\,\hat{p}_i^2+\frac{m\omega^2}{2}\,\hat{q}_i^2 
\right) 
+
 \sum_{\langle i,j \rangle}\frac{\kappa}{2}(\hat{q}_{i} -\hat{q}_j)^2
 \,=\,
 \frac{1}{2}\, \hat{\boldsymbol{r}}^{\textrm t} H^{\textrm{\tiny phys}} \, \hat{\boldsymbol{r}}
\ee
where the second sum is performed over the nearest neighbour sites. 
The position and momentum operators $\hat{q}_i$ and $\hat{p}_i$ 
are hermitian and  satisfy the canonical commutation relations 
$[\hat{q}_i, \hat{q}_j] = [\hat{p}_i, \hat{p}_j]  =0$ and $[\hat{q}_i, \hat{p}_j] = \textrm{i} \delta_{ij}$ 
(we set $\hbar =1$ throughout this manuscript).
The boundary conditions do not change the following discussion,
although  they  are crucial to determine the explicit expressions of the correlators.
Collecting the position and momentum operators into the vector 
$\hat{\boldsymbol{r}} \equiv (\hat{q}_1 , \dots , \hat{q}_N, \hat{p}_1, \dots, \hat{p}_N)^{\textrm{t}}$,
the canonical commutation relations can be written in the form $[\hat{r}_i, \hat{r}_j] = \textrm{i} J_{ij}$,
where $J$ is the standard symplectic matrix 
\be
\label{Jmat}
J \equiv
\bigg(  \begin{array}{cc}
 \boldsymbol{0} &  \; \boldsymbol{1} \\
 - \boldsymbol{1} &  \; \boldsymbol{0} \\
\end{array}  \, \bigg)
\ee
 and we have denoted by $\boldsymbol{1}$ the $N \times N$ identity matrix 
and $ \boldsymbol{0}$ the matrix with the proper size having all its elements equal to zero. 
Notice that $J^2 = - \boldsymbol{1}$ and $J^{\textrm{t}} = J^{-1} = -J$.

The real symplectic group $\textrm{Sp}(2N,\mathbb{R})$ is
made by the real $2N \times 2N$ matrices $S$ characterising
the linear transformations $\hat{\boldsymbol{r}} \to \hat{\boldsymbol{r}}' = S\, \hat{\boldsymbol{r}} $
that preserve the canonical commutation relations
\cite{Simon87,Simon94,Arvind:1995ab,Simon99,deGosson06book}.
This condition is equivalent to $S J S^{\textrm t} = J$.  
Given $S\in \textrm{Sp}(2N,\mathbb{R})$, it can be shown that
$\textrm{det}(S) =  1$, $S^{\textrm{t}} \in \textrm{Sp}(2N,\mathbb{R})$ and $S^{-1} = J S^{\textrm t} J^{-1}$, hence $S^{-\textrm{t}} = J^{\textrm t}  S J$
(we have adopted the notation $M^{-\textrm{t}} \equiv  ( M^{\textrm{t}} )^{-1}$).
The real dimension of $\textrm{Sp}(2N,\mathbb{R})$ is $N(2N+1)$.

The density matrix $\hat{\rho}$, that characterises
a state of the quantum system described by the hamiltonian (\ref{HC ham}), 
is a positive definite, hermitean operator whose trace is normalised to one. 
When the state is pure, the operator $\hat{\rho}$ is a projector.

A useful way to characterise a density matrix is based on the Wigner function $w(\boldsymbol{r})$,
that depends on the vector $\boldsymbol{r}$ made by $2N$ real components.
The Wigner function is defined through
the Wigner characteristic function associated to $\hat{\rho}$, that is \cite{Weedbrook12b,Adesso14,Serafini17book}
\be
\label{CharFuncDef}
\chi (\mathbf{\boldsymbol{\xi}})
\,\equiv\,
\mathrm{Tr}\big(
\hat{\rho}\,
e^{ \textrm{i}\, \hat{\boldsymbol{r}}^{\textrm{t}} J\, \boldsymbol{\xi}}
\big)
=
\mathrm{Tr}\big(
\hat{\rho}\,\widehat{D}_{\boldsymbol{\xi}}
\big)
\;\;\qquad \;\;
\boldsymbol{\xi} \in \mathbb{R}^{2N}
\ee
where in the last step we have introduced the displacement operator as
\be
\label{displacement-op-def-0}
\widehat{D}_{\boldsymbol{a}}
\equiv \,
e^{-\textrm{i}\boldsymbol{a}^{\textrm{t}} J \,\hat{\boldsymbol{r}}}
\;\;\qquad\;\;
\boldsymbol{a} \in \mathbb{R}^{2N}\,.
\ee
The Fourier transform of the Wigner characteristic function provides 
the Wigner function
\be
\label{WignerFuncDef}
w(\boldsymbol{r})
\,\equiv\,
\frac{1}{(2\pi)^{2N}}
 \int\! 
\chi(\mathbf{\boldsymbol{\xi}})\,
e^{- \textrm{i}\, \boldsymbol{r}^{\textrm{t}} J \,\boldsymbol{\xi}} \,
d\boldsymbol{\xi}
\ee
where $d\boldsymbol{\xi}=\prod_{i=1}^{2N} d\xi_i$ 
denotes the integration over the $2N$ real components of $\boldsymbol{\xi}$.

In this manuscript we focus on the Gaussian states of the harmonic lattices,
which are the states whose Wigner function is Gaussian 
 \cite{Weedbrook12b,Eisert:2008ur, Audenaert:2002xfl, Plenio:2004he,Cramer:2005mx,
 Holevo75, Holevo11book}
\be
\label{WignerGaussian}
w_{\textrm{\tiny G}}(\boldsymbol{r};\gamma,\langle\hat{\boldsymbol{r}}\rangle)
\equiv
\frac{e^{-\frac{1}{2}
(\boldsymbol{r}-\langle\hat{\boldsymbol{r}}\rangle)^{\textrm{t}}
\,\gamma^{-1}\,
(\boldsymbol{r}-\langle\hat{\boldsymbol{r}}\rangle)}
}{
(2\pi)^{N} \, \sqrt{\textrm{det}(\gamma)}}\,.
\ee
The $2N \times 2N$ real, symmetric and positive definite matrix $\gamma$
is the covariance matrix of the Gaussian state,
whose elements can be defined in terms of the anticommutator of the operators $\hat{r}_i$ as follows
\be
\label{def_CM_gen}
\gamma_{i,j}
=
\frac{1}{2} \langle \{ \hat{r}_i - \langle\hat{r}_i\rangle\,, \hat{r}_j - \langle\hat{r}_j\rangle\} \rangle
\,=\,
\frac{1}{2} \langle \{ \hat{r}_i \,, \hat{r}_j \} \rangle
- \langle  \hat{r}_i \rangle \,\langle  \hat{r}_j \rangle 
=
\langle \hat{r}_i\, \hat{r}_j \rangle 
- \langle  \hat{r}_i \rangle \,\langle  \hat{r}_j \rangle 
- \frac{\textrm{i}}{2} \, J_{i,j}\,.
\ee
The covariance matrix $\gamma$ is determined by $N(2N+1)$ real parameters.
The expressions (\ref{WignerGaussian}) and (\ref{def_CM_gen})
tell us that the Gaussian states
are completely characterised by the one-point correlators (first moments) 
and by the two-points correlators (second moments) of the 
position and momentum operators collected into the vector $\hat{\boldsymbol{r}}$.
It is important to remark that the validity of the uncertainty principle imposes the following condition 
on the covariance matrix
\cite{Simon94,Weedbrook12b}
\be
\label{UncertaintyPrinc}
\gamma + \frac{\textrm{i}}{2} \, J \,\geqslant \,0\,.
\ee
In \cite{Bathia15} a real, positive matrix with an even size and 
satisfying (\ref{UncertaintyPrinc}) is called Gaussian matrix.
Thus, every symmetric Gaussian matrix provides the covariance matrix of a Gaussian state.

A change of base $\hat{\boldsymbol{r}} \to \hat{\boldsymbol{r}}' = S\, \hat{\boldsymbol{r}} $
characterised by $S\in \textrm{Sp}(2N,\mathbb{R})$
induces  the transformation $\gamma \to \gamma' = S\, \gamma \,S^{\textrm t}$
on the covariance matrix.

In this manuscript we mainly consider Gaussian states with vanishing first moments,
i.e. having $\langle \hat{r}_i \rangle=0$
(pure states that do not fulfil this condition are discussed in Sec.\,\ref{sec-coherent-states}).
In this case the generic element of covariance matrix (\ref{def_CM_gen}) becomes
\be
\label{def_CM_zero_mom}
\gamma_{i,j} 
=
\frac{1}{2} \langle \{ \hat{r}_i \,, \hat{r}_j \} \rangle
= 
\textrm{Re} \big[ \langle \hat{r}_i\, \hat{r}_j\rangle \big]
\ee
and the Wigner function (\ref{WignerGaussian}) slightly simplifies to
\be
\label{WignerGaussian-zero-moments}
w_{\textrm{\tiny G}}(\boldsymbol{r};\gamma)
=
\frac{e^{-\frac{1}{2}\,
\boldsymbol{r}^{\textrm{t}}
\,\gamma^{-1}\,
\boldsymbol{r}}
}{
(2\pi)^{N} \, \sqrt{\textrm{det}(\gamma)}}\,.
\ee
where we have lightened the notation with respect to
(\ref{WignerGaussian}) by setting 
$w_{\textrm{\tiny G}}(\boldsymbol{r};\gamma)\equiv 
w_{\textrm{\tiny G}}(\boldsymbol{r};\gamma,\boldsymbol{0})$.
The quantities introduced above characterise generic mixed Gaussian states. 
The subclass made by the pure states is discussed in Sec.\,\ref{subsec-pure-states}.

The most familiar way to describe the Hilbert space is the Schr\"odinger representation, 
which employs 
the wave functions $\psi(\boldsymbol{q})=\langle \boldsymbol{q} | \psi \rangle$
on $\mathbb{R}^N$ 
(elements of $L^2(\mathbb{R}^N)$ depending on
$\boldsymbol{q} \equiv (q_1 , \dots , q_N)^{\textrm{t}}$) 
for the vectors of the Hilbert space
and the kernels 
$O(\boldsymbol{q},  \tilde{\boldsymbol{q}}) 
= \langle \boldsymbol{q} |\,\widehat{O}\,| \tilde{\boldsymbol{q}} \rangle$
for the linear operators $\widehat{O}$ acting on the Hilbert space \cite{Simon87}.
In the Appendix\;\ref{app:ww-transform} we relate the kernel 
$\rho(\boldsymbol{q},  \tilde{\boldsymbol{q}}) 
= \langle \boldsymbol{q} |\,\hat{\rho}\,| \tilde{\boldsymbol{q}} \rangle$ of the density matrix 
to the corresponding Gaussian Wigner function (\ref{WignerGaussian-zero-moments}).
In the Appendix\;\ref{app:rdm-schrod} 
we express the kernel $\rho_A(\boldsymbol{q}_A,  \tilde{\boldsymbol{q}}_A) $ for the reduced density matrix 
of a spatial subsystem $A$ in terms of the parameters defining the wave function of the pure state 
describing the entire bipartite system.

\subsection{Fisher-Rao distance}
\label{subsec:FRdist}

The set made by the probability density functions (PDF's) parameterised by the quantities $\gamma$
is a manifold.
In information geometry, the distinguishability between PDF's characterised
by two different sets of parameters $\gamma_1$ and $\gamma_2$  is described through
a scalar quantity $D(\gamma_1,\gamma_2)$ called divergence \cite{Amari16book,nielsen2018elementary},
a function such that $D(\gamma_1,\gamma_2)\geqslant 0$ and 
$D(\gamma_1,\gamma_2)= 0$ if and only if $\gamma_1=\gamma_2$ and 
\be
\label{divergence-def2}
D(\gamma,\gamma+d\gamma)
=\frac{1}{2}\sum_{i,j} g_{ij}\,d y_i d y_j + O\big((dy)^3\big)
\ee
where $g_{ij}$ is symmetric and positive definite and 
$\boldsymbol{y}$ denotes the vector collecting the  
independent parameters that determine $\gamma=\gamma(\boldsymbol{y})$.
In general $D(\gamma_1,\gamma_2) \neq D(\gamma_2,\gamma_1)$;
nonetheless, notice that the terms that could lead to the loss of this symmetry
are subleading in the expansion (\ref{divergence-def2}).
Thus, every divergence $D$ introduces a metric tensor $g_{ij}$ that makes
$\mathcal{M}$ a Riemannian manifold.

A natural requirement for a measure of distinguishability between states
is the  information monotonicity \cite{Amari16book,nielsen2018elementary}.
Let us denote by $\boldsymbol{s}=\boldsymbol{s}(\boldsymbol{r})$
a change of variables in the PDF's
and by $\bar{D}(\gamma_1,\gamma_2)$ the 
result obtained from $D(\gamma_1,\gamma_2)$ after this change of variables.
If $\boldsymbol{s}(\boldsymbol{r})$ is not invertible, 
a loss of information occurs because we cannot reconstruct 
$\boldsymbol{r}$ from $\boldsymbol{s}$. 
This information loss leads to a less distinguishability between PDF's, 
namely $\bar{D}(\gamma_1,\gamma_2)< D(\gamma_1,\gamma_2)$. 
Instead, when $\boldsymbol{s}(\boldsymbol{r})$ is invertible, 
information is not lost and the distinguishability of the two functions is preserved, 
i.e. $\bar{D}(\gamma_1,\gamma_2)= D(\gamma_1,\gamma_2)$. 
Thus, it is naturally to require that 
any change of variables must lead to \cite{Amari16book,nielsen2018elementary}
\be
\label{info monotonicity}
\bar{D}(\gamma_1,\gamma_2)\leqslant D(\gamma_1,\gamma_2)\,.
\ee
This property is called information monotonicity for the divergence $D$.

Let us consider a geometric structure on $\mathcal{M}$ 
induced by a metric tensor $g_{ij}$ associated to a divergence satisfying (\ref{info monotonicity}).
An important theorem in information geometry due to Chentsov 
claims that, 
considering any set of the PDF's, 
a unique metric satisfying (\ref{info monotonicity}) exists up to multiplicative constants \cite{Chentsov82book,Amari16book}.

The Wigner functions of the bosonic Gaussian states
 (\ref{WignerGaussian}) 
 with vanishing first moments
 are PDF's that provide a manifold $\mathcal{M}_{\textrm{\tiny G}}$
 parameterised by the covariance matrices $\gamma$.
The Chentsov's theorem for these PDF's
leads to introduce the
{\it Fisher information matrix}
\cite{Fisher21,Rao45,Chentsov82book,Amari16book, Felice17}
\be
\label{FisherRao for Wigner}
g_{ij}
=
\int w_{\textrm{\tiny G}}(\boldsymbol{r},\gamma) 
\;\frac{\partial \log [w_{\textrm{\tiny G}}(\boldsymbol{r};\gamma)]}{\partial y_i}
\; \frac{\partial \log [w_{\textrm{\tiny G}}(\boldsymbol{r};\gamma)]}{\partial y_j} 
\;d\boldsymbol{r}
\ee
which provides the {\it Fisher-Rao distance} 
between two bosonic Gaussian states with vanishing first moments.
Denoting by $\gamma_1$ and $\gamma_2$ 
the covariance matrices of these states, 
their Fisher-Rao distance reads
\cite{Atkinson81,Forstner2003, Bhatia07book, Bathia15,Lawson01,PineleCosta19}
\be
\label{deltaAB_sec2}
d(\gamma_1, \gamma_2)
\,\equiv\,
\sqrt{
\textrm{Tr} \big[ (\log \Delta)^2 \big] 
}
\,\equiv\,
\big|\!\big|  \log\! \big(\gamma_1^{-1/2} \, \gamma_2 \, \gamma_1^{-1/2} \big)  \big|\!\big|_2
\;\;\qquad\;\;
\Delta \equiv \gamma_{\textrm{\tiny 2}} \,\gamma_{\textrm{\tiny 1}}^{-1} \,.
\ee
This is the main formula employed throughout this manuscript to
study the complexity of Gaussian mixed states.

In Appendix \ref{app:first-moments} we report known results about the Fisher-Rao 
distance between Gaussian PDF's with non vanishing first moments
\cite{Fisher21,Hotelling30,Mahalanobis36,Bhattacharyya43,Rao45,Atkinson81}.
We remark that (\ref{deltaAB_sec2}) is the Fisher-Rao distance also when 
the reference state and the target state have the same first moments,
that can be non vanishing \cite{Atkinson81,PineleCosta19,PineleCosta20}.
Although an explicit expression for the Fisher-Rao distance 
in the most general case of different covariance matrices and different first moments
is not available in the literature, 
interesting classes of Gaussian PDF's have been identified 
where explicit expressions for this distance have been found \cite{COSTA201559,Strapasson16,PineleCosta19,PineleCosta20}.

The distance between two states can be evaluated also through the distance between the corresponding density matrices.
Various expressions for distances have been constructed and it is natural to ask whether 
they satisfy a property equivalent to the information monotonicity (\ref{info monotonicity}),
that is known as contractivity \cite{ NielsenChuang00book,GeomQuantumStates_book,spehner2016geometric}.
A quantum operation $\Theta$ is realised by a completely positive operator 
which acts on the density matrix $\hat{\rho}$, providing another quantum state $\Theta(\hat{\rho})$ \cite{NielsenChuang00book,GeomQuantumStates_book,Weedbrook12b} (see also Sec.\,\ref{sec:gauss-channel}).
A distance $d$ between two states characterised by 
their density matrices $\hat{\rho}_1$ and $\hat{\rho}_2$  
is contractive when the action of a quantum operation $\Theta$
reduces the distance between any two given states \cite{NielsenChuang00book,spehner2016geometric},
namely\footnote{In \cite{GeomQuantumStates_book} both the properties (\ref{info monotonicity}) and (\ref{contractive distance}) are called monotonicity.}
\be 
\label{contractive distance}
d(\hat{\rho}_1,\hat{\rho}_2)
\, \geqslant \,
d\big(\Theta(\hat{\rho}_1),\Theta(\hat{\rho}_2)\big)\,.
\ee
This is a crucial property imposed to a distance in quantum information theory. 

The main contractive distances 
are the Bures distance, defined in terms of the fidelity $\mathcal{F}$ as follows
\be
\label{Bures distance Hilbert spacev1}
d^2_\textrm{\tiny B}(\hat{\rho}_1 , \hat{\rho}_2)
\equiv
2\left(
1-
\mathcal{F}(\hat{\rho}_1 , \hat{\rho}_2)
\right)
\qquad
\mathcal{F}(\hat{\rho}_1 , \hat{\rho}_2)
\,\equiv\,
\textrm{Tr}\!\left(\sqrt{\sqrt{\hat{\rho}_1} \; \hat{\rho}_2 \, \sqrt{\hat{\rho}_1}} \;\right)
\ee
the Hellinger distance
\be
\label{Hellinger distance Hilbert space}
d^2_\textrm{\tiny H}(\hat{\rho}_1 , \hat{\rho}_2)
=
2\left[
1-
\mathrm{Tr} \Big(\! \sqrt{\hat{\rho}_1} \;\sqrt{\hat{\rho}_2}\,\Big)
\right]
\ee
and the trace distance
\be
d_{L^1}(\hat{\rho}_1 , \hat{\rho}_2)
\,\equiv\,
\mathrm{Tr}\big| \hat{\rho}_1 - \hat{\rho}_2 \big|\,.
\ee
The trace distance is the $L^p$-distance with $p=1$ and it is the only contractive distance among the
$L^p$-distances.
For $p=2$ we have the Hilbert-Schmidt distance \cite{GeomQuantumStates_book}
\be
\label{HS distance hilbert v1}
d_{\textrm{\tiny HS}}(\hat{\rho}_1 , \hat{\rho}_2) 
=
\sqrt{\,\textrm{Tr} \big(\hat{\rho}_1 - \hat{\rho}_2\big)^2}
\ee
which is non contractive. 
In Appendix\;\ref{app:other_dist} we further discuss the Bures distance 
and the Hilbert-Schmidt distance specialised to the bosonic Gaussian states.

The Bures distance and the Hellinger distance are Riemannian\footnote{In \cite{PETZ199681} Petz has classified all the contractive Riemannian metrics, 
finding a general formula that provides (\ref{Bures distance Hilbert spacev1}) 
and (\ref{Hellinger distance Hilbert space}) as particular cases.},
being induced by a metric tensor, while the trace distance is not. 
Another difference occurs when we restrict to the subset of the pure states. 
It is well known that the only Riemannian distance between pure states is the
the Fubini-Study distance $d^2_{\textrm{\tiny FS}}=2\left(1- |\langle \psi_1|\psi_2\rangle| \right)$,
where $ \hat{\rho}_1=|\psi_1\rangle \langle\psi_1|$ 
and $\hat{\rho}_2=|\psi_2\rangle \langle\psi_2|$.
Restricting to pure states, the Bures distance becomes exactly the Fubini-Study distance,
while the Hellinger distance and  trace distance become
$d^2_{\textrm{\tiny H}}=2\left(1- |\langle \psi_1|\psi_2\rangle|^2\right)$
and $d^2_{L^1}=4\left(1- |\langle \psi_1|\psi_2\rangle|^2\right)$ respectively, namely
a  function of the Fubini-Study distance \cite{spehner2016geometric}.

\subsection{Williamson's decomposition}
\label{sec:williamson}

The {\it Williamson's theorem}
is a very important tool to study Gaussian states
\cite{Williamson36}:
it provides a decomposition for the covariance matrix $\gamma$
that is crucial throughout our analysis. 

The Williamson's theorem holds for any real, symmetric and positive matrix with even size;
hence also for the covariance matrices.
Given a  covariance matrix $\gamma$, the Williamson's theorem guarantees that
a symplectic matrix $W\in \textrm{Sp}(2N,\mathbb{R})$ can be constructed such that
\be
\label{williamson th gammaA}
\gamma = W^{\textrm t} \,\mathcal{D} \,W
\ee
where $\mathcal{D} \equiv \textrm{diag} (\sigma_1 , \dots , \sigma_N)  \oplus \textrm{diag} (\sigma_1 , \dots , \sigma_N)$ and $\sigma_k > 0$.
The set $\{ \sigma_k \}$ is the {\it symplectic spectrum} of $\gamma$ and its elements are the symplectic eigenvalues
(we often call $\mathcal{D}$ the symplectic spectrum throughout this manuscript, with a slight abuse of notation).
The symplectic spectrum is uniquely determined up to permutations of 
the symplectic eigenvalues and it is invariant under symplectic transformations. 
Throughout this manuscript 
we refer to (\ref{williamson th gammaA}) as the Williamson's decomposition\footnote{It is often called normal modes decomposition \cite{Serafini17book}.} 
of $\gamma$,
choosing a decreasing ordering for the symplectic eigenvalues.
The real dimension of the set made by the covariance matrices is $N(2N+1)$ \cite{Serafini17book}.

Combining (\ref{UncertaintyPrinc}) and (\ref{williamson th gammaA}), it can be shown that
$\sigma_k \geqslant \frac{1}{2}$ \cite{Weedbrook12b}. 
A diagonal matrix is symplectic when it has the form $\Upsilon \oplus \Upsilon ^{-1}$.
This implies that a generic covariance matrix is not symplectic
because of the occurrence of the diagonal matrix $\mathcal{D}$ 
in the Williamson's decomposition (\ref{williamson th gammaA}).

Another important tool for our analysis is the 
{\it Euler decomposition} of a symplectic matrix $S$ (also known as Bloch-Messiah decomposition) 
\cite{Arvind:1995ab}. It reads
\be
\label{EulerDec}
S=L \, \mathcal{X}  \, R 
\;\;\qquad\;\; 
\mathcal{X}=e^\Lambda \oplus e^{-\Lambda}
 \;\;\qquad\;\; 
 L,R \in K(N)\equiv \textrm{Sp}(2N, \mathbb{R}) \cap O(2N)
\ee
where $ \Lambda=\textrm{diag}(\Lambda_1,\dots,\Lambda_N)$
with $\Lambda_j \geqslant 0$.
The non-uniqueness of the decomposition (\ref{EulerDec}) is due only to the freedom to order the elements along the diagonal of $\Lambda$. 
By employing the Euler decomposition (\ref{EulerDec}) and
that the real dimension of $K(N)$ is $N^2$, it is straightforward to realise that
the real dimension of the symplectic group $\textrm{Sp}(2N, \mathbb{R})$ is $2N^2 + N$,
as already mentioned in Sec.\,\ref{subsec:phase-space}.
The simplest case corresponds to the one-mode case, i.e. $N=1$,
where a $2\times 2$ real symplectic matrix can be parameterised by two rotation angles and a squeezing parameter $\Lambda_1$.

The quantities explored in this manuscript provide important tools
to study the entanglement quantifiers in harmonic lattices. 
For instance, the symplectic spectrum in (\ref{williamson th gammaA})
for the reduced density matrix allows to evaluate 
the entanglement spectrum and therefore the entanglement entropies 
\cite{Audenaert:2002xfl,Plenio:2004he,Cramer:2005mx,PeschelChung99,Peschel03,Peschel_2009}
and  the Euler decomposition (\ref{EulerDec})
applied to the symplectic matrix occurring in the 
Williamson's decomposition of the covariance matrix of a subsystem
has been employed in \cite{Coser:2017dtb}
to construct a contour function for the entanglement entropies
\cite{Botero04,Chen_2014}.
The Williamson's decomposition is also crucial to study the 
entanglement negativity 
\cite{VidalWerner-neg,Audenaert:2002xfl,Calabrese:2012ew,Calabrese:2012nk,Calabrese:2014yza}
a measure of the bipartite entanglement for mixed states.

\subsubsection{Covariance matrix of a pure state}
\label{subsec-pure-states}

A Gaussian state is pure if and only if all the symplectic eigenvalues equal to $\frac{1}{2}$,
i.e. $\mathcal{D} = \tfrac{1}{2} \, \boldsymbol{1}$. 
Thus, the Williamson's decomposition of the covariance matrix characterising a pure state reads 
\be
\label{gamma-ps-generic}
\gamma
=\frac{1}{2}\,  W^{\textrm t} \, W 
=  \frac{1}{2}\,  R^{\textrm t}\, \mathcal{X}^2 R
\;\;\qquad\;\;
W=L \, \mathcal{X} \,  R \,.
\ee
The last expression, which has been found by employing the Euler decomposition (\ref{EulerDec})
for the symplectic matrix $W$, tells us that the covariance matrix of a pure state
can be determined by fixing $N^2 + N$ real parameters. 

The covariance matrix of a pure state satisfies the following constraint \cite{HolevoWerner01}
\be
\label{pure-state-condition}
\big( \textrm{i} J  \gamma \big)^2 = \frac{1}{4}\, \boldsymbol{1}\,.
\ee

After a change of basis characterised by the symplectic matrix $S$,
the covariance matrix (\ref{gamma-ps-generic}) becomes 
$\gamma'= \tfrac{1}{2} \,S \,W^{\textrm t}  \,W  S^{\textrm{t}}$.
Choosing $S= K\,W^{-\textrm t}$, 
where $K \in K(N)$,
the covariance matrix drastically simplifies to 
$\gamma'=\frac{1}{2}\, \boldsymbol{1}$.

In the Schr\"odinger representation, the wave function of a pure Gaussian state reads \cite{Arvind:1995ab}
\be
\label{wf-ps}
\psi (\boldsymbol{q}) \,
=
\left( \frac{\textrm{det}(E)}{\pi^N}\right)^{1/4}
e^{-\frac{1}{2} \, \boldsymbol{q}^{\textrm t}  (E + \textrm{i} F)\, \boldsymbol{q}}
\ee
where $E$ and $F$ are $N \times N$ real symmetric matrices
and $E$ is also positive definite;
hence  the pure state is parameterised by $N(N+1)$ real coefficients,
in agreement with the counting of the real parameters discussed above. 
The $L^2$ norm of (\ref{wf-ps}) is equal to one.

The covariance matrix corresponding to the pure state (\ref{wf-ps}) 
can be written in terms of the matrices $E$ and $F$ introduced in the wave function 
(\ref{wf-ps}) as follows \cite{Arvind:1995ab}
\be
\label{gamma-pure-wave-parameters}
\gamma \,=\,
\frac{1}{2}\,
\bigg(  \begin{array}{cc}
E^{-1} \;&  \; -\,E^{-1} F \\
-\,F\,E^{-1} \;&  \; E + F\,E^{-1}F \\
\end{array}  \, \bigg)
=
\frac{1}{2}\,  W^{\textrm t} \, W 
\ee
where the symplectic matrix $W$ and its inverse are given respectively by 
\be
\label{WWinvEF}
W = 
\bigg(  \begin{array}{cc}
E^{-1/2} \;&  \; -\,E^{-1/2} F \\
\boldsymbol{0}\;& \; E^{1/2} \\
\end{array}  \, \bigg)
\;\;\qquad\;\;
W^{-1} = 
\bigg(  \begin{array}{cc}
E^{1/2} \;&  \;  F\, E^{-1/2} \\
\boldsymbol{0}\;& \; E^{-1/2} \\
\end{array}  \, \bigg)
\ee
The expression (\ref{wf-ps}) is employed in 
the Appendix\;\ref{app:rdm-schrod} to provide the kernel of
a reduced density matrix in the Schr\"odinger representation.


\subsection{Mixed states}
\label{subsec:mixed}

Considering the set $\mathbb{P}(N)$
made by the $2N \times 2N$ real and positive definite matrices,
the covariance matrices provide the proper subset of $\mathbb{P}(N)$
made by those matrices that also satisfy the inequality (\ref{UncertaintyPrinc}).

The set $\mathbb{P}(N)$ equipped with the Fisher-Rao distance is a Riemannian manifold 
where the length of a generic path $\gamma : [a,b] \to \mathbb{P}(N)$ is given by\footnote{An explicit computation that relates
(\ref{FisherRao for Wigner}) to (\ref{length curve}) can be found e.g. 
in appendix\;A of \cite{Felice17}.}
\cite{Forstner2003, Bhatia07book, Bathia15, Lawson01,Felice17} 
\be
\label{length curve}
L[\gamma(\tau)]
=
\int_a^b \!\sqrt{\,\textrm{Tr}\, \Big\{ \big[\gamma(\tau)^{-1} \, \dot{\gamma}(\tau)\big]^2 \Big\}}\; d\tau\,.
\ee
The unique geodesic connecting two matrices in the manifold $\mathbb{P}(N)$ has been constructed \cite{Bhatia07book}.
In our analysis we restrict to the subset made by the covariance matrices $\gamma$.
Considering the covariance matrix  $\gamma_{\textrm{\tiny R}}$ 
and the covariance matrix $\gamma_{\textrm{\tiny T}}$, 
that correspond to the reference state and to the target state respectively,
the unique geodesic that connects $\gamma_{\textrm{\tiny R}}$ to $\gamma_{\textrm{\tiny T}}$ is
\cite{Bhatia07book}
\be
\label{optimal circuit}
G_s(\gamma_{\textrm{\tiny R}} \, , \gamma_{\textrm{\tiny T}})
\,\equiv \,
\gamma_{\textrm{\tiny R}}^{1/2} 
\Big(  \gamma_{\textrm{\tiny R}}^{- 1/2}  \,\gamma_{\textrm{\tiny T}} \,\gamma_{\textrm{\tiny R}}^{-1/2}  \Big)^s
\gamma_{\textrm{\tiny R}}^{1/2} 
\;\; \qquad \;\;
0 \leqslant s \leqslant 1
\ee
where $s$ parameterises the generic matrix along the geodesic 
(we always assume $0 \leqslant s \leqslant 1$ throughout this manuscript)
and it is straightforward to verify that
\be
\label{optimal-circuit-bc}
G_0(\gamma_{\textrm{\tiny R}} \, , \gamma_{\textrm{\tiny T}}) = \gamma_{\textrm{\tiny R}} 
\;\;\;\qquad\;\;\;
G_1(\gamma_{\textrm{\tiny R}} \, , \gamma_{\textrm{\tiny T}}) = \gamma_{\textrm{\tiny T}} \,.
\ee
The geodesic (\ref{optimal circuit}) provides the optimal circuit 
connecting $\gamma_{\textrm{\tiny R}}$ to $\gamma_{\textrm{\tiny T}}$.
In the mathematical literature, the matrix (\ref{optimal circuit})
is also known as the 
{\it $s$-geometric mean} of $\gamma_{\textrm{\tiny R}} $ and $\gamma_{\textrm{\tiny T}}$.
The matrix associated to $s=1/2$  provides the {\it geometric mean} 
of $\gamma_{\textrm{\tiny R}} $ and $\gamma_{\textrm{\tiny T}}$.
We remark that,
since $\gamma_{\textrm{\tiny R}} $ and $\gamma_{\textrm{\tiny T}}$ are symmetric Gaussian matrices,
it can be shown that also the matrices belonging to the geodesic (\ref{optimal circuit}) are symmetric and Gaussian \cite{Bathia15}.

By employing (\ref{f(MN) identity app}), we find that
the geodesic (\ref{optimal circuit}) can be written in the following form
\be
\label{optimal circuit compact}
G_s(\gamma_{\textrm{\tiny R}} \, , \gamma_{\textrm{\tiny T}})
= 
\left(
\gamma_{\textrm{\tiny T}}\,\gamma_{\textrm{\tiny R}}^{-1} 
\right)^s
\gamma_{\textrm{\tiny R}}
= \,
\gamma_{\textrm{\tiny R}}
\left(
\gamma_{\textrm{\tiny R}}^{-1} \gamma_{\textrm{\tiny T}}
\right)^s.
\ee

The Fisher-Rao distance between $\gamma_{\textrm{\tiny R}}$ and $\gamma_{\textrm{\tiny T}}$
is the length of the geodesic (\ref{optimal circuit}) evaluated through (\ref{length curve}).
It is given by 
\be
\label{deltaAB}
d(\gamma_{\textrm{\tiny R}}, \gamma_{\textrm{\tiny T}})
\,\equiv\,
\sqrt{
\textrm{Tr} \big[ (\log \Delta_{\textrm{\tiny TR}})^2 \big] 
}
\,\equiv\,
\big|\!\big|  \log\! \big(\gamma_{\textrm{\tiny R}}^{-1/2} \, \gamma_{\textrm{\tiny T}} \, \gamma_{\textrm{\tiny R}}^{-1/2} \big)  \big|\!\big|_2
\ee
where\footnote{The expression (\ref{deltaAB}) cannot be written as 
$|\!|  \log (\gamma_{\textrm{\tiny T}} \,\gamma_{\textrm{\tiny R}}^{-1} )  |\!|_2$
(see Appendix\;\ref{app:matrix_id}).}
\be
\label{def relative CM}
\Delta_{\textrm{\tiny TR}} \equiv \gamma_{\textrm{\tiny T}} \,\gamma_{\textrm{\tiny R}}^{-1} \,.
\ee
This distance provides the following definition of complexity 
\be
\label{c2 complexity}
\mathcal{C}_2
\,=\,
\frac{1}{2\sqrt{2}}\;
d
(\gamma_{\textrm{\tiny R}} \,, \gamma_{\textrm{\tiny T}}  )
\ee
It is straightforward to realise that, in the special case where both $\gamma_{\textrm{\tiny R}}$ 
and $\gamma_{\textrm{\tiny T}}$ correspond to pure states, 
the complexity (\ref{c2 complexity}) becomes the result obtained in \cite{Chapman:2018hou} 
for the $F_2$ complexity, based on the $F_2$ cost function;
hence we refer to (\ref{c2 complexity}) also as $F_2$ complexity in the following. 
The matching with \cite{Chapman:2018hou} 
 justifies the introduction of the numerical factor $\tfrac{1}{2\sqrt{2}}$ in (\ref{c2 complexity})
with respect to the distance (\ref{deltaAB}). 
Equivalently, also the $\kappa =2$ complexity given by 
$\mathcal{C}_{\kappa=2} \equiv \mathcal{C}_2^2$ can be considered.

We remark that the complexity (\ref{c2 complexity}) and the optimal circuit (\ref{optimal circuit})
can be applied also for circuits where 
the reference state and the target state have the same first moments \cite{Atkinson81,PineleCosta19,PineleCosta20}.

The symmetry $d(\gamma_{\textrm{\tiny R}} \,, \gamma_{\textrm{\tiny T}}  ) = d(\gamma_{\textrm{\tiny T}} \,, \gamma_{\textrm{\tiny R}} )$,  
imposed on any proper distance, can be verified for the Fisher-Rao distance (\ref{deltaAB}) by observing that $\Delta_{\textrm{\tiny TR}} \, \leftrightarrow  \,\Delta_{\textrm{\tiny TR}}^{-1}$ under the exchange $\gamma_{\textrm{\tiny R}}  \leftrightarrow \gamma_{\textrm{\tiny T}}$.

Evaluating the distance (\ref{deltaAB}) between $\gamma$ and $\gamma+\delta \gamma$, 
which are infinitesimally close, one obtains \cite{Bhatia07book,LinkStrunz18}
\be
\label{ds FR squared}
d(\gamma, \gamma+\delta \gamma)^2
=
\textrm{Tr} \,\Big\{ \big[ \log\! \big(\boldsymbol{1}+\delta \gamma\, \gamma^{-1} \big) \big]^2  \Big\}
=
\textrm{Tr} \,\Big\{ \big[ \,\delta \gamma\, \gamma^{-1} + \dots\,\big]^2  \Big\}
=
\textrm{Tr}  \big[( \gamma^{-1}\delta \gamma)^2 \big] + \dots
\ee
where the dots correspond to $O((\delta \gamma)^3) $ terms.

Performing a change of basis characterised by the symplectic matrix $S$, the matrix $\Delta_{\textrm{\tiny TR}} $ changes as follows
\be
\label{Delta trans}
\Delta'_{\textrm{\tiny TR}} = 
\gamma'_{\textrm{\tiny T}} (\gamma'_{\textrm{\tiny R}})^{-1} 
=
S\,
\Delta_{\textrm{\tiny TR}}
\,S^{-1} \,.
\ee
From this expression it is straightforward to observe that the Fisher-Rao distance (\ref{deltaAB}), 
and therefore the complexity (\ref{c2 complexity}) as well,
is invariant under a change of basis.
We remark that (\ref{c2 complexity}) is invariant under any transformation that induces 
on $\Delta_{\textrm{\tiny TR}}$ the transformation (\ref{Delta trans})
for any matrix $S$ (even complex and not necessarily symplectic).

From the expression (\ref{optimal circuit}) of the geodesic connecting
$\gamma_{\textrm{\tiny R}}$ to $\gamma_{\textrm{\tiny T}}$,
one can show that the change $s\to 1-s$ provides the geodesic connecting $\gamma_{\textrm{\tiny T}}$ to $\gamma_{\textrm{\tiny R}}$;
indeed, we have  that\footnote{This result can be found by considering e.g. the last expression in (\ref{optimal circuit compact}), 
that gives $G_{1-s}(\gamma_{\textrm{\tiny R}} \, , \gamma_{\textrm{\tiny T}}) = 
\gamma_{\textrm{\tiny T}} \left( \gamma_{\textrm{\tiny T}}^{-1} \gamma_{\textrm{\tiny R}} \right)^{s}$
and becomes (\ref{optimal circuit reversed}), once (\ref{f(MN) identity app}) with $M=\gamma_{\textrm{\tiny T}}^{1/2} $ is employed.}
\be
\label{optimal circuit reversed}
G_{1-s}(\gamma_{\textrm{\tiny R}} \, , \gamma_{\textrm{\tiny T}})
\,=\,
\gamma_{\textrm{\tiny T}}^{1/2} 
\Big(  \gamma_{\textrm{\tiny T}}^{- 1/2}  \,\gamma_{\textrm{\tiny R}} \,\gamma_{\textrm{\tiny T}}^{-1/2}  \Big)^s
\gamma_{\textrm{\tiny T}}^{1/2} 
\,=\,
G_{s}(\gamma_{\textrm{\tiny T}} \, , \gamma_{\textrm{\tiny R}}) \,.
\ee
Another interesting result is the Fisher-Rao distance between the initial matrix 
$\gamma_{\textrm{\tiny R}}$ and the generic symmetric Gaussian matrix along the geodesic  (\ref{optimal circuit})
reads \cite{Bhatia07book}
\be
\label{distance along geodesic}
d\big(\gamma_{\textrm{\tiny R}}\,, G_s(\gamma_{\textrm{\tiny R}} \, , \gamma_{\textrm{\tiny T}}) \big) 
\,=\, 
s\;  d(\gamma_{\textrm{\tiny R}} \, , \gamma_{\textrm{\tiny T}}) \,.
\ee

The derivation of some results reported in the forthcoming sections
are based on the geodesic (\ref{optimal circuit}) written in the following form\footnote{The expression (\ref{optimal circuit Myers}) 
can be found by first writing (\ref{optimal circuit}) as
\be
\label{newoptimalcircuitderivation}
G_s(\gamma_{\textrm{\tiny R}} \, , \gamma_{\textrm{\tiny T}})
\,= \,
\bigg[\,
\gamma_{\textrm{\tiny R}}^{1/2} 
\Big(  \gamma_{\textrm{\tiny R}}^{-1/2}  \,\gamma_{\textrm{\tiny T}} \,\gamma_{\textrm{\tiny R}}^{-1/2}  \Big)^{s/2}
\,\bigg]
\,\bigg[
\Big(  \gamma_{\textrm{\tiny R}}^{- 1/2}  \,\gamma_{\textrm{\tiny T}} \,\gamma_{\textrm{\tiny R}}^{-1/2}  \Big)^{s/2}
\gamma_{\textrm{\tiny R}}^{1/2} 
\,\bigg]
\ee
and then employing (\ref{f(MN) identity app}) in both the expressions within the square brackets of (\ref{newoptimalcircuitderivation}) with $f(x)=x^{s/2}$.}
\be
\label{optimal circuit Myers}
G_s(\gamma_{\textrm{\tiny R}} \, , \gamma_{\textrm{\tiny T}})
\,=\,
U_s\,
\gamma_{\textrm{\tiny R}} \,
U_s^{\textrm{t}}
\;\;\qquad\;\;
U_s \equiv\Delta_{\textrm{\tiny TR}}^{s/2} \,.
\ee
This expression is interesting because the generic matrix of the optimal circuit is written 
in a form that reminds a symplectic transformation of $\gamma_{\textrm{\tiny R}} $ through the $U_s$.
Nonetheless, we remark that in general $U_s$ is not symplectic because the covariance matrices are not symplectic matrices. 
The steps performed to obtain (\ref{optimal circuit Myers}) lead to write (\ref{optimal circuit reversed}) as follows
\be
G_{s}(\gamma_{\textrm{\tiny T}} \, , \gamma_{\textrm{\tiny R}})
\,=\,
U_s^{-1}\,
\gamma_{\textrm{\tiny T}} \,
U_s^{-\textrm{t}} .
\ee

\begin{figure}[t!]
\vspace{-.2cm}
\hspace{-1.1cm}
 \begin{center}
\includegraphics[width=0.8\textwidth]{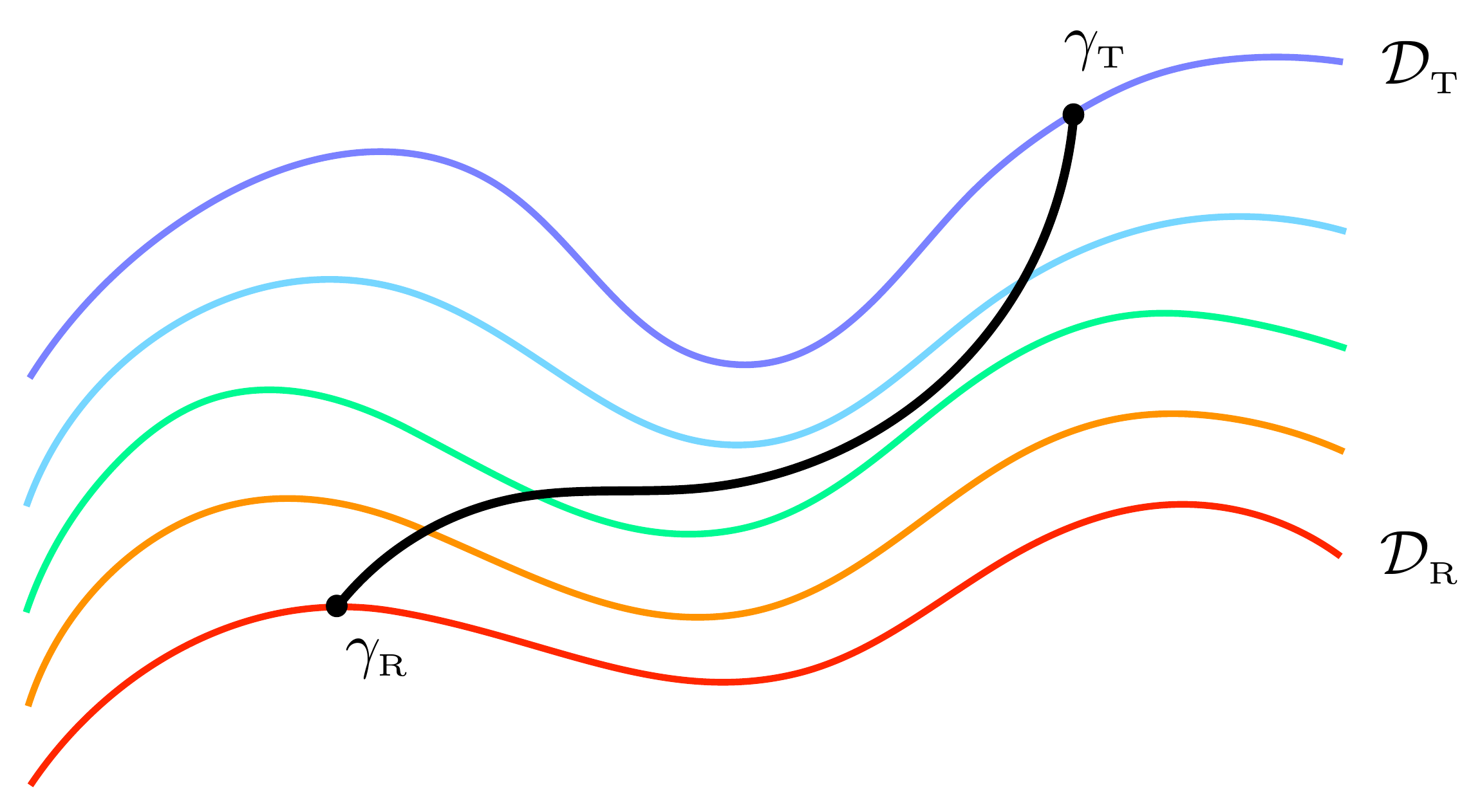}
 \end{center}
\vspace{-.0cm}
\caption{ 
Pictorial representation of the optimal circuit (\ref{optimal circuit}) connecting $\gamma_{\textrm{\tiny R}}$ to $\gamma_{\textrm{\tiny T}}$ (solid black curve).
Coloured solid curves represent the sets made by symmetric  Gaussian matrices having the same symplectic spectrum. 
The red curve corresponds to $\mathcal{D}_{\textrm{\tiny R}}$
and the blue curve to $\mathcal{D}_{\textrm{\tiny T}}$.
}
\label{fig:different_D}
\end{figure}

It is enlightening to exploit the Williamson's decomposition of the covariance matrices discussed in Sec.\,\ref{sec:williamson} 
in the expressions for the complexity and for the optimal circuit.
The Williamson's decomposition (\ref{williamson th gammaA})
allows to write $\gamma_{\textrm{\tiny R}} $ and $\gamma_{\textrm{\tiny T}} $ as follows
\be
\label{gamma R and T williamson}
\gamma_{\textrm{\tiny R}} =  W^{\textrm{t}}_{\textrm{\tiny R}} \, \mathcal{D}_{\textrm{\tiny R}} \,W_{\textrm{\tiny R}}
\;\;\qquad \;\;
\gamma_{\textrm{\tiny T}} = W^{\textrm{t}}_{\textrm{\tiny T}} \, \mathcal{D}_{\textrm{\tiny T}} \,W_{\textrm{\tiny T}}
\;\;\;\;\qquad \;\;\;\;
W_{\textrm{\tiny R}} \,, W_{\textrm{\tiny T}} \in \textrm{Sp}(2N, \mathbb{R})
\ee
where $\mathcal{D}_{\textrm{\tiny R}}$ and $\mathcal{D}_{\textrm{\tiny T}}$ 
contain the symplectic spectra of $\gamma_{\textrm{\tiny R}}$ and $\gamma_{\textrm{\tiny T}}$ respectively.
Let us introduce also the Williamson's decomposition of the generic matrix along the geodesic (\ref{optimal circuit}), namely
\be
\label{williamson-geodesic-generic}
G_s(\gamma_{\textrm{\tiny R}} \, , \gamma_{\textrm{\tiny T}})
=  W^{\textrm{t}}_s \, \mathcal{D}_s \,W_s
\;\;\qquad \;\;
W_s \in \textrm{Sp}(2N, \mathbb{R}) \,.
\ee
It would be insightful to find analytic expressions for $W_s $ and $ \mathcal{D}_s$ 
in terms of $\gamma_{\textrm{\tiny R}}$ and $\gamma_{\textrm{\tiny T}}$.
This has been done later in the manuscript for some particular optimal circuits.

In Fig.\,\ref{fig:different_D} we show a pictorial representation of the optimal circuit 
(\ref{optimal circuit}), which corresponds to the solid black curve.
The figure displays that the symplectic spectrum changes along the geodesic because
the black curve crosses solid curves having different colours, which
correspond to the sets of matrices having the same symplectic spectrum.

In order to write the complexity (\ref{c2 complexity})  in a convenient form
depending on the symplectic spectra 
and on the symplectic matrices $W_{\textrm{\tiny R}}$ and $W_{\textrm{\tiny T}}$,
let us employ that, after a canonical transformation characterised by the symplectic matrix $S$, 
the covariance matrices in (\ref{gamma R and T williamson}) become
\be
\label{gamma-prime-SW-gen}
\gamma'_{\textrm{\tiny R}} = S\, W^{\textrm{t}}_{\textrm{\tiny R}} \,\mathcal{D}_{\textrm{\tiny R}} 
\,W_{\textrm{\tiny R}} S^{\textrm{t}}
\;\;\qquad \;\;
\gamma'_{\textrm{\tiny T}} =  S\,W^{\textrm{t}}_{\textrm{\tiny T}}  \,\mathcal{D}_{\textrm{\tiny T}} 
\, W_{\textrm{\tiny T}} S^{\textrm{t}} \,.
\ee
By choosing $S = K_{\mathcal{D}_{\textrm{\tiny R}}} W^{-\textrm{t}}_{\textrm{\tiny R}}$
where $K_{\mathcal{D}_{\textrm{\tiny R}}}$ is symplectic and such that 
$K_{\mathcal{D}_{\textrm{\tiny R}}}
\mathcal{D}_{\textrm{\tiny R}}\, K^{\textrm{t}}_{\mathcal{D}_{\textrm{\tiny R}}} 
= \mathcal{D}_{\textrm{\tiny R}} $
(the set of matrices made by $K_{\mathcal{D}_{\textrm{\tiny R}}}$ 
is a subgroup of $\textrm{Sp}(2N, \mathbb{R})$ called 
stabilizer \cite{Chapman:2018hou}),
we have that (\ref{gamma-prime-SW-gen}) become respectively 
\be
\label{gammaprime-general}
\gamma'_{\textrm{\tiny R}} = \mathcal{D}_{\textrm{\tiny R}} 
\;\;\qquad \;\;
\gamma'_{\textrm{\tiny T}} =  
\big(W_{\textrm{\tiny TR}} \, K^{\textrm{t}}_{\mathcal{D}_{\textrm{\tiny R}}}\big)^{\textrm{t}}
\,\mathcal{D}_{\textrm{\tiny T}} 
\big(W_{\textrm{\tiny TR}} \, K^{\textrm{t}}_{\mathcal{D}_{\textrm{\tiny R}}}\big)
\ee
where we have introduced the symplectic matrix $W_{\textrm{\tiny TR}}$ defined as follows
\be
\label{W_TR def}
W_{\textrm{\tiny TR}} \equiv W_{\textrm{\tiny T}} \,W_{\textrm{\tiny R}}^{-1} .
\ee
For later convenience, let us consider the Euler decomposition (defined in  Sec.\,\ref{sec:williamson}) of the symplectic matrix $W_{\textrm{\tiny TR}}$, namely
\be
\label{euler dec TR}
W_{\textrm{\tiny TR}} = L_{\textrm{\tiny TR}} \, \mathcal{X}_{\textrm{\tiny TR}} \,R_{\textrm{\tiny TR}}
\ee
where
\be
\label{LRchi-TR-def}
L_{\textrm{\tiny TR}}\, , R_{\textrm{\tiny TR}} \in K(N)
\;\;\qquad\;\;
\mathcal{X}_{\textrm{\tiny TR}} = e^{\Lambda_{\textrm{\tiny TR}}} \oplus e^{-\Lambda_{\textrm{\tiny TR}}} \in \textrm{Sp}(2N, \mathbb{R})
\ee
and $\Lambda_{\textrm{\tiny TR}}$ is a diagonal matrix with positive entries.
By specifying (\ref{Delta trans}) to (\ref{gammaprime-general}), 
we find that
\be
\label{gammagammaDTRcase}
\Delta'_{\textrm{\tiny TR}} 
= 
K_{\mathcal{D}_{\textrm{\tiny R}}} 
W^{\textrm{t}}_{\textrm{\tiny TR}}\,\mathcal{D}_{\textrm{\tiny T}}\, W_{\textrm{\tiny TR}}
\,\mathcal{D}_{\textrm{\tiny R}}^{-1}
K_{\mathcal{D}_{\textrm{\tiny R}}}^{-1}
\ee
which allows to write the $F_2$ complexity (\ref{c2 complexity}) 
as\footnote{The expression (\ref{C-2-complexity}) can be obtained also by first plugging (\ref{gamma R and T williamson}) 
into (\ref{c2 complexity}) and then employing the cyclic property of the trace.}
\be
\label{C-2-complexity}
\mathcal{C}_2
=
\frac{1}{2\sqrt{2}}\;
\sqrt{\,
\textrm{Tr} \,\Big\{ \big[\log\! \big(
\mathcal{D}_{\textrm{\tiny T}}\, W_{\textrm{\tiny TR}}\,\mathcal{D}_{\textrm{\tiny R}}^{-1}\, W^{\textrm{t}}_{\textrm{\tiny TR}}
\big)\big]^2  \Big\}
} \;.
\ee
This expression is independent of $K_{\mathcal{D}_{\textrm{\tiny R}}} $
and tells us that, 
in order to evaluate the $F_2$ complexity (\ref{c2 complexity}) we need
the symplectic spectra  $\mathcal{D}_{\textrm{\tiny R}}$ and $\mathcal{D}_{\textrm{\tiny T}}$ 
and the symplectic matrix (\ref{W_TR def}).

By employing the Euler decomposition (\ref{euler dec TR}), 
the second covariance matrix in (\ref{gammaprime-general}) can be decomposed as follows
\be
\gamma'_{\textrm{\tiny T}} = 
 K_{\mathcal{D}_{\textrm{\tiny R}}} 
 R^{\textrm{t}}_{\textrm{\tiny TR}} \,\mathcal{X}_{\textrm{\tiny TR}}\, L^{\textrm{t}}_{\textrm{\tiny TR}}
\, \mathcal{D}_{\textrm{\tiny T}}\,
L_{\textrm{\tiny TR}}\,\mathcal{X}_{\textrm{\tiny TR}}\,  R_{\textrm{\tiny TR}} 
K_{\mathcal{D}_{\textrm{\tiny R}}}^{\textrm{t}}
\ee
which cannot be further simplified in the general case. 
Similarly, the Euler decomposition (\ref{euler dec TR})
does not simplify (\ref{C-2-complexity}) in a significant way.

From (\ref{optimal circuit}), one finds that
the geodesic $G_s(\gamma'_{\textrm{\tiny R}}  , \gamma'_{\textrm{\tiny T}})$ 
 connecting $\gamma'_{\textrm{\tiny R}}$ to $\gamma'_{\textrm{\tiny T}}$ 
 defined in (\ref{gammaprime-general}) reads
\bea
\label{GsDTRcase}
G_s(\gamma'_{\textrm{\tiny R}}\, , \gamma'_{\textrm{\tiny T}}  ) 
&=&
\mathcal{D}_{\textrm{\tiny R}}^{1/2} 
\Big(  \mathcal{D}_{\textrm{\tiny R}}^{- 1/2}  \,
K_{\mathcal{D}_{\textrm{\tiny R}}} W_{\textrm{\tiny TR}}^{\textrm{t}} 
\,\mathcal{D}_{\textrm{\tiny T}}\, 
W_{\textrm{\tiny TR}} \, K^{\textrm{t}}_{\mathcal{D}_{\textrm{\tiny R}}}
\,\mathcal{D}_{\textrm{\tiny R}}^{-1/2}  \Big)^s
\mathcal{D}_{\textrm{\tiny R}}^{1/2} 
\nonumber
\\
&=&
\Big(  
K_{\mathcal{D}_{\textrm{\tiny R}}} W_{\textrm{\tiny TR}}^{\textrm{t}} 
\,\mathcal{D}_{\textrm{\tiny T}}\, 
W_{\textrm{\tiny TR}} \, K^{\textrm{t}}_{\mathcal{D}_{\textrm{\tiny R}}}
\,\mathcal{D}_{\textrm{\tiny R}}^{-1}  \Big)^s
\mathcal{D}_{\textrm{\tiny R}}
\eea
which is simpler to compute than (\ref{optimal circuit}) because $\gamma'_{\textrm{\tiny R}}$ is diagonal.
Let us remark that $G_s(\gamma'_{\textrm{\tiny R}}  , \gamma'_{\textrm{\tiny T}})$ is different from
$G_s(\gamma_{\textrm{\tiny R}}  , \gamma_{\textrm{\tiny T}})$ but they have the same length given by (\ref{C-2-complexity}).
Furthermore, while the optimal circuit (\ref{GsDTRcase}) depends on the matrix 
$K_{\mathcal{D}_{\textrm{\tiny R}}}$,
its length (\ref{C-2-complexity}) does not.

For pure states, both  (\ref{C-2-complexity}) and (\ref{GsDTRcase}) simplify in a significant way, 
as discussed in Sec.\,\ref{subsec:pure}.

\subsubsection{One-mode mixed states}
\label{subsec-one-mode-mixed}

For mixed states defined by a single mode (i.e. $N=1$),
the results discussed above significantly simplify 
because the diagonal matrices $ \mathcal{D}_{\textrm{\tiny R}} $ 
and $ \mathcal{D}_{\textrm{\tiny T}} $ are proportional to the $2 \times 2$ identity matrix;
hence the covariance matrices of the reference state and of the target state become respectively
\be
\label{gamma R and T williamson one-mode}
\gamma_{\textrm{\tiny R}} =  \sigma_{\textrm{\tiny R}} \,W^{\textrm{t}}_{\textrm{\tiny R}}  \,W_{\textrm{\tiny R}}
\;\;\qquad \;\;
\gamma_{\textrm{\tiny T}} = \sigma_{\textrm{\tiny T}} \, W^{\textrm{t}}_{\textrm{\tiny T}} \, W_{\textrm{\tiny T}}
\;\;\;\;\qquad \;\;\;\;
W_{\textrm{\tiny R}} \,, W_{\textrm{\tiny T}} \in \textrm{Sp}(2, \mathbb{R})
\ee
where $\sigma_{\textrm{\tiny R}} \geqslant 1/2$ and $\sigma_{\textrm{\tiny T}} \geqslant 1/2$.

In this case the Williamson's decomposition for the optimal circuit (\ref{optimal circuit}) can be explicitly written.
Indeed, from (\ref{gamma R and T williamson one-mode}) one finds that 
$\Delta_{\textrm{\tiny TR}} = \sigma_{\textrm{\tiny T}}  \,\sigma_{\textrm{\tiny R}}^{-1} \,
W^{\textrm{t}}_{\textrm{\tiny T}}  \, W_{\textrm{\tiny TR}}\,  W^{-\textrm{t}}_{\textrm{\tiny R}}$
and this leads to write the expression (\ref{optimal circuit Myers}) for the optimal circuit
as follows
\be
\label{G_s decomposition one-mode}
G_s(\gamma_{\textrm{\tiny R}} \, , \gamma_{\textrm{\tiny T}})
\,=\,
\sigma_s\, W^{\textrm{t}}_s \, W_s
\ee
where
\be
\label{Williamson D and W one mode}
\sigma_s = \sigma_{\textrm{\tiny T}}^s  \,\sigma_{\textrm{\tiny R}}^{1-s} 
\hspace{.6cm} \qquad \hspace{.6cm}
W_s \,=\,
W_{\textrm{\tiny R}}
\Big[
\big(  W^{\textrm{t}}_{\textrm{\tiny T}}  \, W_{\textrm{\tiny TR}}\,  W^{-\textrm{t}}_{\textrm{\tiny R}}\big)^{s/2}\,
\Big]^{\textrm{t}}
\ee
which provide the Williamson's decomposition of the generic matrix along the optimal circuit. 

By specialising  the complexity (\ref{C-2-complexity}) 
to the one-mode mixed states in (\ref{gamma R and T williamson one-mode})
we get
\be
\label{complexity-onemode}
\mathcal{C}_2
=
\frac{1}{2\sqrt{2}}\;
\sqrt{
\textrm{Tr}\, \Big\{ \big[\log (
\sigma_\textrm{\tiny T}\,\sigma_\textrm{\tiny R}^{-1}\, W_{\textrm{\tiny TR}}\, W^{\textrm{t}}_{\textrm{\tiny TR}}
)\,\big]^2  \Big\}
} \;.
\ee
Thus, the formal expression for the complexity does not simplify 
significantly for the one-mode mixed states
with respect to the general case with $N\geqslant 1$.

\subsection{Pure states}
\label{subsec:pure}

It is very insightful to specialise the results presented in Sec.\,\ref{subsec:mixed} to pure states.

When both the reference state $| \psi_{\textrm{\tiny R}}\rangle $ and the target state $| \psi_{\textrm{\tiny T}}\rangle $ are pure states, the corresponding density matrices are the projectors 
$ \hat{\rho}_{\textrm{\tiny R}} = | \psi_{\textrm{\tiny R}}\rangle \langle \psi_{\textrm{\tiny R}}|$ 
and $ \hat{\rho}_{\textrm{\tiny T}} = | \psi_{\textrm{\tiny T}}\rangle \langle \psi_{\textrm{\tiny T}}|$ respectively. 
In this case the symplectic spectra drastically simplify to 
\be
\label{D-pure}
\mathcal{D}_{\textrm{\tiny R}} =\mathcal{D}_{\textrm{\tiny T}} = \frac{1}{2}\,\boldsymbol{1}
\ee
where $\boldsymbol{1}$ is the $2N \times 2N$ identity matrix.
This implies that the Williamson's decompositions in (\ref{gamma R and T williamson}) become respectively 
\be
\label{gamma R and T williamson pure}
\gamma_{\textrm{\tiny R}} =  \frac{1}{2}\,W^{\textrm{t}}_{\textrm{\tiny R}}  \,W_{\textrm{\tiny R}}
\;\;\;\;\qquad \;\;\;\;
\gamma_{\textrm{\tiny T}} =  \frac{1}{2}\,W^{\textrm{t}}_{\textrm{\tiny T}}  \,W_{\textrm{\tiny T}} \,.
\ee

The complexity of pure states can be easily found by specialising (\ref{C-2-complexity}) to (\ref{D-pure}).
The resulting expression can be further simplified 
by employing (\ref{euler dec TR}), (\ref{LRchi-TR-def}) and the cyclic property of the trace.
This gives the result obtained in \cite{Chapman:2018hou} 
\be
\label{C-2-complexity-pure}
\mathcal{C}_2
=
\frac{1}{2\sqrt{2}}\;
\sqrt{\,
\textrm{Tr} \,\Big\{ \big[\log\! \big(
W_{\textrm{\tiny TR}}\, W^{\textrm{t}}_{\textrm{\tiny TR}}
\big)\big]^2  \Big\}
}
\,=\,
\frac{1}{2\sqrt{2}}\;
\sqrt{
\textrm{Tr} \,\Big\{\big[\log\! \big( \mathcal{X}_{\textrm{\tiny TR}}^2 \big)\big]^2 \Big\} 
}
\,=\,
\sqrt{\;
\sum_{i} (\Lambda_{\textrm{\tiny TR}})_{i}^2
}
\ee
which can be also obtained through the proper choice of the base described below.

Since we are considering pure states, 
(\ref{WWinvEF}) can be employed to write $W_{\textrm{\tiny R}}$ and $W_{\textrm{\tiny T}}$
in terms of the pairs of symmetric matrices $(E_{\textrm{\tiny R}}, F_{\textrm{\tiny R}})$ 
and $(E_{\textrm{\tiny T}}, F_{\textrm{\tiny T}})$  occurring in  
the wave functions  (\ref{wf-ps}) of the reference state and of the target state respectively.
The matrix $W_{\textrm{\tiny TR}}$  in (\ref{W_TR def}), that provides the complexity 
(\ref{C-2-complexity-pure}),
can be written as follows
\be
\label{W-TR-EF}
W_{\textrm{\tiny TR}}
= 
\left(  \begin{array}{cc}
E_{\textrm{\tiny T}}^{-1/2}  E_{\textrm{\tiny R}}^{1/2} \;\;
&  
\;\;  E_{\textrm{\tiny T}}^{-1/2} F_{\textrm{\tiny R}}\, E_{\textrm{\tiny R}}^{-1/2} 
- E_{\textrm{\tiny T}}^{-1/2} F_{\textrm{\tiny T}}\, E_{\textrm{\tiny R}}^{-1/2}
\\
\rule{0pt}{.5cm}
\boldsymbol{0}\;& \; E_{\textrm{\tiny T}}^{1/2}  E_{\textrm{\tiny R}}^{-1/2} 
\end{array}  \, \right)
\ee
which becomes block diagonal for real wave functions 
(i.e. when $F_{\textrm{\tiny R}}= F_{\textrm{\tiny T}}=\boldsymbol{0}$).

As for the optimal circuit (\ref{optimal circuit}),
by specialising the form (\ref{optimal circuit compact})
to the covariance matrices of pure states in (\ref{gamma R and T williamson pure}), we obtain
\be
G_s(\gamma_{\textrm{\tiny R}} \, , \gamma_{\textrm{\tiny T}})
\,=\, 
\frac{1}{2}\,
W_{\textrm{\tiny R}}^{\textrm{t}}
\left(
W_{\textrm{\tiny TR}}^{\textrm{t}}\,W_{\textrm{\tiny TR}}
\right)^s
W_{\textrm{\tiny R}} \,.
\ee
We find it instructive also to specialise the expression (\ref{optimal circuit Myers})
for the optimal circuit to pure states. 
Indeed, in this case $\Delta_{\textrm{\tiny TR}}$ is symplectic and the result reads
\be
\label{Ws-circuit-pure-states}
G_s(\gamma_{\textrm{\tiny R}} \, , \gamma_{\textrm{\tiny T}})
=
\frac{1}{2}\, W_s^{\textrm{t}} \, W_s
\;\;\qquad\;\;
W_s = W_{\textrm{\tiny R}} \,U_s^{\textrm{t}} 
\ee
This expression provides the Williamson's decomposition of the optimal circuit made by pure states,
given that $W_s \in \textrm{Sp}(2N, \mathbb{R})$.

A proper choice of the basis leads to a simple expression for the optimal circuit. 
Since $\mathcal{D}_{\textrm{\tiny R}} =\tfrac{1}{2}\,\boldsymbol{1}$,
we have that $ K_{\mathcal{D}_{\textrm{\tiny R}}} $ introduced in the text below (\ref{gamma-prime-SW-gen})
is an orthogonal matrix.
For pure states the convenient choice is $K_{\mathcal{D}_{\textrm{\tiny R}}} = R_{\textrm{\tiny TR}}$.
Indeed, by specifying (\ref{gammaprime-general}) to this case we obtain
that in this basis the covariance matrices $\gamma_{\textrm{\tiny R}}$ and $\gamma_{\textrm{\tiny T}}$ become
the following diagonal matrices
\be
\label{gamma mat pure simple form}
\gamma'_{\textrm{\tiny R}} = \frac{1}{2}\, \boldsymbol{1}
\;\;\;\qquad \;\;\;
\gamma'_{\textrm{\tiny T}} = \frac{1}{2}\, \mathcal{X}_{\textrm{\tiny TR}}^2 \,.
\ee
We remark that this result has been obtained by exploiting the peculiarity of the pure states mentioned in Sec.\,\ref{sec:williamson}, namely that,
after a change of basis that brings the covariance matrix into the diagonal form $\tfrac{1}{2} \boldsymbol{1}$,
another change of basis characterised by a symplectic matrix that is also orthogonal
leaves the covariance matrix invariant. 
The occurrence of non trivial symplectic spectra considerably complicates this analysis
(see (\ref{C-2-complexity}) and (\ref{GsDTRcase})).

Specialising the form (\ref{optimal circuit compact}) of the optimal circuit
to the covariance matrices in (\ref{gamma mat pure simple form}), the following simple expression is obtained
\cite{Chapman:2018hou}
\be
\label{geodesic_pure}
G_s(\gamma'_{\textrm{\tiny R}}\, , \gamma'_{\textrm{\tiny T}}  ) 
=
\frac{1}{2}\, \mathcal{X}_{\textrm{\tiny TR}}^{2s} \,.
\ee
This expression tells us that, 
for pure states, this basis is very convenient because the optimal circuit is determined by 
the diagonal matrix $\mathcal{X}_{\textrm{\tiny TR}}$.
\\

\subsection{Thermal states}
\label{subsec:thermal}

The thermal states provide an important class of Gaussian mixed states.
The density matrix of a thermal state at temperature $T \equiv 1/\beta$ is  
$\hat{\rho}_{\textrm{\tiny th}}= e^{- \beta\hat{H}}/\mathcal{Z}$,
where $\widehat{H}$ is the hamiltonian (\ref{HC ham}) 
for the harmonic lattices that we are considering 
and the constant $\mathcal{Z}=\textrm{Tr} \big( e^{-\beta\hat{H}}\big)$ 
guarantees the normalisation condition $\textrm{Tr} \hat{\rho}_{\textrm{\tiny th}} =1$.

In order to study the Williamson's decomposition of the covariance matrix 
associated to a thermal state,
let us  observe  the matrix $H^{\textrm{\tiny phys}}$ in (\ref{HC ham}) can be written as
\be
\label{Hphys_blocks}
H^{\textrm{\tiny phys}}=Q^{\textrm{\tiny phys}}\oplus P^{\textrm{\tiny phys}}
\ee
where $P^{\textrm{\tiny phys}}=\frac{1}{m}\boldsymbol{1}$ and $Q^{\textrm{\tiny phys}}$ is a $N\times N $ real, symmetric and positive definite matrix 
whose explicit expression is not important for the subsequent discussion.

Denoting by $\widetilde{V}$ the real orthogonal matrix that diagonalises $Q^{\textrm{\tiny phys}}$
(for the special case of the harmonic chain with periodic boundary conditions, 
$\widetilde{V}$ has been written in  (\ref{Vtilde-def-even}) e (\ref{Vtilde-def-odd})),
it is straightforward to notice that (\ref{Hphys_blocks}) can be diagonalised as follows
\be
\label{Hphysgen_diag}
H^{\textrm{\tiny phys}}= \,V\,
\bigg[\, \frac{1}{m}\;\textrm{diag}\big( (m\Omega)^2_1,\dots,(m\Omega_N)^2,1,\dots,1\big) \bigg] \, V^{\textrm{t}}
\;\;\qquad\;\;
V\equiv\widetilde{V}\oplus \widetilde{V}
\ee
where $\Omega^2_k$ are the real eigenvalues of $ Q^{\textrm{\tiny phys}}/m$.
It is worth remarking that the $2N\times 2N$ matrix $V$ is symplectic and orthogonal, given that $\widetilde{V}$ is orthogonal.
By employing the argument that leads to (\ref{W-chi-O}), 
the r.h.s. of (\ref{Hphysgen_diag}) can be written as
\be
\label{Hphys-Wdec}
H^{\textrm{\tiny phys}}= \,V\,\mathcal{X}_{\textrm{\tiny phys}}\,
\Big[ \textrm{diag}\big(\Omega_1,\dots,\Omega_N,\Omega_1,\dots,\Omega_N\big) \Big] \, \mathcal{X}_{\textrm{\tiny phys}}\,V^{\textrm{t}}
\ee
where we have introduced the following symplectic and diagonal matrix 
\be
\mathcal{X}_{\textrm{\tiny phys}}=\textrm{diag}\Big( (m\Omega_1)^{1/2},\dots, (m\Omega_N)^{1/2}, (m\Omega_1)^{-1/2},\dots,(m\Omega_N)^{-1/2}\Big)
 \,.
\ee

The expression (\ref{Hphys-Wdec}) provides
the Williamson's decomposition of the matrix $H^{\textrm{\tiny phys}}$ 
entering in the hamiltonian (\ref{HC ham}). It reads
\be
\label{williamson-Hphys-gen}
H^{\textrm{\tiny phys}}
\,= \,
W^{\textrm{t}}_{\textrm{\tiny phys}}  \, \mathcal{D}_{\textrm{\tiny phys}}  \, W_{\textrm{\tiny phys}}
\ee
where
\be
\label{Wphys-mat-def}
\mathcal{D}_{\textrm{\tiny phys}}  =
\textrm{diag}\big(\Omega_1,\dots,\Omega_N,\Omega_1,\dots,\Omega_N\big)
\;\;\qquad\;\;
W_{\textrm{\tiny phys}} = \mathcal{X}_{\textrm{\tiny phys}}\,V^{\textrm{t}} \,.
\ee

The Williamson's decomposition (\ref{williamson-Hphys-gen}) 
suggests to write the physical hamiltonian (\ref{HC ham}) 
in terms of the canonical variables defined through $W_{\textrm{\tiny phys}}$.
The result is
\be
\label{Williamson-basis-Hphys}
\widehat{H}=\frac{1}{2}\, \hat{\boldsymbol{s}}^{\textrm t}\, \mathcal{D}_{\textrm{\tiny phys}} \, \hat{\boldsymbol{s}}
\;\;\qquad\;\;
\hat{\boldsymbol{s}} 
\equiv
W_{\textrm{\tiny phys}}\,\hat{\boldsymbol{r}}
 \equiv
\bigg( 
\begin{array}{c}
\hat{\boldsymbol{\mathfrak{q}}} \\  \hat{\boldsymbol{\mathfrak{p}}}
\end{array}  \bigg) \,.
\ee

Following the standard quantisation procedure, one introduces 
the annihilation operators $\hat{\mathfrak{b}}_k$ and the creation operators $\hat{\mathfrak{b}}_k^\dagger$
as 
\be
\label{b_operators def}
\hat{\boldsymbol{b}}
\equiv 
\big(\,
\hat{\mathfrak{b}}_1, \dots ,  \hat{\mathfrak{b}}_N, \, 
\hat{\mathfrak{b}}_1^\dagger,   \dots, \hat{\mathfrak{b}}_N^\dagger\,\big)^{\textrm t}
\equiv
\Theta^{-1} \hat{\boldsymbol{s}}
\;\qquad\;
\hat{\mathfrak{b}}_k \equiv \frac{\hat{\mathfrak{q}}_k +\textrm{i}\,\hat{\mathfrak{p}}_k}{\sqrt{2}}
\;\qquad\;
\Theta \equiv 
\frac{1}{\sqrt{2}}
\bigg(  \begin{array}{cc}
 \boldsymbol{1} &  \boldsymbol{1} \\
 -\textrm{i} \boldsymbol{1} \; & \textrm{i} \boldsymbol{1} \\
\end{array}   \bigg)
\ee
which satisfy the well known algebra given by $[\hat{\boldsymbol{b}}_i,\hat{\boldsymbol{b}}_j]=J_{ij}$.
In terms of these operators, the  hamiltonian (\ref{Williamson-basis-Hphys}) assumes the standard form
\be
\label{Hphys-op-Omega}
\widehat{H}
=
\sum_{k=1}^N 
\Omega_{k} \! \left( \hat{\mathfrak{b}}_k^\dagger\, \hat{\mathfrak{b}}_k +\frac{1}{2}\, \right) .
\ee
Thus, the symplectic spectrum in (\ref{Wphys-mat-def}) provides the dispersion relation of the model. 

The operator (\ref{Hphys-op-Omega}) leads us to introduce the eigenstates $|n_k\rangle$ of the occupation number operator $  \hat{\mathfrak{b}}_k^\dagger \hat{\mathfrak{b}}_k $,
whose eigenvalues are given by non negative integers $n_k$, and the states $|\boldsymbol{n}\rangle\equiv\bigotimes_{k=1}^N| n_k \rangle$.
The expectation value of an operator $\widehat{O}$ on the thermal state reads 
\be
\label{average_thermalstate}
\langle \hat{O} \rangle
=
\textrm{Tr}\big(
\hat{\rho}_{\textrm{\tiny th}}
\widehat{O}
\big)
=
\frac{1}{\mathcal{Z}}
\sum_{\boldsymbol{n}}
\langle\boldsymbol{n}|\widehat{O} |\boldsymbol{n}\rangle \,
e^{-\beta\sum_{k=1}^N \Omega_{k} (n_k+1/2)}  
\,.
\ee

Considering the covariance matrix of a Gaussian state defined in
(\ref{def_CM_zero_mom}), by employing (\ref{Williamson-basis-Hphys}),
where $W_{\textrm{\tiny phys}}$ is a real matrix,
one finds that the covariance matrix of the thermal state can be written as
\be
\label{CM_Williamsonbasis}
\gamma_{\textrm{\tiny th}}
=
W_{\textrm{\tiny phys}}^{-1}\; 
\textrm{Re}\,\langle \hat{\boldsymbol{s}}\, \hat{\boldsymbol{s}}^{\textrm{t}} \rangle\, 
W_{\textrm{\tiny phys}}^{-\textrm{t}}
\ee
in terms of the covariance matrix in the canonical variables collected into $\hat{\boldsymbol{s}}$,
whose elements are given by  the correlators 
$\langle \hat{\mathfrak{q}}_k \,\hat{\mathfrak{q}}_{k'} \rangle$,  
$\langle \hat{\mathfrak{p}}_k \,\hat{\mathfrak{p}}_{k'} \rangle$ 
and  $\langle \hat{\mathfrak{q}}_k \,\hat{\mathfrak{p}}_{k'} \rangle$.
These correlators can be evaluated by first using (\ref{b_operators def}) to write $\textrm{Re}\,\langle \hat{\boldsymbol{s}}\, \hat{\boldsymbol{s}}^{\textrm{t}} \rangle=\textrm{Re}\,\big(\Theta\langle \hat{\boldsymbol{b}}\, \hat{\boldsymbol{b}}^{\textrm{t}}\rangle\Theta^{\textrm{t}}\big) $,
where we remark that $\Theta$ is not a symplectic matrix 
because it does not preserve the canonical commutation relations. 
Then, by exploiting (\ref{average_thermalstate}) and the action of 
$\hat{\mathfrak{b}}_k$ and $\hat{\mathfrak{b}}_k^{\dagger}$ 
onto the Fock states, one computes 
$\langle \hat{\boldsymbol{b}}\, \hat{\boldsymbol{b}}^{\textrm{t}}\rangle $.
This leads to a diagonal matrix
$\textrm{Re}\,\langle \hat{\boldsymbol{s}}\, \hat{\boldsymbol{s}}^{\textrm{t}} \rangle$
whose  non vanishing elements are given by \cite{Serafini17book, Weedbrook12b}
\be
\label{tfd-corr-diag}
\textrm{Re}\langle \hat{\mathfrak{q}}_k \hat{\mathfrak{q}}_{k} \rangle=
 \textrm{Re}\langle \hat{\mathfrak{p}}_k \hat{\mathfrak{p}}_{k} \rangle=
 \frac{1}{2}
\coth(\beta \Omega_{k}/2)
\;\;\qquad\;\;
 \textrm{Re}\langle \hat{\mathfrak{q}}_k \hat{\mathfrak{p}}_{k} \rangle=0 \,.
\ee
Combining these results with (\ref{CM_Williamsonbasis}), 
for the Williamson's decomposition of the
covariance matrix of the thermal state one obtains 
\be
\label{Williamson_CM_th}
\gamma_{\textrm{\tiny th}}
=W_{\textrm{\tiny th}}^{\textrm{t}} \,\mathcal{D}_\textrm{\tiny th}\,W_{\textrm{\tiny th}}
\ee
where the symplectic eigenvalues entering in the diagonal matrix $\mathcal{D}_\textrm{\tiny th}$ 
and the symplectic matrix $W_{\textrm{\tiny th}}$ are given respectively by 
\be
\label{sympspec_CM_th}
\sigma_{\textrm{\tiny th},k}=\frac{1}{2} \coth (\beta\Omega_{k}/2)
\;\;\qquad\;\;
W_{\textrm{\tiny th}}=W_{\textrm{\tiny phys}}^{-\textrm{t}} \,.
\ee
We remark that $W_{\textrm{\tiny th}}$ is independent of the temperature. 

Taking the zero temperature limit $\beta \to +\infty$ of (\ref{Williamson_CM_th}), 
one obtains the Williamson's decomposition of the
covariance matrix of the ground state. 
This limit gives $\sigma_{\textrm{\tiny th},k} \to 1/2$, 
as expected from the fact that the ground state is a pure state,
while $W_{\textrm{\tiny th}}$ does not change, being independent of the temperature.
Thus, the Williamson's decomposition of the covariance matrix of the ground state reads
\be
\label{Williamson_CM_gs}
\gamma_{\textrm{\tiny gs}}=\frac{1}{2} \,W_{\textrm{\tiny phys}}^{-1} \,W_{\textrm{\tiny phys}}^{-\textrm{t}} 
\ee
where $W_{\textrm{\tiny phys}}$ has been defined in (\ref{Wphys-mat-def}).

It is worth considering the complexity when the reference state and the target state are thermal states
having the same physical hamiltonian $\widehat{H}$ but different temperatures
(we denote respectively by $\beta_{\textrm{\tiny R}}$ and $\beta_{\textrm{\tiny T}}$ their inverse temperatures).
From (\ref{Williamson_CM_th}), we have that 
the Williamson's decomposition of the covariance matrices of 
the reference state and the target state read respectively 
\be
\label{Williamson_target_reference_th}
\gamma_\textrm{\tiny th,R}= W_{\textrm{\tiny th}}^{\textrm{t}}\,\mathcal{D}_\textrm{\tiny th,R}\,W_{\textrm{\tiny th}}
\;\;\qquad\;\;
\gamma_\textrm{\tiny th,T}=W_{\textrm{\tiny th}}^{\textrm{t}}\,\mathcal{D}_\textrm{\tiny th,T}\,W_{\textrm{\tiny th}}
\ee
where $W_{\textrm{\tiny th}}$ is independent of the temperature;
hence $W_{\textrm{\tiny R}}=W_{\textrm{\tiny T}}$.
This means that $W_{\textrm{\tiny TR}}=\boldsymbol{1}$  in this case (see (\ref{W_TR def}));
hence the expression (\ref{C-2-complexity}) for the complexity significantly simplifies to
\be
\label{C-2-complexity-th}
\mathcal{C}_2
=
\frac{1}{2\sqrt{2}}\;
\sqrt{\,
\textrm{Tr} \,\Big\{ \big[\log\! \big(
\mathcal{D}_{\textrm{\tiny th,T}}\,\mathcal{D}_{\textrm{\tiny th,R}}^{-1}
\big)\big]^2  \Big\}}
=\,
\frac{1}{2}\,
\sqrt{\,
\sum_{k=1}^N \,\Bigg\{ \bigg[\log\! \bigg(
\frac{\coth(\beta_{\textrm{\tiny T}} \Omega_{k}/2)}{\coth(\beta_{\textrm{\tiny R}}  \Omega_{k}/2)}
\bigg)\bigg]^2  \Bigg\}
}
 \;.
\ee

The optimal path connecting these particular thermal states is obtained by 
plugging (\ref{Williamson_target_reference_th}) into (\ref{optimal circuit Myers}). 
Furthermore, by exploiting (\ref{f(MN) identity app}) and some straightforward matrix manipulations, 
we find that the Williamson's decomposition of the generic covariance matrix belonging to this optimal path 
reads
\be
\label{optimal_circ_thermal}
G_s(\gamma_\textrm{\tiny th,R} \, , \gamma_\textrm{\tiny th,T})
=  W_{\textrm{\tiny th}}^{\textrm{t}} \, \mathcal{D}_s \,W_{\textrm{\tiny th}}
\;\;\qquad \;\;
\mathcal{D}_s= \mathcal{D}_\textrm{\tiny th,T}^s \,\mathcal{D}_\textrm{\tiny th,R}^{1-s}
\;\; \qquad \;\;
0 \leqslant s \leqslant 1
\ee
where the same symplectic matrix $W_{\textrm{\tiny th}}$ of the reference state and of the target state occurs
and only the symplectic spectrum depends on the parameter $s$ labelling the 
covariance matrices along the optimal path.

It is worth asking whether, for any given value of $s$, 
the covariance matrix 
$G_s(\gamma_\textrm{\tiny th,R} \, , \gamma_\textrm{\tiny th,T})$
in (\ref{optimal_circ_thermal}) can be associated to a thermal state of the system characterised by the same
physical hamiltonian underlying the reference and the target states.
Denoting by $\sigma_{s,k}$ the symplectic eigenvalues of (\ref{optimal_circ_thermal})
this means to find a temperature $T_s \equiv \beta_s^{-1}$  such that 
$\sigma_{s,k}=\tfrac{1}{2}\coth(\beta_s \Omega_{k}/2)$.
This equation can be written more explicitly as follows
\be
\label{eq-beta-s-th}
\coth(\beta_s\, \Omega_{k}/2)
=
\bigg[
\frac{\coth(\beta_\textrm{\tiny T}\, \Omega_{k}/2)}{\coth(\beta_\textrm{\tiny R}\, \Omega_{k}/2)}
\bigg]^s
\coth(\beta_\textrm{\tiny R}\, \Omega_{k}/2) \,.
\ee
We checked numerically that a solution 
$T_s=T_s(T_\textrm{\tiny R},T_\textrm{\tiny T})$ for any $1\leqslant k\leqslant N$
does not exist.

The quantities discussed above are further explored in Sec.\,\ref{sec-thermal-states-examples},
where the thermal states of the harmonic chain are considered.


\subsection{Coherent states}
\label{sec-coherent-states}

The coherent states are pure states with non vanishing first moments \cite{Serafini17book}.
They can be introduced through the displacement operator defined in (\ref{displacement-op-def-0}),
where the real vector $\boldsymbol{a}\in \mathbb{R}^{2N}$ can be parameterised
in terms of the complex vector $\boldsymbol{\alpha} \in \mathbb{C}^N$ as
$\boldsymbol{a}^\textrm{t}=\sqrt{2}\,
( \textrm{Re}(\boldsymbol{\alpha})^\textrm{t}, \textrm{Im}(\boldsymbol{\alpha})^\textrm{t} )$.

The displacement operator (\ref{displacement-op-def-0}), which is unitary and satisfies $\widehat{D}_{\boldsymbol{a}}^{-1}=\widehat{D}_{-\boldsymbol{a}}$,
shifts the position and the momentum operators as follows
\be
\label{displacement}
\widehat{D}_{-\boldsymbol{a}} \,\hat{\boldsymbol{r}} \,
\widehat{D}_{\boldsymbol{a}}
=
\hat{\boldsymbol{r}}+\boldsymbol{a} \,.
\ee
The coherent state $|\boldsymbol{\alpha}\rangle$ is the pure state
obtained by applying the displacement operator to the ground state
\be
\label{coherentstate def}
|\boldsymbol{\alpha}\rangle\equiv \widehat{D}_{\boldsymbol{a}} |0\rangle \,.
\ee
This state is Gaussian and,
from (\ref{displacement-op-def-0}), we have that the ground state corresponds
to the coherent state with vanishing $\boldsymbol{\alpha}$ \cite{Serafini17book}.
From (\ref{displacement}), (\ref{coherentstate def})
and the fact that  $ \langle0|\hat{\boldsymbol{r}}|0\rangle=0$, 
for the first moments of the coherent state (\ref{coherentstate def}) one finds
\be 
\label{coherentstate first moment}
\langle\boldsymbol{\alpha}|\hat{\boldsymbol{r}}|\boldsymbol{\alpha}\rangle
=
\langle0|\widehat{D}_{-\boldsymbol{a}}\,\hat{\boldsymbol{r}}\,
\widehat{D}_{\boldsymbol{a}}|0\rangle 
=
 \langle0|\hat{\boldsymbol{r}}+\boldsymbol{a}|0\rangle=\boldsymbol{a} \,.
\ee
By employing this property, 
from the definition (\ref{def_CM_gen}) 
for the covariance matrix $\gamma_{\textrm{\tiny cs}}$ of a coherent state we find that
\be
\label{CM_Coherent}
\gamma_{\textrm{\tiny cs}}
+\frac{\textrm{i}}{2}\,J
\,=\,
\langle\boldsymbol{\alpha}|(\hat{\boldsymbol{r}}-\boldsymbol{a})\,(\hat{\boldsymbol{r}}-\boldsymbol{a})^{\textrm{t}}|\boldsymbol{\alpha}\rangle
\,=\,
\langle0|\widehat{D}_{-\boldsymbol{a}}(\hat{\boldsymbol{r}}-\boldsymbol{a})\widehat{D}_{\boldsymbol{a}}\,
\widehat{D}_{-\boldsymbol{a}}(\hat{\boldsymbol{r}}-\boldsymbol{a})^{\textrm{t}}\widehat{D}_{\boldsymbol{a}}|0\rangle
\,=\,
\langle0|\hat{\boldsymbol{r}} \, \hat{\boldsymbol{r}}^{\textrm{t}}|0\rangle
\ee
where (\ref{displacement}) has been used in the last step.
Thus, the coherent states have the same covariance matrix of the ground state, but
their first moments (\ref{coherentstate first moment}) are non vanishing.
Combining this observation with (\ref{Williamson_CM_gs}), 
for $\gamma_{\textrm{\tiny cs}}$ we find 
\be
\label{Williamson_CM_coherent}
\gamma_{\textrm{\tiny cs}}=\frac{1}{2} \, W_{\textrm{\tiny phys}}^{-1} \,W_{\textrm{\tiny phys}}^{-\textrm{t}}  \,.
\ee

The distance (\ref{deltaAB}), that is mainly used throughout this manuscript to study the circuit complexity of mixed states, 
is valid for states having the same first moments \cite{Atkinson81,Bhatia07book,PineleCosta19},
as reported in the Appendix\;\ref{app:first-moments}.

In the Appendix\;\ref{app:first-moments} it is also mentioned that an explicit expression for 
the complexity between coherent states is  available in the literature if we  restrict to the coherent states 
having a diagonal covariance matrix 
and  $\boldsymbol{a}=(\sqrt{2}\,\alpha,0,\dots,0)$ \cite{COSTA201559, Strapasson16, PineleCosta19}.
These states provide a manifold parametrised by $(\alpha,v_1,\dots,v_{2N})$,
where $v_k^2$ is the $k$-th entry of the diagonal covariance matrix,
and whose metric is given by (\ref{productFRmetric_v2}) with $n=2N$ and $\mu=\sqrt{2}\alpha$, namely
\be
\label{productFRmetric_coherent}
ds^2=\frac{2\,d\alpha^2+2 \,d v_1^2 }{v_1^2}+2\sum_{k=2}^{2N} \frac{ d v_k^2 }{v_k^2} \,.
\ee

Let us remind that the covariance matrices that we are considering must satisfy the constraint $(\ref{UncertaintyPrinc})$, 
which is equivalent $\sigma_k\geqslant1/2$ for the symplectic eigenvalues, 
where $k=1,\dots,N$.
By using (\ref{W-chi-O}), one finds that the symplectic eigenvalues of
the diagonal covariance matrix  $\textrm{diag}(v^{2}_1,\dots,v^{2}_{2N})$
are $\sigma_k=v_k \,v_{k+N}$ for $k=1,\dots,N$.
Thus, in our case the manifold equipped with the metric (\ref{productFRmetric_coherent}) 
must be constrained by the conditions $v_k \,v_{k+N}\geqslant1/2$ for $k=1,\dots,N$.

The coherent states are pure states, hence their covariance matrices must satisfy the condition (\ref{pure-state-condition}),
which holds also when the first moments are non vanishing \cite{HolevoWerner01,Serafini17book}.
For the class of coherent states that we are considering, the constraint (\ref{pure-state-condition}) leads to
\be
\label{puritycondition_diagonalCM}
v_{k+N}^2=\frac{1}{4 v_{k}^2}
\,\,\qquad\,\,
k=1,\dots ,N
\ee
which saturate the constraints $v_k \,v_{k+N}\geqslant1/2$ introduced above. 
By imposing the conditions (\ref{puritycondition_diagonalCM}),  
the metric (\ref{productFRmetric_coherent}) becomes
\be
\label{metricFR_pure_mu}
ds^2=\frac{2\,d\alpha^2+4 \,d v_1^2 }{v_1^2}+4\sum_{k=2}^{N} \frac{d v_k^2 }{v_k^2}
=
2\,\bigg(\frac{d\alpha^2+2 \,d v_1^2 }{v_1^2}+2\sum_{k=2}^{N} \frac{d v_k^2 }{v_k^2}\bigg)
\ee
which is twice (\ref{productFRmetric_v2}) with $n=N$ and $\mu=\alpha$.
The geometry given by  (\ref{metricFR_pure_mu}) has been found also in \cite{Guo:2018kzl}.
The constraint (\ref{puritycondition_diagonalCM})  
tells us that the metric (\ref{metricFR_pure_mu}) is defined on a set of pure states,
but we are not guaranteed that the resulting manifold is totally geodesic.
This is further discussed in the final part of this subsection.

Given a reference state and a target state parametrised\footnote{The vector $\boldsymbol{\phi}$ corresponds to the vector $\boldsymbol{\theta}$ used in Appendix \ref{app:first-moments} restricted by the condition (\ref{puritycondition_diagonalCM}).}
by $\boldsymbol{\phi}_{\textrm{\tiny R}}=(\alpha_{\textrm{\tiny R}},v_{\textrm{\tiny R},1},\dots,v_{\textrm{\tiny R},N})$ and $\boldsymbol{\phi}_{\textrm{\tiny T}}=(\alpha_{\textrm{\tiny T}},v_{\textrm{\tiny T},1},\dots,v_{\textrm{\tiny T},N})$ respectively,
the square of the complexity of the circuit corresponding to the geodesic connecting these states in the manifold equipped with (\ref{metricFR_pure_mu})
is easily obtained from (\ref{FR-mu-diag}). The result reads
\be
\label{distanceFR_pure_mu_v2}
d_{\textrm{\tiny cs}} (\boldsymbol{\phi}_{\textrm{\tiny R}}, \boldsymbol{\phi}_{\textrm{\tiny T}}) 
=
2\sqrt{2}\;\sqrt{ \bigg[\,\textrm{arccosh}\bigg(1+\frac{(\alpha_{\textrm{\tiny R}}-\alpha_{\textrm{\tiny T}})^2/2+(v_{\textrm{\tiny R},1}-v_{\textrm{\tiny T},1})^2}{2 \,v_{\textrm{\tiny R},1} v_{\textrm{\tiny T},1}}\bigg)\bigg]^2
\!\!
+\sum_{k=2}^N \,\bigg[\log\!\bigg(\frac{v_{\textrm{\tiny T},k}}{v_{\textrm{\tiny R},k}}\bigg) \bigg]^2}
 \,.
\ee
By adopting the normalisation in (\ref{c2 complexity}), which is consistent with \cite{Jefferson:2017sdb,Guo:2018kzl}, 
one can introduce the complexity between coherent states as  follows
\be
\label{c2 complexity cs}
\mathcal{C}_2
\,=\,
\frac{1}{2\sqrt{2}}\;
d_{\textrm{\tiny cs}} (\boldsymbol{\phi}_{\textrm{\tiny R}}, \boldsymbol{\phi}_{\textrm{\tiny T}})  \,.
\ee
Setting $\alpha_{\textrm{\tiny R}}=0$ (or $\alpha_{\textrm{\tiny T}}=0$, equivalently) in (\ref{c2 complexity cs}),
one obtains the complexity between a coherent state in the particular set introduced above and the ground state.
As consistency check, we observe that, by setting $\alpha_{\textrm{\tiny R}}=\alpha_{\textrm{\tiny T}}$ in (\ref{c2 complexity cs}),
the complexity (\ref{C-2-complexity-pure}) between pure states is recovered. 

It is instructive to compare (\ref{c2 complexity cs}) with the results reported in \cite{Guo:2018kzl},
where the complexity $\mathcal{C}_{\kappa=2}=\mathcal{C}_{2}^2$ between the ground state and a bosonic
coherent state has been studied through the Nielsen's approach \cite{Nielsen06, NielsenDowling06, DowlingNielsen08}.
The analytic expression for the complexity in \cite{Guo:2018kzl} has been obtained 
for reference and target states with diagonal covariance matrices 
and first moments with at most one non vanishing component. 
Since these are the assumptions under which (\ref{c2 complexity cs}) has been obtained, 
we can compare the two final results for the complexity. 
The analysis of \cite{Guo:2018kzl} allows to write the complexity $\mathcal{C}_{\kappa=2}$ in terms of a free parameter $x_0$ 
which does not occur neither in the reference state nor in the target state. 
We observe that  the square of (\ref{c2 complexity cs}) with $\alpha_\textrm{\tiny R}=0$ 
coincides with the result in \cite{Guo:2018kzl}\footnote{In Eqs.\,(4.23) and (4.24) of \cite{Guo:2018kzl}, set $i=1$, $\sqrt{2} \,\omega_{\textrm{\tiny R,there}} =1/v_{\textrm{\tiny R},k}$, $\sqrt{2 \,m_{\textrm{there}}\,\omega_{k,\textrm{\tiny there}}}=1/v_{\textrm{\tiny T},k}$ and $\mathfrak{a}_{i,\textrm{\tiny there}}=a_\textrm{\tiny there}/x_0= \alpha_\textrm{\tiny T}/(\sqrt{2} \,v_{\textrm{\tiny R,1}})$.} with $x_0=2 v_{\textrm{\tiny R,1}}$.

In the following we consider circuits in the space of the Gaussian states with non vanishing first moments
such that the reference and the target states 
are given by two coherent states (\ref{coherentstate def}) originating from the same ground state,
denoting their first moments by $\boldsymbol{a}_\textrm{\tiny R} $ and $\boldsymbol{a}_\textrm{\tiny T} $ respectively. 
These states have the same covariance matrix $\gamma_0$ (see (\ref{CM_Coherent})),
whose symplectic eigenvalues are equal to $1/2$, 
given that the coherent states are pure states.
Parametrising the reference state by $\boldsymbol{\theta}_\textrm{\tiny R}=(\boldsymbol{a}_\textrm{\tiny R},\gamma_0)$
and the target state by $\boldsymbol{\theta}_\textrm{\tiny T}=(\boldsymbol{a}_\textrm{\tiny T},\gamma_0)$,
a recent result obtained in \cite{PineleCosta20} and discussed in Appendix\;\ref{app:first-moments}
allows us to write the circuit complexity as follows
\be
\label{c2 complexity cs same gamma0}
\mathcal{C}_2
\,=\,
\frac{1}{2\sqrt{2}}\;
d_\textrm{\tiny FR}(\boldsymbol{\theta}_\textrm{\tiny R},\boldsymbol{\theta}_\textrm{\tiny T})
\ee
where $d_\textrm{\tiny FR}$ has been defined in (\ref{FR-distance-same-sigma}). 
We are not able to prove that $\sigma_k \geqslant 1/2$ for the symplectic eigenvalues of 
the symmetric and positive matrices making the geodesic whose length is 
(\ref{c2 complexity cs same gamma0}).


It is worth remarking that the expressions (\ref{c2 complexity cs}) and (\ref{c2 complexity cs same gamma0}) 
for the complexity are defined for different sets of Gaussian Wigner functions with a non vanishing intersection. 
Indeed, (\ref{c2 complexity cs same gamma0}) holds between PDF's with the same covariance matrix 
(that can be also non diagonal), 
while (\ref{c2 complexity cs}) is valid for diagonal covariance matrices that can be different.
Moreover in (\ref{c2 complexity cs}), $\boldsymbol{a}_\textrm{\tiny R}$ and $\boldsymbol{a}_\textrm{\tiny T}$ can have only one non vanishing components, while in (\ref{c2 complexity cs same gamma0}) they are generic.
Thus, in order to compare (\ref{c2 complexity cs}) and (\ref{c2 complexity cs same gamma0}) we have to consider reference and target states 
which have the same diagonal covariance matrix and 
and whose first moments have only one non vanishing component.
Setting
$v_{\textrm{\tiny R},k}=v_{\textrm{\tiny T},k}\equiv v_k$ with $k=1,\dots,N$ in (\ref{c2 complexity cs}), we obtain
\be
\label{c2 complexity cs particular}
\mathcal{C}_2
\,=\,
\textrm{arccosh}\bigg(1+\frac{(\alpha_{\textrm{\tiny R}}-\alpha_{\textrm{\tiny T}})^2}{4 \,v_{1}^2} \bigg) \,.
\ee
Plugging $\gamma_0=\diag(v_1^2,\dots,v^2_N,(2v_1)^{-2},\dots, (2v_N)^{-2})$ 
and $\boldsymbol{a}_{\textrm{\tiny S}}=(\sqrt{2}\,\alpha_{\textrm{\tiny S}},0,\dots,0)$ 
for $\textrm{S}=\textrm{R}$ and $\textrm{S}=\textrm{T}$ in (\ref{c2 complexity cs same gamma0}), 
one finds
\be
\label{c2 complexity cs same gamma0 particular}
\mathcal{C}_2
\,=\,
\frac{1}{\sqrt{2}}\;
\textrm{arccosh}\bigg(1+\frac{(\alpha_{\textrm{\tiny R}}-\alpha_{\textrm{\tiny T}})^2}{2 \,v_{1}^2} \bigg) \,.
\ee

A simple numerical inspection shows that (\ref{c2 complexity cs same gamma0 particular}) 
is always smaller than (\ref{c2 complexity cs particular}). 
This example allows to conclude that 
the submanifold of pure states with diagonal covariance matrix 
equipped with the metric (\ref{metricFR_pure_mu}) is not totally geodesics.

\section{Spectrum complexity and basis complexity}
\label{sec:spectrum-basis}

In this section we discuss the spectrum complexity and the basis complexity 
for mixed  Gaussian states in harmonic lattices.

\begin{figure}[t!]
\vspace{-.2cm}
\hspace{-1.1cm}
 \begin{center}
\includegraphics[width=0.8\textwidth]{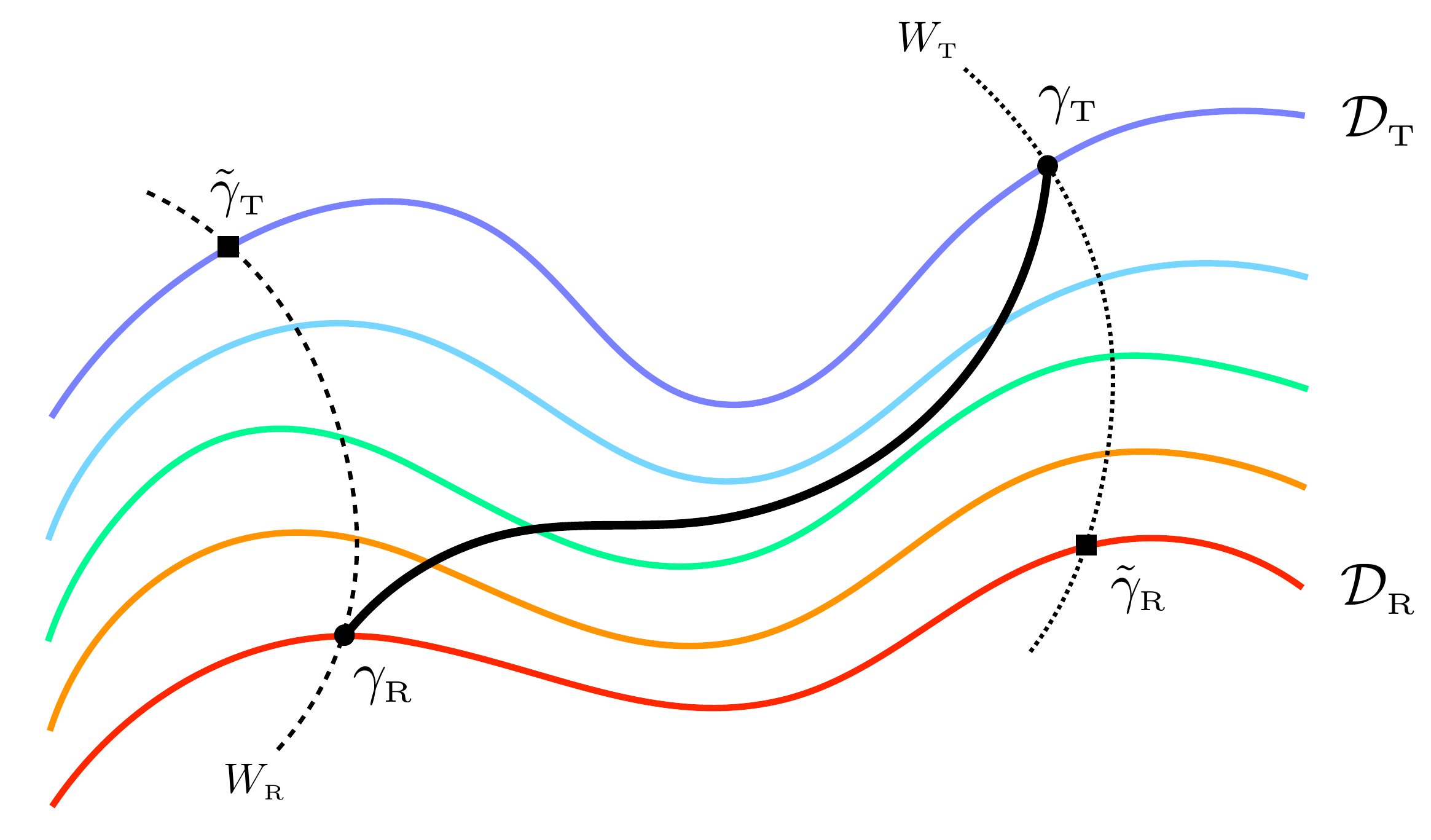}
 \end{center}
\vspace{-.0cm}
\caption{ 
The solid black curve and the solid coloured curves have been defined in Fig\;\ref{fig:different_D}.
Eq.\,(\ref{W-path-williamson}) tells us that
the dashed black curves represent the $W_{\textrm{\tiny R}}$ path and the 
$W_{\textrm{\tiny T}}$ path that pass through 
$\gamma_{\textrm{\tiny R}}$ and $\gamma_{\textrm{\tiny T}}$ respectively
(the auxiliary covariance matrices $\tilde{\gamma}_{\textrm{\tiny R}}$ and $\tilde{\gamma}_{\textrm{\tiny T}}$ 
have been defined in (\ref{tilded gammas})).
The arcs of the dashed curves that connect the blue curve to the red curve
have the same length given by (\ref{compl_same_W}).
}
\label{fig:different_D_sp}
\end{figure}

By exploiting the Williamson's decomposition
we introduce the {\it $W$ path} as the optimal circuit 
connecting two covariance matrices with
$W_{\textrm{\tiny R}}=W_{\textrm{\tiny T}}\equiv W$
and the 
 {\it $\mathcal{D}$ path} as the optimal circuit 
connecting two covariance matrices  having
$\mathcal{D}_{\textrm{\tiny R}}=\mathcal{D}_{\textrm{\tiny T}}\equiv \mathcal{D}$.
In order to study these circuits,
in Sec.\,\ref{sec_first_law} we discuss the first law of complexity 
for the Gaussian states that we are considering. 
The lengths of a $W$ path and of a $\mathcal{D}$ path 
are employed to study 
the spectrum complexity 
(Sec.\,\ref{subsec:spectrum-comp})
and the basis complexity (Sec.\,\ref{subsec:basis-comp}) respectively.
In Fig.\;\ref{fig:different_D_sp} the dashed curves correspond to 
$W$ paths (see (\ref{W-path-williamson})).

\subsection{First law of complexity}
\label{sec_first_law}

It is worth investigating the first law of complexity \cite{Bernamonti:2019zyy, Bernamonti:2020bcf} 
for the states described in Sec.\,\ref{sec:FR-distance}.
The derivations of the results reported below are given in the Appendix \ref{app:sec_first_law}.

Let us consider the following functional
\be
\label{lagrangian action}
S[q(t)] = \int_{t_0}^{t_1} \!\! \mathcal{L}[q(t) , \dot{q}(t), t] \, dt
\ee
where  $q_0 = q(t_0)$ and  $q_1 = q(t_1)$
are the initial and final configurations respectively.

It is well known that the first variation of (\ref{lagrangian action})
under an infinitesimal change of the boundary conditions
$q_i \to q_i + \delta q_i$ for $i=0,1$
evaluated on a solution of the equations of motion is 
\be
\label{variation action}
\delta S[q(t)] \,=  \,
\frac{\partial \mathcal{L}}{\partial \dot{q}}\, \delta q\, \bigg|_{t_1} 
\!\! - \frac{\partial \mathcal{L}}{\partial \dot{q}}\, \delta q \,\bigg|_{t_0}  \,.
\ee

The functional we are interested in is the length functional (\ref{length curve}) 
and the solution of its equations of motion is given by
the optimal circuit (\ref{optimal circuit}), that satisfies the boundary conditions (\ref{optimal-circuit-bc}).
In order to apply (\ref{variation action}), 
one considers the infinitesimal variations 
$ \gamma_{\textrm{\tiny T}} \to  \gamma_{\textrm{\tiny T}}+ \delta  \gamma_{\textrm{\tiny T}}$ 
and $ \gamma_{\textrm{\tiny R}} \to  \gamma_{\textrm{\tiny R}}+ \delta  \gamma_{\textrm{\tiny R}}$
of the covariance matrices of the reference and of the target states
that preserve the properties of these matrices.
In other words, these variations are such that also the resulting matrices are covariance matrices.

The length functional (\ref{length curve}) leads to introduce the following cost function
\be
\label{FR cost function}
F(\gamma,\dot{\gamma})\,
=\,
\sqrt{\textrm{Tr}
\big[
\left(
\gamma^{-1} \,\dot{\gamma}
\right)^2
\big]} \,.
\ee
By applying (\ref{variation action}) to the length functional (\ref{length curve}),
one obtains the first law of complexity
\be
\label{first law complexity}
\delta d=\sum_{i j} 
\frac{\partial F}{\partial \dot{\gamma}_{i j}}\, \delta \gamma_{i j}\,\bigg |_{s=0}^{s=1}
\ee
where the r.h.s. is evaluated on the geodesic (\ref{optimal circuit}).

Equivalent expressions for the variation (\ref{first law complexity})  
have been derived in Appendix\;\ref{app:sec_first_law}.
For instance, it can be written as 
\be
\label{1st law final text}
\delta d
\,=\,
 \frac{1}{d}
 \bigg(
 \textrm{Tr}\, \Big\{ 
\partial_s G_s^{-1} \big|_{s=0} \, \delta \gamma_{\textrm{\tiny R}}
\Big\}
 -
\textrm{Tr} \,\Big\{ 
\partial_s G_s^{-1} \big|_{s=1} \, \delta \gamma_{\textrm{\tiny T}}
\Big\}
\bigg)
\ee
where $G_s$ is the geodesic (\ref{optimal circuit}).
Another useful expression for  (\ref{first law complexity}),
which is simpler to evaluate than (\ref{1st law final text}),  reads
\be
\label{first variation d-squared v1}
\delta d
\,=\,
\frac{1}{d}\;
\textrm{Tr} \,\Big\{
\big( 
\delta \gamma_{\textrm{\tiny T}} \, \gamma_{\textrm{\tiny T}}^{-1}
-
\delta \gamma_{\textrm{\tiny R}} \, \gamma_{\textrm{\tiny R}}^{-1}
\big)
\log\!\big(\Delta_{\textrm{\tiny TR}}\big)
\Big\} 
\ee
where $\Delta_{\textrm{\tiny TR}}$ has been defined in (\ref{def relative CM}).

We find it worth providing also an expression for the variation (\ref{first law complexity})
that is based on the Williamson's decompositions (\ref{gamma R and T williamson}).
It is given by 
\bea
\label{first-law-williamson}
\delta d
&=&
\frac{1}{d}\;\bigg[\,
2\,\textrm{Tr} \,\Big\{
\log\!\big(
\gamma_\textrm{\tiny T}^{-1} \gamma_\textrm{\tiny R}
\big)\,
\big( 
W_{\textrm{\tiny R}}^{-1}\,\delta W_{\textrm{\tiny R}} 
-
W_{\textrm{\tiny T}}^{-1}\,\delta W_{\textrm{\tiny T}}
\big)
\Big\} 
\\
& & \hspace{.8cm}
+\,
\textrm{Tr}\, \Big\{ 
\big[\,
W_{\textrm{\tiny T}}
\log\!\big(
\gamma_\textrm{\tiny R}^{-1} \gamma_\textrm{\tiny T}
\big)
W_{\textrm{\tiny T}}^{-1}
\,\big]\,
\mathcal{D}_\textrm{\tiny T}^{-1}\, \delta \mathcal{D}_\textrm{\tiny T}
-
\big[\,
W_{\textrm{\tiny R}}
\log\!\big(
\gamma_\textrm{\tiny R}^{-1} \gamma_\textrm{\tiny T}
\big)
W_{\textrm{\tiny R}}^{-1}
\,\big]\,
\mathcal{D}_\textrm{\tiny R}^{-1}\, \delta \mathcal{D}_\textrm{\tiny R}\,
\,\Big\}
\bigg]
\nonumber
\eea
where, by using (\ref{f(MN) identity app}), one can write 
the matrices within the square brackets 
in terms of the matrix
$W_{\textrm{\tiny TR}}$ introduced in (\ref{W_TR def})
as follows
\bea
\label{WT log WTinv-1}
& &
W_{\textrm{\tiny T}}
\log\!\big(
\gamma_\textrm{\tiny R}^{-1} \gamma_\textrm{\tiny T}
\big)
W_{\textrm{\tiny T}}^{-1}
= 
\log\!\big( 
W_{\textrm{\tiny TR}}\,\mathcal{D}_{\textrm{\tiny R}}^{-1}\, W_{\textrm{\tiny TR}}^{\textrm{t}}\,\mathcal{D}_{\textrm{\tiny T}}
\big)
\\
\label{WT log WTinv-2}
\rule{0pt}{.5cm}
& &
W_{\textrm{\tiny R}}
\log\!\big(
\gamma_\textrm{\tiny R}^{-1} \gamma_\textrm{\tiny T}
\big)
W_{\textrm{\tiny R}}^{-1}
=
\log\!\big(
\mathcal{D}_{\textrm{\tiny R}}^{-1}\, W_{\textrm{\tiny TR}}^{\textrm{t}}\,\mathcal{D}_{\textrm{\tiny T}}\,W_{\textrm{\tiny TR}}
\big) \,.
\label{WR log WRinv}
\eea

The form (\ref{first-law-williamson}) for $\delta d$ tells us that 
this variation can be written as the sum of four contributions:
two terms from the variations $\delta W_{\textrm{\tiny R}}  $ and $\delta W_{\textrm{\tiny T}}  $
of the symplectic matrices in (\ref{gamma R and T williamson})
and two terms 
from the diagonal and non negative variations $\delta\mathcal{D}_{\textrm{\tiny R}}$ and $\delta\mathcal{D}_{\textrm{\tiny T}}$
of the symplectic spectra.

\subsection{Solving $\delta d=0$}
\label{sec:solving}

It is natural to look for relations
between $\gamma_{\textrm{\tiny R}}$  and $\gamma_{\textrm{\tiny T}}$ 
that lead to $\delta d=0$
and, in order to find them, 
let us consider the first law of complexity written in the form
(\ref{first-law-williamson}).

First we focus on
the variations of $W_{\textrm{\tiny R}} $ and $W_{\textrm{\tiny T}} $.
When $\delta \mathcal{D}_{\textrm{\tiny R}} = \delta \mathcal{D}_{\textrm{\tiny T}} = \boldsymbol{0}$ 
in (\ref{first-law-williamson}), the equation $\delta d=0$ becomes 
\be
\label{delta spectrum comp final}
\textrm{Tr} \,\Big\{
\log\!\big(
\gamma_\textrm{\tiny T}^{-1} \gamma_\textrm{\tiny R}
\big)\,
\big( 
W_{\textrm{\tiny R}}^{-1}\,\delta W_{\textrm{\tiny R}} 
-
W_{\textrm{\tiny T}}^{-1}\,\delta W_{\textrm{\tiny T}}
\big)
\Big\} 
=0 \,.
\ee
A trivial solution of this equation is given by 
\be
\label{WR=WT}
W_{\textrm{\tiny R}} = W_{\textrm{\tiny T}} \,.
\ee

Another solution of (\ref{delta spectrum comp final}) is $W_{\textrm{\tiny R}} = M\,W_{\textrm{\tiny T}}$,
where $M$ is a constant symplectic matrix whose elements are just real numbers,
i.e. it does not contain parameters to vary. 
Notice that these two simple solutions require that 
both $W_{\textrm{\tiny R}}$ and $W_{\textrm{\tiny T}}$ are allowed to vary.

In order to find solutions to (\ref{delta spectrum comp final})
for any choice of the independent variations 
$\delta W_{\textrm{\tiny R}} $ and $\delta W_{\textrm{\tiny T}} $,
let us first observe that, by taking the first variation of the relation $W^{\textrm t} J \,W = J$,
that characterises a symplectic matrix $W^{\textrm t}$,
it is not difficult to realise that 
$\delta X \equiv W^{\textrm t} J \,\delta W   $ must be a symmetric matrix.
By using that $W^{-1} = J^{-1} \,W^{\textrm t} J$, 
for the two terms in the r.h.s. of (\ref{delta spectrum comp final}) one obtains
$ \textrm{Tr}[A\,W^{-1}\,\delta W] = \textrm{Tr} (Y \delta X) $,
where $Y \equiv A \,J^{-1}$.
Since $\delta X$ corresponds to a generic infinitesimal real and symmetric matrix, 
$\textrm{Tr} (Y \delta X) =0$ is satisfied for every $\delta X$ 
e.g. when $Y$ is a real antisymmetric matrix. 
These observations lead us to write (\ref{delta spectrum comp final}) as 
\be
\label{delta spectrum comp final X-version}
\textrm{Tr} \,\Big\{
\log\!\big(
\gamma_\textrm{\tiny T}^{-1} \gamma_\textrm{\tiny R}
\big)\,J^{-1}
\big[
\delta X_{\textrm{\tiny R}} 
-
\delta X_{\textrm{\tiny T}}
\big]
\Big\} 
=0
\ee
where $\delta X_{\textrm{\tiny R}}  \equiv W_{\textrm{\tiny R}} ^{\textrm t} \,J \,\delta W_{\textrm{\tiny R}} $ 
and $\delta X_{\textrm{\tiny T}}  \equiv W_{\textrm{\tiny T}} ^{\textrm t} \,J \,\delta W_{\textrm{\tiny T}} $
are real and symmetric matrices. 
Thus, from (\ref{delta spectrum comp final X-version}), 
we have that (\ref{delta spectrum comp final}) is satisfied for 
$ \gamma_\textrm{\tiny T}^{-1} \gamma_\textrm{\tiny R} = \, e^{Y J} $, i.e.
\be
\label{exp-Y}
 \gamma_{\textrm{\tiny T}} \,= \, \gamma_{\textrm{\tiny R}} \,e^{-Y J} 
\ee
where $Y$ is a real antisymmetric matrix that can depend on 
$ \gamma_{\textrm{\tiny R}} $ or $ \gamma_{\textrm{\tiny T}}$.
We remark that (\ref{exp-Y}) solves (\ref{delta spectrum comp final})
for any choice of the independent variations 
$\delta W_{\textrm{\tiny R}} $ and $\delta W_{\textrm{\tiny T}} $.

It is worth asking when the case $W_{\textrm{\tiny R}} = M\,W_{\textrm{\tiny T}}$ 
mentioned above, where $M$ is a symplectic matrix made of real numbers,
becomes a special case of (\ref{exp-Y}). 
The Williamson's decompositions (\ref{gamma R and T williamson}) 
and $W_{\textrm{\tiny R}} = M\,W_{\textrm{\tiny T}}$
lead to
$\log(\gamma_\textrm{\tiny T}^{-1} \gamma_\textrm{\tiny R})J^{-1}
=
\log(W_\textrm{\tiny T}^{-1} \mathcal{D}_\textrm{\tiny T}^{-1} M^{\textrm{t}} \mathcal{D}_\textrm{\tiny R} M W_\textrm{\tiny T})J^{-1}$.
Then, (\ref{f(MN) identity app}) allows us to write the transpose of this matrix as
$J \log(W_\textrm{\tiny T}^{\textrm{t}} M^{\textrm{t}} \mathcal{D}_\textrm{\tiny R} M \mathcal{D}_\textrm{\tiny T}^{-1} W^{-\textrm{t}}_\textrm{\tiny T})$.
By inserting the identity $\boldsymbol{1} = J^{-1} J$ between all the factors within the argument of the logarithm occurring in this matrix, 
using (\ref{f(MN) identity app}) again 
and exploiting the properties of the symplectic matrices, 
one arrives to
$
\big[\log(\gamma_\textrm{\tiny T}^{-1} \gamma_\textrm{\tiny R})J^{-1}\big]^{\textrm{t}}
=
-\log( W_\textrm{\tiny T}^{-1}  M^{-1}  \mathcal{D}_\textrm{\tiny R}  M^{-\textrm{t}}  \mathcal{D}_\textrm{\tiny T}^{-1} W_\textrm{\tiny T})J^{-1}
$.
Comparing this expression with the one reported above, we conclude that
the matrix $\log(\gamma_\textrm{\tiny T}^{-1} \gamma_\textrm{\tiny R})J^{-1}$
is antisymmetric when $W_{\textrm{\tiny R}} = M\,W_{\textrm{\tiny T}}$
and $ M^{-1}  \mathcal{D}_\textrm{\tiny R}  M^{-\textrm{t}}
=M^{\textrm{t}} \mathcal{D}_\textrm{\tiny R} M $. 
This is the case for a symplectic matrix that is also orthogonal,
i.e. $M\in K(N)$.
In particular, the special case given by $M=\boldsymbol{1} \in K(N)$
corresponds to (\ref{WR=WT}).
Summarising, in the special case given by $W_{\textrm{\tiny R}} = M\,W_{\textrm{\tiny T}}$,
the matrix $Y$ introduced in (\ref{exp-Y}) can be written as $Y=\log(W_\textrm{\tiny T}^{-1} \mathcal{D}_\textrm{\tiny T}^{-1} M^{\textrm{t}} \mathcal{D}_\textrm{\tiny R} M W_\textrm{\tiny T})J^{-1}$ with 
$ M^{-1}  \mathcal{D}_\textrm{\tiny R}  M^{-\textrm{t}} =M^{\textrm{t}} \mathcal{D}_\textrm{\tiny R} M $.

As for the terms corresponding to the 
variations of $ \mathcal{D}\textrm{\tiny R}$ and $ \mathcal{D}\textrm{\tiny T}$ 
in (\ref{first-law-williamson}), let us observe that,
for a diagonal matrix $\Lambda>0$,
we have that
$\Lambda^{-1} \delta \Lambda = \delta \log\Lambda$
and that 
$\textrm{Tr} ( H \,\delta \log\Lambda ) = 0$ holds for a generic $\delta \Lambda$
when all the elements along the diagonal of $H$ vanish.
The matrices having vanishing elements on their main diagonal are called hollow matrices.
By employing these observations in the equation $\delta d=0$
with $\delta d$ given by (\ref{first-law-williamson}),
where the variations of  $ \mathcal{D}\textrm{\tiny R}$ and $ \mathcal{D}\textrm{\tiny T}$ 
 are independent, we conclude that the  main diagonals of the matrices 
 within the square brackets in (\ref{first-law-williamson}) must vanish.
By introducing two non vanishing hollow matrices $Z$ and $\widetilde{Z}$, this gives
\be
\label{ZtildeZ}
\log\!\big(
\gamma_\textrm{\tiny R}^{-1} \gamma_\textrm{\tiny T}
\big)
= W_{\textrm{\tiny T}}^{-1} Z\, W_{\textrm{\tiny T}}
\;\;\qquad\;\;
\log\!\big(
\gamma_\textrm{\tiny R}^{-1} \gamma_\textrm{\tiny T}
\big)
= W_{\textrm{\tiny R}}^{-1} \widetilde{Z}\, W_{\textrm{\tiny R}}
\ee
which correspond to the terms containing 
 $\delta \mathcal{D}\textrm{\tiny T}$ and $\delta \mathcal{D}\textrm{\tiny R}$  
 respectively in (\ref{first-law-williamson}).
 By employing (\ref{WT log WTinv-1}) and (\ref{WT log WTinv-2}), 
 one finds that the relations in (\ref{ZtildeZ})
 can be written respectively as
 \be
\label{ZtildeZ-exp}
W_{\textrm{\tiny TR}}\,\mathcal{D}_{\textrm{\tiny R}}^{-1}\, W_{\textrm{\tiny TR}}^{\textrm{t}}\,\mathcal{D}_{\textrm{\tiny T}}
=e^{Z}
\;\;\qquad\;\;
\mathcal{D}_{\textrm{\tiny R}}^{-1}\, W_{\textrm{\tiny TR}}^{\textrm{t}}\,\mathcal{D}_{\textrm{\tiny T}}\,W_{\textrm{\tiny TR}}
=e^{\widetilde{Z}} \,.
\ee

These observations tell us that $\delta d=0$
for generic variations of $\gamma_\textrm{\tiny R}$ and $\gamma_\textrm{\tiny T}$
when these covariance matrices are related by (\ref{exp-Y})
with $Y$ constrained by the conditions that
the elements on the diagonals of 
$W_{\textrm{\tiny T}} \,YJ \,W_{\textrm{\tiny T}}^{-1}$
and
$W_{\textrm{\tiny R}} \,YJ \,W_{\textrm{\tiny R}}^{-1}$
vanish.

A rough analysis shows that this problem is too constrained for $N=1$ and $N=2$.
Indeed, the antisymmetric matrix $Y$ depends on $N(2N-1)$ parameters
and imposing that the diagonals of $Z$ and $\widetilde{Z}$ vanish provides  
$4N$ constraints. 
In particular, when $N=1$ the $2\times 2$ antisymmetric matrix $Y$ has only one non vanishing off diagonal element $a$ in the top right position
and it is straightforward to check that $Y J=-a\,\boldsymbol{1}$. 
By using also (\ref{exp-Y}) and (\ref{ZtildeZ}) specialised to this case, 
we obtain that the above procedure leads to impose that $W_{\textrm{\tiny S}} \,YJ \,W_{\textrm{\tiny S}}^{-1}$,
with $\textrm{S}=\{\textrm{R},\textrm{T}\}$, 
must have vanishing elements along the diagonal.
This is possible only for $a=0$, i.e. $Y=\boldsymbol{0}$.
Thus, when $N=1$ we cannot find a solution of the form (\ref{exp-Y}) for the equation $\delta d=0$ with $\delta d$ given by (\ref{first-law-williamson}).

\subsection{Spectrum complexity}
\label{subsec:spectrum-comp}

It is worth exploring the possibility to define the 
circuit complexity associated to the change of the symplectic spectrum. 

Let us consider a reference state and a target state 
such that in the Williamson's decompositions of their
covariance matrices $\gamma_{\textrm{\tiny R}}$ and $\gamma_{\textrm{\tiny T}}$
(see (\ref{gamma R and T williamson}))
the same symplectic matrix occurs, namely
\be
\label{gamma R and T williamson same W}
\gamma_{\textrm{\tiny R}} =  W^{\textrm{t}} \, \mathcal{D}_{\textrm{\tiny R}} \,W
\;\;\qquad \;\;
\gamma_{\textrm{\tiny T}} = W^{\textrm{t}} \, \mathcal{D}_{\textrm{\tiny T}} \,W
\;\;\;\;\qquad \;\;\;\;
W  \in \textrm{Sp}(2N, \mathbb{R}) \,.
\ee
We call {\it $W$ path} the optimal circuit (\ref{optimal circuit}) connecting 
these two covariance matrices.

In order to study the Williamson's decomposition of a matrix belonging to a $W$ path, 
we consider the expression (\ref{optimal circuit Myers}) for the optimal circuit. 
When (\ref{gamma R and T williamson same W}) holds,
from (\ref{def relative CM}) it is straightforward to find that 
$\Delta_{\textrm{\tiny TR}} 
= W^{\textrm{t}} \, \mathcal{D}_{\textrm{\tiny T}} \,\mathcal{D}_{\textrm{\tiny R}}^{-1} \,W^{-\textrm{t}} $.
Then, by employing (\ref{f(MN) identity app}) both in $U_s$ and in $U_s^{\textrm{t}}$
occurring in (\ref{optimal circuit Myers}), we obtain 
\be
\label{W type path}
G_s(\gamma_{\textrm{\tiny R}} \, , \gamma_{\textrm{\tiny T}})
=  W^{\textrm{t}} \big(\, \mathcal{D}_{\textrm{\tiny R}}^{1-s} \, \mathcal{D}_{\textrm{\tiny T}}^s \,\big)  W
\ee
which tells us that the Williamson's decomposition 
of the matrix along the $W$ path is (\ref{williamson-geodesic-generic}) with
\be
\label{W-path-williamson}
\mathcal{D}_s\, =\, \mathcal{D}_{\textrm{\tiny R}}^{1-s} \, \mathcal{D}_{\textrm{\tiny T}}^s
\;\;\qquad\;\;
W_s=W \,.
\ee

It is remarkable that the symplectic matrix $W_s$ is independent of $s$.
This means that in the 
Williamson's decomposition of a matrix belonging to a $W$ path
the same symplectic matrix $W$ occurs.  
In Fig.\;\ref{fig:different_D_sp} the dashed curves correspond to 
the $W_{\textrm{\tiny R}}$ path and to the $W_{\textrm{\tiny T}}$ path.
Considering e.g. the $W_{\textrm{\tiny R}}$ path in Fig.\;\ref{fig:different_D_sp},
from (\ref{W-path-williamson}) we have  that
the Williamson's decomposition of a generic matrix $\gamma$ belonging to 
this $W_{\textrm{\tiny R}}$ path is given by the symplectic matrix $W_{\textrm{\tiny R}}$
and by the symplectic spectrum corresponding to the coloured line
intersecting the dashed line of the $W_{\textrm{\tiny R}}$ path at $\gamma$.

An interesting example of $W$ path is given by the thermal states 
of a given model at different temperatures
(see Sec.\,\ref{subsec:thermal}).
Indeed, in the Williamson's decomposition (\ref{Williamson_CM_th}), 
the symplectic matrix $W_{\textrm{\tiny th}}$ is independent of the temperature.

For a $W$ path we have $\delta d=0$ (see (\ref{WR=WT}));
hence the $W$ paths provide a preferred way to connect the 
set of covariance matrices with symplectic spectrum
$\mathcal{D}_{\textrm{\tiny R}} $ 
to the set of covariance matrices with symplectic spectrum
$\mathcal{D}_{\textrm{\tiny T}} $.

We find it natural to define the spectrum complexity as the length of a $W$ path
because this quantity is independent of the choice of $W$.
In particular, 
from (\ref{gamma R and T williamson same W}), 
we have that $W_{\textrm{\tiny TR}}  =\boldsymbol{1}$, 
hence (\ref{C-2-complexity}) simplifies to
\be
\label{compl_same_W}
d_{\textrm{\tiny spectrum}}(\gamma_{\textrm{\tiny R}} \, , \gamma_{\textrm{\tiny T}})
\equiv
\sqrt{
\textrm{Tr} \Big\{ \big[\log\! \big( \mathcal{D}_{\textrm{\tiny T}} \,  \mathcal{D}_{\textrm{\tiny R}}^{-1} \big)\big]^2  \Big\}}
=\,
\sqrt{2
\sum_{k=1}^N \left[ \,\log\! \left( \frac{\sigma_{\textrm{\tiny T}, k}}{\sigma_{\textrm{\tiny R}, k}} \right)\right]^2
}
\ee
which is independent of $W$.
This implies that 
$d_{\textrm{\tiny spectrum}}(\gamma_{\textrm{\tiny R}} \, , \gamma_{\textrm{\tiny T}})
=
d_{\textrm{\tiny spectrum}}(\mathcal{D}_{\textrm{\tiny R}} \, , \mathcal{D}_{\textrm{\tiny T}})$.
Thus, in Fig.\,\ref{fig:different_D_sp}
the arcs of the dashed curves that connect the blue curve to the red curve
have the same length given by (\ref{compl_same_W}).

Another natural definition for the spectrum complexity
is the distance between the set of covariance matrices whose symplectic spectrum is 
$\mathcal{D}_{\textrm{\tiny R}} $ (red curve in Fig.\,\ref{fig:different_D_sp})
and the set of covariance matrices whose symplectic spectrum is 
$\mathcal{D}_{\textrm{\tiny T}} $ (blue curve in Fig.\,\ref{fig:different_D_sp}).
It reads
\be
\label{spectrum complexity tilded}
\tilde{d}_{\textrm{\tiny spectrum}}( \mathcal{D}_{\textrm{\tiny R}}  \, , \mathcal{D}_{\textrm{\tiny T}} )
\equiv\,
\textrm{min}
\big[\,
d\big( W^{\textrm{t}} \, \mathcal{D}_{\textrm{\tiny R}} \,W 
 \,, 
\widetilde{W}^{\textrm{t}}  \, \mathcal{D}_{\textrm{\tiny T}} \,\widetilde{W}\big) 
\,\big]
\;\;\qquad \;\;
W , \,  \widetilde{W} \in \textrm{Sp}(2N, \mathbb{R})
\ee
where the minimisation over the symplectic matrices
$W$ and $\widetilde{W}$ is difficult to perform. 
It is straightforward to realise that
$ \tilde{d}_{\textrm{\tiny spectrum}}( \mathcal{D}_{\textrm{\tiny R}}  \, , \mathcal{D}_{\textrm{\tiny T}} ) 
\leqslant
d_{\textrm{\tiny spectrum}}( \mathcal{D}_{\textrm{\tiny R}}  \, , \mathcal{D}_{\textrm{\tiny T}} )$.

In the simplest case of one-mode mixed states (i.e. when $N=1$),
the optimal circuit (\ref{W type path}) simplifies to
\be
G_s(\gamma_{\textrm{\tiny R}} \, , \gamma_{\textrm{\tiny T}})
\,=\,
\sigma_{\textrm{\tiny R}}^{1-s}  \sigma_{\textrm{\tiny T}}^s
\;W^{\textrm{t}}\,  W
\,=\,
\bigg( \frac{\sigma_{\textrm{\tiny T}}}{\sigma_{\textrm{\tiny R}}} \bigg)^s \gamma_{\textrm{\tiny R}} 
\ee
which tells us that the $2\times 2$ matrix belonging to the $W$ path is a 
proper rescaling of the covariance matrix of the reference state.

\subsection{Basis complexity}
\label{subsec:basis-comp}

In order to study the circuit complexity associated to a change of basis,
let us consider the Williamson's decompositions of two
covariance matrices $\gamma_{\textrm{\tiny R}} $ and $\gamma_{\textrm{\tiny T}} $ 
having the same symplectic spectrum, i.e.
\be
\label{gamma R and T williamson same D}
\gamma_{\textrm{\tiny R}} =  W^{\textrm{t}}_{\textrm{\tiny R}}  \, \mathcal{D}\,W_{\textrm{\tiny R}} 
\;\;\qquad \;\;
\gamma_{\textrm{\tiny T}} = W^{\textrm{t}}_{\textrm{\tiny T}}  \, \mathcal{D}\,W_{\textrm{\tiny T}} 
\ee
that have been obtained by setting $\mathcal{D}_{\textrm{\tiny R}}=\mathcal{D}_{\textrm{\tiny T}}=\mathcal{D}$
in (\ref{gamma R and T williamson}).
An important example is given by  states whose density matrices $\hat{\rho}_{\textrm{\tiny T}} $ and $\hat{\rho}_{\textrm{\tiny R}} $ are related through a unitary transformation $U$, namely 
$\hat{\rho}_{\textrm{\tiny T}} = U \hat{\rho}_{\textrm{\tiny R}}  U^\dagger$.
Indeed, this means that the corresponding covariance matrices are related through a symplectic matrix
(that does not change the symplectic spectrum) \cite{Weedbrook12b,Serafini17book}.

We denote as {\it $\mathcal{D}$ path} the optimal circuit connecting the covariance matrices 
having the same symplectic spectrum, identifying its length as a  basis complexity.
This basis complexity can be found by
specifying (\ref{C-2-complexity}) to (\ref{gamma R and T williamson same D}) and
the result is\footnote{The result (\ref{compl_same_D}) can be obtained also by employing (\ref{gammagammaDTRcase}) with $ \mathcal{D}_{\textrm{\tiny R}} =  \mathcal{D}_{\textrm{\tiny T}}$.}
\be
\label{compl_same_D}
d_{\textrm{\tiny basis}}(\gamma_{\textrm{\tiny R}} \, , \gamma_{\textrm{\tiny T}})
=\,
\sqrt{
\textrm{Tr}\, \Big\{ \big[\log\! \big( 
 \mathcal{D}\,
W_{\textrm{\tiny TR}} 
\, \mathcal{D}^{-1}\,
W_{\textrm{\tiny TR}}^{\textrm{t}}
\big)\big]^2  \Big\}}
\ee
where $W_{\textrm{\tiny TR}} $ has been defined in (\ref{W_TR def}).
Notice that we have not required that all the matrices along a $\mathcal{D}$ path 
have the same symplectic spectrum.

We find it reasonable to introduce also another definition of basis complexity
as the minimal length of an optimal circuit that connects a covariance matrix
whose Williamson's decomposition contains the symplectic matrix $W_{\textrm{\tiny R}}$
(i.e. that lies on the dashed curve on the left in Fig.\,\ref{fig:different_D_sp})
to a covariance matrix
having the symplectic matrix $W_{\textrm{\tiny T}}$ in its Williamson's decomposition
(i.e. that belongs to the dashed curve on the right in Fig.\,\ref{fig:different_D_sp}).
This basis complexity is defined as follows
\be
\label{basis complexity tilded}
\tilde{d}_{\textrm{\tiny basis}}( W_{\textrm{\tiny R}}  \, , W_{\textrm{\tiny T}} )
\equiv\,
\textrm{min}
\big[\,
d\big( W_{\textrm{\tiny R}}^{\textrm{t}} \, \mathcal{D} \,W_{\textrm{\tiny R}} 
\, , \, 
W_{\textrm{\tiny T}}^{\textrm{t}}  \, \widetilde{\mathcal{D}} \,W_{\textrm{\tiny T}}\big) 
\,\big]
\;\;\qquad \;\;
\mathcal{D} , \,  \widetilde{\mathcal{D}} \in \textrm{Diag}(N, \mathbb{R})
\ee
where the minimisation is performed over the set $\textrm{Diag}(2N, \mathbb{R})$
made by  the diagonal matrices of the form
 $\textrm{diag}(\boldsymbol{\sigma}) \oplus \textrm{diag}(\boldsymbol{\sigma})$,
with $\boldsymbol{\sigma}$ vector of $N$ real numbers $\sigma_i \geqslant 1/2$.
It is immediate to notice that (\ref{basis complexity tilded}) is a lower bound for (\ref{compl_same_D}),
i.e. $ \tilde{d}_{\textrm{\tiny basis}}( W_{\textrm{\tiny R}}  \, , W_{\textrm{\tiny T}} ) 
\leqslant
d_{\textrm{\tiny basis}}( \gamma_{\textrm{\tiny R}}  \, , \gamma_{\textrm{\tiny T}} )$.

Specifying the form (\ref{optimal circuit Myers}) for the optimal circuit to (\ref{gamma R and T williamson same D}), it is straightforward to find that the $\mathcal{D}$ path is given by 
\be
\label{geod-D-path-symp}
G_s(\gamma_{\textrm{\tiny R}} \, , \gamma_{\textrm{\tiny T}})
\,=\,
\widetilde{W}^{\textrm{t}}_s \, \mathcal{D}\,  \widetilde{W}_s
\;\;\qquad\;\;
\widetilde{W}_s \equiv W_{\textrm{\tiny R}} \,U_s^{\textrm{t}}
\ee
where we remark that $\widetilde{W}_s $ is not symplectic in general.

It is worth asking when $\widetilde{W}_s $ is symplectic because in these cases 
(\ref{geod-D-path-symp}) provides the Williamson's decomposition of the $\mathcal{D}$ path.
The requirement $U_s \in \textrm{Sp}(2N,\mathbb{R})$ leads to
\be
\label{sympl_cond_Us}
\big[\,W_{\textrm{\tiny TR}}\,, \mathcal{D}\,\big] \,=\,0 \,.
\ee
When this condition holds, (\ref{compl_same_D}) simplifies to
the following expression 
\be
\label{compl_same_D_comm}
d_{\textrm{\tiny basis}}(\gamma_{\textrm{\tiny R}} \, , \gamma_{\textrm{\tiny T}})
=\,
\sqrt{
\textrm{Tr} \,\Big\{ \big[\log\! \big( 
W_{\textrm{\tiny TR}} 
\,W_{\textrm{\tiny TR}}^{\textrm{t}}
\big)\big]^2  \Big\}}
\ee
which is independent of $\mathcal{D}$.

For pure states, which have $\mathcal{D} = \tfrac{1}{2}\,\boldsymbol{1}$,
the condition (\ref{sympl_cond_Us}) is trivially verified.
Another interesting example where (\ref{sympl_cond_Us}) holds is given by the one-mode states,
where $\mathcal{D}$ is proportional to the $2\times 2$ identity matrix.
In this case we can always connect two covariance matrices with the same symplectic spectrum through the optimal circuit (\ref{G_s decomposition one-mode}), that can be written as
\be
G_s(\gamma_{\textrm{\tiny R}} \, , \gamma_{\textrm{\tiny T}})
\,=\,\sigma \;W^{\textrm{t}}_s \,W_s
\ee
where  $\sigma_{\textrm{\tiny R}}=\sigma_{\textrm{\tiny T}}\equiv\sigma$ and $W_s$ 
is defined in (\ref{Williamson D and W one mode});
hence from (\ref{geod-D-path-symp}) we have that $W_s =\widetilde{W}_s$.

When $N>1$ the condition (\ref{sympl_cond_Us}) is a non trivial requirement.
For instance, when $W_{\textrm{\tiny TR}}$ is diagonal, 
(\ref{sympl_cond_Us}) is verified and
(\ref{geod-D-path-symp}) holds with
$\widetilde{W}_s = \mathcal{X}_{\textrm{\tiny TR}}^s W_{\textrm{\tiny R}}$.
The basis complexity (\ref{compl_same_D}) simplifies to
$d^2_{\textrm{\tiny basis}}(\gamma_{\textrm{\tiny R}} \, , \gamma_{\textrm{\tiny T}})
= \textrm{Tr}\big\{[\log( \mathcal{X}_{\textrm{\tiny TR}}^2 )]^2 \big\} $,
that is independent of $\mathcal{D}$.

Writing $W_{\textrm{\tiny TR}}$ as a block matrix made by four $N \times N$ matrices,
it is straightforward to find that the condition (\ref{sympl_cond_Us}) holds 
whenever every block of $W_{\textrm{\tiny TR}}$ commutes with 
$\textrm{diag} (\sigma_1 , \dots , \sigma_N)$.
Then, we can exploit the fact that a diagonal matrix with distinct elements commutes with another matrix 
only when the latter one is diagonal\footnote{Consider the diagonal matrix $\Lambda=\textrm{diag}(\lambda_1,\dots,\lambda_N)$ with $\lambda_i\neq \lambda_j$ and a matrix $M$ such that $[\Lambda,M]=0$.
The generic element of this relation reads $M_{i,j} \lambda_j =\lambda_i M_{i,j} $, i.e. $M_{i,j} (\lambda_i-\lambda_j)=0$.
Since $\lambda_i\neq \lambda_j$ when $i\neq j$, we have $M_{i,j}=0$ for $i\neq j$.
}.
Thus, if the symplectic spectrum is non degenerate, all the blocks of $W_{\textrm{\tiny TR}}$ 
must be diagonal to fulfil the condition (\ref{sympl_cond_Us}).
We remark that the non-degeneracy condition for the symplectic spectrum is not guaranteed;
indeed, the symplectic spectrum has some degeneracy in several interesting cases.
For instance, for pure states all the symplectic eigenvalues are equal to $\frac{1}{2}$.
Another important example is
the reduced covariance matrix of an interval 
in an infinite harmonic chain with non vanishing mass \cite{PeschelChung99}.

We find it worth discussing the relation between the optimal circuits considered above to study the basis complexity
and the solutions of the equation $\delta d= 0$ described in Sec.\,\ref{sec:solving}.
For the set of paths occurring in (\ref{basis complexity tilded}),
which includes the $\mathcal{D}$ paths, we have 
$\delta W_{\textrm{\tiny R}}=\delta W_{\textrm{\tiny T}}=0$ in (\ref{first-law-williamson}).
In this case,  in Sec.\,\ref{sec:solving}
we found that a solution of $\delta d= 0$ is given by
(\ref{ZtildeZ-exp}), where $Z$ and $\widetilde{Z}$ are non vanishing hollow matrices. 
Restricting to the cases of $\mathcal{D}$ paths that satisfy also (\ref{sympl_cond_Us}),
these relations simplify respectively to
$W_{\textrm{\tiny TR}} \,W_{\textrm{\tiny TR}}^{\textrm{t}} =e^{Z}$
and
$W_{\textrm{\tiny TR}}^{\textrm{t}} \,W_{\textrm{\tiny TR}} =e^{\widetilde{Z}}$,
whose solution is non trivial because
a matrix does not commute with its transpose in general 
(the matrices satisfying this property are called normal matrices).
Notice that $W_{\textrm{\tiny TR}} \in K(N)$ are not admissible solutions because
$Z$ and $\widetilde{Z}$ are non vanishing. 

A slightly more general solution can be obtained by restricting to the $\mathcal{D}$ paths
(see (\ref{gamma R and T williamson same D})).
In this case, from (\ref{basis complexity tilded}) with 
$\delta W_{\textrm{\tiny R}}=\delta W_{\textrm{\tiny T}}=0$
and
$\delta \mathcal{D}_{\textrm{\tiny R}}=\delta \mathcal{D}_{\textrm{\tiny T}} \equiv \delta \mathcal{D}$,
we have that $\delta d=0$ becomes
\be
\label{d=0-basis-Dpath}
\textrm{Tr}\, \Big\{ 
\big[\,
W_{\textrm{\tiny T}}
\log\!\big(
\gamma_\textrm{\tiny R}^{-1} \gamma_\textrm{\tiny T}
\big)
W_{\textrm{\tiny T}}^{-1}
-
W_{\textrm{\tiny R}}
\log\!\big(
\gamma_\textrm{\tiny R}^{-1} \gamma_\textrm{\tiny T}
\big)
W_{\textrm{\tiny R}}^{-1}
\,\big]\,
\mathcal{D}^{-1}\, \delta \mathcal{D}\,
\Big\}
=0 \,.
\ee
By using (\ref{WT log WTinv-1}), (\ref{WT log WTinv-2}) 
and the discussion made in Sec.\,\ref{sec:solving},
one finds that (\ref{d=0-basis-Dpath}) is solved 
when
$\log( W_{\textrm{\tiny TR}}\,\mathcal{D}^{-1}\, W_{\textrm{\tiny TR}}^{\textrm{t}}\,\mathcal{D})
- \log(\mathcal{D}^{-1}\, W_{\textrm{\tiny TR}}^{\textrm{t}}\,\mathcal{D}\,W_{\textrm{\tiny TR}})$
is a non vanishing hollow matrix.
When (\ref{sympl_cond_Us}) also holds,
this condition simplifies to the requirement that 
$\log( W_{\textrm{\tiny TR}}\,W_{\textrm{\tiny TR}}^{\textrm{t}})
- \log(W_{\textrm{\tiny TR}}^{\textrm{t}} W_{\textrm{\tiny TR}})$
is a non vanishing hollow matrix,
which is independent of $\mathcal{D}$.

\section{Purification through the $W$ path}
\label{sec:purification}

The purification of a mixed state is a process that provides a pure state starting from a mixed state.
This procedure is not unique. 
Considering the context of the bosonic Gaussian states that we are exploring,
in this section we discuss the purification of a mixed state by employing the results
reported in Sec.\,\ref{sec:spectrum-basis}.

Given a mixed state that is not pure and that is 
characterised by the covariance matrix $\gamma_{\textrm{\tiny R}}$,
any circuit connecting $\gamma_{\textrm{\tiny R}}$ to a pure state 
provides a {\it purification path}. 
A purification path connects
the covariance matrices $\gamma_{\textrm{\tiny R}} $ to $\gamma_{\textrm{\tiny T}}$
whose Williamson's decompositions are given
respectively by 
\be
\label{purification gamma choice}
\gamma_{\textrm{\tiny R}} = 
W^{\textrm{t}}_{\textrm{\tiny R}}\,\mathcal{D}\, W_{\textrm{\tiny R}}
\;\;\qquad \;\;
\gamma_{\textrm{\tiny T}} = \frac{1}{2}\,W^{\textrm{t}}_{\textrm{\tiny T}} \, W_{\textrm{\tiny T}}
\ee
where $W_{\textrm{\tiny R}} \in \textrm{Sp}(2N,\mathbb{R})$ and $\mathcal{D}\neq \tfrac{1}{2}\,\boldsymbol{1}$ are assigned, 
while $W_{\textrm{\tiny T}} \in \textrm{Sp}(2N,\mathbb{R})$ is not.
Among all the possible paths, the optimal circuit is obtained
by specifying (\ref{optimal circuit}) to (\ref{purification gamma choice}).
The result is
\be
\label{purification path}
G^{(0)}_s(\gamma_{\textrm{\tiny R}} , W_{\textrm{\tiny T}})
\equiv\,
\frac{1}{2^s}\;
\gamma_{\textrm{\tiny R}}^{1/2} 
\Big(
\gamma_{\textrm{\tiny R}}^{-1/2} 
\,W^{\textrm{t}}_{\textrm{\tiny T}} \, W_{\textrm{\tiny T}} \,
\gamma_{\textrm{\tiny R}}^{-1/2} 
\,\Big)^s
\gamma_{\textrm{\tiny R}}^{1/2} 
\ee
which depends on the symplectic matrix $W_{\textrm{\tiny T}}$ that determines the final pure state. 
The length of the purification path (\ref{purification path}) can be found by 
evaluating (\ref{deltaAB}) for the special case described by (\ref{purification gamma choice}).
It reads
\be
\label{length purification path}
d_0(\gamma_{\textrm{\tiny R}}  , W_{\textrm{\tiny T}})
\equiv
\sqrt{
\textrm{Tr} \,\Big\{ \big[\log \! \big(
2\,\gamma_{\textrm{\tiny R}} \, 
W^{-1}_{\textrm{\tiny T}} W_{\textrm{\tiny T}}^{-\textrm{t}}
\big)\big]^2  \Big\}
} \,.
\ee

\begin{figure}[t!]
\vspace{-.2cm}
\hspace{-1.1cm}
 \begin{center}
\includegraphics[width=0.8\textwidth]{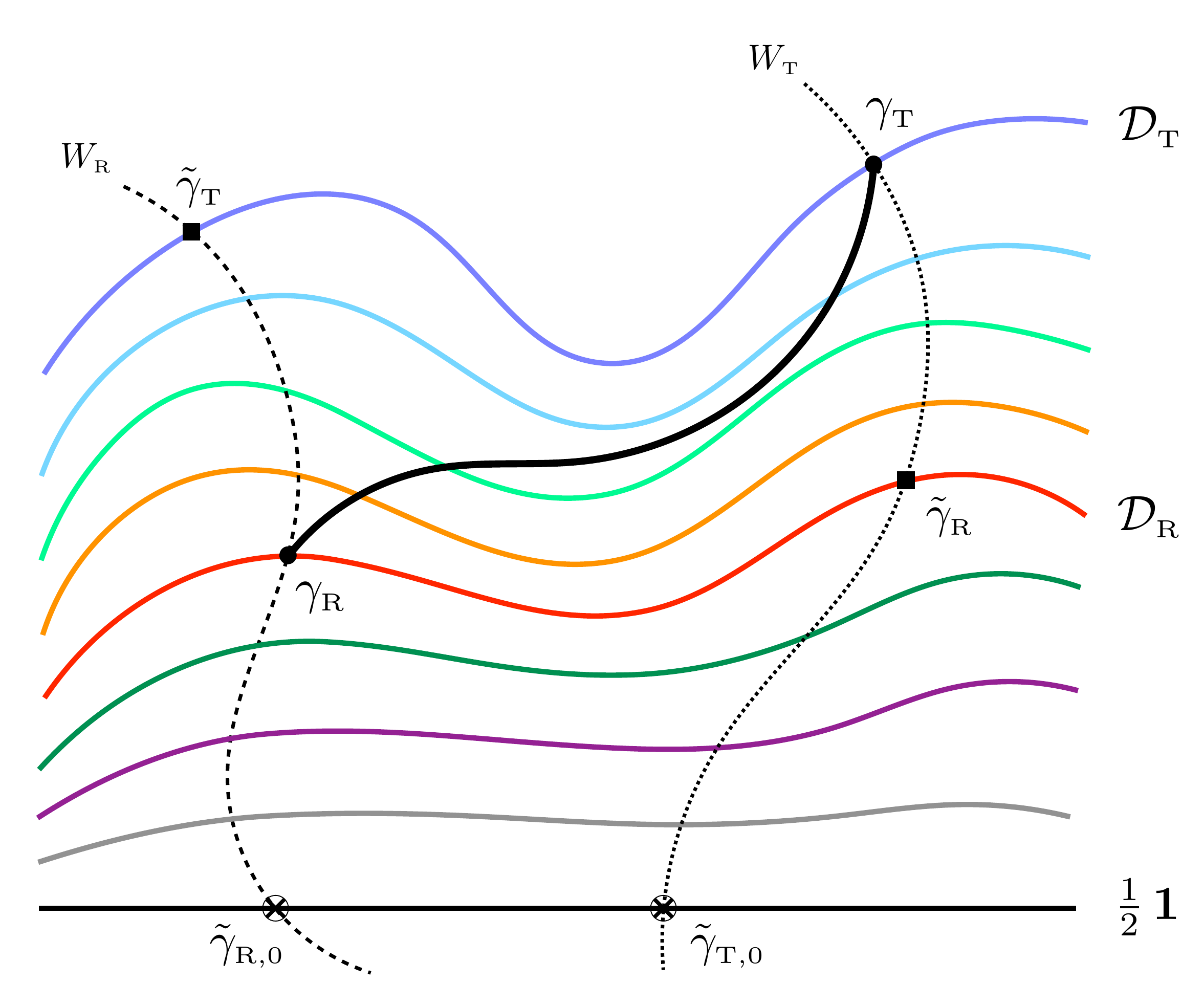}
 \end{center}
\vspace{-.0cm}
\caption{ 
The optimal purification paths for $\gamma_{\textrm{\tiny R}}$ and $\gamma_{\textrm{\tiny T}}$
correspond respectively to the $W_{\textrm{\tiny R}}$ path and to the $W_{\textrm{\tiny T}}$ path,
that are represented through dashed lines.
The straight black solid line represent the set of the pure states,
whose symplectic spectrum is given by $\mathcal{D} = \tfrac{1}{2} \boldsymbol{1}$.
}
\label{fig:different_D_pure}
\end{figure}

The {\it optimal purification path} is the 
purification path with minimal length,
which can be found by minimising (\ref{length purification path})
as $W_{\textrm{\tiny T}}\in \textrm{Sp}(2N, \mathbb{R})$ varies
within the symplectic group. 
This extremization procedure 
selects a symplectic matrix $W_0$
that determines the pure state through its covariance matrix
$\tfrac{1}{2} \,W^{\textrm{t}}_0 \, W_0$.
The matrix $W_0$ is obtained by solving $\delta d_0 = 0$,
where $d_0$ is defined in (\ref{length purification path}).
This is a special case 
of the analysis performed in Sec.\,\ref{sec:solving}
corresponding to 
$\delta \mathcal{D}_{\textrm{\tiny R}} = \delta \mathcal{D}_{\textrm{\tiny T}} 
= \delta W_{\textrm{\tiny R}} =\boldsymbol{0}$.

In Sec.\,\ref{sec:solving} we have shown that a $W$ path provides 
a solution to this equation, namely 
\be
\label{optim purif}
W_0 = W_{\textrm{\tiny R}}
\ee
which is the trivial solution corresponding to $\delta W_{\textrm{\tiny R}} =\boldsymbol{0} $.
In the following we focus on the purification process based on the $W$ paths. 
We cannot prove that, among all the solution of  $\delta d_0 = 0$ 
(see Sec.\,\ref{sec:solving}), the $W$ path corresponds to the one having minimal length.

The  $W_{\textrm{\tiny R}}$ path connects the mixed state
$\gamma_{\textrm{\tiny R}}= W^{\textrm{t}}_{\textrm{\tiny R}}\,\mathcal{D}\, W_{\textrm{\tiny R}}$ 
to the pure state $\gamma_0=  \tfrac{1}{2}\,W^{\textrm{t}}_{\textrm{\tiny R}} \, W_{\textrm{\tiny R}}$.
By specialising (\ref{W type path}) to $\mathcal{D}_{\textrm{\tiny T}} =\tfrac{1}{2}\,\boldsymbol{1} $,
we find that this  $W_{\textrm{\tiny R}}$ path is given by 
\be
\label{optimal purification path}
G_s(\gamma_{\textrm{\tiny R}} \, , \gamma_0)
= G^{(0)}_s(\gamma_{\textrm{\tiny R}} , W_{\textrm{\tiny R}})
=  \frac{1}{2^s}\;W_{\textrm{\tiny R}}^{\textrm{t}} \, \mathcal{D}^{1-s} \,  W_{\textrm{\tiny R}}
\ee
and its length can be easily obtained by setting 
$\mathcal{D}_{\textrm{\tiny T}} =\tfrac{1}{2}\,\boldsymbol{1} $ in (\ref{compl_same_W}),
finding an expression that depends only on $\mathcal{D}$
\be
\label{length-optimal-purification}
d_0(\gamma_{\textrm{\tiny R}}  , W_{\textrm{\tiny R}})
\equiv
d_{\textrm{\tiny spectrum}}(\gamma_{\textrm{\tiny R}} \, , \gamma_0 )
=
\sqrt{
\textrm{Tr} \Big\{ \big[\log\! \big( 2\,  \mathcal{D}\big)\big]^2  \Big\}}
=
\sqrt{2
\sum_{k=1}^N \big[ \log ( 2\,\sigma_{k} )\big]^2
} \,.
\ee

It is instructive to focus on the one-mode mixed states,
when the covariance matrices (\ref{purification gamma choice}) become
$\gamma_{\textrm{\tiny R}} =  \sigma \,W^{\textrm{t}}_{\textrm{\tiny R}}  \,W_{\textrm{\tiny R}}$
and $\gamma_{\textrm{\tiny T}} = \tfrac{1}{2} \, W^{\textrm{t}}_{\textrm{\tiny T}} \, W_{\textrm{\tiny T}}$.
Any purification path corresponding to a geodesic can be written as in (\ref{G_s decomposition one-mode})
with $\sigma_s=\frac{1}{2}\left(2 \sigma\right)^{1-s}$ and $ W_s$ defined in (\ref{Williamson D and W one mode}).
In particular, for the $W_{\textrm{\tiny R}}$ path we have $W_s = W_{\textrm{\tiny R}}$ and its length
is given by $d_0(\gamma_{\textrm{\tiny R}} , W_{\textrm{\tiny R}}) =\sqrt{2}\; |\log(2\sigma)\,|$.

The thermal states are interesting examples of mixed states to explore. 
The Williamson's decomposition of the covariance matrix of a thermal state is given by 
(\ref{Williamson_CM_th}). 
By specialising (\ref{optimal purification path}) to this case,
we obtain the $W$ path that purifies a thermal state.
It reads
\be
\label{optimal purification path thermal gen}
G_s(\gamma_{\textrm{\tiny th}}  \, , \gamma_0)
=  \frac{1}{2^s}\;W^{\textrm{t}}_{\textrm{\tiny th}}  \, \mathcal{D}_{\textrm{\tiny th}}^{1-s} \,  W_{\textrm{\tiny th}} 
\ee
where it is worth reminding that the symplectic matrix  $W_{\textrm{\tiny th}}$,
given in (\ref{sympspec_CM_th}), does not depend on the temperature of the thermal state,
but only on the parameters occurring in the hamiltonian.

It is natural to ask whether the $W_{\textrm{\tiny th}} $ path (\ref{optimal purification path thermal gen}) is made by thermal states. 
This is the case if, for any given $s\in [0,1]$, the symplectic spectrum of (\ref{optimal purification path thermal gen}) 
is thermal at some inverse temperature $\beta_s$ determined by the inverse temperature $\beta$ of the thermal state
that plays the role of the reference state in this purification path. 
Using (\ref{sympspec_CM_th}), this requirement leads to the following condition
\be
\label{thermalstate along Geod condition}
\big[  \coth(\beta\Omega_k/2)\,\big]^{1-s}
\! =\,
\coth(\beta_s\Omega_k/2)
\ee
which corresponds to (\ref{eq-beta-s-th}) when $\beta_\textrm{\tiny T} \to \infty$, as expected. 
This condition depends on the dispersion relation of the model. 
A straightforward numerical inspection for the periodic chains (see Sec.\,\ref{subsec-examples-hc})
shows that (\ref{thermalstate along Geod condition})  cannot be solved by $\beta_s = \beta_s(\beta)$ for any $1 \leqslant k \leqslant N$;
hence we conclude that the purification path (\ref{optimal purification path thermal gen}) 
is not entirely made by thermal states.

The $W$ paths provide a natural alternative way to connect two generic mixed states
$\gamma_{\textrm{\tiny R}}$ and $\gamma_{\textrm{\tiny T}}$ 
by using a path that passes through the set of pure states.
In particular, by exploiting
the Williamson's decompositions given in (\ref{gamma R and T williamson}),
one first considers
the $W_{\textrm{\tiny R}}$ path that connects 
$\gamma_{\textrm{\tiny R}}$ to the pure state $ \tilde{\gamma}_{\textrm{\tiny R,$0$}} $ 
and the $W_{\textrm{\tiny T}}$ path that connects 
$\gamma_{\textrm{\tiny T}}$ to the pure state $ \tilde{\gamma}_{\textrm{\tiny T,$0$}} $.
From (\ref{optimal purification path}), these $W$ paths are given respectively by 
\be
G_s(\gamma_{\textrm{\tiny R}} \, , \tilde{\gamma}_{\textrm{\tiny R,$0$}} )
=  \frac{1}{2^s}\;W_{\textrm{\tiny R}}^{\textrm{t}} \, \mathcal{D}_{\textrm{\tiny R}}^{1-s} \,  W_{\textrm{\tiny R}}
\;\;\qquad\;\;
G_s(\gamma_{\textrm{\tiny T}} \, , \tilde{\gamma}_{\textrm{\tiny T,$0$}} )
=  \frac{1}{2^s}\;W_{\textrm{\tiny T}}^{\textrm{t}} \, \mathcal{D}_{\textrm{\tiny T}}^{1-s} \,  W_{\textrm{\tiny T}}
\ee 
where 
\be
\label{pur-path-aux-gamma}
\tilde{\gamma}_{\textrm{\tiny R,$0$}} \equiv \frac{1}{2}\,W^{\textrm{t}}_{\textrm{\tiny R}} \, W_{\textrm{\tiny R}}
\;\;\qquad\;\;
\tilde{\gamma}_{\textrm{\tiny T,$0$}} \equiv  \frac{1}{2}\,W^{\textrm{t}}_{\textrm{\tiny T}} \, W_{\textrm{\tiny T}} \,.
\ee
Then, within the set of the pure states, we consider the geodesic connecting 
$\tilde{\gamma}_{\textrm{\tiny R,$0$}} $ to $\tilde{\gamma}_{\textrm{\tiny T,$0$}} $.
Our preferred path to connect 
$\gamma_{\textrm{\tiny R}}$ and $\gamma_{\textrm{\tiny T}}$ passing through the set of pure states
is obtained by combining these three paths as follows
\be
\label{purification complexity path}
\gamma_{\textrm{\tiny R}} 
\longrightarrow  \tilde{\gamma}_{\textrm{\tiny R,$0$}} 
\longrightarrow    \tilde{\gamma}_{\textrm{\tiny T,$0$}} 
\longrightarrow   \gamma_{\textrm{\tiny T}}
\ee

The length $d_{\textrm{\tiny pur}}(\gamma_{\textrm{\tiny R}} \, , \gamma_{\textrm{\tiny T}})$ of this path 
can be found by summing the lengths of its three components. 
From (\ref{compl_same_D}) and (\ref{length-optimal-purification}), we get
\bea
\label{purif-complexity-formal}
d_{\textrm{\tiny pur}}(\gamma_{\textrm{\tiny R}} \, , \gamma_{\textrm{\tiny T}})
&\equiv &
d_0(\gamma_{\textrm{\tiny R}}  , W_{\textrm{\tiny R}})
+
d_{\textrm{\tiny basis}}(\tilde{\gamma}_{\textrm{\tiny R,$0$}}\, , \tilde{\gamma}_{\textrm{\tiny T,$0$}})
+
d_0(\gamma_{\textrm{\tiny T}}  , W_{\textrm{\tiny T}})
\eea
which can be written more explicitly as follows
\be
\label{purif-complexity-explicit}
d_{\textrm{\tiny pur}}(\gamma_{\textrm{\tiny R}} \, , \gamma_{\textrm{\tiny T}})
\,=\,
\sqrt{
\textrm{Tr} \Big\{ \big[\log\! \big( 2\,\mathcal{D}_{\textrm{\tiny R}}  \big)\big]^2  \Big\}}
+
\sqrt{
\textrm{Tr} \Big\{ \big[\log\! \big( 2\,\mathcal{D}_{\textrm{\tiny T}}  \big)\big]^2  \Big\}}
+
\sqrt{
\textrm{Tr} \Big\{ \big[\log\! \big( 
W_{\textrm{\tiny TR}} 
\,W_{\textrm{\tiny TR}}^{\textrm{t}}
\big)\big]^2  \Big\}}
\,.
\ee
This expression provides an upper bound 
$d(\gamma_{\textrm{\tiny R}} \, , \gamma_{\textrm{\tiny T}})
\leqslant d_{\textrm{\tiny pur}}(\gamma_{\textrm{\tiny R}} \, , \gamma_{\textrm{\tiny T}})$ 
on (\ref{deltaAB}).

\section{Bounding complexity}
\label{sec:bounds}

Explicit formulas to evaluate the circuit complexity for mixed states are difficult to obtain;
hence it is worth finding calculable expressions which provide either higher or lower bounds to this quantity.
In this section we construct some bounds in the setup of the bosonic Gaussian states that we are exploring. 
In Sec.\;\ref{sec-bound-vanish} we focus on the states with vanishing first moments, 
while in Sec.\;\ref{sec-bound-non-vanish} the most general case of states with non vanishing first moments is considered.

\subsection{States with vanishing first moments}
\label{sec-bound-vanish}

The complexity (\ref{c2 complexity}), 
which holds for states with vanishing first moments,
is proportional to 
the length of the optimal circuit (\ref{optimal circuit}) connecting 
$\gamma_{\textrm{\tiny R}}$ to $\gamma_{\textrm{\tiny T}}$;
hence it is straightforward to observe that the length of any other path 
connecting these two covariance matrices provides an upper bound
on the complexity. 
The analysis reported in Sec.\,\ref{sec:spectrum-basis} and in Sec.\,\ref{sec:purification} 
naturally leads to consider some particular paths.

The simplest choice is a path made by two geodesics that connect 
$\gamma_{\textrm{\tiny R}}$ and $\gamma_{\textrm{\tiny T}}$
to an auxiliary covariance matrix $\gamma_{\textrm{\tiny aux}}$
that does not belong to the optimal circuit (\ref{optimal circuit})
(i.e. that does not lie on the black solid curve in Fig.\,\ref{fig:different_D_sp}).
Natural candidates for $\gamma_{\textrm{\tiny aux}}$ are the 
covariance matrices whose Williamson's decompositions contain either 
$\mathcal{D}_{\textrm{\tiny R}}$ or $\mathcal{D}_{\textrm{\tiny T}}$ 
or $W_{\textrm{\tiny R}}$ or $W_{\textrm{\tiny T}}$.
For instance, we can choose for $\gamma_{\textrm{\tiny aux}}$
a covariance matrix whose symplectic spectrum is $\mathcal{D}_{\textrm{\tiny R}}$
or a covariance matrix whose symplectic spectrum is $\mathcal{D}_{\textrm{\tiny T}}$
(that lie respectively on the red solid curve and on the blue solid curve  in Fig.\,\ref{fig:different_D_sp}).
Different choices for $\gamma_{\textrm{\tiny aux}}$ lead to different bounds; 
hence it is worth asking whether a particular choice provides the best bound. 
The answer depends on the set where $\gamma_{\textrm{\tiny aux}}$ is allowed to vary.

Let us consider some natural paths where only a single auxiliary covariance matrix 
$\gamma_{\textrm{\tiny aux}}$ is involved. 
In Sec.\,\ref{sec:solving} we have shown that the $W$ paths satisfy the first law of complexity 
with $\delta d=0$. 
Thus, natural candidates to consider for $\gamma_{\textrm{\tiny aux}}$ are
\be
\label{tilded gammas}
\tilde{\gamma}_{\textrm{\tiny R}} \equiv  W^{\textrm{t}}_{\textrm{\tiny T}}  \, \mathcal{D}_{\textrm{\tiny R}} W_{\textrm{\tiny T}} 
\;\;\qquad \;\;
\tilde{\gamma}_{\textrm{\tiny T}} \equiv  W^{\textrm{t}}_{\textrm{\tiny R}}  \, \mathcal{D}_{\textrm{\tiny T}} W_{\textrm{\tiny R}} 
\ee
that have been represented by black squares in Fig.\,\ref{fig:different_D_sp} and Fig.\,\ref{fig:different_D_pure}.

By first applying the triangle inequality for the paths
$\gamma_{\textrm{\tiny R}} \to \tilde{\gamma}_{\textrm{\tiny R}}  \to \gamma_{\textrm{\tiny T}}$ 
and $\gamma_{\textrm{\tiny R}} \to \tilde{\gamma}_{\textrm{\tiny T}}  \to \gamma_{\textrm{\tiny T}}$,
and then picking the one that provides the best constraint between the $\mathcal{D}$ paths, 
we obtain
\be
d(\gamma_{\textrm{\tiny R}} \, , \gamma_{\textrm{\tiny T}})
\,\leqslant\,
d_{\textrm{\tiny spectrum}}(\gamma_{\textrm{\tiny R}} \, , \gamma_{\textrm{\tiny T}})
+
\textrm{min}
\big[\,
d_{\textrm{\tiny basis}}(\gamma_{\textrm{\tiny R}} \, , \tilde{\gamma}_{\textrm{\tiny R}}) 
\, ,\,
d_{\textrm{\tiny basis}}( \tilde{\gamma}_{\textrm{\tiny T}} \, , \gamma_{\textrm{\tiny T}}) 
\,\big]\,.
\ee
Denoting by $\tilde{d}(\gamma_{\textrm{\tiny R}} \, , \gamma_{\textrm{\tiny T}})$
the r.h.s. of this inequality,
by using  (\ref{compl_same_W}) and (\ref{compl_same_D})
we find that
\bea
\label{d_tilde_RT def}
\tilde{d}(\gamma_{\textrm{\tiny R}} \, , \gamma_{\textrm{\tiny T}})
& = &
\sqrt{
\textrm{Tr} \Big\{ \big[\log\! \big( \mathcal{D}_{\textrm{\tiny T}} \,  \mathcal{D}_{\textrm{\tiny R}}^{-1} \big)\big]^2  \Big\}}\;
\\
& &
+\;
\textrm{min} \Bigg[\,
\sqrt{
\textrm{Tr} \Big\{ \big[\log\! \big( 
 \mathcal{D}_{\textrm{\tiny R}}\,
W_{\textrm{\tiny TR}} 
\, \mathcal{D}_{\textrm{\tiny R}}^{-1}\,
W_{\textrm{\tiny TR}}^{\textrm{t}}
\big)\big]^2  \Big\}}
\,,\,
\sqrt{
\textrm{Tr} \Big\{ \big[\log\! \big( 
 \mathcal{D}_{\textrm{\tiny T}}\,
W_{\textrm{\tiny TR}} 
\, \mathcal{D}_{\textrm{\tiny T}}^{-1}\,
W_{\textrm{\tiny TR}}^{\textrm{t}}
\big)\big]^2  \Big\}}
\; \Bigg]\,.
\nonumber
\eea
The path $\gamma_{\textrm{\tiny R}} \to
\tilde{\gamma}_{\textrm{\tiny T}} \to
\gamma_{\textrm{\tiny T}}$
corresponds to an explicit realisation
of the proposal made in Fig.\,6 of \cite{Agon:2018zso}
within the approach that we are considering, 
that does not require the addition of ancillary degrees of freedom.

Better bounds could be obtained by
constructing paths 
that involve more 
auxiliary covariance matrices $\gamma_{\textrm{\tiny aux}}$.
For instance, one can consider paths
$\gamma_{\textrm{\tiny R}} \to
\gamma_{\textrm{\tiny aux},1} \to 
\gamma_{\textrm{\tiny aux},2}\to
\gamma_{\textrm{\tiny T}}$
that involve two auxiliary covariance matrices
$\gamma_{\textrm{\tiny aux},1}$ and $\gamma_{\textrm{\tiny aux},2}$.
Referring to Fig.\,\ref{fig:different_D_sp},
natural paths to consider within this class
are the ones where 
$\gamma_{\textrm{\tiny aux},1}$ belongs to the $W_{\textrm{\tiny R}}$ path 
and $\gamma_{\textrm{\tiny aux},2}$ to the $W_{\textrm{\tiny T}}$ path,
or the ones where
$\gamma_{\textrm{\tiny aux},1}$ belongs to the red curve
(i.e. its symplectic spectrum is $\mathcal{D}_{\textrm{\tiny R}}$)
and $\gamma_{\textrm{\tiny aux},2}$ belongs to the blue curve
(i.e. its symplectic spectrum is $\mathcal{D}_{\textrm{\tiny T}}$).

Another interesting path to consider is the one constructed in (\ref{purification complexity path}):
it involves the two auxiliary matrices 
$\gamma_{\textrm{\tiny aux},1}=\tilde{\gamma}_{\textrm{\tiny R,$0$}} $
and $\gamma_{\textrm{\tiny aux},2}=\tilde{\gamma}_{\textrm{\tiny T,$0$}} $
and its length is (\ref{purif-complexity-formal}) 
(see Fig.\,\ref{fig:different_D_pure}).
It is straightforward to observe that
$d(\gamma_{\textrm{\tiny R}} \, , \gamma_{\textrm{\tiny T}})
\leqslant d_{\textrm{\tiny pur}}(\gamma_{\textrm{\tiny R}} \, , \gamma_{\textrm{\tiny T}})$,
but it is non trivial to find the best bound between 
$\tilde{d}(\gamma_{\textrm{\tiny R}} \, , \gamma_{\textrm{\tiny T}})$ and 
$d_{\textrm{\tiny pur}}(\gamma_{\textrm{\tiny R}} \, , \gamma_{\textrm{\tiny T}})$.
Since we cannot provide a general solution to this problem, 
in the following we focus on simple special cases where we can show that
$ d(\gamma_{\textrm{\tiny R}} \, , \gamma_{\textrm{\tiny T}})
\leqslant
\tilde{d}(\gamma_{\textrm{\tiny R}} \, , \gamma_{\textrm{\tiny T}})
\leqslant
d_{\textrm{\tiny pur}}(\gamma_{\textrm{\tiny R}} \, , \gamma_{\textrm{\tiny T}})$.

When $ \gamma_{\textrm{\tiny T}}$ is pure, from (\ref{d_tilde_RT def}) it is straightforward to observe that
$\tilde{d}(\gamma_{\textrm{\tiny R}} \, , \gamma_{\textrm{\tiny T}}) \leqslant 
d_{\textrm{\tiny pur}}(\gamma_{\textrm{\tiny R}} \, , \gamma_{\textrm{\tiny T}})$.
Another class of special cases that we find interesting to consider is given by the pairs $(\gamma_{\textrm{\tiny R}} ,\gamma_{\textrm{\tiny T}} )$
such that all the matrices along the $\mathcal{D}_{\textrm{\tiny R}}$ path connecting 
$\gamma_{\textrm{\tiny R}} $ to $\tilde{\gamma}_{\textrm{\tiny R}}$ have the same symplectic spectrum
$\mathcal{D}_{\textrm{\tiny R}}$ and, similarly, 
all the matrices along the $\mathcal{D}_{\textrm{\tiny T}}$ path connecting 
$\gamma_{\textrm{\tiny T}} $ to $\tilde{\gamma}_{\textrm{\tiny T}}$ have the same symplectic spectrum
$\mathcal{D}_{\textrm{\tiny T}}$. This means that (\ref{sympl_cond_Us}) holds for 
both $\mathcal{D}=\mathcal{D}_{\textrm{\tiny R}}$ and $\mathcal{D}=\mathcal{D}_{\textrm{\tiny T}}$;
hence (\ref{d_tilde_RT def}) simplifies to 
\be
\label{d_tilde_RT def_particularcase}
\tilde{d}(\gamma_{\textrm{\tiny R}} \, , \gamma_{\textrm{\tiny T}})
 = 
\sqrt{
\textrm{Tr} \Big\{ \big[\log\! \big( \mathcal{D}_{\textrm{\tiny T}} \,  \mathcal{D}_{\textrm{\tiny R}}^{-1} \big)\big]^2  \Big\}}
+\sqrt{
\textrm{Tr} \,\Big\{ \big[\log\! \big( 
W_{\textrm{\tiny TR}} 
\,W_{\textrm{\tiny TR}}^{\textrm{t}}
\big)\big]^2  \Big\}}\,.
\ee
The first square root in the r.h.s. can be bounded as follows
\bea
\label{ineq-dtilde-D}
\sqrt{
\textrm{Tr} \Big\{ \big[\log\! \big( \mathcal{D}_{\textrm{\tiny T}} \,  \mathcal{D}_{\textrm{\tiny R}}^{-1} \big)\big]^2  \Big\}}
&=&
\sqrt{
\textrm{Tr} \Big\{ \big[\log\! \big( 2\mathcal{D}_{\textrm{\tiny T}} \big)-\log\! \big( 2\mathcal{D}_{\textrm{\tiny R}}\big)\big]^2  \Big\}}
\nonumber
\\
\rule{0pt}{.7cm}
& \leqslant &
\sqrt{
\textrm{Tr} \Big\{ \big[\log\! \big( 2\,\mathcal{D}_{\textrm{\tiny R}}  \big)\big]^2  \Big\}}
+
\sqrt{
\textrm{Tr} \Big\{ \big[\log\! \big( 2\,\mathcal{D}_{\textrm{\tiny T}}  \big)\big]^2  \Big\}}
\eea
where we have employed first that all the elements of $2\mathcal{D}$ are larger than or equal to $1$ 
(in order to discard a positive term under the square root)
and then the inequality $\sqrt{a+b} \leqslant \sqrt{a} +\sqrt{b}$, that holds for any $a$ and $b$.
By employing (\ref{ineq-dtilde-D}) in (\ref{d_tilde_RT def_particularcase})
and comparing the result against (\ref{purif-complexity-explicit}),
we can conclude that 
$\tilde{d}(\gamma_{\textrm{\tiny R}} \, , \gamma_{\textrm{\tiny T}}) \leqslant 
d_{\textrm{\tiny pur}}(\gamma_{\textrm{\tiny R}} \, , \gamma_{\textrm{\tiny T}})$.

\subsection{States with non vanishing first moments}
\label{sec-bound-non-vanish}

In the most general case where the first moments are non vanishing
 $\langle\hat{\boldsymbol{r}}\rangle\equiv \boldsymbol{a} \neq \boldsymbol{0}$,
a closed expression for the Fisher-Rao distance is not known,
as also remarked in the Appendix\;\ref{app:first-moments},
where the notation $\boldsymbol{\mu}= \boldsymbol{a}$ and $\Sigma=\gamma$ has been adopted.
Nonetheless, lower and upper bounds on the complexity can be written 
by employing some known results  about the Gaussian PDF's
\cite{CalvoOller90,Strapasson15,Strapasson16,
PineleCosta19,PineleCosta20}.

Given a reference state and a target state, 
that can be parameterised by 
$\boldsymbol{\theta}_{\textrm{\tiny R}}=(\boldsymbol{a}_{\textrm{\tiny R}},\gamma_{\textrm{\tiny R}}) $ 
and $\boldsymbol{\theta}_{\textrm{\tiny T}}=(\boldsymbol{a}_{\textrm{\tiny T}},\gamma_{\textrm{\tiny T}}) $
respectively, 
let us introduce the following $(2N+1)\times(2N+1)$ matrix
\be
\label{aux Gamma lowerbound}
\Gamma_{\textrm{\tiny S}}
=
\bigg( \,\begin{array}{cc}
 \gamma_{\textrm{\tiny S}}
 +  \boldsymbol{a}_{\textrm{\tiny S}}\, \boldsymbol{a}_{{\textrm{\tiny S}}}^{\textrm{t}} & \quad \boldsymbol{a}_{\textrm{\tiny S}}
\\
 \boldsymbol{a}_{\textrm{\tiny S}}^{\textrm{t}} &\quad 1
\end{array}\bigg)
\;\;\qquad\;\;
\textrm{S}\in \big\{\textrm{R},\textrm{T}\big\}\,.
\ee
A lower bound  for the Fisher-Rao distance, first obtained in \cite{CalvoOller90}, is given by 
\be
\label{delta-lower-bound}
d_{\textrm{\tiny lower}}(\boldsymbol{\theta}_{\textrm{\tiny R}},\boldsymbol{\theta}_{\textrm{\tiny T}})
\equiv
\sqrt{
\textrm{Tr} \big[ (\log \Delta_{\Gamma,\textrm{\tiny TR}})^2 \big] 
}
=
\big|\!\big|  \log\! 
\Big(\Gamma_{\textrm{\tiny R}}^{-1/2} \, \Gamma_{\textrm{\tiny T}} \, \Gamma_{\textrm{\tiny R}}^{-1/2}\, \Big)  
\big|\!\big|_2
=
\Bigg[ \, \sum_{i=1}^{2N+1}\!\!\big(\log (\tilde{\lambda}_i)\big)^2\,\Bigg]^{1/2}
\ee
where
$\Delta_{\Gamma,\textrm{\tiny TR}} \equiv \Gamma_{\textrm{\tiny T}} \,\Gamma_{\textrm{\tiny R}}^{-1} $
and 
$\tilde{\lambda}_i$  are the eigenvalues of 
$\Gamma_{\textrm{\tiny R}}^{-1/2}\,\Gamma_{\textrm{\tiny T}}\,\Gamma_{\textrm{\tiny R}}^{-1/2}$.

Upper bounds for the Fisher-Rao distance have been also found for non vanishing first moments
\cite{Strapasson15,Strapasson16, PineleCosta19,PineleCosta20}.
An upper bound 
can be written through $d^{(1)}_{\textrm{\tiny FR}} $ 
defined in (\ref{univariateGaussian_FR}) as follows \cite{Strapasson15}
\be
\label{d_upper_1}
d_{\textrm{\tiny upper},1}(\boldsymbol{\theta}_{\textrm{\tiny R}},\boldsymbol{\theta}_{\textrm{\tiny T}})
\equiv
\bigg[\,
\sum_{i=1}^{2N} d^{(1)}_{\textrm{\tiny FR}}
\big((0,1),(\tilde{a}_{i}, \lambda_i)\big)^2 
\,\bigg]^{1/2}
\ee
where $\lambda_i$ is the $i$-th eigenvalue of 
$\gamma_{\textrm{\tiny R}}^{-1/2}\gamma_{\textrm{\tiny T}}\,\gamma_{\textrm{\tiny R}}^{-1/2}$,
$\tilde{a}_i$ is the $i$-th component of 
$\tilde{\boldsymbol{a}}_{\textrm{\tiny T,R}}\equiv \widetilde{O}^{\textrm{t}} \,\boldsymbol{a}_{\textrm{\tiny T,R}} $, 
\textcolor{red}{being}
$\boldsymbol{a}_{\textrm{\tiny T,R}}
\equiv \gamma_{\textrm{\tiny R}}^{-1/2} (\boldsymbol{a}_{\textrm{\tiny T}}-\boldsymbol{a}_{\textrm{\tiny R}}) $\textcolor{red}{,}
and $\widetilde{O}$ is the orthogonal matrix whose columns are the eigenvectors of  
$\gamma_{\textrm{\tiny R}}^{-1/2}\gamma_{\textrm{\tiny T}}\,\gamma_{\textrm{\tiny R}}^{-1/2}$.

Another upper bound has been found in \cite{Strapasson16}.
It has been written by introducing 
the $2N\times 2N$ orthogonal matrix $O$ such that 
$O \,\boldsymbol{a}_{\textrm{\tiny T,R}}=(|\boldsymbol{a}_{\textrm{\tiny T,R}}|,0,\dots,0) $
and the following $2N\times 2N$ matrices
\be
D_{\textrm{\tiny T,R}}\equiv\textrm{diag}\big(\sqrt{(|\boldsymbol{a}_{\textrm{\tiny T,R}}|+2)/2}\,,1,\dots,1\big)
\;\;\qquad\;\;
\gamma_{\textrm{\tiny T,R}} \equiv O^{-1} D_{\textrm{\tiny T,R}}\, O^{-\textrm{t}}\,.
\ee
These matrices are employed to identify the states corresponding to the following vectors
\be
\label{vectors upperbound 2}
\boldsymbol{\theta}_0 \equiv (0,\boldsymbol{1})
\qquad
\boldsymbol{\theta}_{O}\equiv (O \,\boldsymbol{a}_{\textrm{\tiny T,R}} \,,D_{\textrm{\tiny T,R}})
\qquad
\boldsymbol{\theta}_{*}\equiv 
\big(\boldsymbol{a}_{\textrm{\tiny T,R}} \,,
\gamma_{\textrm{\tiny R}}^{-1/2}\gamma_{\textrm{\tiny T}}\,\gamma_{\textrm{\tiny R}}^{-1/2}\,\big)
\qquad
\boldsymbol{\theta}_\gamma\equiv (\boldsymbol{a}_{\textrm{\tiny T,R}}\,,\gamma_{\textrm{\tiny T,R}})\,.
\ee 
The upper bound reads
\be
\label{d_upper_2}
d_{\textrm{\tiny upper},2}(\boldsymbol{\theta}_{\textrm{\tiny R}},\boldsymbol{\theta}_{\textrm{\tiny T}})
\equiv
d_{\textrm{\tiny diag}}(\boldsymbol{\theta}_0,\boldsymbol{\theta}_{O})
+
d_{\boldsymbol{a}_{\textrm{\tiny T,R}}}(\boldsymbol{\theta}_{*},\boldsymbol{\theta}_\gamma)
\ee
where $d_{\boldsymbol{a}_{\textrm{\tiny T,R}}}$ is defined in (\ref{FR-distance-mu}) 
and $d_{\textrm{\tiny diag} }$ in (\ref{d_diag-def}).
Since an inequality between the two upper bounds in (\ref{d_upper_1}) and (\ref{d_upper_2})
cannot be found for any value of $\boldsymbol{\theta}_{\textrm{\tiny R}} $ and $\boldsymbol{\theta}_{\textrm{\tiny T}}$
\cite{PineleCosta20}, we pick the minimum between them.

Combining the above results, one obtains 
\be
\label{pc-bounds}
d_{\textrm{\tiny lower}}(\boldsymbol{\theta}_{\textrm{\tiny R}},\boldsymbol{\theta}_{\textrm{\tiny T}})
\leqslant 
d(\boldsymbol{\theta}_{\textrm{\tiny R}},\boldsymbol{\theta}_{\textrm{\tiny T}}) 
\leqslant
\min\!\big[
d_{\textrm{\tiny upper},1}(\boldsymbol{\theta}_{\textrm{\tiny R}} ,\boldsymbol{\theta}_{\textrm{\tiny T}}),
d_{\textrm{\tiny upper},2}(\boldsymbol{\theta}_{\textrm{\tiny R}} ,\boldsymbol{\theta}_{\textrm{\tiny T}})\big]
\ee

In order to provide a consistency check for these bounds,
let us consider the case $\boldsymbol{a}_{\textrm{\tiny R}}=\boldsymbol{a}_{\textrm{\tiny T}}=\boldsymbol{a}$. 
From (\ref{aux Gamma lowerbound}) we obtain
\be
\label{gamma-gamma-TR}
\Gamma_{\textrm{\tiny T}} \,\Gamma_{\textrm{\tiny R}}^{-1}=
\bigg( \,\begin{array}{cc}
 \gamma_{\textrm{\tiny T}}\, \gamma_{\textrm{\tiny R}}^{-1}
 &\;\;
 \big(\boldsymbol{1}-\gamma_{\textrm{\tiny T}} \,\gamma_{\textrm{\tiny R}}^{-1}\big)\boldsymbol{a}
\\
\boldsymbol{0} &\;\; 1
\end{array}\bigg)\,.
\ee
By employing a formula for the determinant of a block matrix reported below (see (\ref{det-block-mat})), 
one finds that the  first $2N$ eigenvalues of (\ref{gamma-gamma-TR})
are the eigenvalues of $\gamma_{\textrm{\tiny T}} \gamma_{\textrm{\tiny R}}^{-1}$, 
while the last eigenvalue is equal to $1$.
Thus, $d_{\textrm{\tiny lower}}$ in (\ref{delta-lower-bound}) becomes (\ref{deltaAB})  in this case, saturating the lower bound.
As for the upper bound in (\ref{pc-bounds}), we have (\ref{vectors upperbound 2}) 
simplify to $\boldsymbol{\theta}_0=\boldsymbol{\theta}_O =\boldsymbol{\theta}_\gamma= (0,\boldsymbol{1}) $
in this case.
This implies that $d_{\textrm{\tiny upper},2}$ in (\ref{d_upper_2}) becomes (\ref{deltaAB}); 
hence also the upper bound is saturated.

\section{Optimal path for entanglement hamiltonians}
\label{sec:comp-eh}

The density matrix of a mixed state can be written as follows
\be
\hat{\rho}\,\propto\, e^{-\widehat{K}}
\ee
where the proportionality constant determines the normalisation of $\hat{\rho}$.
We  denote the operator $\widehat{K}$  as {\it entanglement hamiltonian},
with a slight abuse of notation. 
Indeed, the operator $\widehat{K}$ is the entanglement hamiltonian
when $\hat{\rho}= \hat{\rho}_A = \textrm{Tr}_{\mathcal{H}_B} \hat{\rho}_0$ 
is the reduced density matrix obtained by tracing out the part $\mathcal{H}_B$
of a bipartite Hilbert space $\mathcal{H} = \mathcal{H}_A \otimes \mathcal{H}_B$.
Instead, for instance, the thermal states are mixed states
that do not correspond to a bipartition of the Hilbert space.
For these states $\widehat{K} = \beta \,\widehat{H}$,
where $\widehat{H}$ is the hamiltonian of the system 
and $\beta$ the inverse temperature.

The entanglement hamiltonians associated to some particular reduced density matrices 
have been largely studied for simple models, both in quantum field theories
\cite{Bisognano:1975ih,  Casini:2009sr, Casini:2011kv, Wong:2013gua, Cardy:2016fqc,  
Arias:2016nip, Arias:2017dda,Tonni:2017jom, Roy:2020frd}
and on the lattice
\cite{Peschel_2009, PeschelChung00,Peschel03, Peschel_2004,BanchiPirandola-15,
Eisler:2017cqi, Eisler:2019rnr, DiGiulio:2019lpb,DiGiulio:2019cxv}.
The spectrum of the entanglement hamiltonian,
that is usually called entanglement spectrum \cite{Li:2008kda},
is rich in information. 
For instance, in conformal field theories the entanglement spectrum provides
both the central charge \cite{PhysRevA.78.032329}
and the conformal spectrum of the underlying model
\cite{Lauchli:2013jga,Cardy:2016fqc,Tonni:2017jom,Alba:2017bgn,
Surace:2019mft,DiGiulio:2019cxv,DiGiulio:2019lpb,Roy:2020frd}

For the bosonic Gaussian states that we are considering, the entanglement hamiltonians are
quadratic operators in terms of the position and momentum operators; 
hence they can be written as follows
\be
\label{ent-ham HC}
\widehat{K}
= \frac{1}{2}\, \hat{\boldsymbol{r}}^{\textrm t} H\, \hat{\boldsymbol{r}}
\hspace{.5cm} \qquad \hspace{.5cm} 
\hat{\boldsymbol{r}} = 
\bigg( \hspace{-.0cm} 
\begin{array}{c}
\hat{\boldsymbol{q}} \\  \hat{\boldsymbol{p}}
\end{array} \hspace{-.0cm}  \bigg)
\ee
where $H$ is a $2N\times 2N$ symmetric and positive definite matrix
that characterises the underlying mixed state. 
We denote $H$ as the entanglement hamiltonian matrix.
It can be written in terms of the corresponding covariance matrix $\gamma$
as follows
\cite{PeschelChung00, Peschel03, Peschel_2004,BanchiPirandola-15, DiGiulio:2019cxv, Casini:2009sr}
\be
\label{EHfromgamma}
H
=
 2\,\textrm{i} \, J  \,  \textrm{arccoth}(2\,\textrm{i}\gamma\, J) 
 \equiv 
 h(\gamma)
\ee
where $J$ is the standard symplectic matrix (\ref{Jmat}). 
The expression (\ref{EHfromgamma}) holds for covariance matrices that are not associated to pure states.
Thus, in particular, the purification procedure reported in Sec.\,\ref{sec:purification} cannot be described 
through the entanglement hamiltonian matrices $H$  defined by (\ref{ent-ham HC}).

Since the matrix $H$ is symmetric and positive definite, 
we can adapt to the entanglement hamiltonian matrices 
many results reported in the previous discussions for the covariance matrices.

Given the matrices $H_{\textrm{\tiny R}}$ and $H_{\textrm{\tiny T}}$
corresponding to the reference state $\gamma_{\textrm{\tiny R}}$ and 
to the target state $\gamma_{\textrm{\tiny T}}$ respectively, 
we can consider the optimal path connecting $H_{\textrm{\tiny R}}$ to $H_{\textrm{\tiny T}}$,
namely
\be
\label{optimal circuit H}
\widetilde{G}_s(H_{\textrm{\tiny R}} \, , H_{\textrm{\tiny T}})
\,\equiv \,
H_{\textrm{\tiny R}}^{1/2} 
\Big(  H_{\textrm{\tiny R}}^{- 1/2}  \,H_{\textrm{\tiny T}} \,H_{\textrm{\tiny R}}^{-1/2}  \Big)^s
H_{\textrm{\tiny R}}^{1/2} 
\;\; \qquad \;\;
0 \leqslant s \leqslant 1
\ee
whose boundary conditions are given by
$\widetilde{G}_0(H_{\textrm{\tiny R}} \, , H_{\textrm{\tiny T}}) = H_{\textrm{\tiny R}}$ 
and $\widetilde{G}_1(H_{\textrm{\tiny R}} \, , H_{\textrm{\tiny T}}) = H_{\textrm{\tiny T}}$.
The length of the geodesic (\ref{optimal circuit H}) measured through the 
Fisher-Rao metric reads
\be
\label{H-distance}
d(H_{\textrm{\tiny R}}, H_{\textrm{\tiny T}})
\equiv
\sqrt{\,
\textrm{Tr} \,\Big\{ \big[ \log\! \big(H_{\textrm{\tiny T}} \,H_{\textrm{\tiny R}}^{-1} \big) \big]^2  \Big\}
}\,.
\ee

The Williamson's decomposition of the entanglement hamiltonian matrix $H$ is given by
\be
\label{williamson-dec-H}
H =  \widetilde{W}^{\textrm{t}}  \, \mathcal{E}\, \widetilde{W}
\ee
where $\mathcal{E} \equiv \textrm{diag} (\varepsilon_1 , \dots , \varepsilon_N)  \oplus \textrm{diag} (\varepsilon_1 , \dots , \varepsilon_N)$ with $\varepsilon_k > 0$.
The symplectic spectrum  of $H$ can be determined from the symplectic spectrum of $\gamma$ as follows
\cite{PeschelChung00, Peschel03, Peschel_2004}
\be
\label{Ediag_Ddiag}
\mathcal{E} 
=
2 \,\textrm{arccoth}(2\,\mathcal{D})
=
\log\!\bigg( \frac{\mathcal{D} +1/2}{\mathcal{D} - 1/2} \bigg) \,.
\ee
This formula cannot be applied for pure states, which have $\mathcal{D} =\tfrac{1}{2}\boldsymbol{1}$.
The symplectic matrices $W$ and $\widetilde{W}$,
introduced in (\ref{williamson th gammaA}) and (\ref{williamson-dec-H}) respectively, 
are related as follows \cite{BanchiPirandola-15, DiGiulio:2019cxv}
\be
\label{W widetildeW relation}
\widetilde{W} \equiv J^{\textrm t}  \,W J
= 
W^{-\textrm t}\,.
\ee

We find it worth expressing the distance (\ref{H-distance}) 
 in terms of the matrices occurring in the Williamson's decompositions
 of $H_{\textrm{\tiny R}}$ and  $H_{\textrm{\tiny T}}$,
 as done in Sec.\,\ref{subsec:mixed} for the covariance matrices. 
 These decompositions read 
 \be
\label{EH R and T williamson}
H_{\textrm{\tiny R}} =  \widetilde{W}^{\textrm{t}}_{\textrm{\tiny R}}  \, \mathcal{E}_{\textrm{\tiny R}}\, \widetilde{W}_{\textrm{\tiny R}} 
\;\;\qquad \;\;
H_{\textrm{\tiny T}} =  \widetilde{W}^{\textrm{t}}_{\textrm{\tiny T}}  \, \mathcal{E}_{\textrm{\tiny T}}\, \widetilde{W}_{\textrm{\tiny T}} 
\ee
where 
$\widetilde{W}_{\textrm{\tiny R}} $ and $\widetilde{W}_{\textrm{\tiny T}} $ 
are related respectively to 
$W_{\textrm{\tiny R}} $ and $W_{\textrm{\tiny T}} $ 
through (\ref{W widetildeW relation}).
By using (\ref{C-2-complexity}) and the following relation
\be
\label{W widetildeW relation TR}
\widetilde{W}_{\textrm{\tiny TR}} 
\equiv 
\widetilde{W}_{\textrm{\tiny T}} \,\widetilde{W}_{\textrm{\tiny R}}^{-1}
= J^{\textrm t}  \,W_{\textrm{\tiny TR}} \,J
\ee
we can write the distance (\ref{H-distance}) as
\be
\label{C-2-complexity-eh}
d(H_{\textrm{\tiny R}}, H_{\textrm{\tiny T}})
=
\sqrt{\,
\textrm{Tr} \,\Big\{ \big[\log \big(
\mathcal{E}_{\textrm{\tiny T}}\, W_{\textrm{\tiny TR}}\,\mathcal{E}_{\textrm{\tiny R}}^{-1}\, W^{\textrm{t}}_{\textrm{\tiny TR}}
\big)\big]^2  \Big\}
}\,.
\ee

The expression (\ref{EHfromgamma}) (or equivalently (\ref{Ediag_Ddiag}) and (\ref{W widetildeW relation})) 
provides a highly non trivial relation between the set made by the covariance matrices $\gamma$ 
that are associated to the mixed states that are not pure states
and the set of the entanglement hamiltonian matrices $H$.
The map $h$ in (\ref{EHfromgamma}) is not an isometry, 
hence the distances are not preserved
and geodesics are not sent into geodesics.
Thus, we find it worth comparing the distance 
$d(\gamma_{\textrm{\tiny R}}, \gamma_{\textrm{\tiny T}}) = d(h^{-1}(H_{\textrm{\tiny R}}), h^{-1}(H_{\textrm{\tiny T}}))$ 
from (\ref{C-2-complexity}) 
and the distance $d(H_{\textrm{\tiny R}}, H_{\textrm{\tiny T}})$ in (\ref{C-2-complexity-eh}).

For the sake of simplicity, let us explore the case of one-mode mixed states,
where $\mathcal{D} =\sigma \,\boldsymbol{1}$ 
and $\mathcal{E} =\varepsilon \,\boldsymbol{1}$ are proportional to the $2\times 2$ identity matrix. 
In this simple case the expressions for $d(\gamma_{\textrm{\tiny R}}, \gamma_{\textrm{\tiny T}})^2$ from (\ref{C-2-complexity}) 
and for $d(H_{\textrm{\tiny R}}, H_{\textrm{\tiny T}})^2$ from (\ref{C-2-complexity-eh})
take the form\footnote{The l.h.s. of (\ref{step-1-onemode-a}) comes from Baker-Campbell-Hausdorff formula \cite{Hall15book}.}
\be
\label{step-1-onemode-a}
\textrm{Tr}\,\Big\{\big[ a\,\boldsymbol{1} + \log( W_{\textrm{\tiny TR}}\, W^{\textrm{t}}_{\textrm{\tiny TR}})\big]^2\Big\}
\,=\,
2\,a\, \Big( a + \textrm{Tr}\big[\log( W_{\textrm{\tiny TR}}\, W^{\textrm{t}}_{\textrm{\tiny TR}})\big]\Big)
+ 
\textrm{Tr}\Big\{\big[ \log( W_{\textrm{\tiny TR}}\, W^{\textrm{t}}_{\textrm{\tiny TR}})\big]^2\Big\}
\ee
with $a= \log (\sigma_\textrm{\tiny T}/\sigma_\textrm{\tiny R})\equiv a_\sigma$  
and  $a= \log (\varepsilon_\textrm{\tiny T}/\varepsilon_\textrm{\tiny R})\equiv a_\varepsilon$ 
respectively, which can take any real value. 
Since $d(\gamma_{\textrm{\tiny R}}, \gamma_{\textrm{\tiny T}})$ is symmetric under the exchange 
$\gamma_{\textrm{\tiny R}} \leftrightarrow \gamma_{\textrm{\tiny T}}$, we can assume 
$\sigma_\textrm{\tiny R} \geqslant \sigma_{\textrm{\tiny T}}$ without loss of generality.
Then, since the function $\tfrac{2\textrm{arccoth}(2x)}{x}$ is a properly decreasing function when $x>0$,
we have that 
$\tfrac{2\textrm{arccoth}(2\sigma_\textrm{\tiny R})}{\sigma_\textrm{\tiny R}}\leqslant
\tfrac{2\textrm{arccoth}(2\sigma_\textrm{\tiny T})}{\sigma_\textrm{\tiny T}}$,
i.e.  $\sigma_\textrm{\tiny T}/\sigma_\textrm{\tiny R} \leqslant \varepsilon_\textrm{\tiny T}/\varepsilon_\textrm{\tiny R}$, 
once (\ref{Ediag_Ddiag}) has been used;
hence $a_\sigma \leqslant a_\varepsilon$.
This does not provide a relation between
$d(\gamma_{\textrm{\tiny R}}, \gamma_{\textrm{\tiny T}})$ and $d(H_{\textrm{\tiny R}}, H_{\textrm{\tiny T}})$
because the r.h.s. of (\ref{step-1-onemode-a}) does not have a well defined monotonicity as function of $a$,
given that $\log( W_{\textrm{\tiny TR}}\, W^{\textrm{t}}_{\textrm{\tiny TR}})$ is  non vanishing in general.
 
Thus, the one-mode case teaches us that $W_{\textrm{\tiny TR}}$ plays a major role in finding a possible relation 
between $d(H_{\textrm{\tiny R}}, H_{\textrm{\tiny T}})$ 
and $d(\gamma_{\textrm{\tiny R}}, \gamma_{\textrm{\tiny T}})$.
In order to find this relation in some simple cases, 
the expression (\ref{step-1-onemode-a}) naturally leads us
to consider the special cases of one-mode mixed states such that 
$\log( W_{\textrm{\tiny TR}}\, W^{\textrm{t}}_{\textrm{\tiny TR}}) = \boldsymbol{0}$.
In this cases (\ref{step-1-onemode-a}) tells us that
$d(\gamma_{\textrm{\tiny R}}, \gamma_{\textrm{\tiny T}}) = \sqrt{2}\, |a_\sigma|$ 
and $d(H_{\textrm{\tiny R}}, H_{\textrm{\tiny T}}) = \sqrt{2}\, |a_\varepsilon|$.
Since $a^2_\sigma \leqslant a^2_\varepsilon$ is equivalent to $(a_\sigma  -a_\varepsilon)(a_\sigma +a_\varepsilon)\leqslant 0$,
we observe that the latter inequality is satisfied because $a_\sigma \leqslant a_\varepsilon$
and\footnote{This inequality comes from the fact that the function $x \,\textrm{arccoth}(2x) $  is properly decreasing for $ x>0$ and 
that $\sigma_\textrm{\tiny R} \geqslant \sigma_{\textrm{\tiny T}}$ has been assumed.}
$a_\sigma +a_\varepsilon=\log\big(\tfrac{\sigma_{\textrm{\tiny T}} \textrm{arccoth}(2\sigma_{\textrm{\tiny T}})}{\sigma_{\textrm{\tiny R}} \textrm{arccoth}(2\sigma_{\textrm{\tiny R}})}\big)\geqslant 0$.
Thus, for one-mode states such that $W_{\textrm{\tiny TR}}\, W^{\textrm{t}}_{\textrm{\tiny TR}} = \boldsymbol{1}$
we have that
$d(\gamma_{\textrm{\tiny R}}, \gamma_{\textrm{\tiny T}}) \leqslant d(H_{\textrm{\tiny R}}, H_{\textrm{\tiny T}})$.

When  $N\geqslant 1$ and $W_{\textrm{\tiny T}}=W_{\textrm{\tiny R}}$, 
i.e. $W_{\textrm{\tiny TR}}=\boldsymbol{1}$
(this includes the thermal states originating from the same physical hamiltonian),
the distance (\ref{C-2-complexity-eh}) simplifies to 
\be
\label{C-2-complexity-eh-sameW}
d(H_{\textrm{\tiny R}}, H_{\textrm{\tiny T}})
=
\sqrt{\,
\textrm{Tr} \,\Big\{ \big[\log \big(
\mathcal{E}_{\textrm{\tiny T}}\,\mathcal{E}_{\textrm{\tiny R}}^{-1}
\big)\big]^2  \Big\} 
}= 
\,
\sqrt{2
\sum_{k=1}^N \left[ \,\log\! \left( \frac{\varepsilon_{\textrm{\tiny T},k}}{\varepsilon_{\textrm{\tiny R}, k}} \right)\right]^2}
\ee
while $d(\gamma_{\textrm{\tiny R}}, \gamma_{\textrm{\tiny T}})$ 
is given by (\ref{compl_same_W}).
By applying the above analysis made for the one-mode case to the $k$-th mode, we can conclude that
$[\log(\sigma_{\textrm{\tiny T}, k}/\sigma_{\textrm{\tiny R}, k})]^2 
\leqslant [\log(\varepsilon_{\textrm{\tiny T}, k}/\varepsilon_{\textrm{\tiny R}, k})]^2 $
for any given $k$; hence
$d(\gamma_{\textrm{\tiny R}}, \gamma_{\textrm{\tiny T}})^2 \leqslant d(H_{\textrm{\tiny R}}, H_{\textrm{\tiny T}})^2$ 
is obtained after summing over the modes.

By using the decompositions (\ref{EH R and T williamson}),
one can draw a pictorial representation similar to 
Fig.\,\ref{fig:different_D} and Fig.\,\ref{fig:different_D_sp} also for the 
entanglement hamiltonian matrices $H$,
just by replacing each $\gamma$ with the corresponding $H$,
each $W$ with the corresponding $\widetilde{W}$
and where the solid coloured lines are labelled 
by the corresponding symplectic spectra $\mathcal{E}$.

We find it worth discussing further the set of thermal states
through the approach based on the entanglement hamiltonian matrices 
because the simplicity of these matrices in this case 
allows to write analytic results. 
For a thermal state $H = \beta H^{\textrm{\tiny phys}}$, 
where $H^{\textrm{\tiny phys}}$ is the matrix characterising the
physical hamiltonian (\ref{HC ham})
and $\beta$ is the inverse temperature. 
This implies that the symplectic eigenvalues of $H$ are 
$\varepsilon_{\textrm{\tiny th},k} = \beta \,\sigma_{\textrm{\tiny phys},k}$,
where $\sigma_{\textrm{\tiny phys},k}$ are
the symplectic eigenvalues of $H^{\textrm{\tiny phys}}$.

We denote by $\beta_{\textrm{\tiny R}}$ and $\beta_{\textrm{\tiny T}}$ the inverse temperatures
of the reference state and of the target state respectively.
An interesting special case is given by 
thermal states of the same system, which have the same $H^{\textrm{\tiny phys}}$.
In this case 
$H_{\textrm{\tiny T}} H_{\textrm{\tiny R}}^{-1} = (\beta_{\textrm{\tiny T}} / \beta_{\textrm{\tiny R}}) \,\boldsymbol{1}$;
hence (\ref{H-distance}) simplifies to
\be
\label{Complexity through EH Thermal}
d(H_{\textrm{\tiny R}}, H_{\textrm{\tiny T}})
=
\big| \log(\beta_{\textrm{\tiny T}} / \beta_{\textrm{\tiny R}})\big|\,
\sqrt{2V}
=
\sqrt{\,
\textrm{Tr} \,\Big\{ \big[\log \big(
\mathcal{E}_{\textrm{\tiny T}}\, \mathcal{E}_{\textrm{\tiny R}}^{-1}\, \big)\big]^2  \Big\}
}
\ee
where $V$ is the number of sites in the harmonic lattice
and the last expression has been obtained by specialising (\ref{C-2-complexity-eh}) to this case,
where $\widetilde{W}_{\textrm{\tiny TR}}  = \boldsymbol{1}$.
Furthermore, from (\ref{optimal circuit H}) it is straightforward to observe that in this case
the entire optimal circuit is made by thermal states having the same $H^{\textrm{\tiny phys}}$.
The optimal circuit (\ref{optimal circuit H}) significantly simplifies to 
\be
\label{optimal-circuit-th-general}
\widetilde{G}_s(H_{\textrm{\tiny R}} \, , H_{\textrm{\tiny T}})
= 
\beta_s\,
H^{\textrm{\tiny phys}}
\;\; \qquad \;\;
\beta_s \equiv \,\beta_{\textrm{\tiny R}} \left( \frac{\beta_{\textrm{\tiny T}}}{\beta_{\textrm{\tiny R}}} \right)^s
\;\; \qquad \;\;
0 \leqslant s \leqslant 1\,.
\ee
By employing (\ref{williamson-Hphys-gen}), one finds that the Williamson's decomposition of this optimal circuit reads
\be
\label{optimal-circuit-th-general-williamson}
\widetilde{G}_s(H_{\textrm{\tiny R}} \, , H_{\textrm{\tiny T}})
\,= \,
W^{\textrm{t}}_{\textrm{\tiny phys}}  \, \mathcal{D}_s \, W_{\textrm{\tiny phys}}
 \;\; \qquad \;\;
\mathcal{D}_s = \beta_s \, \mathcal{D}_{\textrm{\tiny phys}}
\;\; \qquad \;\;
0 \leqslant s \leqslant 1
\ee
where $W_{\textrm{\tiny phys}}$ is independent of $s$.
Thus, (\ref{optimal-circuit-th-general}) tells us that $\beta_s$
is the inverse temperature of the thermal state labelled by $s$ along this optimal circuit.

In Sec.\,\ref{subsec-eh-num} the above results are applied to the thermal states of the harmonic chain with periodic boundary conditions. 
\\

\section{Gaussian channels}
\label{sec:gauss-channel}

Quantum operations are described by completely positive operators acting on a
quantum state, which can be either pure or mixed, 
and they are classified in {\it quantum channels} and {\it quantum measurements}
\cite{GeomQuantumStates_book,Benatti:2005uq}.
The quantum channels are trace preserving quantum operations, while quantum measurements are not trace preserving \cite{Giedke02}.

The output $\Theta(\hat{\rho})$ of a quantum channel 
applied to the density matrix $\hat{\rho}$ of a system is obtained 
by first extending the system through an ancillary system (the environment)
in a pure state $|\Phi_E\rangle$,
then by allowing an interaction characterised by a unitary transformation $U$
and finally by tracing out the degrees of freedom of the environment
\cite{Stinespring55,NielsenChuang00book, Weedbrook12b}, namely
\be
\Theta(\hat{\rho})
=
\textrm{Tr}_E \big[\,U^\dagger\big(\hat{\rho}\otimes|\Phi_E\rangle\langle\Phi_E|\big) \,U\,\big]\,.
\ee

While within the set of the pure states
the unitary transformations are the only operations that allow to pass from a state to another,
within the general set of mixed states also non unitary operations must be taken into account.

In this manuscript we consider circuits in the space made by quantum Gaussian states; 
hence only quantum operations between Gaussian states (also called Gaussian operations) 
can be considered \cite{Giedke02}. 
The quantum channels and the quantum measurements restricted to the set of the Gaussian states
are often called  Gaussian channels \cite{HolevoWerner01,Weedbrook12b}
and Gaussian measurements \cite{Adesso14} respectively.

In the following we focus only on the Gaussian channels.
A Gaussian state with vanishing first moments is completely described by its covariance matrix;
hence the action of a Gaussian channel on a Gaussian state
can be defined through its effect on the covariance matrix of the Gaussian state. 
This effect can be studied by  introducing two real matrices $T$ and $N$ as \cite{HolevoWerner01}
\be
\label{gaussian channel}
\gamma \,\to\, T\, \gamma\, T^{\textrm{t}} + N
\;\;\qquad\;\;
N = N^{\textrm{t}}
\;\;\qquad\;\;
N + \textrm{i}\, \frac{J}{2}- \textrm{i}\, T \,\frac{J}{2} \,T^{\textrm{t}} \geqslant 0
\ee
where $T$ is unconstrained and  the last inequality corresponds to the complete positivity condition.
The Gaussian unitary transformations are the Gaussian channels with $N=0$ and symplectic $T$.
In this case the inequality in (\ref{gaussian channel}) is saturated. 
Further interesting results for Gaussian operations have been reported e.g. in \cite{Caruso08, Adesso14}.

We find it worth asking whether a matrix along the optimal circuit (\ref{optimal circuit})
can be obtained by acting with a Gaussian channel on the reference state.
This means  finding $T_s$ and $N_s$ that fulfil (\ref{gaussian channel}) 
for any $0 \leqslant s \leqslant 1$ and such that
\be
\label{geodesics gaussian channel}
G_s(\gamma_{\textrm{\tiny R}} \, , \gamma_{\textrm{\tiny T}})
\,=\,
U_s\,\gamma_{\textrm{\tiny R}} \,U_s^{\textrm{t}}
\,=\,
T_s\,\gamma_{\textrm{\tiny R}} \,T_s^{\textrm{t}} + N_s
\;\;\; \qquad \;\;\;
0 \leqslant s \leqslant 1
\ee
where $U_s$ is defined in (\ref{optimal circuit Myers}).
Unfortunately, we are not able to determine $T_s$ and $N_s$
as functions of $U_s$ in full generality. 
In the following we provide some simple particular solutions.

A simple possibility reads
\be
\label{TGs sol1}
T_s = 0 \;\;\qquad \;\; N_s = G_s
\ee
which satisfies the inequality in (\ref{gaussian channel}), 
since $G_s$ is a symmetric Gaussian matrix (see Sec.\,\ref{subsec:mixed}). 

Another, less trivial, solution is given by 
\be
\label{TGs sol4}
T_s = U_s
\;\;\qquad \;\; 
N_s =\boldsymbol{0}
\ee
where the complete positivity condition in (\ref{gaussian channel}) becomes 
$ \textrm{i}\, \frac{J}{2}- \textrm{i}\, T_s \,\frac{J}{2} \,T_s^{\textrm{t}} \geqslant 0$. 
We have considered numerically some cases, finding that $U_s$ satisfies
the complete positivity condition only when it is symplectic
(in this case the complete positivity inequality is saturated).

An explicit example belonging to the class identified by (\ref{TGs sol4})
can be constructed by considering a particular $\mathcal{D}$ path where $W_{\textrm{\tiny T}} = \mathcal{X}_{\textrm{\tiny TR}} W_{\textrm{\tiny R}}$
(see Sec.\,\ref{subsec:basis-comp}). In this case (\ref{geod-D-path-symp}) holds, 
hence (\ref{TGs sol4}) is realised with\footnote{
The last expression in (\ref{channel-4}) is obtained by observing that
$W_{\textrm{\tiny T}} = \mathcal{X}_{\textrm{\tiny TR}} W_{\textrm{\tiny R}}$ 
and $\mathcal{D}_\textrm{\tiny T} = \mathcal{D}_{\textrm{\tiny R}}$ into  
(\ref{optimal circuit Myers}) give
$U_s=(W_{\textrm{\tiny R}}^{\textrm{t}}\, \mathcal{X}^2_{\textrm{\tiny TR}} W_{\textrm{\tiny R}}^{-\textrm{t}})^{s/2}$,
that becomes (\ref{channel-4}) once (\ref{f(MN) identity app}) is employed
with $M=W_{\textrm{\tiny R}}^{\textrm{t}}\, \mathcal{X}^2_{\textrm{\tiny TR}} $ and $N=W_{\textrm{\tiny R}}^{-\textrm{t}}$.
}
\be
\label{channel-4}
T_s= U_s= W_{\textrm{\tiny R}}^{\textrm{t}} \;\mathcal{X}_{\textrm{\tiny TR}}^s \,W_{\textrm{\tiny R}}^{-\textrm{t}}\,.
\ee

A more general solution where both $T_s$ and $N_s$ can be non vanishing 
is obtained by imposing the following relation
\be
\label{TGs sol3}
T_s \,=\, \left[\big( G_s - N_s \big)\, \gamma_{\textrm{\tiny R}}^{-1}\right]^{1/2}
\ee
which solves (\ref{geodesics gaussian channel}) for any symmetric $N_s$. 
The solution (\ref{TGs sol1}) is recovered from (\ref{TGs sol3}) with $T_s=0$.
When $N_s=0$, the relation (\ref{TGs sol3}) gives  
$T_s=\left(G_s\gamma_{\textrm{\tiny R}}^{-1}\right)^{1/2}=U_s$,
where the last equality is obtained from (\ref{optimal circuit compact}) and (\ref{optimal circuit Myers}).
Plugging (\ref{TGs sol3}) into the complete positivity condition in (\ref{gaussian channel}),
we obtain
\be
\label{cp-ineq-sol3}
N_s + \textrm{i}\, \frac{J}{2}
- \textrm{i}\, \big[\big( G_s - N_s \big)\, \gamma_{\textrm{\tiny R}}^{-1}\big]^{1/2}\,  \frac{J}{2} \,\big[\gamma_{\textrm{\tiny R}}^{-1} \,\big( G_s - N_s \big)\big]^{1/2}  \,\geqslant\, 0
\ee
Thus, for any $N_s$ fulfilling  this inequality, by using (\ref{TGs sol3})
we can implement our optimal circuit (\ref{optimal circuit}) through Gaussian channels.

An interesting class of $N_s$ that saturates  (\ref{cp-ineq-sol3}) has been constructed in \cite{Lindblad00}.
It is given by 
\be
\label{LindbladSolution}
N_s=\sqrt{K_s^{\textrm{t}} \,K_s}
\;\;\qquad\;\;
K_s=T_s^{\textrm{t}} \, \frac{J}{2}\, T_s - \frac{J}{2}\,.
\ee
By plugging (\ref{LindbladSolution}) into (\ref{TGs sol3}) first and then employing (\ref{optimal circuit compact}), 
we find the following equation for $T_s$
\be
\label{eq for T_s general}
T_s^2=\Delta_\textrm{\tiny TR}^s-\sqrt{K_s^{\textrm{t}} K_s}\;\gamma_\textrm{\tiny  R}^{-1}
\ee
whose solutions provide realisations of the optimal circuit (\ref{optimal circuit}) through Gaussian channels.

Plugging the definition of $ K_s$ given in (\ref{LindbladSolution}) into (\ref{eq for T_s general}) we find
\be
\label{eq for T_s}
T_s^2+ \textrm{i}\, T_s^{\textrm{t}} \, \frac{J}{2} \, T_s\gamma_\textrm{\tiny  R}^{-1}
=
\Delta_\textrm{\tiny TR}^s+ \textrm{i}\,\frac{J}{2} \,\gamma_\textrm{\tiny  R}^{-1}\,.
\ee
The real part of this relation tells us that $T_s=\Delta_\textrm{\tiny TR}^{s/2}=U_s$,
while from the imaginary part we find that
$T_s^{\textrm{t}}\,J\, T_s=J$, i.e. that $T_s$ is symplectic.
The latter result and (\ref{LindbladSolution}) lead to $K_s=N_s=0$.

Let us conclude by emphasising that all the explicit expressions for the
Gaussian channels given above saturate the complete positivity condition in (\ref{gaussian channel}) 
(more details can be found in \cite{Lindblad00}). 
It would be interesting to explore also Gaussian channels where 
this inequality is not saturated, as done e.g. in (\ref{TGs sol3}) and (\ref{cp-ineq-sol3}).


\section{Complexity of mixed states through ancillae}
\label{sec:purification-approaches}

In this section we discuss the approach to the complexity of mixed states 
explored in \cite{Caceres:2019pgf}, which is based on the introduction of ancillary degrees of freedom.

Consider a quantum system in a mixed state characterised by the density matrix $\hat{\rho}$.
A pure state can be always constructed from $\hat{\rho}$ by adding ancillary degrees of freedom. 
This purification procedure consists in first extending the 
Hilbert space of the system  to a larger Hilbert space
$\mathcal{H}_{\textrm{\tiny extended}} \equiv \mathcal{H} \otimes \mathcal{H}_{\textrm{\tiny anc}}$ 
through an auxiliary Hilbert space $\mathcal{H}_{\textrm{\tiny anc}}$,
and then finding a pure state $ |\Omega\rangle \in \mathcal{H}_{\textrm{\tiny extended}}$ 
such that the original mixed state is obtained as the reduced density matrix given by 
\be
\label{tracing-aux}
\hat{\rho} \,=\,\textrm{Tr}_{\mathcal{H}_{\textrm{\tiny anc}}}
 |\Omega\rangle\! \langle \Omega | 
\ee
where the ancillary degrees of freedom have been traced out. 
We remark that the purifications discussed in Sec.\,\ref{sec:purification} do not involve 
ancillary degrees of freedom. 

There are infinitely many ways to construct $\mathcal{H}_{\textrm{\tiny extended}}$ 
and $|\Omega\rangle$ such that  (\ref{tracing-aux}) is satisfied;
hence a purification criterion must be introduced. 
Different purtification criteria have been considered in the literature to study different quantities.
An important example is the entanglement of purification \cite{Terhal_2002, Takayanagi:2017knl, Nguyen:2017yqw, Bhattacharyya:2018sbw}.
In this manuscript we are interested in the purification complexity \cite{Agon:2018zso}, 
that has been employed in \cite{Caceres:2019pgf} to study the complexity of mixed states.

\subsection{Covariance matrix of the extended system}
\label{sec:cov-mat-pure-extended}

We are interested in a generic harmonic lattice made by $N$ sites in the Gaussian mixed state
characterised by the covariance matrix $\gamma$ and by vanishing first moments. 
The covariance matrix $\gamma$ can be decomposed in blocks
as follows
\be
\label{gamma-block-QPM}
\gamma
\equiv
\bigg(\begin{array}{cc}
Q & M
\\
M^{\textrm t} &P
\end{array}\bigg)
\ee
where $Q$ and $P$ are $N \times N$ symmetric matrices, while $M$ is a generic 
$N \times N$ real matrix; hence $N(2N+1)$ real parameters
must be fixed to determine $\gamma$.

We consider a simplification of the purification process 
by focussing only on Gaussian purifications.
This means that a mixed state characterised by the covariance matrix (\ref{gamma-block-QPM}) 
is purified by introducing ancillary degrees of freedom
and construcing a $2N_{\textrm{\tiny ext}} \times 2N_{\textrm{\tiny ext}} $ 
covariance matrix $\gamma_{\textrm{\tiny ext}}$ 
that corresponds to a Gaussian pure state $|\Omega\rangle$ for the extended lattice
having $N_{\textrm{\tiny ext}}\equiv N+N_{\textrm{\tiny anc}}$ sites. 
For the sake of simplicity, we assume also that 
$ | \Omega \rangle$ has vanishing first moments, i.e.
$\langle \Omega | \hat{\boldsymbol{r}}_{\textrm{\tiny ext}}  | \Omega \rangle = 0$, where
$\hat{\boldsymbol{r}}_{\textrm{\tiny ext}}^{\textrm t} \equiv
(\hat{\boldsymbol{q}}^{\textrm t}, \hat{\boldsymbol{q}}_{\textrm{\tiny anc}}^{\textrm t} , 
\hat{\boldsymbol{p}}^{\textrm t}, \hat{\boldsymbol{p}}_{\textrm{\tiny anc}}^{\textrm t})$ and we have separated the ancillary degrees of freedom 
from the ones associated to the physical system. 

By writing also $\gamma_{\textrm{\tiny ext}}$ through 
the block decomposition (\ref{gamma-block-QPM}) we have
\be
\label{gamma-block-QPM-ext}
\gamma_{\textrm{\tiny ext}}
\equiv
\bigg(\,\begin{array}{cc}
Q_{\textrm{\tiny ext}} & M_{\textrm{\tiny ext}}
\\
M_{\textrm{\tiny ext}}^{\textrm t} &P_{\textrm{\tiny ext}}
\end{array}\bigg)
\ee
where $Q_{\textrm{\tiny ext}}$ and $P_{\textrm{\tiny ext}}$ are 
$N_{\textrm{\tiny ext}}\times N_{\textrm{\tiny ext}}$ 
symmetric matrices.
Since the covariance matrix (\ref{gamma-block-QPM-ext}) corresponds to
a pure state, the condition (\ref{pure-state-condition}) must hold.
This tells us that the blocks occurring in (\ref{gamma-block-QPM-ext})  
are related by the following constraints 
\be
\label{pure-state-exteded-conds}
Q_{\textrm{\tiny ext}}\, P_{\textrm{\tiny ext}} - M_{\textrm{\tiny ext}}^2
 = \,\frac{1}{4}\, \boldsymbol{1}
\;\;\;\qquad\;\;\;
M_{\textrm{\tiny ext}}\, Q_{\textrm{\tiny ext}} 
= Q_{\textrm{\tiny ext}}\, M_{\textrm{\tiny ext}}^{\textrm t}
\;\;\;\qquad\;\;\;
P_{\textrm{\tiny ext}}\,M_{\textrm{\tiny ext}} 
= M_{\textrm{\tiny ext}}^{\textrm t} \,P_{\textrm{\tiny ext}}\,.
\ee
The first relation tells us that $M_{\textrm{\tiny ext}}$ is determined by the product 
$Q_{\textrm{\tiny ext}}\, P_{\textrm{\tiny ext}} $,
while the remaining two relations mean that 
$M_{\textrm{\tiny ext}}\, Q_{\textrm{\tiny ext}} $ 
and $P_{\textrm{\tiny ext}}\,M_{\textrm{\tiny ext}} $ are symmetric. 
Thus, (\ref{gamma-block-QPM-ext}) is determined by the symmetric matrices $Q_{\textrm{\tiny ext}}$ and $P_{\textrm{\tiny ext}}$,
that depend on $2\,\tfrac{N_{\textrm{\tiny ext}}(N_{\textrm{\tiny ext}}+1)}{2}$ real parameters,
as expected also from Sec.\,\ref{subsec-pure-states} (see (\ref{wf-ps})).

We can impose that $\gamma_{\textrm{\tiny ext}} $ is the covariance matrix of a pure state
also by using (\ref{gamma-ps-generic}),
i.e. by requiring that the Williamson's decomposition of (\ref{gamma-block-QPM-ext}) reads
\be
\label{gamma_extended pure}
\gamma_{\textrm{\tiny ext}} 
=\frac{1}{2}\, 
W_{\textrm{\tiny ext}}^{\textrm t} W_{\textrm{\tiny ext}}
=\frac{1}{2}\, 
R_{\textrm{\tiny ext}}^{\textrm t}\, \mathcal{X}^2_{\textrm{\tiny ext}}\, R_{\textrm{\tiny ext}}
\ee
where $W_{\textrm{\tiny ext}} \in \textrm{Sp}(2N_{\textrm{\tiny ext}},\mathbb{R})$
and 
the last expression has been obtained from the Euler decomposition of $W_{\textrm{\tiny ext}}$,
that includes $R_{\textrm{\tiny ext}} \in K(N_{\textrm{\tiny ext}})$.
The symplectic matrix $W_{\textrm{\tiny ext}}$ can be partitioned 
through $N_{\textrm{\tiny ext}}\times N_{\textrm{\tiny ext}}$ matrices as follows
\be
\label{W-b}
W_{\textrm{\tiny ext}} 
=
\bigg( \,\begin{array}{cc}
U_{\textrm{\tiny ext}}  & Y_{\textrm{\tiny ext}} 
\\
Z_{\textrm{\tiny ext}}  & V_{\textrm{\tiny ext}} 
\end{array}\bigg)
\;\;\qquad\;\;
\left\{
\begin{array}{l}
\textrm{$U_{\textrm{\tiny ext}} Y_{\textrm{\tiny ext}}^{\textrm t}$ 
and $V_{\textrm{\tiny ext}} Z_{\textrm{\tiny ext}}^{\textrm t}$ are symmetric}
\\
\rule{0pt}{.5cm}
U_{\textrm{\tiny ext}} V_{\textrm{\tiny ext}}^{\textrm t}-Y_{\textrm{\tiny ext}} Z_{\textrm{\tiny ext}}^{\textrm t} = \boldsymbol{1}\,.
\end{array}
\right.
\ee
The relation (\ref{gamma_extended pure}) provides
the blocks in (\ref{gamma-block-QPM-ext}) through to the ones in (\ref{W-b}). The result reads
\be
Q_{\textrm{\tiny ext}} = \frac{1}{2} \big( U_{\textrm{\tiny ext}}^{\textrm t} U_{\textrm{\tiny ext}} + Z_{\textrm{\tiny ext}}^{\textrm t} Z_{\textrm{\tiny ext}}  \big)
\qquad
P_{\textrm{\tiny ext}} = \frac{1}{2} \big(  V_{\textrm{\tiny ext}}^{\textrm t} V_{\textrm{\tiny ext}} + Y_{\textrm{\tiny ext}}^{\textrm t} Y_{\textrm{\tiny ext}}   \big)
\qquad
M_{\textrm{\tiny ext}} = \frac{1}{2} \big( U_{\textrm{\tiny ext}}^{\textrm t} Y_{\textrm{\tiny ext}} + Z_{\textrm{\tiny ext}}^{\textrm t} V_{\textrm{\tiny ext}}  \big)
\,.
\ee

Another useful way to impose the purity condition on the final state of this purification process 
exploits the general form (\ref{wf-ps}) for the wave function of a pure state
and the corresponding covariance matrix (\ref{gamma-pure-wave-parameters}).
This allows us to write the covariance matrix of the extended system as\footnote{The special case 
$F_{\textrm{\tiny ext}}  = \boldsymbol{0}$ 
has been considered e.g.  in \cite{Caceres:2019pgf, Bhattacharyya:2018sbw}.}
\be
\label{gamma-pure-wave-parameters-ext}
\gamma_{\textrm{\tiny ext}} \,=\,
\frac{1}{2}\,
\bigg(  \begin{array}{cc}
E^{-1}_{\textrm{\tiny ext}} \;
&  \; -\,E^{-1}_{\textrm{\tiny ext}} \,F_{\textrm{\tiny ext}} 
\\
\rule{0pt}{.4cm}
-F_{\textrm{\tiny ext}}\,E^{-1}_{\textrm{\tiny ext}} \;\;
&  \; E_{\textrm{\tiny ext}} + F_{\textrm{\tiny ext}}\,E^{-1}_{\textrm{\tiny ext}}\, F_{\textrm{\tiny ext}} 
\end{array}  \, \bigg)
\ee
where the $N_{\textrm{\tiny ext}} \times N_{\textrm{\tiny ext}}$ symmetric matrices 
$E_{\textrm{\tiny ext}}$ and $F_{\textrm{\tiny ext}}$ are related to the blocks occurring
in (\ref{gamma-block-QPM-ext}) as follows
\be
\label{EFext-from-QMext}
E_{\textrm{\tiny ext}} = \frac{1}{2}\, Q^{-1}_{\textrm{\tiny ext}}
\;\;\qquad\;\;
F_{\textrm{\tiny ext}} = -\,Q^{-1}_{\textrm{\tiny ext}}\, M_{\textrm{\tiny ext}}
\,.
\ee
The second relation in (\ref{pure-state-exteded-conds}) ensures that
$F_{\textrm{\tiny ext}}$ is symmetric.
We remark that (\ref{EFext-from-QMext}) also tell us that the relation
$P_{\textrm{\tiny ext}}
= \tfrac{1}{2} (E_{\textrm{\tiny ext}} + F_{\textrm{\tiny ext}}\,E^{-1}_{\textrm{\tiny ext}}\, F_{\textrm{\tiny ext}}  )$
coming from the second block on the diagonal in (\ref{gamma-block-QPM-ext})
becomes the first relation in (\ref{pure-state-exteded-conds}).

In order to relate $\gamma$ in (\ref{gamma-block-QPM}) 
to $\gamma_{\textrm{\tiny ext}}$ in (\ref{gamma-block-QPM-ext}),
one observes that, since $\hat{\boldsymbol{r}}_{\textrm{\tiny ext}}^{\textrm t} \equiv
(\hat{\boldsymbol{q}}^{\textrm t}, \hat{\boldsymbol{q}}_{\textrm{\tiny anc}}^{\textrm t} , 
\hat{\boldsymbol{p}}^{\textrm t}, \hat{\boldsymbol{p}}_{\textrm{\tiny anc}}^{\textrm t})$,
we have that the $N_{\textrm{\tiny ext}}\times N_{\textrm{\tiny ext}}$ blocks 
occurring in (\ref{gamma-block-QPM-ext}) 
can be partitioned in blocks as follows 
\be
\label{gamma_extended blocks-dec}
Q_{\textrm{\tiny ext}} 
\equiv
\bigg(\,\begin{array}{cc}
Q & \Gamma_Q
\\
\Gamma_Q^{\textrm t} \; & Q_{\textrm{\tiny anc}}
\end{array}\bigg)
\;\;\qquad\;\;
P_{\textrm{\tiny ext}} 
\equiv
\bigg(\,\begin{array}{cc}
P & \Gamma_P
\\
\Gamma_P^{\textrm t} \; & P_{\textrm{\tiny anc}}
\end{array}\bigg)
\;\;\qquad\;\;
M_{\textrm{\tiny ext}} 
\equiv
\bigg(\,\begin{array}{cc}
M & \Gamma_M
\\
\widetilde{\Gamma}_M^{\textrm t} \; & M_{\textrm{\tiny anc}}
\end{array}\bigg)
\ee
where $Q$, $P$ and $M$ are the $N \times N$ blocks of $\gamma$ in (\ref{gamma-block-QPM}),
while $Q_{\textrm{\tiny anc}}$ and $P_{\textrm{\tiny anc}}$ are $N_{\textrm{\tiny anc}}  \times N_{\textrm{\tiny anc}} $ symmetric matrices. 
Instead, $M_{\textrm{\tiny anc}}$ is a  generic $N_{\textrm{\tiny anc}}  \times N_{\textrm{\tiny anc}} $ real matrix. 
Indeed, by plugging (\ref{gamma_extended blocks-dec}) into (\ref{gamma-block-QPM-ext}), 
it is straightforward to observe that the covariance matrix (\ref{gamma-block-QPM}) is obtained
by restricting $\gamma_{\textrm{\tiny ext}} $ 
to the sites corresponding to the original degrees of freedom.
Instead, by restricting $\gamma_{\textrm{\tiny ext}} $ to the ancillary sites,
one gets the following $2N_{\textrm{\tiny anc}}\times 2N_{\textrm{\tiny anc}}$  symmetric matrix
\be
\label{gamma_anc_def}
\gamma_{\textrm{\tiny anc}}
\equiv
\bigg(\,\begin{array}{cc}
Q_{\textrm{\tiny anc}} & M_{\textrm{\tiny anc}}
\\
M_{\textrm{\tiny anc}}^{\textrm t} &P_{\textrm{\tiny anc}}
\end{array}\bigg)\,.
\ee
By changing the order of the rows and the columns, the matrix in (\ref{gamma-block-QPM-ext})
becomes
\be
\label{gamma_extended def-anc}
\bigg(\,\begin{array}{cc}
\gamma \,& \Gamma
\\
 \Gamma^{\textrm t} \,& \gamma_{\textrm{\tiny anc}}
\end{array}\bigg)
\ee
where $\gamma$ is the covariance matrix (\ref{gamma-block-QPM}),
$\gamma_{\textrm{\tiny anc}}$ is the symmetric matrix defined in   (\ref{gamma_anc_def})
and
\be
\Gamma
\equiv
\bigg(\,\begin{array}{cc}
\Gamma_Q \;& \Gamma_M
\\
\widetilde{\Gamma}_M \;&\Gamma_P
\end{array}\bigg)
\ee
By using that (\ref{gamma_extended def-anc}) is positive definite,
it can be shown that also $\gamma_{\textrm{\tiny anc}}$ is positive definite\footnote{By employing the following formula for the determinant of a block matrix
\be
\label{det-block-mat}
\textrm{det}\bigg(
\begin{array}{cc}
A & B \\
C & D
\end{array}\bigg)
=
\textrm{det}(A - B \, D^{-1} C)\, \textrm{det}(D)
\ee
where it is assumed that $D$ is invertible, one finds that the eigenvalues of $\gamma_{\textrm{\tiny anc}}$ 
are also eigenvalues of $\gamma_{\textrm{\tiny ext}}$.
If $A$ is invertible, a formula similar to (\ref{det-block-mat}) can be written where $\textrm{det}(A)$ is factorised and this result can be used to show that 
the eigenvalues of $\gamma$ are eigenvalues of $\gamma_{\textrm{\tiny ext}}$ as well.};
hence $\gamma_{\textrm{\tiny anc}}$  
can be interpreted  as the covariance matrix of the ancillary system made by $N_{\textrm{\tiny anc}}$ sites.

An alternative approach exploits the expressions in the Schr\"odinger representation discussed in
Appendix\;\ref{app:wwm}.
In particular, given the covariance matrix $\gamma$ in the block matrix form (\ref{gamma-block-QPM}),
we can construct the $N \times N$ complex matrices $\Theta$ and $\Phi$
by using  (\ref{ThetaQPM}) and (\ref{PhiQPM}). 
Then, (\ref{ThetaPhi-A}) provide the constraints for the blocks of
$E_{\textrm{\tiny ext}} $ and  $F_{\textrm{\tiny ext}} $ 
in terms of the complex matrices $\Theta$ and $\Phi$.

There are many ways to construct the pure state $|\Omega \rangle$.
They correspond to the freedom to fix $N_{\textrm{\tiny anc}}$ first 
and then to choose e.g. the blocks in (\ref{gamma_extended blocks-dec})
that are different from $Q$, $P$ and $M$,
provided that the constraints (\ref{pure-state-exteded-conds}) 
are satisfied.

\subsubsection{One-mode mixed states}
\label{sec-purification-ext-1mode}

We find it instructive to consider explicitly the simplest case of
a one-mode mixed state, i.e.  $N =1$.
The minimal choice for the number of ancillae is  $N_{\textrm{\tiny anc}} =1$.

When $N=1$, only a non trivial symplectic eigenvalue $\sigma$ occurs;
hence the Williamson's decomposition 
(\ref{williamson th gammaA}) and the Euler decomposition (\ref{EulerDec}) of a symplectic matrix
provide the $2\times 2$ covariance matrix given by 
\be
\label{gamma-one-mode-3forms}
\gamma 
= \sigma \, W^{\textrm t} \, W
 = \sigma \, R^{\textrm t}  \eta^2  R 
 =  R^{\textrm t} \, \textrm{diag}\big( \sigma  e^{2\lambda} ,\sigma e^{-2\lambda}  \big)  R 
 =
 \bigg(\begin{array}{cc}
Q & M
\\
M &P
\end{array}\bigg)
\ee
where $\lambda$ is the squeezing parameter and
$R$ is a $2\times 2$ rotation matrix, which is completely fixed by the rotation angle $\theta$.
Notice that $Q$, $P$ and $M$ are real parameters in  (\ref{gamma-one-mode-3forms}).
Le us remark that
the pure state condition (\ref{pure-state-condition})
for (\ref{gamma-one-mode-3forms}) gives $1-4d_\gamma =0$,
where we have introduced $d_\gamma \equiv \textrm{det}(\gamma) = QP -M^2$.
This implies that $1-4d_\gamma\neq 0$ for the covariance matrices (\ref{gamma-one-mode-3forms})
that correspond to the mixed states that are not pure.

When $N_{\textrm{\tiny anc}} =1$, the covariance matrix (\ref{gamma_extended pure}) of the pure state 
for the extended system reads
\be
\gamma_{\textrm{\tiny ext}}
= \frac{1}{2} \, W^{\textrm t}_{\textrm{\tiny ext}}  W_{\textrm{\tiny ext}}
 =  \frac{1}{2} \,R_{\textrm{\tiny ext}}^{\textrm t}  \,\eta_{\textrm{\tiny ext}}^2  \,R_{\textrm{\tiny ext}}
 =  \frac{1}{2} \,R_{\textrm{\tiny ext}}^{\textrm t} \, 
 \textrm{diag}\big( e^{2\lambda_1} , e^{2\lambda_2} , e^{-2\lambda_1} , e^{-2\lambda_2}  \big)  R_{\textrm{\tiny ext}} \,.
\ee
This $4\times 4$ covariance matrix corresponds to a pure state, hence
it depends on $N_{\textrm{\tiny ext}} (N_{\textrm{\tiny ext}} +1)=6$ real parameters 
($2^2$ from $R_{\textrm{\tiny ext}}$ and two squeezing parameters $\lambda_i$), since $N_{\textrm{\tiny ext}} =2$.
Writing the $4\times 4$ covariance matrix $\gamma_{\textrm{\tiny ext}}$ in the form (\ref{gamma-block-QPM-ext}),
it is straightforward to realise that $3$ elements are given by the real parameters 
$Q$, $P$ and $M$.
Thus, we are left with three real parameters to construct the pure state for the extended system.

We find it instructive to write explicit expressions for the elements of the 
covariance matrix $\gamma_{\textrm{\tiny ext}}$. 
The constraints (\ref{pure-state-exteded-conds}) for the $2\times 2$ matrices 
$Q_{\textrm{\tiny ext}}$, $P_{\textrm{\tiny ext}}$ and $M_{\textrm{\tiny ext}}$
provide six equations: 
four from the first relation and one from each one of the
other two relations
(that can be written in the form $X=0$, where $X$ is a $2\times 2$ antisymmetric matrix).

When $ \Gamma_M\neq 0$, 
the solution of this system can be written in terms of 
$\Gamma_Q$, $\Gamma_P$ and $\Gamma_M$
as 
\bea
\label{onemodepure_solexpli_GammaMnon0}
& &
Q_\textrm{\tiny anc}
= -
\frac{1- 4 d_\gamma}{4 \Gamma_M^2  }
\left(
Q
+ \frac{8\Gamma_Q (M\, \Gamma_M-Q\, \Gamma_P)}{1- 4 d_\gamma}
+\frac{16\Gamma_Q^2 \big[ P\, \Gamma_M^2-\Gamma_P(2 M\, \Gamma_M - Q\, \Gamma_P) \big]}{(1- 4 d_\gamma)^2} 
\right) 
\nonumber
\\
\rule{0pt}{.7cm}
& &
M_\textrm{\tiny anc}
=
- \frac{1}{\Gamma_M}
\left(M\, \Gamma_M - Q\, \Gamma_P
+ \frac{4\Gamma_Q\big[P\, \Gamma_M^2-\Gamma_P(2 M\, \Gamma_M - Q\, \Gamma_P)\big]}{1- 4 d_\gamma} 
\right)
\nonumber
\\
\rule{0pt}{.7cm}
& &
P_\textrm{\tiny anc}
=
-\frac{4\big[P\, \Gamma_M^2-\Gamma_P(2 M\, \Gamma_M - Q\, \Gamma_P)\big]}{1- 4 d_\gamma}
\nonumber
\\
\rule{0pt}{.7cm}
& &
\widetilde{\Gamma}_M
=
\frac{4\Gamma_Q\,\Gamma_P+ 4d_\gamma-1}{4 \Gamma_M}\,.
\eea
When $ \Gamma_M = 0$ and $ \Gamma_P\neq 0$, we find
\bea
\label{onemodepure_solexpli_GammaM0}
& &\hspace{-1cm}
Q_\textrm{\tiny anc}
=
\frac{ P(4d_\gamma-1)}{4\Gamma_P^2}
+\frac{2 M\,\widetilde{\Gamma}_M}{\Gamma_P}
-\frac{4 Q \,\widetilde{\Gamma}_M^2}{1-4d_\gamma}
\hspace{1.5cm}
\Gamma_Q
=
\frac{1-4d_\gamma}{4\Gamma_P}
\\
\rule{0pt}{.7cm}
& &\hspace{-1cm}
P_\textrm{\tiny anc}
=-\frac{4 Q\, \Gamma_P^2}{1-4d_\gamma}
\hspace{5.4cm}
M_\textrm{\tiny anc}
=M-\frac{4Q\,\Gamma_P\,\widetilde{\Gamma}_M}{1-4d_\gamma}
\nonumber
\eea
while a solution does not exist for $ \Gamma_M =  \Gamma_P=0$.
Notice that $1-4d_\gamma \neq 0$ in these expressions because 
$\gamma$ does not corresponds to a pure state. 

We remark that also the analysis 
based on the Schr\"odinger representation
reported in the Appendix \ref{app:rdm-schrod} 
allows to conclude that
the purification of a one-mode mixed state can be realised 
through a pure state in an extended lattice with $N_{\textrm{\tiny ext}} =2$ 
that depends on three real parameters.

\subsubsection{Block diagonal covariance matrices}
\label{subsec:blockdiagCM}

Many interesting mixed states are described by a 
block diagonal covariance matrix $\gamma = Q\oplus P$.
In this cases $M=\boldsymbol{0}$ in (\ref{gamma-block-QPM}).

It is worth considering a pure state for the extended system
such that $M_{\textrm{\tiny ext}}=\boldsymbol{0}$
in the corresponding covariance matrix (\ref{gamma-block-QPM-ext}).
In this case (\ref{pure-state-exteded-conds}) reduce to
\be
\label{pure-state-exteded-conds-M0}
Q_{\textrm{\tiny ext}}\, P_{\textrm{\tiny ext}} 
 = \,\frac{1}{4}\, \boldsymbol{1}
 \qquad \Longleftrightarrow \qquad
 \left\{\begin{array}{l}
 \displaystyle
Q \,P = \frac{1}{4}\, \boldsymbol{1} - \Gamma_Q \,\Gamma_P^{\textrm t}
\\
\rule{0pt}{.6cm}
 \displaystyle
Q_{\textrm{\tiny anc}} \,P_{\textrm{\tiny anc}} = 
\frac{1}{4}\, \boldsymbol{1} - \Gamma_Q^{\textrm t} \,\Gamma_P
\\
\rule{0pt}{.6cm}
 \displaystyle
Q \, \Gamma_P +  \Gamma_Q\, P_{\textrm{\tiny anc}} = \boldsymbol{0}
\\
\rule{0pt}{.6cm}
 \displaystyle
P\,\Gamma_Q +   \Gamma_P\,Q_{\textrm{\tiny anc}} = \boldsymbol{0}
 \end{array} \right.
\ee
where $\Gamma_Q \,\Gamma_P^{\textrm t} \neq \boldsymbol{0}$ and $\gamma = Q\oplus P$ is the covariance matrix of a state that is not pure.

A common choice consists in considering purifications where the 
extended system has twice the degrees of freedom occurring in the original one,
namely $N_{\textrm{\tiny anc}} = N$.
In these cases $\Gamma_Q$, $\Gamma_P$ and $\Gamma_M$ are $N \times N$ matrices.

Considering the purifications with $N_{\textrm{\tiny anc}} = N$,
a drastic simplification corresponds to require  that $\gamma=\gamma_{\textrm{\tiny anc}}$,
which is equivalent to impose that 
$Q=Q_{\textrm{\tiny anc}}$ and $P=P_{\textrm{\tiny anc}}$.
In this case, a solution is given by 
symmetric and commuting matrices $\Gamma_Q$ and $\Gamma_P$ 
that can be related through the last two equations in (\ref{pure-state-exteded-conds-M0}),
which give
\be
\label{pure-state-exteded-conds-M0-bis}
Q = -\, \Gamma_Q\, P\, \Gamma_P^{-1} 
\;\;\qquad\;\;
\Gamma_Q\, P\, \Gamma_P^{-1} =  \Gamma_P^{-1}\, P\, \Gamma_Q\,.
\ee

Setting $\Gamma_P = \alpha\, \Gamma_Q^{-1}$ with $\alpha \in \mathbb{R}$, the last equality is solved
while the remaining relation $Q = -\, \tfrac{1}{\alpha}\,\Gamma_Q\, P\, \Gamma_Q $,
whose validity is not guaranteed, provides  $\Gamma_Q$.

A different solution for the matrix equations in (\ref{pure-state-exteded-conds-M0-bis}) 
can be written when $Q$ and $P$ can be decomposed 
through three real matrices $A$, $B$ and $\Lambda$ as follows
\be
\label{QP-ABlambda}
Q=A\, \Lambda\, B^{-1}
\;\;\qquad\;\;
P=B\, \Lambda\, A^{-1}\,.
\ee
In this case, we can construct $\Gamma_Q$ and  $\Gamma_P$ as 
\be
\label{gammaQP-ABlambda}
\Gamma_Q=A\, \tilde{\Lambda}\, B^{-1}
\;\;\qquad\;\;
\Gamma_P= -\,B\, \tilde{\Lambda}\, A^{-1}
\;\;\qquad\;\;
\big[\, \tilde{\Lambda}\, , \Lambda \, \big] = 0
\ee
where a new matrix $\tilde{\Lambda}$ that commutes with $\Lambda$ has been introduced. 

It is straightforward to check that (\ref{QP-ABlambda}) and (\ref{gammaQP-ABlambda})
satisfy the matrix equations in (\ref{pure-state-exteded-conds-M0-bis}).
Notice that $\Gamma_P$ is not proportional to $\Gamma_Q^{-1}$ in (\ref{gammaQP-ABlambda}).

An important example where  $N_{\textrm{\tiny anc}} = N$ and $\gamma=\gamma_{\textrm{\tiny anc}}$
is the thermofield double state (TFD).
In Appendix\;\ref{app-tfd} a detailed analysis for this pure state for harmonic lattices is reported. 
The relations (\ref{QQPP_TFD}) and (\ref{GammaPGammaQ_TFD})
tell us that the TFD corresponds to a special case\footnote{In particular, $\Lambda$ and $\tilde{\Lambda}$ are the diagonal matrices in (\ref{LambdaTFD}), 
while $A= \widetilde{V}\, \mathcal{S}^{-1}$ and $B=\widetilde{V}\, \mathcal{S}$,
in terms of the matrices $\widetilde{V}$ and $\mathcal{S}$ introduced in the Appendix\;\ref{app-tfd}.}
of (\ref{QP-ABlambda}) and (\ref{gammaQP-ABlambda}).

The simplest case corresponds to $N=N_{\textrm{\tiny anc}}=1$,
which has been discussed in Sec.\,\ref{sec-purification-ext-1mode} in the most general setting. 
Solving the system (\ref{pure-state-exteded-conds-M0}) for this case, one finds 
\be
\label{onemodepure_solexpli_Mext0}
Q_\textrm{\tiny anc}
= -
\frac{(1-4QP)P}{4\Gamma_P^2}
\,\,\qquad\,\,
P_\textrm{\tiny anc}
=\frac{4 Q \Gamma_P^2}{4QP-1}
\,\,\qquad\,\,
\Gamma_Q
=
\frac{1- 4QP}{4\Gamma_P}\,.
\ee
When $M=0$, the observation in the text below (\ref{gamma-one-mode-3forms}) 
tells us that $4QP-1 \neq 0$ in order to have a mixed state that is not pure to purify. 
Notice also that,
by setting $M=\widetilde{\Gamma}_M=0$ in (\ref{onemodepure_solexpli_GammaM0}),
that holds for $\Gamma_M=0$,
one finds (\ref{onemodepure_solexpli_Mext0}) and $M_{\textrm{\tiny anc}}=0$.
Thus, when $M_{\textrm{\tiny ext}}=\boldsymbol{0}$ and $N=N_{\textrm{\tiny anc}}=1$
we can parameterise the pure state of the extended system through a single parameter.
This is consistent with the analysis reported in \cite{Caceres:2019pgf}.
As final remark about the purifications having $N=N_{\textrm{\tiny anc}}=1$,
let us observe that the second equation in (\ref{pure-state-exteded-conds-M0-bis}) is trivially satisfied,  
while the first one is obtained by setting 
$Q=Q_{\textrm{\tiny anc}}$ and $P=P_{\textrm{\tiny anc}}$ in (\ref{onemodepure_solexpli_Mext0}).

\subsection{Selection criterion for the pure state}
\label{subsec:selectioncriterion}

In the previous discussion we have explored  the constraints 
guaranteeing that the covariance matrix $\gamma_{\textrm{\tiny ext}} $ corresponds to a pure state
under the condition that $\gamma_{\textrm{\tiny ext}} $
provides the covariance matrix $\gamma$ of the given mixed state
once the ancillary degrees of freedom have been traced out. 

These constraints identify the parameter space of the pure states allowed by $\gamma$
for a given value of $N_{\textrm{\tiny anc}}$.
Within this space of parameters, it is natural to introduce a quantity $ \mathcal{F}$
whose minimisation provides a particular pure state with certain properties. 
Thus, $ \mathcal{F}$ characterises the criterion to select the pure state provided by the purification procedure as follows
\be
\label{purif-criterion}
\widetilde{\mathcal{F}} (\gamma)
\,\equiv\,
\min_{\gamma_{\textrm{\tiny ext}}} \! \big[ \mathcal{F} (\gamma_{\textrm{\tiny ext}}) \big]
\ee
where $\gamma_{\textrm{\tiny ext}}$ is the covariance matrix for the extended system,
that is constrained as described in Sec.\,\ref{sec:cov-mat-pure-extended},
and $\widetilde{\mathcal{F}}$ denotes the minimal value of $\mathcal{F}$ 
as $\gamma_{\textrm{\tiny ext}}$ spans all the pure states allowed by $\gamma$.
For the bosonic Gaussian states that we are considering, 
the calculations can be performed by employing 
either the wave functions or the covariance matrices.

For instance, the entanglement of purification for a bipartite mixed state 
\cite{Terhal_2002,Takayanagi:2017knl,Nguyen:2017yqw,Bhattacharyya:2018sbw}
is (\ref{purif-criterion}), with $\mathcal{F}$ given by  the entanglement entropy of a particular bipartition of 
$\mathcal{H}_{\textrm{\tiny extended}}$.

In \cite{Agon:2018zso, Caceres:2019pgf}, 
the purification complexity has been introduced to quantify the complexity of a mixed state. 
The definition of purification complexity is given by (\ref{purif-criterion}) in the special case where
$\mathcal{F}$ is the complexity of the pure state corresponding to $\gamma_{\textrm{\tiny ext}}$ 
with respect to a given fixed pure state in $\mathcal{H}_{\textrm{\tiny extended}}$,
whose covariance matrix is denoted by $ \gamma_{\textrm{\tiny ext},0}$.
This definition of purification complexity requires the choice of a cost function.
The purification complexity explored in \cite{Caceres:2019pgf} reads
\be
\label{purif-comp}
\widetilde{\mathcal{C}}_r (\gamma)
\equiv
\min_{\gamma_{\textrm{\tiny ext}}} \! \big[\, 
\mathcal{C}_r (\gamma_{\textrm{\tiny ext}} , \gamma_{\textrm{\tiny ext},0} )
 \big]
\ee
where either $r=1$ or $r=2$, 
depending on whether the $F_1$ cost function or the $F_2$ cost function is adopted.
In \cite{Caceres:2019pgf} the purification complexity based on the $F_1$ cost function has been mainly studied
because, for the pure states, the divergence structure of the complexity evaluated through the $F_1$ cost function
is closer to the one obtained from holographic calculations \cite{Jefferson:2017sdb,Chapman:2017rqy}. 
The complexity defined through the $F_1$ cost function depends on the choice of the underlying basis,
while the $F_2$ cost function leads to a complexity that is independent of this choice.

This approach to the complexity of mixed states is different from the one considered in this manuscript.
The main difference is due to the fact that in the purification procedure described 
in Sec.\,\ref{sec:purification}
ancillary degrees of freedom have not been introduced.
Moreover, the purification complexity defined in (\ref{purif-comp}) depends on the choice of the pure state 
corresponding to $\gamma_{\textrm{\tiny ext},0}$
(in \cite{Caceres:2019pgf} this pure state has been fixed to the one whose
wave function (\ref{wf-ps}) has $E_{\textrm{\tiny ext}} \propto \boldsymbol{1}$ 
and $F_{\textrm{\tiny ext}}= \boldsymbol{0}$).
Furthermore, in the evaluation of the complexity of a mixed state 
through (\ref{purif-comp}),
no cost is assigned to the purification process of extending the system through ancillary degrees of freedom, given that the circuit considered in (\ref{purif-comp}) is entirely 
made by pure states in  $\mathcal{H}_{\textrm{\tiny extended}}$.

Explicit computations through (\ref{purif-comp}) are technically involved 
and discussing them is beyond the scope of this manuscript.
We refer the interested reader to the detailed analysis performed in \cite{Caceres:2019pgf}.
Focussing on the simple case of one-mode thermal states, 
in Sec.\,\ref{subsec:purif_numerics} we compare the complexity evaluated through the Fisher-Rao distance 
with the results found in  \cite{Caceres:2019pgf} for the $\mathcal{C}_1$ complexity of mixed states based on the purification complexity.
The latter quantity depends on the basis: in Appendix \ref{app:c1} we discuss the 
diagonal basis and the physical basis, that have been introduced in \cite{Caceres:2019pgf} to evaluate this $\mathcal{C}_1$ complexity.

\section{Harmonic chains}
\label{sec:examples}

In this section we further study some of the quantities discussed in the previous sections
by focussing on the one-dimensional case of the harmonic chain,
either on the circle (i.e. with periodic boundary conditions) or on the infinite line. 
In this case we obtain analytic expressions in terms of the parameters of the circuit
for some quantities and provide numerical results for 
the quantities that are more difficult to address analytically. 
After a brief discussion of the model in Sec.\,\ref{subsec-examples-hc},
circuits whose reference and target states are either pure or thermal 
are considered in Sec.\,\ref{sec-pure-states-examples} and Sec.\,\ref{sec-thermal-states-examples}
respectively. 
In Sec.\,\ref{subsec:mutualcomp-TFD} we study the mutual complexity for the 
thermofield double states (TFD's).
Numerical results for the complexity and the mutual complexity associated to subregions
are presented in Sec.\,\ref{sec-subregion-examples}
and Sec.\,\ref{subsec:mutualsubregion}.
Finally, in Sec.\,\ref{subsec:purif_numerics} we consider a simple comparison between the
complexity for mixed states discussed in this manuscript and 
the one based on the purification complexity recently proposed in \cite{Caceres:2019pgf}.

For the sake of simplicity, in this section we consider only examples that
involve states whose covariance matrices are block diagonal.
We remark that the results discussed in the previous sections hold 
also for states characterised by covariance matrices that are not block diagonal. 
For instance, these states typically occur in the out-of-equilibrium 
dynamics of the harmonic lattices \cite{Eisler_2014,Coser:2014gsa,Cotler:2016acd,DiGiulio:2019cxv}.

\subsection{Hamiltonian}
\label{subsec-examples-hc}

The hamiltonian of the periodic harmonic chain 
made by $L$ sites, with frequency $\omega$, mass $m$ and elastic constant $\kappa$
reads
\be
\label{HC ham-1d}
\widehat{H} 
\,=\, 
\sum_{i=1}^{L} \left(\,
\frac{1}{2m}\,\hat{p}_i^2+\frac{m\omega^2}{2}\,\hat{q}_i^2 
+ \frac{\kappa}{2}(\hat{q}_{i+1} -\hat{q}_i)^2
\right)
\ee
where 
$\hat{\boldsymbol{r}} \equiv (\hat{q}_1 , \dots , \hat{q}_L, \hat{p}_1, \dots, \hat{p}_L)^{\textrm{t}}$
collects the position and momentum operators and the periodic boundary condition
$\hat{q}_{L+1} = \hat{q}_{1} $ is imposed. 

Assuming that both $\kappa$ and $m$ are non-vanishing,
the canonical transformation given by 
$\hat{q}_i \to  \hat{q}_i / \sqrt[4]{m\kappa}$ and $\hat{p}_i \to \sqrt[4]{m\kappa}\, \hat{p}_i$ 
allows to write (\ref{HC ham-1d}) as follows
\be
\label{HC ham v2}
\widehat{H} 
= \frac{\sqrt{\kappa/m}}{2}\,
\sum_{i=1}^{L} 
\left(
\hat{p}_i^2+ \frac{\omega^2}{\kappa/m} \, \hat{q}_i^2 +  (\hat{q}_{i+1} -\hat{q}_i)^2
\right)
 \,\equiv\,
 \frac{1}{2}\, \hat{\boldsymbol{r}}^{\textrm t} H^{\textrm{\tiny phys}} \, \hat{\boldsymbol{r}}
\ee
where
\be
\label{Hphys-mat-form}
H^{\textrm{\tiny phys}} = \sqrt{\kappa/m} \,
\Big(
\big[(\tilde{\omega}^2+2)\boldsymbol{1}- T\big]
\oplus \boldsymbol{1}
\Big)
\ee
and we are naturally led to introduce the dimensionless parameter
\be
\label{omega_tilde}
\tilde{\omega}^2 = \frac{\omega^2}{\kappa/m} \,.
\ee
The non vanishing elements of the symmetric matrix $T$ in (\ref{Hphys-mat-form})
are $T_{i,i+1}=T_{i+1,i}=1 $ with $1\leqslant i \leqslant L-1$ and $T_{1,L}=T_{L,1}=1$.

In order to find the Williamson's decomposition (\ref{williamson-Hphys-gen}) for (\ref{Hphys-mat-form}), 
first one observes that the matrix $T$ in (\ref{Hphys-mat-form}) is diagonalised by the 
following unitary matrix 
\be
\label{calU matrix pbc}
\widetilde{U}_{r,s} \equiv
\frac{e^{2\pi \mathrm{i} \,r\, s /L }}{\sqrt{L}}
\ee
that implements the discrete Fourier transform and 
it is independent of the parameters $\omega$, $m$ and $\kappa$.
This implies that $H^{\textrm{\tiny phys}}$ in (\ref{Hphys-mat-form}) is diagonalised by $U\equiv\widetilde{U}\oplus\widetilde{U}$.

Since the symplectic matrix entering in the Williamson's decomposition (\ref{williamson-Hphys-gen}) is real,
let us consider the proper combinations of the eigenvectors entering in (\ref{calU matrix pbc}) 
that correspond to the same eigenvalue.
This leads to introduce the $L \times L$ real and orthogonal matrix $\widetilde{V}$, 
whose generic element for even $L$ is given by
\be
\label{Vtilde-def-even}
\widetilde{V}_{i,k} \equiv
\left\{\begin{array}{ll}
\sqrt{2/L}\; \cos(2\pi i\,k/L)  \hspace{1cm}&   1\leqslant k < L/2
\\
\rule{0pt}{.5cm}
(-1)^i/\sqrt{L} &    k = L/2
\\
\rule{0pt}{.5cm}
\sqrt{2/L}\; \sin(2\pi i\,k/L)  &   L/2+1 \leqslant k < L-1
\\
\rule{0pt}{.5cm}
1/\sqrt{L} &    k = L
\end{array}
\right.
\ee
and for odd $L$ by
\be
\label{Vtilde-def-odd}
\widetilde{V}_{i,k} \equiv
\left\{\begin{array}{ll}
\sqrt{2/L}\; \cos(2\pi i\,k/L)  &   1\leqslant k < (L-1)/2
\\
\rule{0pt}{.5cm}
\sqrt{2/L}\; \sin(2\pi i\,k/L)  \hspace{1cm}&   (L-1)/2+1 \leqslant k < L-1
\\
\rule{0pt}{.5cm}
1/\sqrt{L} &    k = L\,.
\end{array}
\right.
\ee
The matrix $\widetilde{V}$ diagonalises both $T$ and $\boldsymbol{1}$ in (\ref{Hphys-mat-form});
hence, by introducing the orthogonal matrix $V\equiv \widetilde{V} \oplus \widetilde{V}$, that is also symplectic,
we have \cite{Serafini17book}
\be
\label{Hphys-hc-block}
H^{\textrm{\tiny phys}}
=
 \,V\,
\Big[ \sqrt{\kappa/m}\;\textrm{diag}\big( \Omega^2_1,\dots,\Omega^2_L,1,\dots,1\big) \Big] \, V^{-1}
\ee
where $\Omega_{k}$ provides the dispersion relation, which depends on the parameter $\tilde{\omega}$ defined in (\ref{omega_tilde})
as follows
 \be
\label{dispersion relation}
\Omega_{k} \equiv \sqrt{\tilde{\omega}^2 + 4\, (\sin[\pi k/L])^2}
\;\; \qquad\;\;
k=1, \dots, L\,.
\ee

By applying the observation made in the final part of the Appendix\;\ref{app:matrix_id} to (\ref{Hphys-hc-block}),
one obtains the Williamson's decomposition (\ref{williamson-Hphys-gen}) with the symplectic eigenvalues given by
\be
\label{symp-spectrum-phys}
\sigma_{\textrm{\tiny phys},k}= \sqrt{\kappa/m}\;\Omega_k
\ee
and the symplectic matrix $W_{\textrm{\tiny phys}}$ by 
\be
\label{symp-W-phys}
W_{\textrm{\tiny phys}}=
\mathcal{X}_{\textrm{\tiny phys}} \,V^{t}
\;\;\qquad\;\;
\mathcal{X}_{\textrm{\tiny phys}}
\equiv
\textrm{diag}\big(\sqrt{\Omega_1},\dots,\sqrt{\Omega_L},1/\sqrt{\Omega_1},\dots,1/\sqrt{\Omega_L}\,\big)\,.
\ee 
In these expressions the zero mode corresponds to $k=L$
and its occurrence is due to the invariance of the system under translations. 
The comparison between the expressions reported throughout this section
and the corresponding ones in Sec.\,\ref{subsec:thermal} 
can be done once the canonical transformation above (\ref{HC ham v2})
has been taken into account\footnote{This canonical transformation is responsible e.g.
for the different definitions of $\Omega_k$ 
in (\ref{dispersion relation}) and in Sec.\,\ref{subsec:thermal}
and also for the factor between 
(\ref{symp-spectrum-phys}) and (\ref{Wphys-mat-def}),
which is the same prefactor occurring in the hamiltonian (\ref{HC ham v2}).}.

It is worth remarking that the canonical transformation that brings (\ref{HC ham-1d}) into (\ref{HC ham v2}) cannot be defined when $\kappa=0$. 
This implies, for instance, that, 
in order to employ the unentangled product state of the harmonic chain as reference state
(this is often the case in the recent literature on the circuit complexity \cite{Jefferson:2017sdb,Chapman:2018hou,Camargo:2018eof,Guo:2018kzl,Caceres:2019pgf}),
our analysis must be adapted to the hamiltonian (\ref{HC ham-1d}).

\subsection{Pure states}
\label{sec-pure-states-examples}

In this subsection we study the circuit complexity for pure states that are the ground states of periodic 
harmonic chains having different frequencies \cite{Jefferson:2017sdb}.

\subsubsection{Covariance matrix}
\label{subsec:CMHC}

The two-point correlators in the ground state of the hamiltonian (\ref{HC ham v2}),
where periodic boundary conditions are imposed, read
\be
 \label{corrs-periodic}
 \langle \hat{q}_i \hat{q}_j  \rangle =
\frac{1}{2 L} \,\sum_{k=1}^{L} \frac{1}{\Omega_{k}} \, \cos[2 \pi k\,(i-j)/L] 
 \qquad
\langle \hat{p}_i \hat{p}_j  \rangle =
\frac{1}{2 L} \, \sum_{k=1}^{L} \Omega_{k}  \cos[2 \pi k\,(i-j)/L]
 \ee
where $\Omega_{k}$ is the dispersion relation (\ref{dispersion relation}).
The periodic boundary conditions make this system invariant under translations.
The expressions in (\ref{corrs-periodic}) define
the elements of the correlation matrices $Q_{\textrm{\tiny gs}}$ and $P_{\textrm{\tiny gs}}$ respectively.
These matrices provide the block diagonal covariance matrix 
$\gamma_{\textrm{\tiny gs}} = Q_{\textrm{\tiny gs}} \oplus P_{\textrm{\tiny gs}}$.

By introducing the discrete Fourier transform of the operators $\hat{q}_j$ and $\hat{p}_j$ 
in the standard way \cite{AltlandSimons10book}, 
the matrices $Q_{\textrm{\tiny gs}}$ and $P_{\textrm{\tiny gs}}$ can be written as follows
\be
\label{matrix FT}
Q_{\textrm{\tiny gs}}\,=\,\widetilde{U} \,\mathcal{Q}_{\textrm{\tiny gs}} \;\widetilde{U}^{-1}
\qquad
P_{\textrm{\tiny gs}}\,=\,\widetilde{U} \,\mathcal{P}_{\textrm{\tiny gs}} \; \widetilde{U}^{-1}
\ee
where the matrix $\widetilde{U}$ have been defined in (\ref{Hphys-mat-form}),
while $\mathcal{Q}_{\textrm{\tiny gs}}$ and $\mathcal{P}_{\textrm{\tiny gs}}$ are diagonal matrices whose elements read
\be
\label{diag-corr-FT}
(\mathcal{Q}_{\textrm{\tiny gs}})_{k,k}= \frac{1}{2  \,\Omega_{k}}
\;\;\qquad\;\;
(\mathcal{P}_{\textrm{\tiny gs}})_{k,k}=\frac{1}{2}\;\Omega_k
\;\;\;\;\qquad\;\;\;\;
k=1,\dots,L\,.
\ee

In order to find the Williamson's decomposition of $\gamma_{\textrm{\tiny gs}}$, 
we have to consider the symplectic matrix $V$ introduced in Sec.\,\ref{subsec-examples-hc}.
Then, the observation made in the final part of the Appendix\;\ref{app:matrix_id} specified to $\gamma_{\textrm{\tiny gs}}$
leads us to introduce the following symplectic and diagonal matrix
\be
\label{Williamson W for pure HC}
\mathcal{X}_{\textrm{\tiny C}}
\,\equiv\,
\textrm{diag}\big(1/\sqrt{\Omega_1},\dots,1/\sqrt{\Omega_L},\sqrt{\Omega_1},\dots,\sqrt{\Omega_L}\,\big)
=
J^{\textrm{t}}  \,\mathcal{X}_{\textrm{\tiny phys}}\, J
\ee
where $\mathcal{X}_{\textrm{\tiny phys}}$ has been introduced in (\ref{symp-W-phys}).
This matrix is related to (\ref{diag-corr-FT}) as follows
\be
\label{QplusP-diag}
\mathcal{Q}_{\textrm{\tiny gs}} \oplus \mathcal{P}_{\textrm{\tiny gs}}  = \frac{1}{2}\, \mathcal{X}_{\textrm{\tiny C}}^2\,.
\ee
By introducing the symplectic matrix 
\be
\label{W_C matrix def}
W_{\textrm{\tiny C}}
\, \equiv \,
\mathcal{X}_{\textrm{\tiny C}}\,
V^{\textrm{t}} 
\ee
we have that the Williamson's decomposition of $\gamma_{\textrm{\tiny gs}}$ reads
\be
\label{williamson-def-gs}
\gamma_{\textrm{\tiny gs}}
=
\frac{1}{2}\,W^{\textrm{t}}_{\textrm{\tiny C}} \, W_{\textrm{\tiny C}}\,.
\ee

Notice that the symplectic matrices $W_{\textrm{\tiny phys}}$ and $W_{\textrm{\tiny C}}$,
defined in (\ref{symp-W-phys}) and (\ref{W_C matrix def}) respectively, 
are related as follows
\be
\label{WCandWphys}
W_{\textrm{\tiny C}} = 
J^{\textrm{t}}  \,\mathcal{X}_{\textrm{\tiny phys}}\, 
\big(J\, V^{\textrm{t}}  J^{\textrm{t}} \big) J
=
J^{\textrm{t}}  \,\mathcal{X}_{\textrm{\tiny phys}}\,  V^{\textrm{t}}   J
=
J^{\textrm{t}}  \,W_{\textrm{\tiny phys}}  J
=
W_{\textrm{\tiny phys}}^{-\textrm{t}}
\ee
where we have also used (\ref{Williamson W for pure HC}), (\ref{W_C matrix def}), 
the property $S^{-\textrm{t}} = J^{\textrm t}  S J$ of the symplectic matrices $S$
and the fact that symplectic $V$ is also orthogonal (see also (\ref{sympspec_CM_th})).

As consistency check, we can plug (\ref{W_C matrix def}) into (\ref{williamson-def-gs}) first and then use
(\ref{QplusP-diag}), finding that
\be
\label{QP-diag-V-gs}
\gamma_{\textrm{\tiny gs}} 
= 
V \big( \mathcal{Q}_{\textrm{\tiny gs}} \oplus \mathcal{P}_{\textrm{\tiny gs}} \big) \,V^{\textrm{t}}
= 
V \big( \mathcal{Q}_{\textrm{\tiny gs}} \oplus \mathcal{P}_{\textrm{\tiny gs}} \big) \,V^{-1}\,.
\ee
This tells us that $V$ is also the orthogonal matrix that diagonalises the symmetric matrix $\gamma_{\textrm{\tiny gs}}$.
Let us remark that $V$ depends only on the number of sites $L$ of the harmonic chain.

\subsubsection{Complexity}

We consider the circuit complexity where the reference state and the target state 
are the ground states of periodic harmonic chains whose hamiltonians are
characterised by the parameters $(\omega_\textrm{\tiny R}, \kappa_\textrm{\tiny R},m_\textrm{\tiny R})$ 
and  $(\omega_\textrm{\tiny T}, \kappa_\textrm{\tiny T},m_\textrm{\tiny T})$ respectively. 

For the sake of simplicity, in our analysis we set 
$\kappa_\textrm{\tiny R}=\kappa_\textrm{\tiny T}\equiv\kappa$ and $m_\textrm{\tiny R}=m_\textrm{\tiny T}\equiv m$;
hence  only the  parameter $\tilde{\omega}$ distinguishes the reference and the target states.

In this case, from (\ref{QP-diag-V-gs}) and the fact that $V$ is independent of the parameters $\omega$, $\kappa$ and $m$,
it is straightforward to find that (\ref{def relative CM}) becomes
\be
\label{DeltaTR-gs-diag}
\Delta_{\textrm{\tiny TR}}=\, 
V \big( 
\mathcal{Q}_\textrm{\tiny gs,T} \,\mathcal{Q}_\textrm{\tiny gs,R}^{-1}
\oplus 
\mathcal{P}_\textrm{\tiny gs,T} \,\mathcal{P}_\textrm{\tiny gs,R}^{-1}
\big) \,V^{-1}
\ee
where the diagonal matrices $\mathcal{Q}_\textrm{\tiny gs,R} $, $\mathcal{Q}_\textrm{\tiny gs,T} $,
$\mathcal{P}_\textrm{\tiny gs,R} $ and $\mathcal{P}_\textrm{\tiny gs,T} $ can be easily obtained by writing
(\ref{diag-corr-FT}) for the reference and for the target state.
By employing (\ref{DeltaTR-gs-diag}) and (\ref{diag-corr-FT}), it is straightforward to find that,
in this case, 
the complexity given by (\ref{deltaAB}) and (\ref{c2 complexity}) simplifies to \cite{Jefferson:2017sdb}
\be
\label{PureStateComplexityHC}
\mathcal{C}_2
= 
\frac{1}{2\sqrt{2}}\,
\sqrt{\,
\textrm{Tr}\, \Big\{ \big[\log \!\big(\mathcal{Q}_\textrm{\tiny gs,T} \,\mathcal{Q}^{-1}_\textrm{\tiny gs,R} 
\oplus \mathcal{P}_\textrm{\tiny gs,T} \,\mathcal{P}^{-1}_\textrm{\tiny gs,R} \big) \big]^2 \Big\} }
\,=\,
\frac{1}{2}\,
\sqrt{\, \sum_{k=1}^L\!  \Big(\! \log\big[ \Omega_{\textrm{\tiny T},k} /  \Omega_{\textrm{\tiny R},k}\big]\Big)^2}
\ee
where
\be
\label{dispersion relation TR}
\Omega_{\textrm{\tiny S},k} \equiv \sqrt{\tilde{\omega}^2_{\textrm{\tiny S}} + 4 \, (\sin[\pi k/L])^2}
\;\;\;\; \qquad\;\;\;\;
k=1, \dots, L
\; \qquad\;
\textrm{S} \in \big\{\textrm{R}, \textrm{T}\big\}\,.
\ee
Notice that the complexity (\ref{PureStateComplexityHC}) depends on $m$ and $\kappa$
only through the dimensionless parameters $\tilde{\omega}_\textrm{\tiny R}$ and $\tilde{\omega}_\textrm{\tiny T}$.

We remark that, if $\kappa_\textrm{\tiny R}\neq\kappa_\textrm{\tiny T}$, 
the  canonical transformation reported in the text between (\ref{HC ham-1d}) and (\ref{HC ham v2}) is not the same for reference and target states.
This is crucial in the evaluation of the circuit complexity between the ground state of the harmonic chain with $\kappa\neq 0$ and the unentangled product state, where $\kappa=0$.
In these cases we have to consider the hamiltonian (\ref{HC ham-1d}) instead of (\ref{HC ham v2}). 
Adapting our analyses to this hamiltonian is a straightforward exercise whose details will not be reported here. 
For instance, considering $m_{\textrm{\tiny R}}= m_{\textrm{\tiny T}}=m $ but keeping $\kappa_{\textrm{\tiny R}}\neq \kappa_{\textrm{\tiny T}}$, we obtain
\be
\label{complexity different kappa pure}
\mathcal{C}_2
\,=\,
\frac{1}{2}\,
\sqrt{\, \sum_{k=1}^L\!  \bigg(\! \log\bigg[\frac{\sigma_{\textrm{\tiny phys,T},k} }{ \sigma_{\textrm{\tiny phys,R},k}}\bigg]\bigg)^2}
\ee
where $\sigma_{\textrm{\tiny phys,S},k}$ with $\textrm{S} \in \big\{\textrm{R}, \textrm{T}\big\}$ are the symplectic spectra 
corresponding to the reference and the target states, evaluated through the hamiltonian (\ref{symp-spectrum-phys}). Ref. \cite{Jefferson:2017sdb} mainly considered the case
where the reference state is the unentangled product state 
and the target state is the ground state of a harmonic chain (with periodic boundary conditions).
The circuit complexity in this case can be found by taking the limit $\kappa_\textrm{\tiny R}\to 0$ in (\ref{complexity different kappa pure}). 
The result is \cite{Jefferson:2017sdb}
\be
\mathcal{C}_2
\,=\,
\frac{1}{2}\,
\sqrt{\, \sum_{k=1}^L\!  \bigg(\! \log\bigg[\frac{ \sigma_{\textrm{\tiny phys,T},k}}{ \omega_{\textrm{\tiny R}}}\bigg]\bigg)^2}
\ee

It is instructive to obtain (\ref{PureStateComplexityHC}) also as a special case of (\ref{C-2-complexity-pure}),
that is written in terms of the matrix $W_{\textrm{\tiny TR}}$ introduced in (\ref{W_TR def}).
For the pure states that have been chosen, whose covariance matrices have the form (\ref{williamson-def-gs}),
$W_{\textrm{\tiny R}}$ and $W_{\textrm{\tiny T}}$ can be obtained by specialising 
(\ref{W_C matrix def}) to the reference and the target states considered. 
Since the matrix $V$ is the same for both of them, (\ref{W_TR def}) simplifies to 
\be
\label{WTR_thermalstates}
W_{\textrm{\tiny TR}}
=
\mathcal{X}_{\textrm{\tiny C,T}} \, \mathcal{X}^{-1}_{\textrm{\tiny C,R}}
=
\textrm{diag}\!\left[\left(\frac{\Omega_{\textrm{\tiny T},1}}{\Omega_{\textrm{\tiny R},1}}\right)^{-1/2}\!\! \!,
\dots,\left(\frac{\Omega_{\textrm{\tiny T},L}}{\Omega_{\textrm{\tiny R},L}}\right)^{-1/2}\,\right]
\oplus \,
\textrm{diag}\!\left[\left(\frac{\Omega_{\textrm{\tiny T},1}}{\Omega_{\textrm{\tiny R},1}}\right)^{1/2}\!\!\!,
\dots,\left(\frac{\Omega_{\textrm{\tiny T},L}}{\Omega_{\textrm{\tiny R},L}}\right)^{1/2}\,\right]
\ee
where the last expression has been found by using (\ref{Williamson W for pure HC}).
Thus, in this case 
$W_{\textrm{\tiny TR}}=\mathcal{X}_{\textrm{\tiny TR}}=\mathcal{X}_{\textrm{\tiny C,T}} \, \mathcal{X}^{-1}_{\textrm{\tiny C,R}}$
and this leads to obtain (\ref{PureStateComplexityHC}) from (\ref{C-2-complexity-pure}).

In  the thermodynamic limit $L\to \infty$, 
the expression (\ref{PureStateComplexityHC}) for the complexity becomes \cite{Jefferson:2017sdb}
\be
\label{dFR pure thermo-c2}
\mathcal{C}_2
=
a(\tilde{\omega}_{\textrm{\tiny T}}, \tilde{\omega}_{\textrm{\tiny R}}) \,\sqrt{L}
+ \dots
\;\;\qquad\;\;
 \tilde{\omega}_{\textrm{\tiny R}} \neq   \tilde{\omega}_{\textrm{\tiny T}}
\;\;\;\qquad\;\;\;
L\to \infty
\ee
where the  subleading terms have been neglected 
and the coefficient of the leading term reads
\be
\label{dFR pure thermo}
a(\tilde{\omega}_{\textrm{\tiny T}}, \tilde{\omega}_{\textrm{\tiny R}}) 
\equiv
\frac{1}{2}
\;
\sqrt{ \frac{1}{\pi} \int_0^\pi \!\!\Big( \! \log\big[ \Omega_{\textrm{\tiny T},\theta} /  \Omega_{\textrm{\tiny R},\theta}\big]\Big)^2 d\theta}
\;\;\qquad\;\;
 \Omega_{\textrm{\tiny R}} \neq  \Omega_{\textrm{\tiny T}}
\ee
with
\be
\label{dispersion relation TD}
\Omega_{\textrm{\tiny S},\theta} \equiv \sqrt{\tilde{\omega}^2_{\textrm{\tiny S}} +4 \, (\sin\theta)^2}
\;\;\; \qquad\;\;\;
\theta \in (0,\pi)
\ee
which can be easily obtained from (\ref{dispersion relation TR}).
As consistency check, notice that $a(\tilde{\omega}_{\textrm{\tiny R}}, \tilde{\omega}_{\textrm{\tiny R}}) =0$, as expected. 
For  large $\omega_{\textrm{\tiny T}}$, the leading term of (\ref{dFR pure thermo}) is
\be
a(\tilde{\omega}_{\textrm{\tiny T}}, \tilde{\omega}_{\textrm{\tiny R}}) 
= \frac{1}{2}  \log  \tilde{\omega}_{\textrm{\tiny T}} + \dots
\;\;\qquad\;\;
\tilde{\omega}_{\textrm{\tiny T}} \to \infty\,.
\ee

\begin{figure}[t!]
\subfigure
{\hspace{-.8cm}
\includegraphics[width=.5\textwidth]{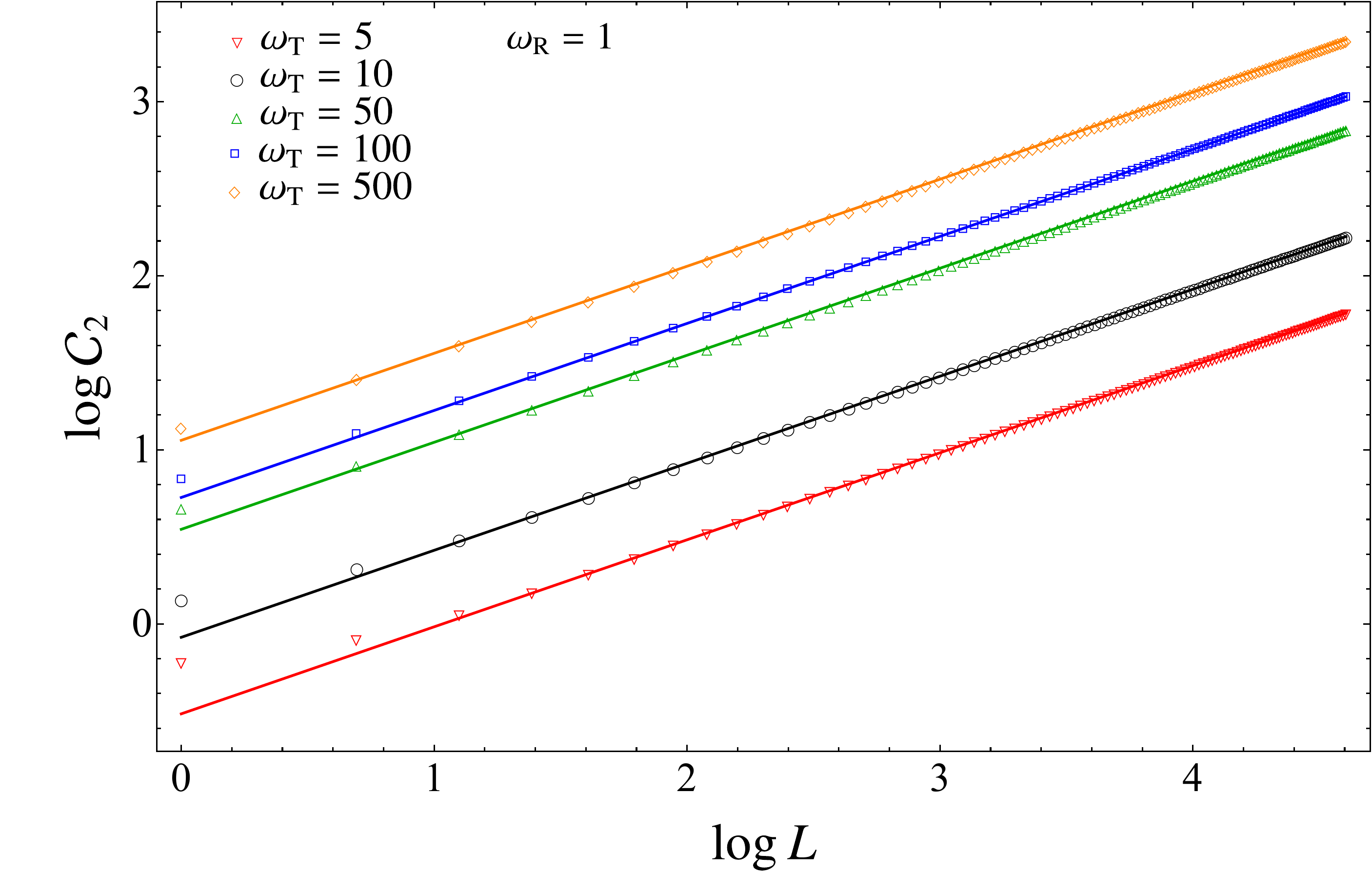}}
\subfigure
{
\hspace{.1cm}\includegraphics[width=.5\textwidth]{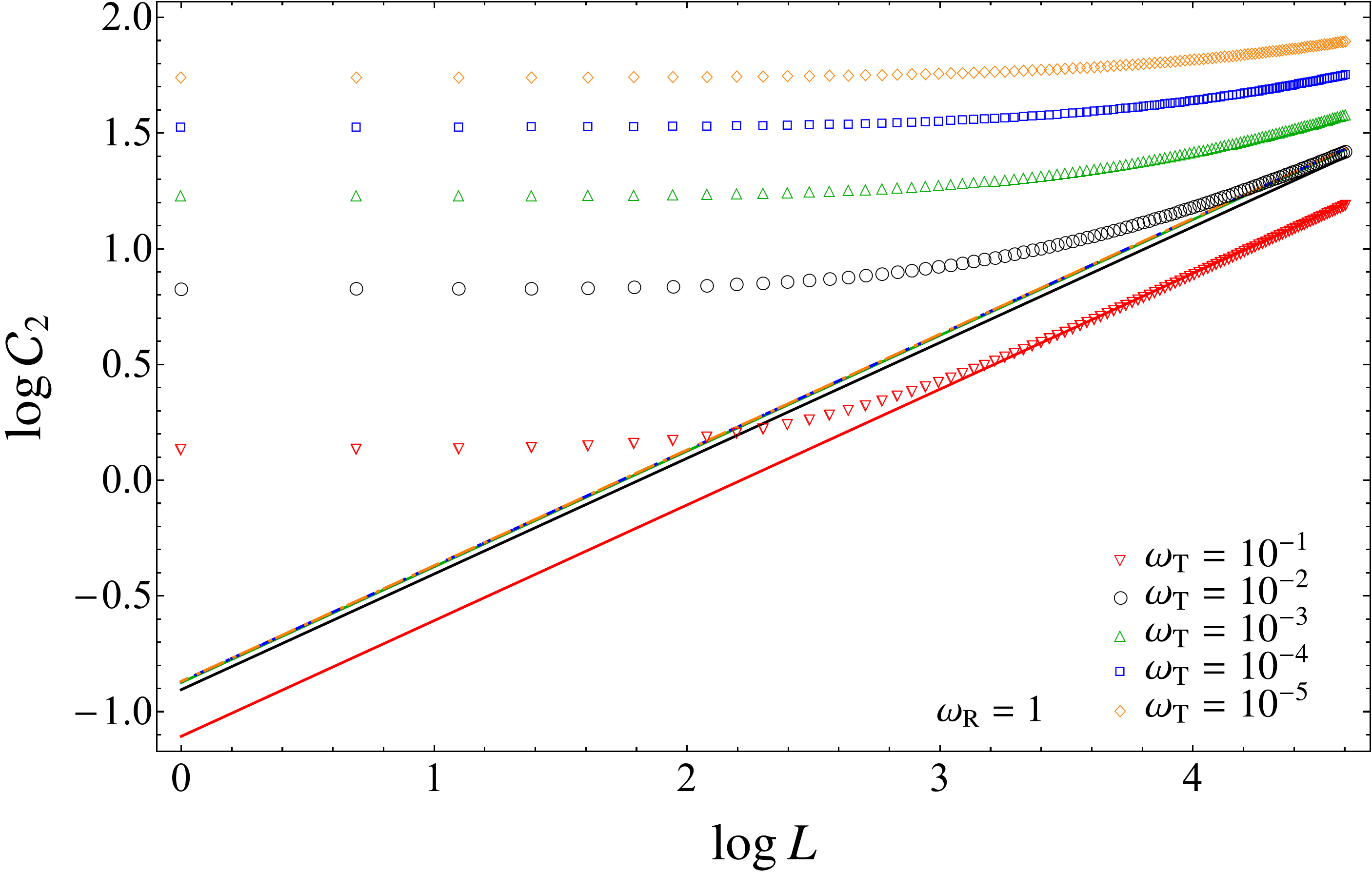}}
\subfigure
{
\hspace{-.8cm}\includegraphics[width=.5\textwidth]{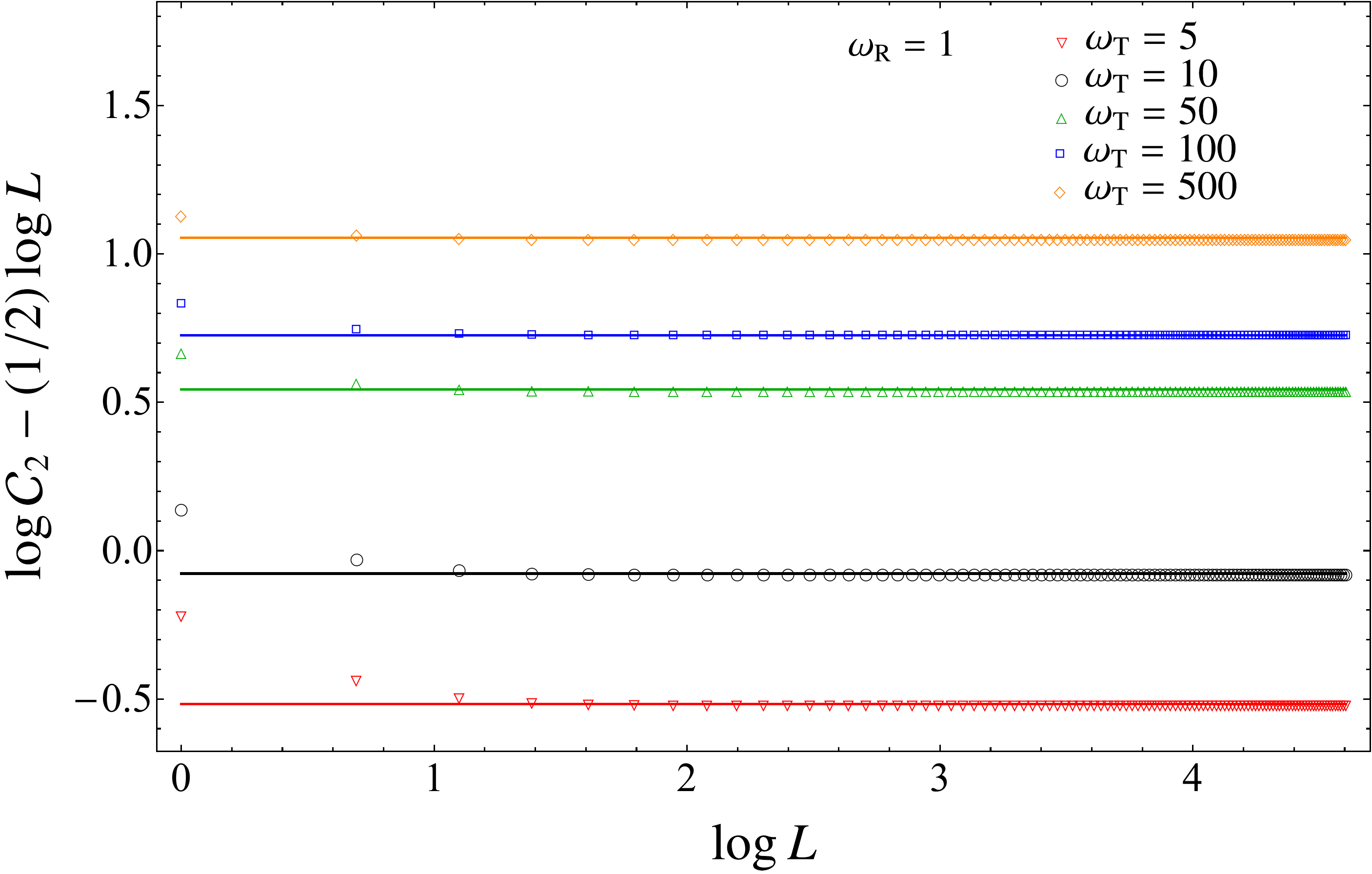}}
\subfigure
{\hspace{.45cm}
\includegraphics[width=.5\textwidth]{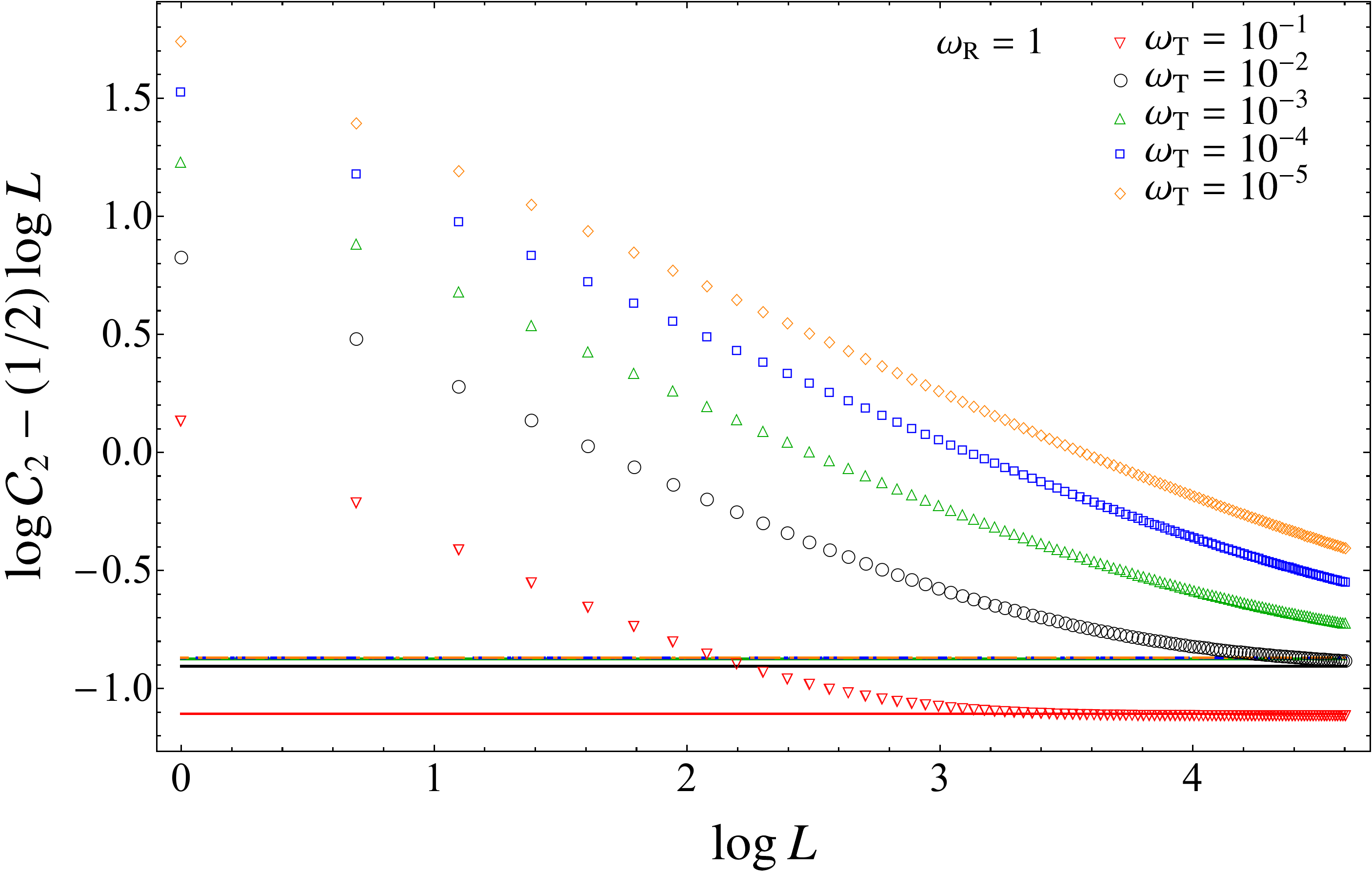}}
\caption{The complexity $\mathcal{C}_2$ in terms of the size $L$ of the periodic harmonic chain. 
The reference and the target states are the ground states with 
 $\omega = \omega_\textrm{\tiny R}$ and  $\omega = \omega_\textrm{\tiny T}$ respectively
 (here $\kappa=m=1$).
The data reported correspond to $\omega_\textrm{\tiny R}=1$ and different $\omega_\textrm{\tiny T}$.
The solid lines in the top panels represent (\ref{dFR pure thermo-c2}),
while the horizontal solid lines in the bottom panels correspond to the constant values 
of $a(\tilde{\omega}_{\textrm{\tiny T}}, \tilde{\omega}_{\textrm{\tiny R}}) $ 
obtained from (\ref{dFR pure thermo}).
} 
\vspace{0.4cm}
\label{fig:FRCompPureStatefunctionofL}
\end{figure}

We find it interesting to observe that, once the limit $L\to\infty$ has been taken, 
either $\tilde{\omega}_{\textrm{\tiny R}}$ or $\tilde{\omega}_{\textrm{\tiny T}}$ can be set to zero. 
For instance, setting $\tilde{\omega}_{\textrm{\tiny T}}=0$ in (\ref{dFR pure thermo}) gives
the following finite result
\be
\label{a-TD limit Massless}
a(\tilde{\omega}_{\textrm{\tiny R}}, \tilde{\omega}_{\textrm{\tiny T}}=0)^2
=
\frac{1}{16\pi} \int_0^\pi\!\!
\left( 
2 \log[\sin \theta] - \log\!\bigg[ \frac{\tilde{\omega}^2_{\textrm{\tiny R}} }{4 }+  (\sin\theta)^2 \bigg]
\,\right)^2
\! d\theta\,.
\ee
On the other hand, it is well known that the correlators $ \langle \hat{q}_i \hat{q}_j  \rangle $ in (\ref{corrs-periodic}) 
diverge when the frequency of the chain vanishes because of the occurrence of the zero mode; 
hence we cannot  evaluate $\mathcal{C}_2$ for a finite chain 
when either $\tilde{\omega}_{\textrm{\tiny T}} =0$ or $\tilde{\omega}_{\textrm{\tiny R}}=0$.
This tells us that the limits $L\to \infty$ and $\tilde{\omega}_{\textrm{\tiny T}} \to 0$ do not commute.

In Fig.\,\ref{fig:FRCompPureStatefunctionofL} we show the complexity $\mathcal{C}_2$ as function of  the size $L$ of the periodic chain.
The numerical results discussed in this manuscript have been obtained for $\kappa=1$ and $m=1$, unless otherwise specified;
hence $\tilde{\omega}_{\textrm{\tiny R}}=\omega_{\textrm{\tiny R}}$ 
and $\tilde{\omega}_{\textrm{\tiny T}}=\omega_{\textrm{\tiny T}}$.
In the left and right panels of Fig.\,\ref{fig:FRCompPureStatefunctionofL} 
we have $\omega_\textrm{\tiny T} > \omega_\textrm{\tiny R}$
and $\omega_\textrm{\tiny T} < \omega_\textrm{\tiny R}$ respectively
(notice that $\omega_\textrm{\tiny R}=1$ for all the panels).
In the top panels the numerical data are compared against the expression (\ref{dFR pure thermo-c2}) (solid lines) obtained in the thermodynamic limit: 
while in the top left panel the agreement is very good at large $L$,  
from the top right panel we conclude that larger values of $L$ 
are needed to observe a reasonable agreement as $\tilde{\omega}_{\textrm{\tiny T}} \to 0$.
In the bottom panels of Fig.\,\ref{fig:FRCompPureStatefunctionofL} we consider the subleading term in (\ref{dFR pure thermo-c2}):
while in the bottom left panel the data agree with the horizontal lines corresponding to 
$a(\tilde{\omega}_{\textrm{\tiny T}}, \tilde{\omega}_{\textrm{\tiny R}})$ given by (\ref{dFR pure thermo}), 
in the bottom right panel the agreement gets worse as $\tilde{\omega}_{\textrm{\tiny T}} \to 0$.
Notice that the solid lines in the right panels of Fig.\,\ref{fig:FRCompPureStatefunctionofL} accumulate on a limiting line as $\omega_{\textrm{\tiny T}}\to 0$.
This line can be found  by plugging (\ref{a-TD limit Massless}) with $\omega_\textrm{\tiny R}=1$ into (\ref{dFR pure thermo-c2}).

\begin{figure}[t!]
\subfigure
{\hspace{-.8cm}
\includegraphics[width=.51\textwidth]{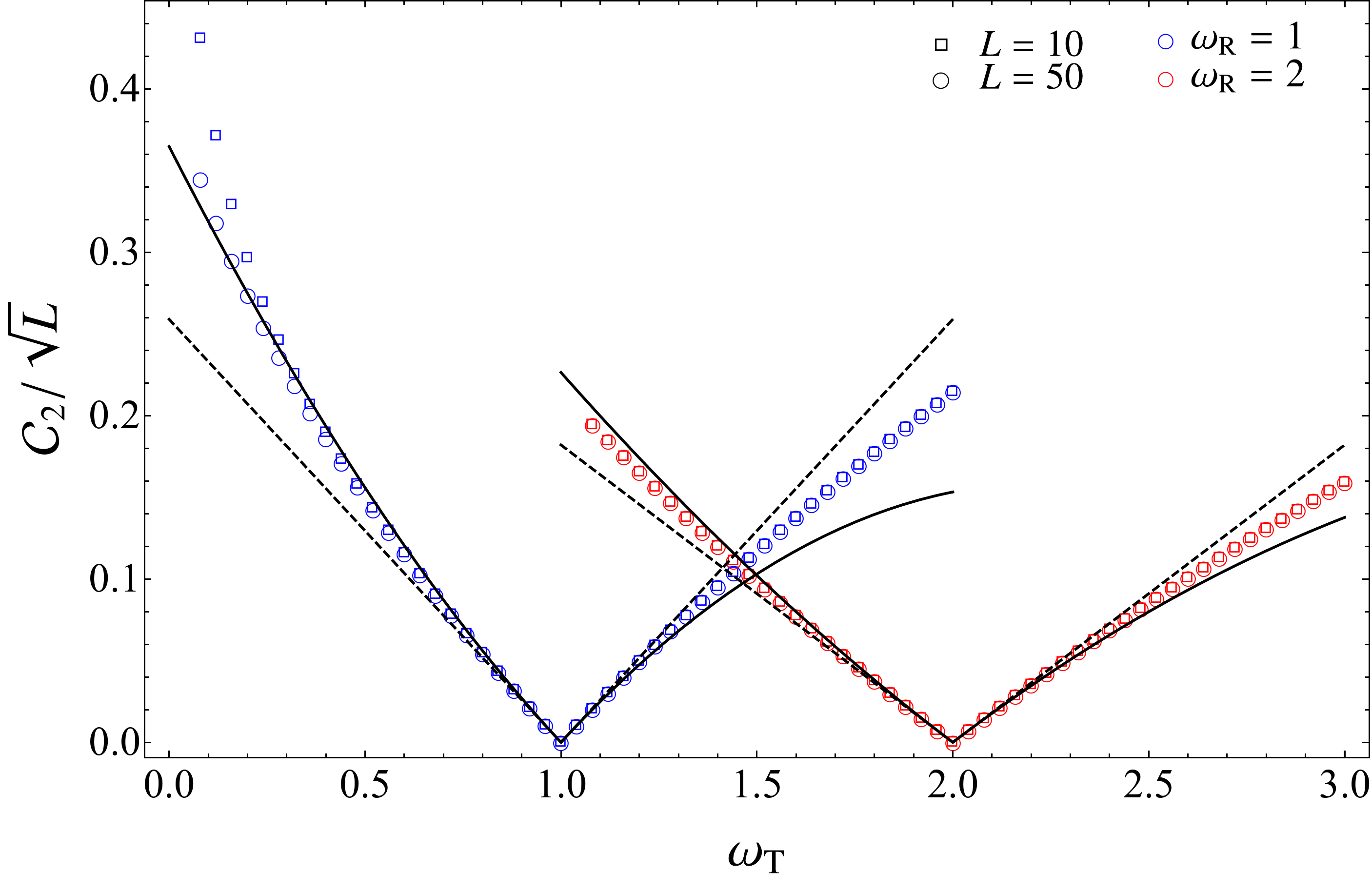}}
\subfigure
{
\hspace{.1cm}\includegraphics[width=.51\textwidth]{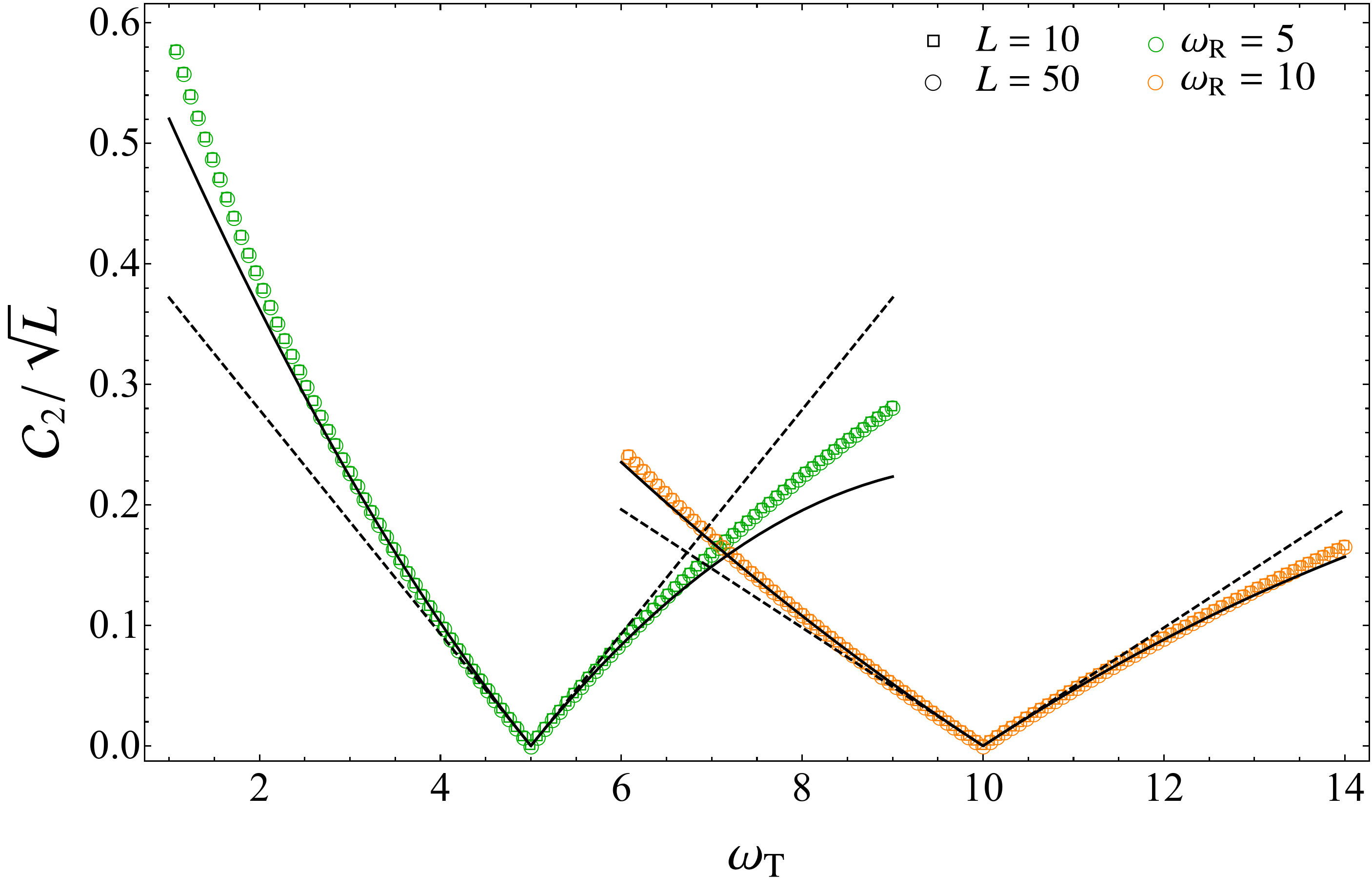}}
\caption{
The complexity $\mathcal{C}_2$ between the ground states of harmonic chains with $\kappa=m=1$
and different frequencies, for a given $\omega_\textrm{\tiny R}$ and as function of $\omega_\textrm{\tiny T}$.
The data points come from (\ref{PureStateComplexityHC}).
The dashed lines correspond to the first order approximation (\ref{complexitypure_smalldeltaomega}) 
and the solid lines to the  second order approximation (\ref{complexitypure_smalldeltaomega_2ndorder}), in the thermodynamic limit
and in the expansion where $\tilde{\omega}_{\textrm{\tiny T}} = \tilde{\omega}_{\textrm{\tiny R}} + \delta \tilde{\omega} $.
} 
\vspace{0.4cm}
\label{fig:FRCompPureStatefunctionofomega}
\end{figure}

We find it worth considering a perturbative expansion of the complexity of these pure states 
when the target state is infinitesimally close to the reference state.
This means that $\tilde{\omega}_{\textrm{\tiny T}} = \tilde{\omega}_{\textrm{\tiny R}} + \delta \tilde{\omega} $ 
with $ | \delta \tilde{\omega} | \ll \tilde{\omega}_{\textrm{\tiny R}}$.
Assuming $\tilde{\omega}_{\textrm{\tiny R}} \neq 0$, we expand (\ref{dFR pure thermo}) as $\delta \tilde{\omega}/\tilde{\omega}_{\textrm{\tiny R}} \to 0$.
The first order of this expansion gives
\be
\label{complexitypure_smalldeltaomega}
\mathcal{C}_2
=
\frac{\sqrt{L}\,  |\delta \tilde{\omega} |}{2\tilde{\omega}_{\textrm{\tiny R}}} \,
\sqrt{ \frac{1}{\pi}
\int_0^\pi  \left[ \,1 + \frac{4 }{ \tilde{\omega}^2_{\textrm{\tiny R}}} \, (\sin\theta)^2\,\right]^2 \!\! d\theta  \; }
=
 \frac{|\delta \tilde{\omega}|}{ 2\,\tilde{\omega}_{\textrm{\tiny R}}} \;
\sqrt{ \frac{1 + 2/\tilde{\omega}^2_{\textrm{\tiny R}}}{(1+ 4/\tilde{\omega}^2_{\textrm{\tiny R}}\,)^{3/2}}
}
\; \sqrt{L}\,.
\ee
Including also the $O((\delta \tilde{\omega})^2)$ term in the expansion of (\ref{dFR pure thermo}), we find 
\be
\label{complexitypure_smalldeltaomega_2ndorder}
\mathcal{C}_2
\,=\,
 \frac{|\delta \tilde{\omega}|}{ 2\,\tilde{\omega}_{\textrm{\tiny R}}} \;
\sqrt{ \frac{1 + 2/\tilde{\omega}^2_{\textrm{\tiny R}}}{(1+ 4/\tilde{\omega}^2_{\textrm{\tiny R}}\,)^{3/2}}
}\,
\left[\,
 1+\frac{\delta\tilde{\omega}}{2 \tilde{\omega}_{\textrm{\tiny R}}}
 \bigg(
 1-\frac{\tilde{\omega}_{\textrm{\tiny R}}^4+4\tilde{\omega}_{\textrm{\tiny R}}^2+6}{4 \tilde{\omega}_{\textrm{\tiny R}}^3 (\tilde{\omega}_{\textrm{\tiny R}}^2+2)(\tilde{\omega}_{\textrm{\tiny R}}^2+4)}
 \bigg) \right]
 \sqrt{L}
\ee

In Fig.\,\ref{fig:FRCompPureStatefunctionofomega} we compare the exact formula (\ref{PureStateComplexityHC}) for finite $L$
against the first order result (\ref{complexitypure_smalldeltaomega}) (dashed lines) 
and against (\ref{complexitypure_smalldeltaomega_2ndorder}), that includes also the second order correction (solid lines).

\subsection{Thermal states}
\label{sec-thermal-states-examples}

The thermal states are the most natural mixed states to consider.
In the following we evaluate the complexity (\ref{c2 complexity})
when both the reference and the target states are thermal states of the harmonic chain.

\subsubsection{Covariance matrix}
\label{cov-mat-thermal-phc}

The two-point correlators of a periodic chain in a thermal state at temperature $T$ read
\bea
\label{corrs-periodic-thermal}
 \langle \hat{q}_i \hat{q}_j  \rangle &=&
\frac{1}{2 L } \sum_{k=1}^{L} \frac{\coth[\Omega_k /(2 \widetilde{T})]}{ \Omega_{k}} \, \cos[2 \pi k\,(i-j)/L] 
 \\
\langle \hat{p}_i \hat{p}_j  \rangle &=&
\frac{1}{2 L} \sum_{k=1}^{L}  \Omega_{k} \coth[\Omega_k /(2 \widetilde{T})] \, \cos[2 \pi k\,(i-j)/L]
\label{corrs-periodic-thermalv2}
\eea
where $\Omega_k$ is given by (\ref{dispersion relation}) and we have introduced the dimensionless parameter
\be
\label{T_tilde}
\widetilde{T} \equiv \frac{T}{\sqrt{\kappa/m}}\,.
\ee
The correlators (\ref{corrs-periodic-thermal}) and (\ref{corrs-periodic-thermalv2}) 
provide the generic elements of the correlation matrices $Q_{\textrm{\tiny th}}$ and $P_{\textrm{\tiny th}}$ respectively,
which are the non vanishing blocks of the covariance matrix $\gamma_{\textrm{\tiny th}} = Q_{\textrm{\tiny th}} \oplus P_{\textrm{\tiny th}}$
of the thermal state.

Following the standard procedure, also for the thermal state one first performs the discrete Fourier transform 
through the matrix $\widetilde{U}$ in (\ref{calU matrix pbc}),
finding that (\ref{matrix FT}) can be written also for
the correlation matrices $Q_{\textrm{\tiny th}}$ and $P_{\textrm{\tiny th}}$, i.e.
\be
\label{matrix FT-th}
Q_{\textrm{\tiny th}}\,=\,\widetilde{U} \,\mathcal{Q}_{\textrm{\tiny th}} \;\widetilde{U}^{-1}
\qquad
P_{\textrm{\tiny th}}\,=\,\widetilde{U} \,\mathcal{P}_{\textrm{\tiny th}} \; \widetilde{U}^{-1}
\ee
with the proper diagonal matrices $\mathcal{Q}_{\textrm{\tiny th}}$ and $\mathcal{P}_{\textrm{\tiny th}}$,
whose elements are given respectively by
 \cite{AltlandSimons10book}
\be
\label{diag-corr-FT-th}
(\mathcal{Q}_{\textrm{\tiny th}})_{k,k}= \frac{\coth\!\big[\Omega_k /(2 \widetilde{T})\big]}{2  \, \Omega_{k}}
\;\;\qquad\;\;
(\mathcal{P}_{\textrm{\tiny th}})_{k,k}=\frac{1}{2}\;
\Omega_{k} \coth\!\big[\Omega_k /(2 \widetilde{T})\big]
\;\;\qquad\;\;
k=1,\dots,L
\ee
which reduce to (\ref{diag-corr-FT}) when $\widetilde{T}\to 0$, as expected. 

By employing the results obtained in Sec.\,\ref{sec-pure-states-examples} for the covariance matrix of the ground state,
it is not difficult to find that the Williamson's decomposition of the covariance matrix $\gamma_{\textrm{\tiny th}}$ 
of the thermal state reads
\be
\label{williamson-dec-hc-thermal}
\gamma_{\textrm{\tiny th}}
=
W^{\textrm{t}}_{\textrm{\tiny C}} \, \mathcal{D}_\textrm{\tiny th} \,W_{\textrm{\tiny C}}\,.
\ee
The matrix $W_{\textrm{\tiny C}} $ is the 
symplectic matrix (\ref{W_C matrix def}) occurring in
the Williamson's decomposition (\ref{williamson-def-gs}) of the ground state
and $\mathcal{D}_\textrm{\tiny th} \equiv \textrm{diag} (\sigma_{\textrm{\tiny th},1} , \dots , \sigma_{\textrm{\tiny th},L}) 
 \oplus \textrm{diag} (\sigma_{\textrm{\tiny th},1} , \dots , \sigma_{\textrm{\tiny th},L})$, 
 with the symplectic eigenvalues given by \cite{Adesso14}
\be
\label{symp-spec-thermal}
\sigma_{\textrm{\tiny th},k} 
=
\frac{1}{2}\, \coth\!\big[\Omega_k/(2\widetilde{T})\big]
=
\frac{1}{2}\, \coth\!\big[ \sigma_{\textrm{\tiny phys},k} /(2T)\big]
\;\;\qquad\;\; k=1,\dots,L \,.
\ee
From these observations, we find that
\be
\label{QP-diag-V-th}
\gamma_{\textrm{\tiny th}} 
= 
V \big( \mathcal{Q}_{\textrm{\tiny th}} \oplus \mathcal{P}_{\textrm{\tiny th}} \big) \,V^{\textrm{t}}
= 
V \big( \mathcal{Q}_{\textrm{\tiny th}} \oplus \mathcal{P}_{\textrm{\tiny th}} \big) \,V^{-1}
\ee
where $V$ is the same symplectic and orthogonal matrix introduced through (\ref{Vtilde-def-even}) and (\ref{Vtilde-def-odd}) 
for the ground state. 
It is straightforward to check that (\ref{QP-diag-V-th}) becomes (\ref{QP-diag-V-gs}) as $T \to 0$, as expected.

\subsubsection{Complexity}
\label{subsec:comp_thermal_HC_Ex}

In order to explore the complexity of two thermal states of a  periodic chain, 
let us consider a reference state characterised by
frequency $\omega_{\textrm{\tiny R}}$ and temperature $T_{\textrm{\tiny R}}$ 
and a target state characterised by frequency $\omega_{\textrm{\tiny T}}$ and temperature $T_{\textrm{\tiny T}}$,
assuming again that $\kappa_\textrm{\tiny R}=\kappa_\textrm{\tiny T}=\kappa$ and $m_\textrm{\tiny R}=m_\textrm{\tiny T}=m$.
Like (\ref{QP-diag-V-gs}), 
we have that also (\ref{DeltaTR-gs-diag}) can be adapted to this case, 
simply by replacing $\mathcal{Q}_\textrm{\tiny gs,M}$ with $\mathcal{Q}_\textrm{\tiny th,M}$
and $\mathcal{P}_\textrm{\tiny gs,M}$ with $\mathcal{P}_\textrm{\tiny th,M}$ taken from (\ref{diag-corr-FT-th}),
with $\textrm{M} \in \{ \textrm{R}, \textrm{T} \}$.
Thus, the complexity given by (\ref{c2 complexity}) and (\ref{deltaAB}) 
for these thermals states becomes
\bea
\label{ThermalComplexityHC-0}
\mathcal{C}_2
&=& 
\frac{1}{2\sqrt{2}}\,
\sqrt{\,
\textrm{Tr}\, \Big\{ \big[\log \!\big(\mathcal{Q}_\textrm{\tiny th,T} \,\mathcal{Q}^{-1}_\textrm{\tiny th,R} 
\oplus \mathcal{P}_\textrm{\tiny th,T} \,\mathcal{P}^{-1}_\textrm{\tiny th,R} \big) \big]^2 \Big\} }
\\
\rule{0pt}{1cm}
\label{ThermalComplexityHC}
&=&
\frac{1}{2\sqrt{2}}\;\sqrt{\,\sum_{k=1}^L
\left\{
 \left[\, \log\!\left( \frac{ \Omega_{{\textrm{\tiny R}},{\textrm{\tiny T}},k}}
 { \Omega_{{\textrm{\tiny T}},{\textrm{\tiny R}},k}  }
 \right)
  \right]^2
\! \!+
\left[\, \log\!\left( \frac{ \Omega_{{\textrm{\tiny T}}, {\textrm{\tiny T}},k}  }
 { \Omega_{{\textrm{\tiny R}},{\textrm{\tiny R}},k}  }
 \right)
  \right]^2\,\right\}
  }
\eea
where we have introduced
\be
\label{OmegaMN-def}
\Omega_{{\textrm{\tiny M}},{\textrm{\tiny N}},k}
\,\equiv\,
\Omega_{{\textrm{\tiny M}},k} \coth\!\big(\Omega_{{\textrm{\tiny N}},k} /(2 \widetilde{T}_{\textrm{\tiny N}})\big) 
\;\;\qquad\;\;
\textrm{M}, \textrm{N} \in \big\{ \textrm{R} \, ,  \textrm{T} \big\}
\ee
with $\Omega_{{\textrm{\tiny S}},k}$ given in (\ref{dispersion relation TR}). 
Notice that $\Omega_{{\textrm{\tiny M}},{\textrm{\tiny N}},k} \to \Omega_{{\textrm{\tiny M}},k}$
as $T_{\textrm{\tiny N}} \to 0$;
hence in the limit given by $\widetilde{T}_{\textrm{\tiny R}} \to 0$ and $\widetilde{T}_{\textrm{\tiny T}} \to 0$, 
the expected expression (\ref{PureStateComplexityHC}) for pure states is recovered.
Notice that the complexity (\ref{ThermalComplexityHC}) depends on the dimensionless parameters 
$\tilde{\omega}_\textrm{\tiny R}$, $\tilde{\omega}_\textrm{\tiny T}$, $\tilde{T}_\textrm{\tiny R}$ and $\tilde{T}_\textrm{\tiny T}$.

As briefly discussed  in Sec.\,\ref{subsec-examples-hc}, 
in order to consider the unentangled product state as reference state,
we have to keep the hamiltonian in the form (\ref{HC ham-1d}) and generalise the above analyses 
by setting  $m_{\textrm{\tiny R}}= m_{\textrm{\tiny T}}=m $ and $\kappa_{\textrm{\tiny R}}\neq \kappa_{\textrm{\tiny T}}$. 
For thermal states, this slight generalisation leads to
\be 
\label{thermal complexity different kappa}
\mathcal{C}_2
\,=\,
\frac{1}{2\sqrt{2}}\;\sqrt{\,\sum_{k=1}^L
\left\{
 \left[\, \log\!\left( \frac{ \sigma_{\textrm{\tiny phys,R,T},k}}
 { \sigma_{\textrm{\tiny phys,T,R},k}  }
 \right)
  \right]^2
\! \!+
\left[\, \log\!\left( \frac{ \sigma_{\textrm{\tiny phys,T,T},k}  }
 { \sigma_{\textrm{\tiny phys,R,R},k}  }
 \right)
  \right]^2\,\right\}
  }
\ee
where $\sigma_{\textrm{\tiny phys,M,N},k}$ is defined in terms of $\sigma_{\textrm{\tiny phys,S},k} $ given in (\ref{symp-spectrum-phys}) as follows
\be
\label{OmegaMN-def new}
\sigma_{\textrm{\tiny phys,M,N},k}
\,\equiv\,
\sigma_{\textrm{\tiny phys,M},k} \coth\!\big(\sigma_{\textrm{\tiny phys,N},k} /(2 T_{\textrm{\tiny N}})\big) 
\;\;\qquad\;\;
\textrm{M}, \textrm{N} \in \big\{ \textrm{R} \, ,  \textrm{T} \big\}
\ee
The complexity between a thermal state at temperature $T_{\textrm{\tiny T}}$ and the unentangled product state
can be found by taking $\kappa_{\textrm{\tiny R}}\to 0$ and $T_{\textrm{\tiny R}}\to 0$ in (\ref{thermal complexity different kappa}).

In the special case where $\tilde{\omega}_\textrm{\tiny R}=\tilde{\omega}_\textrm{\tiny T}\equiv\tilde{\omega}$, 
the expression (\ref{ThermalComplexityHC}) simplifies to
\be
\label{ThermalComplexityHC-eqfreq}
\mathcal{C}_2
\,=\,
\frac{1}{2}\sqrt{\sum_{k=1}^L
 \left[\, \log\!\bigg( \frac{  \coth(\Omega_{k}/(2 \widetilde{T}_{\textrm{\tiny T}}))  }
 { \coth(\Omega_{k}/(2 \widetilde{T}_{\textrm{\tiny R}})) }
 \bigg)
  \right]^2
}
\! =\,
\frac{1}{2\sqrt{2}}\,\sqrt{\textrm{Tr}\left\lbrace\left[\log \left(   \mathcal{D}_\textrm{\tiny th,T} \,  \mathcal{D}_\textrm{\tiny th,R}^{-1}
 \right)
  \right]^2\right\rbrace}\,.
\ee
This result is consistent with 
the general expression (\ref{C-2-complexity}) for the complexity
in the special case where $W_{\textrm{\tiny TR}}=\boldsymbol{1}$.
This relation can be verified by setting 
$\tilde{\omega}_\textrm{\tiny R}=\tilde{\omega}_\textrm{\tiny T}$ 
in (\ref{WTR_thermalstates}).

In the special case of $\tilde{\omega}_\textrm{\tiny R}=\tilde{\omega}_\textrm{\tiny T}$, from (\ref{williamson-dec-hc-thermal}) and (\ref{W_C matrix def}) 
we have that $W_\textrm{\tiny T}=W_\textrm{\tiny R}$, hence (\ref{ThermalComplexityHC-eqfreq}) provides the length of the $W$ path
connecting the reference and the target state that we are considering (see Sec.\,\ref{subsec:spectrum-comp}).
Indeed it is proportional to the proposal (\ref{compl_same_W}) for the spectrum complexity.

As for the basis complexity, from (\ref{symp-spec-thermal}), one observes that,
for a generic number of modes larger than one, the requirement $\mathcal{D}_\textrm{\tiny R}=\mathcal{D}_\textrm{\tiny T}$ 
leads to $\tilde{\omega}_\textrm{\tiny R}=\tilde{\omega}_\textrm{\tiny T}$ and $\tilde{T}_\textrm{\tiny R}=\tilde{T}_\textrm{\tiny T}$.
This implies $W_\textrm{\tiny R}=W_\textrm{\tiny T}$, as just remarked above,
hence  $\gamma_\textrm{\tiny R}=\gamma_\textrm{\tiny T}$ and the basis complexity (\ref{compl_same_D}) vanishes for the thermal states.
Also the corresponding basis complexity (\ref{basis complexity tilded}) vanishes because it is bounded from above by (\ref{compl_same_D}).
We expect that this is a peculiar feature due to the simplicity of the model. 
Notice that, for pure states or one-mode mixed states, the constraint $\mathcal{D}_\textrm{\tiny R}=\mathcal{D}_\textrm{\tiny T}$ 
does not imply that $\gamma_\textrm{\tiny R}=\gamma_\textrm{\tiny T}$, hence a non vanishing basis complexity is obtained.

Another interesting special case to explore is given by a pure reference state,
i.e. $\widetilde{T}_\textrm{\tiny R}=0$. 
In this limit (\ref{ThermalComplexityHC}) becomes
\be
\label{thermal complexity from pure to th}
\mathcal{C}_2
\,=\,
\frac{1}{2\sqrt{2}}\;
\sqrt{\,\sum_{k=1}^L
 \left[\,
 \log\!\left( \frac{ \Omega_{{\textrm{\tiny R}},k}  }
 { \Omega_{{\textrm{\tiny T}},k}} \,\coth(\Omega_{{\textrm{\tiny T}},k}/(2 \widetilde{T}_\textrm{\tiny T}))
 \right)
  \right]^2
\! +
\left[\,
\log\!\left( \frac{ \Omega_{{\textrm{\tiny T}},k} }
 { \Omega_{{\textrm{\tiny R}},k} }\,\coth(\Omega_{{\textrm{\tiny T}},k} /(2 \widetilde{T}_\textrm{\tiny T}))
 \right)
  \right]^2}\,.
\ee
In the low temperature limit
(i.e. when $\widetilde{T}_\textrm{\tiny T} \ll \tilde{\omega}_{\textrm{\tiny R}},\tilde{\omega}_{\textrm{\tiny T}}$),
by using that $\coth x \simeq 1+2 e^{-2x}+\dots$ as $x\to\infty$ and that 
$\log(1+x)\simeq x +\dots$ as $x\to 0$, we find that (\ref{thermal complexity from pure to th}) simplifies to
\be
\label{thermal complexity small T}
\mathcal{C}_2
\,=\,
\frac{1}{2}\;
\sqrt{\,\sum_{k=1}^L
\Big\{
\big[ 
\log\left( \Omega_{{\textrm{\tiny R}},k} /\Omega_{{\textrm{\tiny T}},k}
 \right)
   \big]^2
 +4\, e^{-2\,\Omega_{{\textrm{\tiny T}},k}/\widetilde{T}_\textrm{\tiny T}}
 \Big\} }
\, +\dots
\ee
where the dots denote subleading terms.
This expansion tells us that the first correction to the pure state result (\ref{PureStateComplexityHC}) 
as $\widetilde{T}_\textrm{\tiny T}\to 0$  is exponentially small.
The high temperature regime corresponds to 
$\widetilde{T}_\textrm{\tiny R} \gg \tilde{\omega}_\textrm{\tiny R}$ 
and $\widetilde{T}_\textrm{\tiny T} \gg \tilde{\omega}_\textrm{\tiny T}$.
In this limit, by using  that $\coth x \simeq 1/x$ when $x\to 0$, 
we find that (\ref{ThermalComplexityHC}) becomes
\be
\mathcal{C}_2\,=\,
\frac{1}{2} \;
\sqrt{\,
\sum_{k=1}^L
\left\lbrace
\left[\,
\log \! \big(\widetilde{T}{_\textrm{\tiny T}}/\widetilde{T}{_\textrm{\tiny R}} \big) \right]^2\!
+ 2\,\Big[\log\!\big( \Omega_{{\textrm{\tiny R}},k} / \Omega_{{\textrm{\tiny T}},k} \big)  \Big]^2
\right\rbrace}\,.
\ee

In the thermodynamic limit $L \to \infty$, for the complexity (\ref{ThermalComplexityHC}) we find
\be
\label{Thermal complexity TD}
\mathcal{C}_2=
\sqrt{L}\; 
a\big(\tilde{\omega}_{\textrm{\tiny R}},\tilde{\omega}_{\textrm{\tiny T}},
\widetilde{T}_{\textrm{\tiny R}}, \widetilde{T}_{\textrm{\tiny T}}\big)
+ \dots\,
\ee
where
\be
\label{Thermal complexity TD a-function}
a(\tilde{\omega}_{\textrm{\tiny R}},\tilde{\omega}_{\textrm{\tiny T}},\widetilde{T}_{\textrm{\tiny R}}, \widetilde{T}_{\textrm{\tiny T}})
\,\equiv\,
\frac{1}{2\sqrt{2}}\;
\sqrt{\, 
\int_0^\pi
\!
\left\lbrace
 \left[\, \log\left( \frac{ \Omega_{{\textrm{\tiny R}},{\textrm{\tiny T}},\theta}}{ \Omega_{{\textrm{\tiny T}},{\textrm{\tiny R}},\theta}  }
 \right)
  \right]^2
\! +
\left[\, \log\left( \frac{ \Omega_{{\textrm{\tiny T}},{\textrm{\tiny T}},\theta}}{ \Omega_{{\textrm{\tiny R}},{\textrm{\tiny R}},\theta} }
 \right)
  \right]^2
  \right\rbrace
   \frac{d\theta}{\pi}
} 
\ee
and $\Omega_{{\textrm{\tiny M}},{\textrm{\tiny N}},\theta} \equiv
\Omega_{{\textrm{\tiny M}},\theta} \coth\!\big(\Omega_{{\textrm{\tiny N}},\theta} /(2 \widetilde{T}_{\textrm{\tiny N}})\big) $,
with $\textrm{M}, \textrm{N} \in \{ \textrm{R} \, ,  \textrm{T} \}$
(see (\ref{OmegaMN-def})), 
written in terms of the dispersion relation $\Omega_{{\textrm{\tiny S}},\theta}$ given by (\ref{dispersion relation TD}).
Notice that
$a(\tilde{\omega}_{\textrm{\tiny R}},\tilde{\omega}_{\textrm{\tiny T}},\widetilde{T}_{\textrm{\tiny R}}, \widetilde{T}_{\textrm{\tiny T}}) 
\to\, a(\tilde{\omega}_{\textrm{\tiny R}},\tilde{\omega}_{\textrm{\tiny T}})$ 
when $\widetilde{T}_{\textrm{\tiny R}} \to 0$ and $\widetilde{T}_{\textrm{\tiny T}} \to 0$,
where $a(\tilde{\omega}_{\textrm{\tiny R}},\tilde{\omega}_{\textrm{\tiny T}})$ 
has been defined in (\ref{dFR pure thermo}).

In Sec.\,\ref{sec-pure-states-examples} we have observed that the massless limit of 
the coefficient of the leading term of the complexity of pure states is finite (see (\ref{a-TD limit Massless})).
This happens also for thermal states. 
Indeed, by setting $\tilde{\omega}_{\textrm{\tiny R}}=\tilde{\omega}_{\textrm{\tiny T}}=0$ in (\ref{Thermal complexity TD a-function}), we find
\be
\label{Thermal complexity TD a-function massless}
 a\big(\tilde{\omega}_{\textrm{\tiny R}}=0,\tilde{\omega}_{\textrm{\tiny T}}=0,
 \widetilde{T}_{\textrm{\tiny R}}, \widetilde{T}_{\textrm{\tiny T}}\big)
\, =\,
 \frac{1}{2}\;
\sqrt{\, 
\int_0^\pi\!
 \left[\,
 \log\!\left( 
 \frac{ \coth( (\sin\theta)/\widetilde{T}_{\textrm{\tiny T}}) }
 { \coth((\sin\theta)/\widetilde{T}_{\textrm{\tiny R}}) }
 \right)
  \right]^2
\!  \frac{d\theta}{\pi}
 }\,.
\ee

\begin{figure}[t!]
\subfigure
{\hspace{-.8cm}
\includegraphics[width=.5\textwidth]{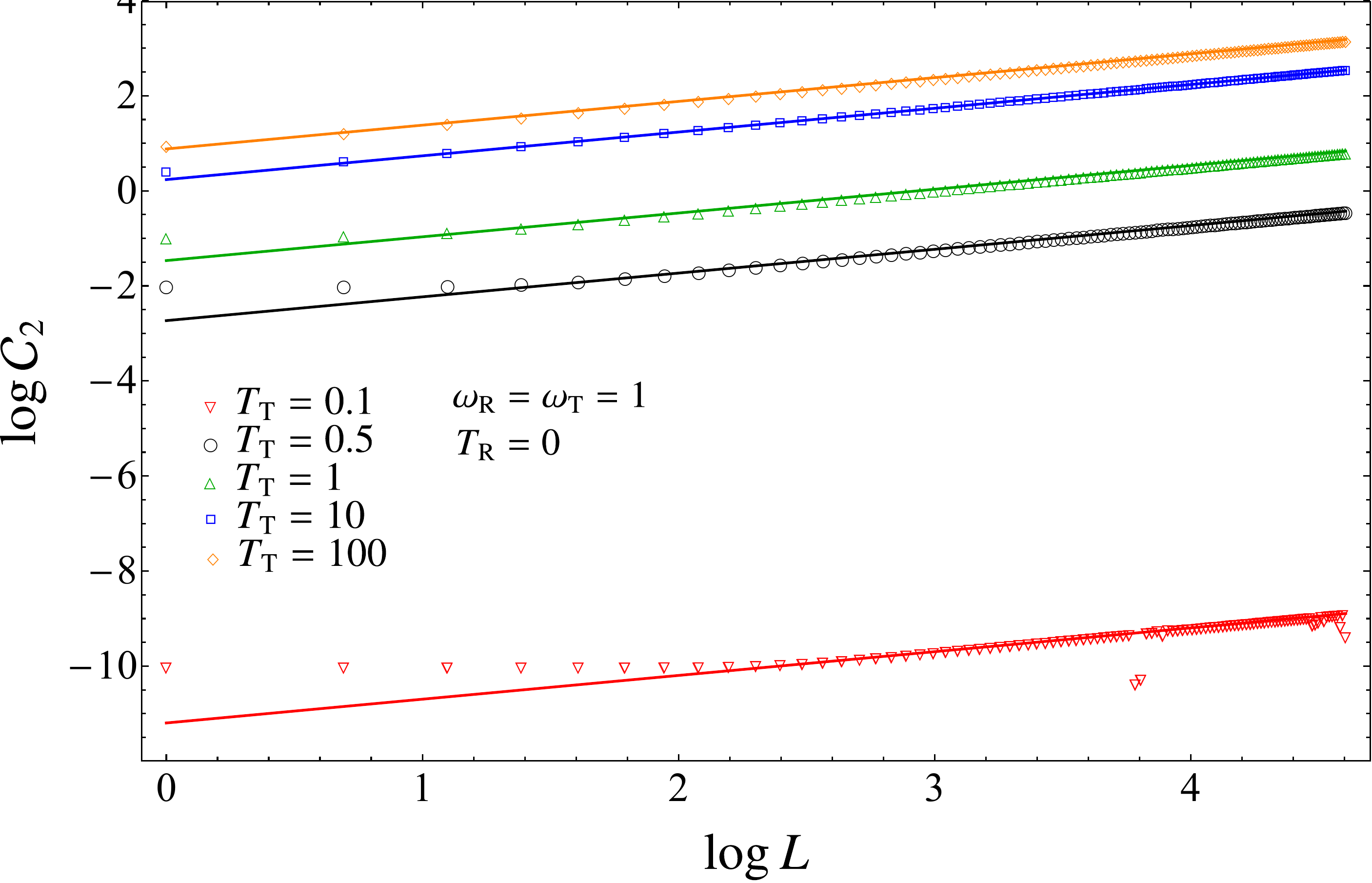}}
\subfigure
{
\hspace{.2cm}\includegraphics[width=.5\textwidth]{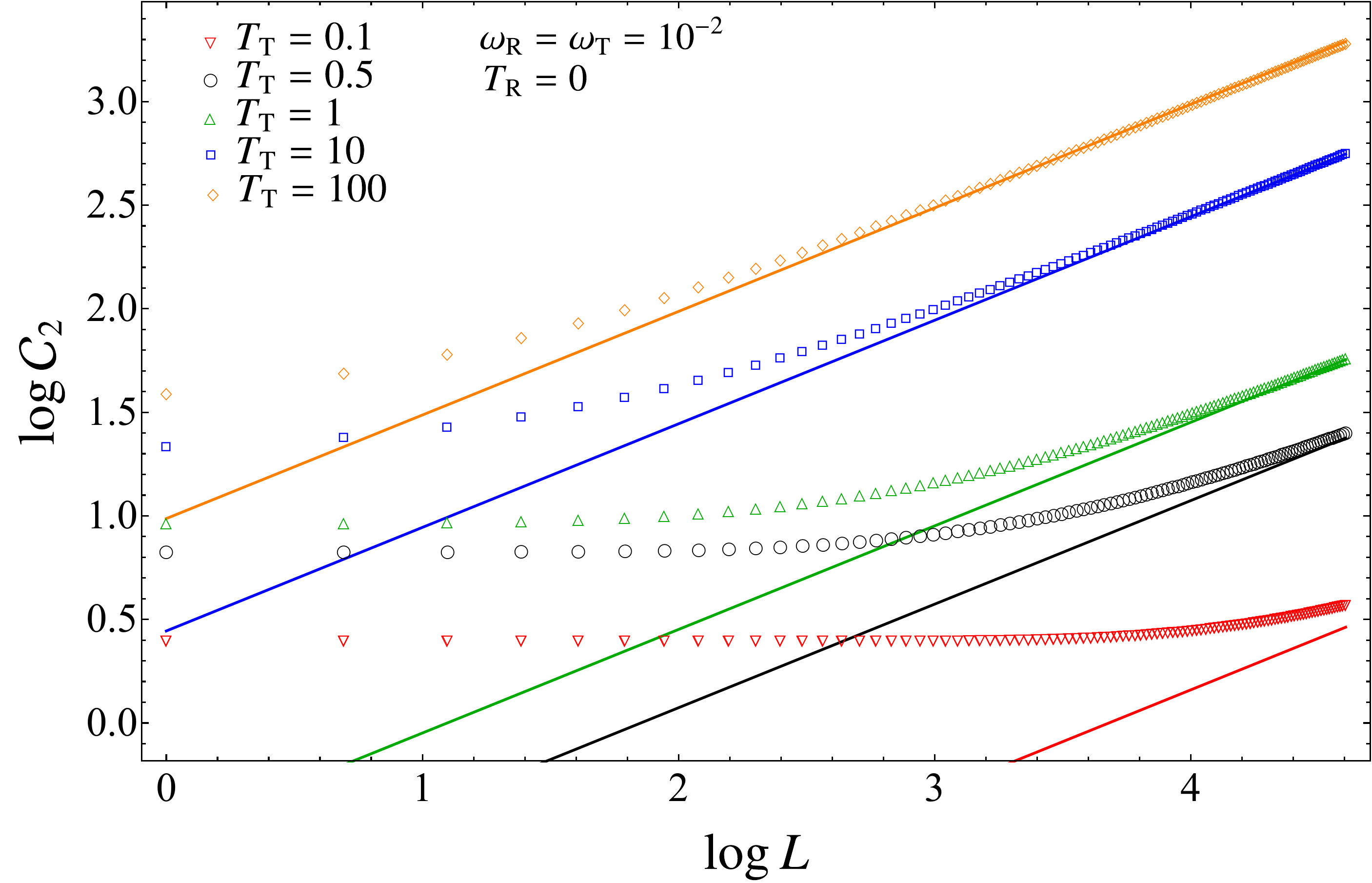}}
\subfigure
{
\hspace{-.8cm}\includegraphics[width=.5\textwidth]{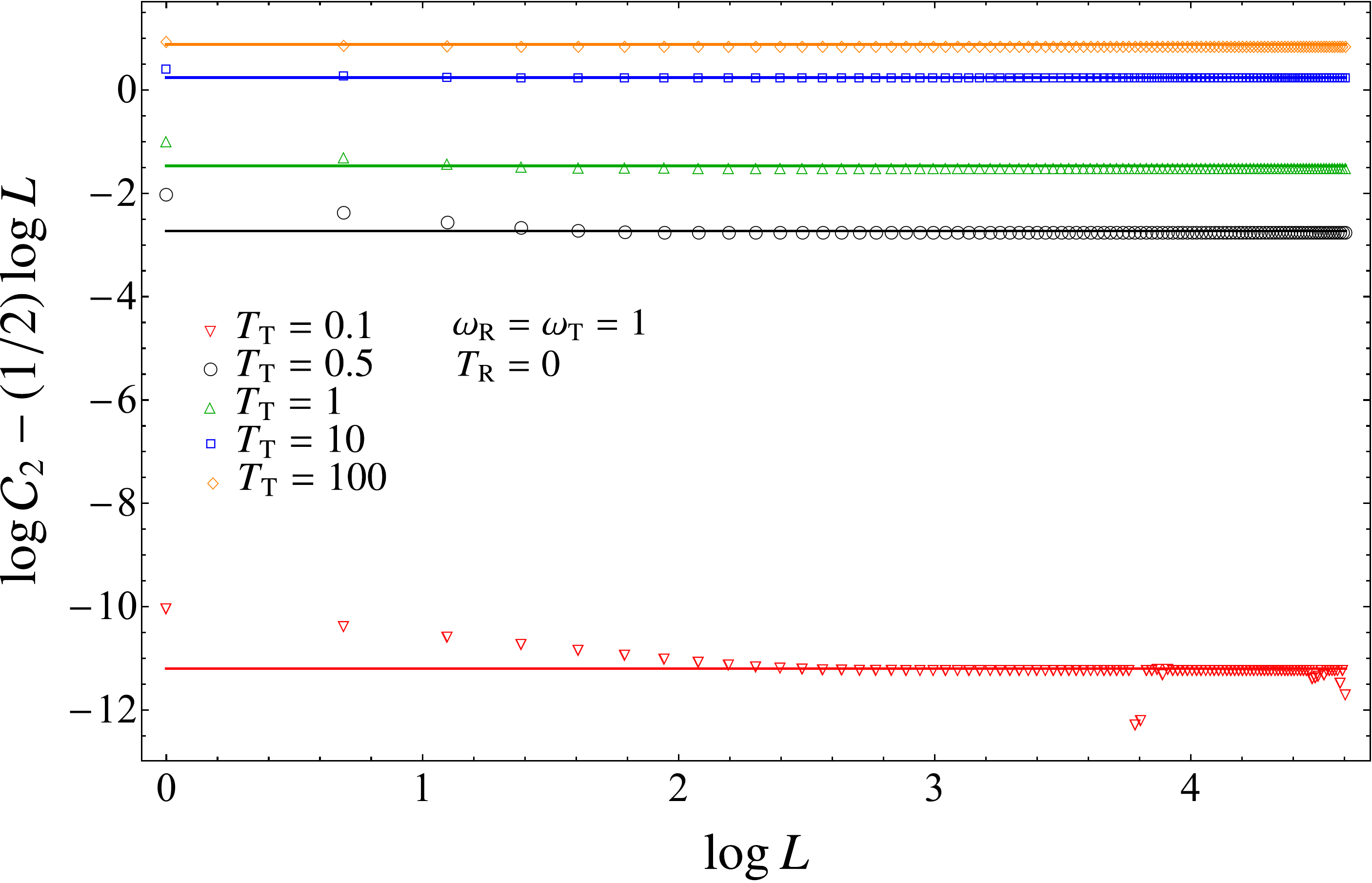}}
\subfigure
{\hspace{.45cm}
\includegraphics[width=.5\textwidth]{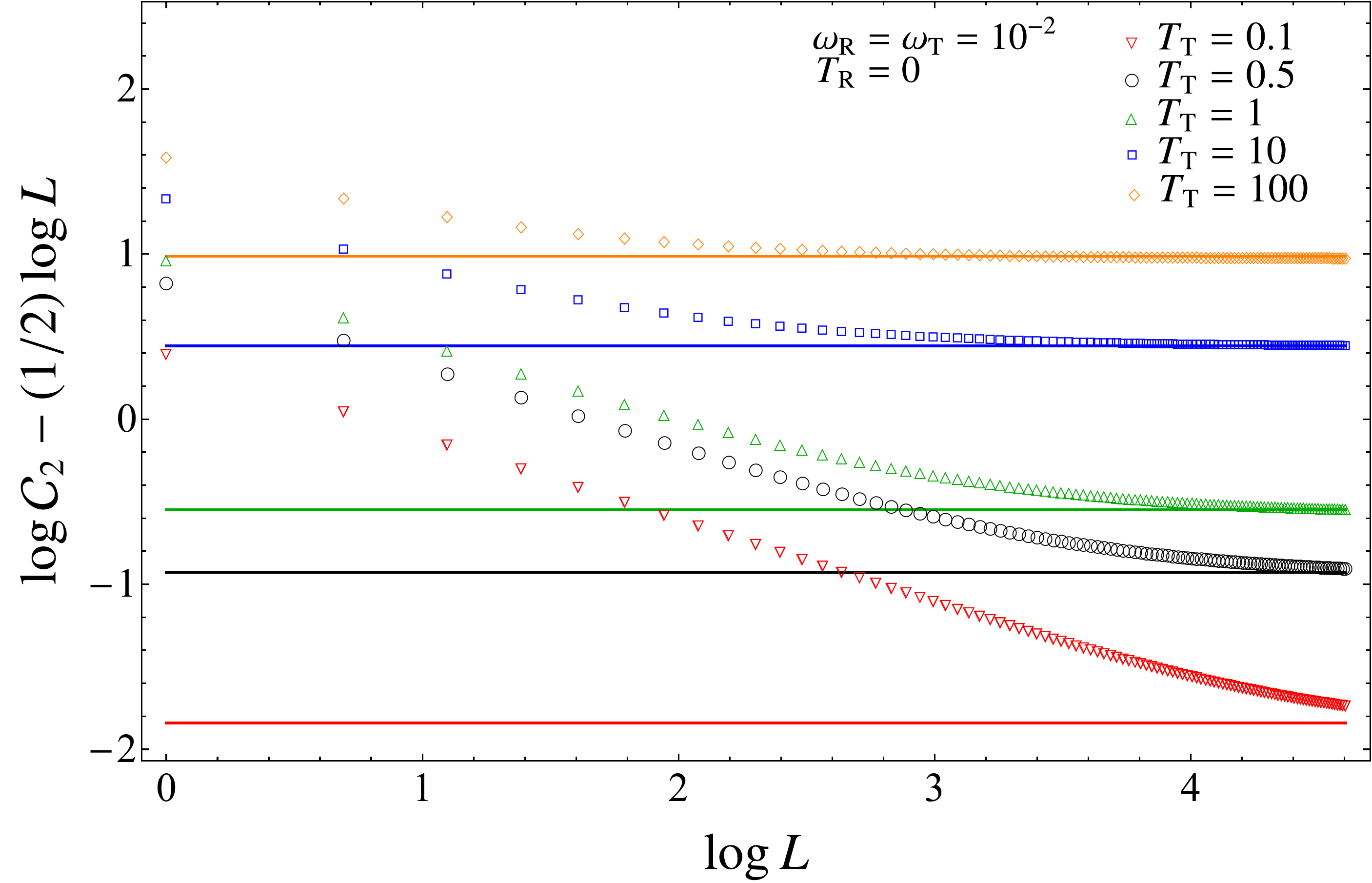}}
\caption{The complexity $\mathcal{C}_2$ for thermal states as function of the size $L$ of the periodic harmonic chain.
Here $\kappa=1$ and $m=1$; hence $\tilde{\omega}_\textrm{\tiny R}=\omega_\textrm{\tiny R}$, $\tilde{\omega}_\textrm{\tiny T}=\omega_\textrm{\tiny T}$, $\widetilde{T}_\textrm{\tiny R}=T_\textrm{\tiny R}$ and $\widetilde{T}_\textrm{\tiny T}=T_\textrm{\tiny T}$.
In all the panels $T_\textrm{\tiny R}=0$ and various values of $T_\textrm{\tiny T}$ are considered. 
Top panels: $\omega_\textrm{\tiny R}=\omega_\textrm{\tiny T}=1$ (left) and 
$\omega_\textrm{\tiny R}=\omega_\textrm{\tiny T}=10^{-2}$ (right). 
The solid lines correspond to (\ref{Thermal complexity TD}). 
Bottom panels: 
subleading term $\log\mathcal{C}_2-\tfrac{1}{2}\log L$ as function of $\log L$, 
for $T_\textrm{\tiny R}=0$  and various values of $T_\textrm{\tiny T}$. 
In the left panel $\omega_\textrm{\tiny R}=\omega_\textrm{\tiny T}=1$  
and $\omega_\textrm{\tiny R}=\omega_\textrm{\tiny T}=10^{-2}$  in the right panel.
The horizontal solid lines correspond to the constant values obtained from (\ref{Thermal complexity TD a-function}).
} 
\vspace{0.4cm}
\label{fig:FRCompThermalStatefunctionofL}
\end{figure}

\begin{figure}[t!]
\subfigure
{\hspace{-.8cm}
\includegraphics[width=.51\textwidth]{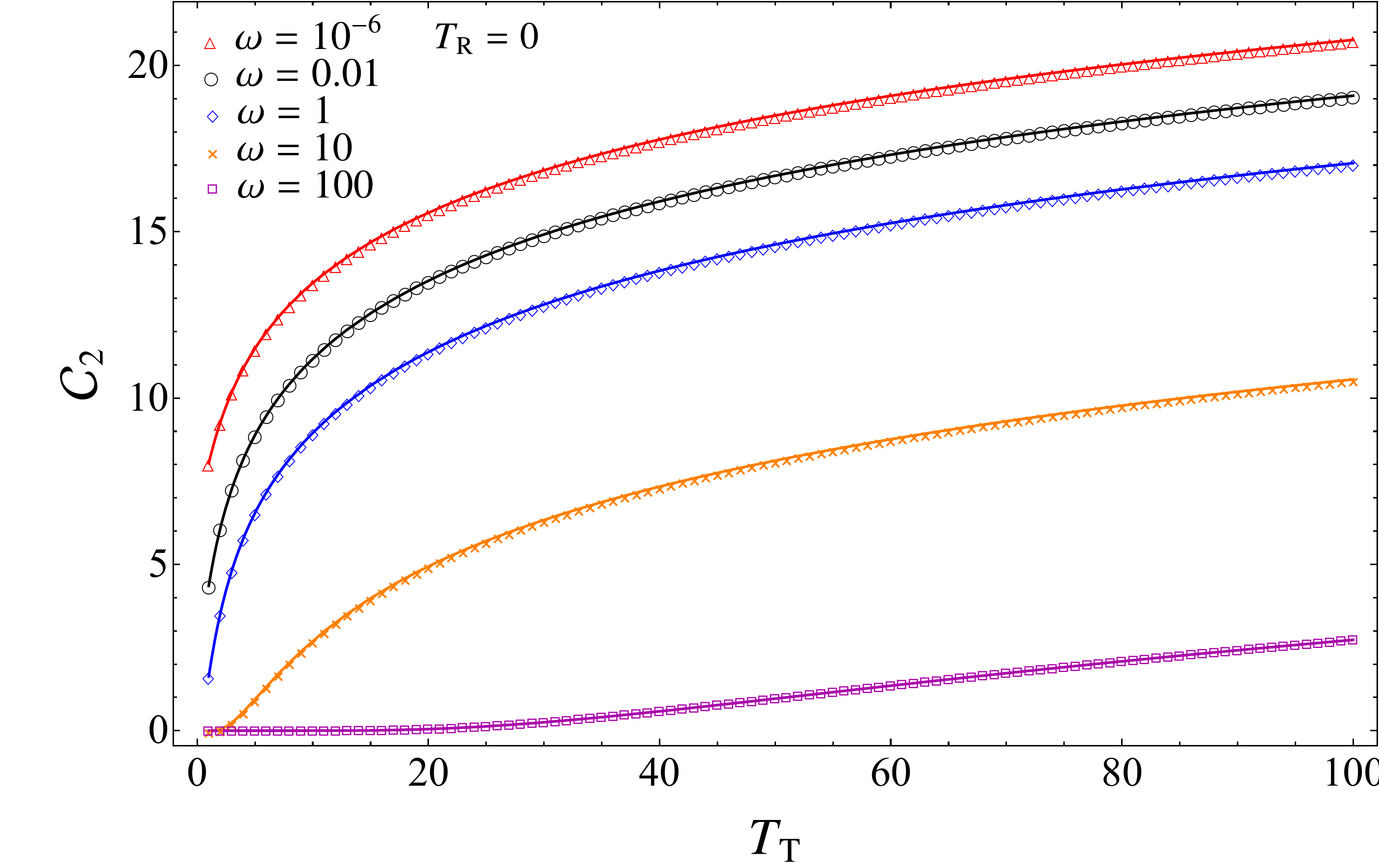}}
\subfigure
{
\hspace{.1cm}\includegraphics[width=.51\textwidth]{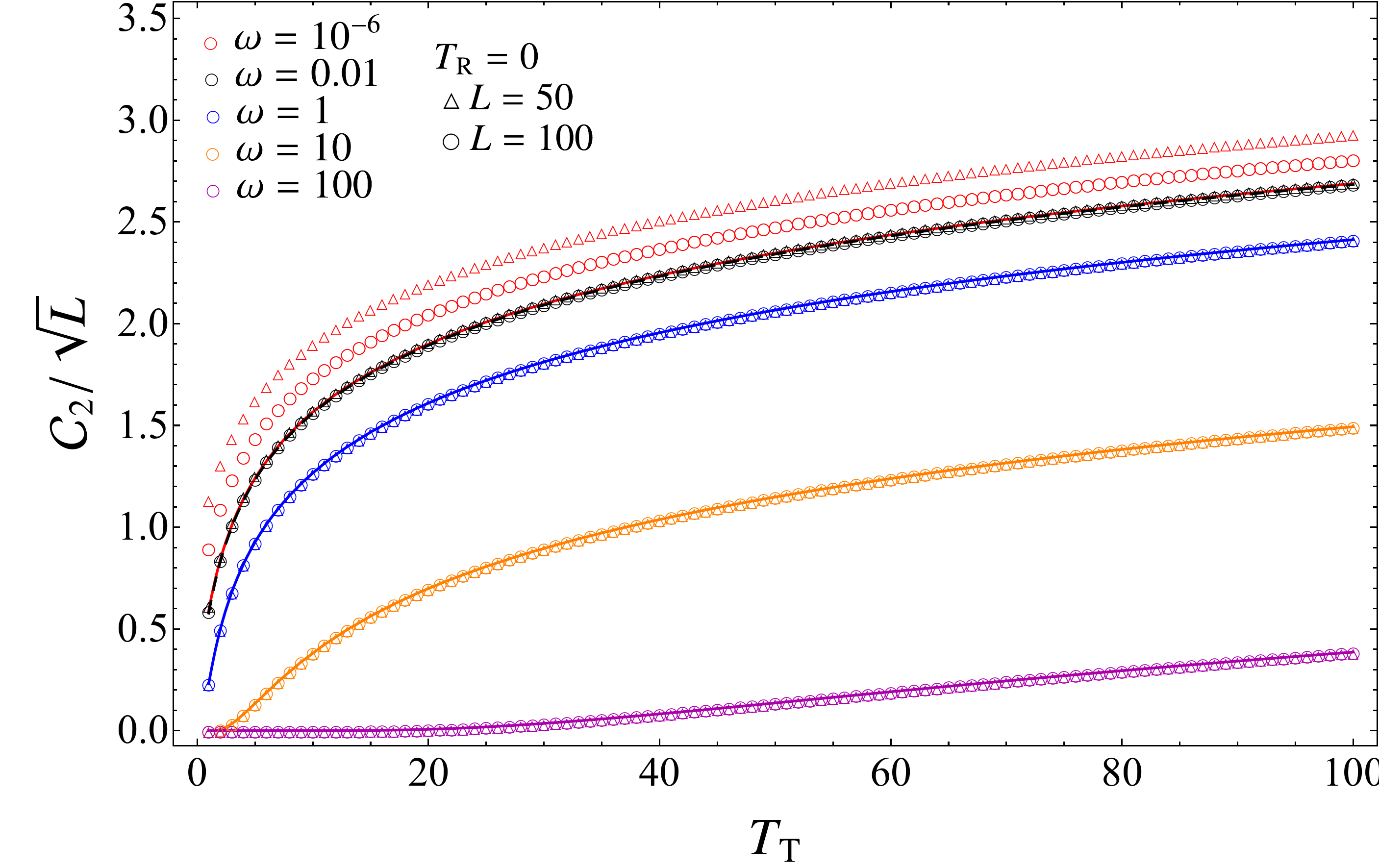}}
\caption{The complexity $\mathcal{C}_2$ for thermal states when $T_\textrm{\tiny R}=0$.
Left panel:
$\mathcal{C}_2$ for a chain of length $L=50$ as function of  $T_\textrm{\tiny T}$, 
reported for different values of $\omega_\textrm{\tiny R}=\omega_\textrm{\tiny T}=\omega$.
The solid lines correspond to (\ref{ThermalComplexityHC}). 
Right panel: $\mathcal{C}_2/\sqrt{L}$ as function of  $T_\textrm{\tiny T}$ for different values of 
$\omega_\textrm{\tiny R}=\omega_\textrm{\tiny T}=\omega$ and two values of $L$. 
The solid lines correspond to (\ref{Thermal complexity TD a-function}). 
The dashed black line and the solid red line are collapsed
on the curve (\ref{Thermal complexity TD a-function massless}).
} 
\vspace{0.4cm}
\label{fig:FRCompThermalStatefunctionofTT}
\end{figure}

In Fig.\,\ref{fig:FRCompThermalStatefunctionofL} and Fig.\,\ref{fig:FRCompThermalStatefunctionofTT} 
we report some numerical results for the complexity (\ref{c2 complexity}) between thermal states with different temperatures
and $\omega_\textrm{\tiny R}=\omega_\textrm{\tiny T}\equiv \omega$. 
The data have been taken for $\kappa=m=1$, 
hence $\widetilde{T}_{\textrm{\tiny R}}=T_{\textrm{\tiny R}}$ and $\widetilde{T}_{\textrm{\tiny T}}=T_{\textrm{\tiny T}}$.
 Notice that these numerical results 
display an example of spectrum complexity for thermal states, as discussed in the text below (\ref{ThermalComplexityHC-eqfreq}).

In Fig.\,\ref{fig:FRCompThermalStatefunctionofL} we consider the complexity 
as function of the length $L$ of the periodic harmonic chain. 
For the sake of simplicity, the reference state is the ground state (i.e. $T_\textrm{\tiny R}=0)$ and the target state is a thermal state with temperature $T_\textrm{\tiny T}$. 
In the left panels $\omega=1$, while in the right panels $\omega=10^{-2}$.
In the top panels the data are compared against the expression (\ref{Thermal complexity TD}) (solid lines), 
while in the bottom panels the subleading term of the same expression is considered (horizontal solid lines).
The data having $\omega=1$ agree very well with the predictions, 
while for the ones with $\omega \ll 1$ the agreement is worse because in these cases the values of $L$ considered are not large enough.

In Fig.\,\ref{fig:FRCompThermalStatefunctionofTT} the same quantity considered in Fig.\,\ref{fig:FRCompThermalStatefunctionofL}
is shown as function of $T_\textrm{\tiny T}$.
The increasing behaviour of the curves tells us that the distance between the states increases with $T_\textrm{\tiny T}$, as expected. 
In the left panel we test numerically the analytic expression (\ref{ThermalComplexityHC}) for $L=50$ and different values of $\omega$.
Instead, in the right panel we test numerically the formula (\ref{Thermal complexity TD a-function}), obtained in the thermodynamic limit $L \to \infty$:
the agreement  is very good when $\omega\gtrsim 1$, while it gets worse when $\omega\ll 1$.
Thus, when $\omega$ is very small,
larger values of $L$ should be explored to observe the expected agreement between 
the numerical data and the curve (\ref{Thermal complexity TD a-function}).
When $T_\textrm{\tiny R}=0$,
in the latter case the curves for (\ref{Thermal complexity TD a-function}) 
collapse onto the limiting curve (\ref{Thermal complexity TD a-function massless}), obtained by setting 
$\omega_\textrm{\tiny R}=\omega_\textrm{\tiny T}=0$.

\subsubsection{Optimal path for entanglement hamiltonians and its complexity}
\label{subsec-eh-num}

In Sec.\,\ref{sec:comp-eh} we have discussed the map that provides the entanglement hamiltonian 
in terms of the covariance matrix of a mixed state.
In the following we explore further the optimal path of entanglement hamiltonian matrices 
for the periodic harmonic chain in the special case where both the 
reference and the target states are thermal states.

The entanglement hamiltonian matrices  $H_{\textrm{\tiny R}}$ and $H_{\textrm{\tiny T}}$
of a reference state and of a target state that are both thermal
can be obtained by applying the map (\ref{EHfromgamma}) to the covariance matrix 
$\gamma_{\textrm{\tiny th}} = Q_{\textrm{\tiny th}} \oplus P_{\textrm{\tiny th}}$ introduced 
in Sec.\,\ref{cov-mat-thermal-phc}, whose Williamson's decomposition is  (\ref{williamson-dec-hc-thermal}).
The symplectic spectrum of the entanglement hamiltonian matrix of a thermal state can be easily obtained 
by plugging (\ref{symp-spec-thermal}) into (\ref{Ediag_Ddiag}), finding
\be
\label{EHspectrum_thermal}
\varepsilon_{\textrm{\tiny th},k}=\frac{\Omega_k}{\widetilde{T}}
\;\;\qquad\;\; k=1,\dots,L \,.
\ee
This provides the elements of the diagonal matrix $\mathcal{E}_\textrm{\tiny th}$ entering in 
the Williamson's decomposition (\ref{williamson-dec-H}) for the thermal state.
Comparing (\ref{EHspectrum_thermal}) with (\ref{T_tilde}) and (\ref{symp-spec-thermal}), 
we get $\varepsilon_{\textrm{\tiny th},k}=\beta \sigma_{\textrm{\tiny phys},k}$, 
as discussed in Sec.\,\ref{sec:comp-eh} for the thermal states in any number of dimensions. 

The distance (\ref{C-2-complexity-eh}) between $H_{\textrm{\tiny R}}$ and $H_{\textrm{\tiny T}}$
can be evaluated by employing (\ref{WTR_thermalstates}) and (\ref{EHspectrum_thermal}).
The result reads
\be
\label{distanceEHthermalgeneral}
d(H_{\textrm{\tiny R}}, H_{\textrm{\tiny T}})
=
\sqrt{\,\sum_{k=1}^L \left\lbrace 
\left[\, \log\!\left(\frac{T_\textrm{\tiny R}}{T_\textrm{\tiny T}}\right)\right]^2
+\left[\,\log\!\left(\frac{T_\textrm{\tiny R} \,\Omega_{\textrm{\tiny T},k}}{T_\textrm{\tiny T} \, \Omega_{\textrm{\tiny R},k}}\right)\right]^2 
\right\rbrace}
\ee
which can be obtained also by replacing 
$\mathcal{D}_\textrm{\tiny th}$ with $\mathcal{E}_\textrm{\tiny th}$  in (\ref{ThermalComplexityHC-eqfreq}).

In the special case of $\widetilde{\omega}_{\textrm{\tiny R}}=\widetilde{\omega}_{\textrm{\tiny T}}$, 
the summation in (\ref{distanceEHthermalgeneral}) can be easily performed, finding
\be
\label{FR-dist-HRT-same-omega}
d(H_{\textrm{\tiny R}}, H_{\textrm{\tiny T}})
=
\sqrt{2 L} \;\big| \log\!\big(\beta_\textrm{\tiny T}/\beta_\textrm{\tiny R}\big) \big|
\ee
which corresponds to (\ref{Complexity through EH Thermal}) specified to the 
one-dimensional harmonic chain. 
In this case we can employ the discussion made in Sec.\,\ref{sec:comp-eh}
for the cases where $W_{\textrm{\tiny TR}}=\boldsymbol{1}$ 
(see (\ref{WTR_thermalstates})) to conclude that 
$d(\gamma_\textrm{\tiny R}, \gamma_\textrm{\tiny T})\leqslant d(H_{\textrm{\tiny R}}, H_{\textrm{\tiny T}})$.

In the periodic harmonic chain the Williamson's decomposition of the optimal circuit 
connecting $H_{\textrm{\tiny R}}$ and $H_{\textrm{\tiny T}}$
is given by (\ref{optimal-circuit-th-general-williamson}), 
with  the symplectic eigenvalues (\ref{symp-spectrum-phys}) 
and the symplectic matrix (\ref{symp-W-phys}). 
Thus, the symplectic eigenvalues for the matrix labeled by $s\in[0,1]$ along this optimal circuit are
\be
\sigma_{k,s}=\beta_{\textrm{\tiny T}}^s \;\beta_{\textrm{\tiny R}}^{1-s}\,\sqrt{\kappa/m}\;\Omega_k
\ee
where $\Omega_k$ is the dispersion relation (\ref{dispersion relation}).
This means that the optimal circuit is made by the entanglement hamiltonian matrices of thermal states,
as also discussed in Sec.\,\ref{sec:comp-eh}. 
This is not a feature of the optimal circuit connecting the covariance matrices of two thermal states,
as discussed in Sec.\,\ref{subsec:thermal}.
This discrepancy is consistent with the fact that the map (\ref{EHfromgamma}) does not send geodesics into geodesics.

\subsection{Mutual complexity of TFD's}
\label{subsec:mutualcomp-TFD}

The thermofield double state (TFD) is a pure state obtained by entangling two equal copies of the harmonic lattice 
and such that a thermal state of the original system is obtained after the partial trace over one of the two copies.
A detailed analysis of the TFD and of the circuit complexity between two TFD's is reported in the Appendix\;\ref{app-tfd}.

It is worth comparing the circuit complexity of two thermal states with the one obtained from the corresponding TFD's.
Following \cite{Caceres:2019pgf}, we introduce the mutual complexity for the TFD's as 
\be
\label{mutualcompTFD}
\mathcal{M}_{\textrm{\tiny TFD}}\big(\tilde{\omega}_{\textrm{\tiny R}},\tilde{\omega}_{\textrm{\tiny T}},
\tilde{\beta}_{\textrm{\tiny R}}, \tilde{\beta}_{\textrm{\tiny T}}\big)=2\,\mathcal{C}^2_{\textrm{\tiny th}}\big(\tilde{\omega}_{\textrm{\tiny R}},\tilde{\omega}_{\textrm{\tiny T}},
\tilde{\beta}_{\textrm{\tiny R}}, \tilde{\beta}_{\textrm{\tiny T}}\big)-\mathcal{C}^2_{\textrm{\tiny TFD}}\big(\tilde{\omega}_{\textrm{\tiny R}},\tilde{\omega}_{\textrm{\tiny T}},
\tilde{\beta}_{\textrm{\tiny R}}, \tilde{\beta}_{\textrm{\tiny T}}\big)
\ee
where $\mathcal{C}_{\textrm{\tiny th}}$ and $\mathcal{C}_{\textrm{\tiny TFD}}$
are given by (\ref{ThermalComplexityHC}) and (\ref{ComplexityTFDHC}) respectively. 
More explicitly, (\ref{mutualcompTFD}) reads
\bea
\label{mutualcompTFDv2}
\mathcal{M}_{\textrm{\tiny TFD}}\big(\tilde{\omega}_{\textrm{\tiny R}},\tilde{\omega}_{\textrm{\tiny T}},
\tilde{\beta}_{\textrm{\tiny R}}, \tilde{\beta}_{\textrm{\tiny T}}\big) =
& &
\\
&&\hspace{-2.5cm}
=\;
\frac{1}{4}\sum_{k=1}^L\bigg\{\bigg[\log\bigg(\frac{\Omega_{\textrm{\tiny R},k}\coth\big(\tilde{\beta}_{\textrm{\tiny R}}\Omega_{\textrm{\tiny R},k}/2\big)}{\Omega_{\textrm{\tiny T},k} \coth\big(\tilde{\beta}_{\textrm{\tiny T}}\Omega_{\textrm{\tiny T},k}/2\big)}\bigg) \bigg]^2-\bigg[\log\bigg(\frac{\Omega_{\textrm{\tiny R},k}\coth\big(\tilde{\beta}_{\textrm{\tiny R}}\Omega_{\textrm{\tiny R},k}/4\big)}{\Omega_{\textrm{\tiny T},k} \coth\big(\tilde{\beta}_{\textrm{\tiny T}}\Omega_{\textrm{\tiny T},k}/4\big)}\bigg) \bigg]^2
\nonumber
\\
&&\hspace{-.7cm}
+\bigg[\log\bigg(\frac{\Omega_{\textrm{\tiny R},k}\coth\big(\tilde{\beta}_{\textrm{\tiny T}}\Omega_{\textrm{\tiny T},k}/2\big)}{\Omega_{\textrm{\tiny T},k}\coth\big(\tilde{\beta}_{\textrm{\tiny R}}\Omega_{\textrm{\tiny R},k}/2\big)}\bigg)\bigg]^2-
\bigg[\log\bigg(\frac{\Omega_{\textrm{\tiny R},k}\coth\big(\tilde{\beta}_{\textrm{\tiny T}}\Omega_{\textrm{\tiny T},k}/4\big)}{\Omega_{\textrm{\tiny T},k}\coth\big(\tilde{\beta}_{\textrm{\tiny R}}\Omega_{\textrm{\tiny R},k}/4\big)}\bigg)\bigg]^2
\bigg
\}
\nonumber
\eea
which can be written also in terms of $\Omega_{{\textrm{\tiny M}},{\textrm{\tiny N}},k}$ defined in (\ref{OmegaMN-def}) as follows
\bea
\mathcal{M}_{\textrm{\tiny TFD}}\big(\tilde{\omega}_{\textrm{\tiny R}},\tilde{\omega}_{\textrm{\tiny T}},
\tilde{\beta}_{\textrm{\tiny R}}, \tilde{\beta}_{\textrm{\tiny T}}\big)
&=&
\frac{1}{4}\sum_{k=1}^L
\Bigg\{
\bigg[\log\bigg(\frac{\Omega_{{\textrm{\tiny R}},{\textrm{\tiny R}},k}}{\Omega_{{\textrm{\tiny T}},{\textrm{\tiny T}},k}}\bigg) \bigg]^2
+
\bigg[\log\bigg(\frac{\Omega_{{\textrm{\tiny R}},{\textrm{\tiny T}},k}}{\Omega_{{\textrm{\tiny T}},{\textrm{\tiny R}},k}}\bigg)\bigg]^2
\\
\rule{0pt}{.7cm}
&& \hspace{-.4cm}
 -\,\bigg[\log\bigg(\frac{\Omega_{{\textrm{\tiny R}},{\textrm{\tiny R}},k}}{\Omega_{{\textrm{\tiny T}},{\textrm{\tiny T}},k}}\bigg)+\log\bigg(\frac{\cosh\big(\tilde{\beta}_{\textrm{\tiny T}}\Omega_{\textrm{\tiny T},k}/2\big)\big(\cosh\big(\tilde{\beta}_{\textrm{\tiny R}}\Omega_{\textrm{\tiny R},k}/2\big)-1\big)}{\cosh\big(\tilde{\beta}_{\textrm{\tiny R}}\Omega_{\textrm{\tiny R},k}/2\big)\big(\cosh\big(\tilde{\beta}_{\textrm{\tiny T}}\Omega_{\textrm{\tiny T},k}/2\big)-1\big)}\bigg) \bigg]^2
\nonumber
\\
\rule{0pt}{.8cm}
&& \hspace{-.4cm}
-\,
\bigg[\log\bigg(\frac{\Omega_{{\textrm{\tiny R}},{\textrm{\tiny T}},k}}{\Omega_{{\textrm{\tiny T}},{\textrm{\tiny R}},k}}\bigg)+\log\bigg(\frac{\cosh\big(\tilde{\beta}_{\textrm{\tiny R}}\Omega_{\textrm{\tiny R},k}/2\big)\big(\cosh\big(\tilde{\beta}_{\textrm{\tiny T}}\Omega_{\textrm{\tiny T},k}/2\big)-1\big)}{\cosh\big(\tilde{\beta}_{\textrm{\tiny T}}\Omega_{\textrm{\tiny T},k}/2\big)\big(\cosh\big(\tilde{\beta}_{\textrm{\tiny R}}\Omega_{\textrm{\tiny R},k}/2\big)-1\big)}\bigg) \bigg]^2
\Bigg\}\,.
\nonumber
\eea
After expanding the squares and a bit of manipulation, one obtains
\bea
\label{mutualcompTFDv4}
\mathcal{M}_{\textrm{\tiny TFD}}\big(\tilde{\omega}_{\textrm{\tiny R}},\tilde{\omega}_{\textrm{\tiny T}},
\tilde{\beta}_{\textrm{\tiny R}}, \tilde{\beta}_{\textrm{\tiny T}}\big)
&=&
\frac{1}{2}\sum_{k=1}^L
F_{\textrm{\tiny TR},k} \,\bigg\{
2 
\log\bigg[\frac{\coth\big(\tilde{\beta}_{\textrm{\tiny T}}\Omega_{\textrm{\tiny T},k}/2\big)}{\coth\big(\tilde{\beta}_{\textrm{\tiny R}}\Omega_{\textrm{\tiny R},k}/2\big)} \bigg]
-F_{\textrm{\tiny TR},k}
\bigg\}
\\
&&\hspace{-2.2cm}
=\;\frac{1}{2}\sum_{k=1}^L
F_{\textrm{\tiny TR},k}\,
\bigg\{
\log\bigg[\frac{\coth^2\big(\tilde{\beta}_{\textrm{\tiny T}}\Omega_{\textrm{\tiny T},k}/2\big)\cosh\big(\tilde{\beta}_{\textrm{\tiny R}}\Omega_{\textrm{\tiny R},k}/2\big)\big(\cosh\big(\tilde{\beta}_{\textrm{\tiny T}}\Omega_{\textrm{\tiny T},k}/2\big)-1\big)}{\coth^2\big(\tilde{\beta}_{\textrm{\tiny R}}\Omega_{\textrm{\tiny R},k}/2\big)\cosh\big(\tilde{\beta}_{\textrm{\tiny T}}\Omega_{\textrm{\tiny T},k}/2\big)\big(\cosh\big(\tilde{\beta}_{\textrm{\tiny R}}\Omega_{\textrm{\tiny R},k}/2\big)-1\big)} \bigg]
\bigg
\}\nonumber
\eea
where
\be
F_{\textrm{\tiny TR},k}=\log\bigg(\frac{\cosh\big(\tilde{\beta}_{\textrm{\tiny T}}\Omega_{\textrm{\tiny T},k}/2\big)\big(\cosh\big(\tilde{\beta}_{\textrm{\tiny R}}\Omega_{\textrm{\tiny R},k}/2\big)-1\big)}{\cosh\big(\tilde{\beta}_{\textrm{\tiny R}}\Omega_{\textrm{\tiny R},k}/2\big)\big(\cosh\big(\tilde{\beta}_{\textrm{\tiny T}}\Omega_{\textrm{\tiny T},k}/2\big)-1\big)}\bigg)\,.
\ee
For fixed $k$, the argument of the sum in (\ref{mutualcompTFDv4}) only depends on $\tilde{\beta}_{\textrm{\tiny T}}\Omega_{\textrm{\tiny T},k} $ and $\tilde{\beta}_{\textrm{\tiny R}}\Omega_{\textrm{\tiny R},k} $ and it is symmetric under the exchange of $\textrm{T}$ and $\textrm{R}$;
hence we can fix $\tilde{\beta}_{\textrm{\tiny T}}\Omega_{\textrm{\tiny T},k} >\tilde{\beta}_{\textrm{\tiny R}}\Omega_{\textrm{\tiny R},k} $ for every $k$
without loss of generality.
This allows to show that every term of the sum (\ref{mutualcompTFDv4}) is negative\footnote{We use that $\frac{\cosh(x/2)}{\cosh(x/2)-1} $ is 
a monotonically decreasing function and that $\frac{\coth^2(x/2)(\cosh(x/2)-1)}{\cosh(x/2)} $ is a monotonically increasing function when $x>0$. 
This implies that $F_{\textrm{\tiny TR},k}<0$,
while the function within the curly brackets 
in the last sum of (\ref{mutualcompTFDv4}) is positive for any value of $k$.} 
and therefore $\mathcal{M}_{\textrm{\tiny TFD}}\big(\tilde{\omega}_{\textrm{\tiny R}},\tilde{\omega}_{\textrm{\tiny T}},
\tilde{\beta}_{\textrm{\tiny R}}, \tilde{\beta}_{\textrm{\tiny T}}\big)$ is always negative.

The formula (\ref{mutualcompTFDv2}) can be generalised to the case where $\kappa_{\textrm{\tiny R}}\neq \kappa_{\textrm{\tiny T}}$ 
as discussed in the final part of Sec.\,\ref{subsec-examples-hc}.
This leads to
\bea
\label{mutualcompTFDv2 differentkappa}
&&\hspace{-.5cm}
\mathcal{M}_{\textrm{\tiny TFD}}\big(\omega_{\textrm{\tiny R}},\omega_{\textrm{\tiny T}},\kappa_{\textrm{\tiny R}},\kappa_{\textrm{\tiny T}}, 
\beta_{\textrm{\tiny R}}, \beta_{\textrm{\tiny T}}\big)
=
\\
&&\hspace{-.3cm}
=\;
\frac{1}{4}\sum_{k=1}^L
\Bigg\{
\bigg[\log\bigg(\frac{\sigma_{\textrm{\tiny phys,R},k}\coth\big(\beta_{\textrm{\tiny R}}\sigma_{\textrm{\tiny phys,R},k}/2\big)}{\sigma_{\textrm{\tiny phys,T},k} \coth\big(\beta_{\textrm{\tiny T}}\sigma_{\textrm{\tiny phys,T},k}/2\big)}\bigg) \bigg]^2-\bigg[\log\bigg(\frac{\sigma_{\textrm{\tiny phys,R},k}\coth\big(\beta_{\textrm{\tiny R}}\sigma_{\textrm{\tiny phys,R},k}/4\big)}{\sigma_{\textrm{\tiny phys,T},k} \coth\big(\beta_{\textrm{\tiny T}}\sigma_{\textrm{\tiny phys,T},k}/4\big)}\bigg) \bigg]^2
\nonumber
\\
&&\hspace{1.4cm}
+\bigg[\log\bigg(\frac{\sigma_{\textrm{\tiny phys,R},k}\coth\big(\beta_{\textrm{\tiny T}}\sigma_{\textrm{\tiny phys,T},k}/2\big)}{\sigma_{\textrm{\tiny phys,T},k}\coth\big(\beta_{\textrm{\tiny R}}\sigma_{\textrm{\tiny phys,R},k}/2\big)}\bigg)\bigg]^2-
\bigg[\log\bigg(\frac{\sigma_{\textrm{\tiny phys,R},k}\coth\big(\beta_{\textrm{\tiny T}}\sigma_{\textrm{\tiny phys,T},k}/4\big)}{\sigma_{\textrm{\tiny phys,T},k}\coth\big(\beta_{\textrm{\tiny R}}\sigma_{\textrm{\tiny phys,R},k}/4\big)}\bigg)\bigg]^2
\Bigg\}
\nonumber
\eea
where $\sigma_{\textrm{\tiny phys,S},k} $ with $\textrm{S}=\{\textrm{R},\textrm{T}\}$ has been defined in (\ref{symp-spectrum-phys}). 
In the special case where the reference state is the unentangled product ground state, {\it i.e.} $\kappa_{\textrm{\tiny R}}=0$ and $\beta_{\textrm{\tiny R}}\to\infty$,
we find that  (\ref{mutualcompTFDv2 differentkappa}) in this limit becomes
\bea
\label{mutualcompTFDv2 refprodstate}
&&\hspace{-.5cm}
\mathcal{M}_{\textrm{\tiny TFD}}\big(\omega_{\textrm{\tiny R}},\omega_{\textrm{\tiny T}},\kappa_{\textrm{\tiny R}}=0,\kappa_{\textrm{\tiny T}}, 
\beta_{\textrm{\tiny R}}\to\infty, \beta_{\textrm{\tiny T}}\big)
=
\\
&&\hspace{-.2cm}
=\,
\frac{1}{4}\sum_{k=1}^L
\Bigg\{
\bigg[\log\!\bigg(\frac{\omega_{\textrm{\tiny R}}}{\sigma_{\textrm{\tiny phys,T},k} \coth\big(\beta_{\textrm{\tiny T}}\sigma_{\textrm{\tiny phys,T},k}/2\big)}\bigg) \bigg]^2
-\bigg[\log\!\bigg(\frac{\omega_{\textrm{\tiny R}}}{\sigma_{\textrm{\tiny phys,T},k} \coth\big(\beta_{\textrm{\tiny T}}\sigma_{\textrm{\tiny phys,T},k}/4\big)}\bigg) \bigg]^2
\nonumber
\\
&&\hspace{3cm}
+\bigg[\log\!\bigg(\frac{\omega_{\textrm{\tiny R}}\coth\big(\beta_{\textrm{\tiny T}}\sigma_{\textrm{\tiny phys,T},k}/2\big)}{\sigma_{\textrm{\tiny phys,T},k}}\bigg)\bigg]^2-
\bigg[\log\!\bigg(\frac{\omega_{\textrm{\tiny R}}\coth\big(\beta_{\textrm{\tiny T}}\sigma_{\textrm{\tiny phys,T},k}/4\big)}{\sigma_{\textrm{\tiny phys,T},k}}\bigg)\bigg]^2
\Bigg\}
\nonumber
\eea

The thermodynamic limit $L \to \infty$ of (\ref{mutualcompTFDv2}) gives
\be
\label{mutualTFD_TD}
\mathcal{M}_{\textrm{\tiny TFD}}\big(\tilde{\omega}_{\textrm{\tiny R}},\tilde{\omega}_{\textrm{\tiny T}},
\tilde{\beta}_{\textrm{\tiny R}}, \tilde{\beta}_{\textrm{\tiny T}}\big)
\,=\,
a_{\textrm{\tiny TFD}}\big(\tilde{\omega}_{\textrm{\tiny R}},\tilde{\omega}_{\textrm{\tiny T}},
\tilde{\beta}_{\textrm{\tiny R}}, \tilde{\beta}_{\textrm{\tiny T}}\big)\, L + \dots
\ee
where the coefficient of the linear divergence can be written in terms 
$\Omega_{{\textrm{\tiny S}},\theta}$ in (\ref{dispersion relation TD}) as
\bea
\label{a-funct_TD_mutualTFDv0}
a_{\textrm{\tiny TFD}}\big(\tilde{\omega}_{\textrm{\tiny R}},\tilde{\omega}_{\textrm{\tiny T}},
\tilde{\beta}_{\textrm{\tiny R}}, \tilde{\beta}_{\textrm{\tiny T}}\big)
&=&
\\
&& \hspace{-3.cm}
=\,
\frac{1}{4\pi}\int_0^\pi
\Bigg\{\bigg[\log\bigg(\frac{\Omega_{\textrm{\tiny R},\theta}\coth\big(\tilde{\beta}_{\textrm{\tiny R}}\Omega_{\textrm{\tiny R},\theta}/2\big)}{\Omega_{\textrm{\tiny T},\theta} \coth\big(\tilde{\beta}_{\textrm{\tiny T}}\Omega_{\textrm{\tiny T},\theta}/2\big)}\bigg) \bigg]^2
-\bigg[\log\bigg(\frac{\Omega_{\textrm{\tiny R},\theta}\coth\big(\tilde{\beta}_{\textrm{\tiny R}}\Omega_{\textrm{\tiny R},\theta}/4\big)}{\Omega_{\textrm{\tiny T},\theta} \coth\big(\tilde{\beta}_{\textrm{\tiny T}}\Omega_{\textrm{\tiny T},\theta}/4\big)}\bigg) \bigg]^2
\nonumber
\\
&& \hspace{-.8cm}
+\, \bigg[\log\bigg(\frac{\Omega_{\textrm{\tiny R},\theta}
\coth\big(\tilde{\beta}_{\textrm{\tiny T}}\Omega_{\textrm{\tiny T},\theta}/2\big)}{\Omega_{\textrm{\tiny T},\theta}\coth\big(\tilde{\beta}_{\textrm{\tiny R}}\Omega_{\textrm{\tiny R},\theta}/2\big)}\bigg)\bigg]^2-
\bigg[\log\bigg(\frac{\Omega_{\textrm{\tiny R},\theta}\coth\big(\tilde{\beta}_{\textrm{\tiny T}}\Omega_{\textrm{\tiny T},\theta}/4\big)}{\Omega_{\textrm{\tiny T},\theta}\coth\big(\tilde{\beta}_{\textrm{\tiny R}}\Omega_{\textrm{\tiny R},\theta}/4\big)}\bigg)\bigg]^2
\Bigg\}  \,d\theta \,.
\nonumber
\eea
We remark that the massless limit of $\mathcal{M}_{\textrm{\tiny TFD}}/L$ diverges when $L<\infty$, 
while it is finite once $L\to\infty$ is considered.
Indeed, by setting $\tilde{\omega}_{\textrm{\tiny R}}=\tilde{\omega}_{\textrm{\tiny T}}=0$ in (\ref{a-funct_TD_mutualTFDv0}) we find
\bea
\label{a_TFD-00}
a_{\textrm{\tiny TFD}}\big(\tilde{\omega}_{\textrm{\tiny R}}=0,\tilde{\omega}_{\textrm{\tiny T}}=0,
\tilde{\beta}_{\textrm{\tiny R}}, \tilde{\beta}_{\textrm{\tiny T}}\big)
&=&
\\
&& \hspace{-1.7cm}
=
\int_0^\pi 
\Bigg\{
\bigg[\log\bigg(\frac{\coth\big(\tilde{\beta}_{\textrm{\tiny R}}\sin\theta\big)}{ \coth\big(\tilde{\beta}_{\textrm{\tiny T}}\sin\theta\big)}\bigg) \bigg]^2-\bigg[\log\bigg(\frac{\coth\big(\tilde{\beta}_{\textrm{\tiny R}}\sin\theta/2\big)}{ \coth\big(\tilde{\beta}_{\textrm{\tiny T}}\sin\theta/2\big)}\bigg) \bigg]^2
\Bigg\}\,  \frac{d\theta}{2\pi}\,.
\nonumber
\eea
This feature has been observed also 
for the complexity of pure states (Sec.\,\ref{sec-pure-states-examples}) 
and for the complexity of thermal states (Sec.\,\ref{sec-thermal-states-examples}).

In the limit $\tilde{\beta}_{\textrm{\tiny R}}\to\infty$ 
both the TFD in (\ref{TFD_Hlattice}) and the thermal reference state 
become the product of two ground states.
In this regime, (\ref{a_TFD-00}) slightly simplifies to 
\bea
\label{a-funct_TD_mutualTFD}
a_{\textrm{\tiny TFD}}\big(\tilde{\omega}_{\textrm{\tiny R}}=0,\tilde{\omega}_{\textrm{\tiny T}}=0,
\tilde{\beta}_{\textrm{\tiny R}}\to	\infty, \tilde{\beta}_{\textrm{\tiny T}}\big)
&=&
\\
&& \hspace{-2.5cm}
=
\int_0^\pi  
\Bigg\{
\bigg[\log\bigg(\!\coth\big(\tilde{\beta}_{\textrm{\tiny T}}\sin\theta\big) \bigg) \bigg]^2
\!-
\bigg[\log\bigg(\!\coth\big(\tilde{\beta}_{\textrm{\tiny T}}\sin\theta/2\big)\bigg) \bigg]^2
\Bigg\}\, \frac{d\theta}{2\pi}
\nonumber
\eea
which depends only on $\tilde{\beta}_{\textrm{\tiny T}}$ and can be easily studied. 
This function is negative for every value of $\tilde{\beta}_{\textrm{\tiny T}}$ and it vanishes when $\tilde{\beta}_{\textrm{\tiny T}}\to\infty$,
as expected. 
When $\tilde{\beta}_{\textrm{\tiny T}} \to 0$ in (\ref{a-funct_TD_mutualTFD}), 
we find the following logarithmic divergence
\be
\label{a-funct-massless-TD-smallbeta}
a_{\textrm{\tiny TFD}}\big(\tilde{\omega}_{\textrm{\tiny R}}=0,\tilde{\omega}_{\textrm{\tiny T}}=0,
\tilde{\beta}_{\textrm{\tiny R}}\to	\infty, \tilde{\beta}_{\textrm{\tiny T}}\big)
\,=\,
\log 2\, \bigg(\! \log\tilde{\beta}_{\textrm{\tiny T}}-\frac{3}{2} \log 2\bigg)
+\dots
\;.
\ee

\begin{figure}[t!]
\subfigure
{\hspace{-.8cm}
\includegraphics[width=.51\textwidth]{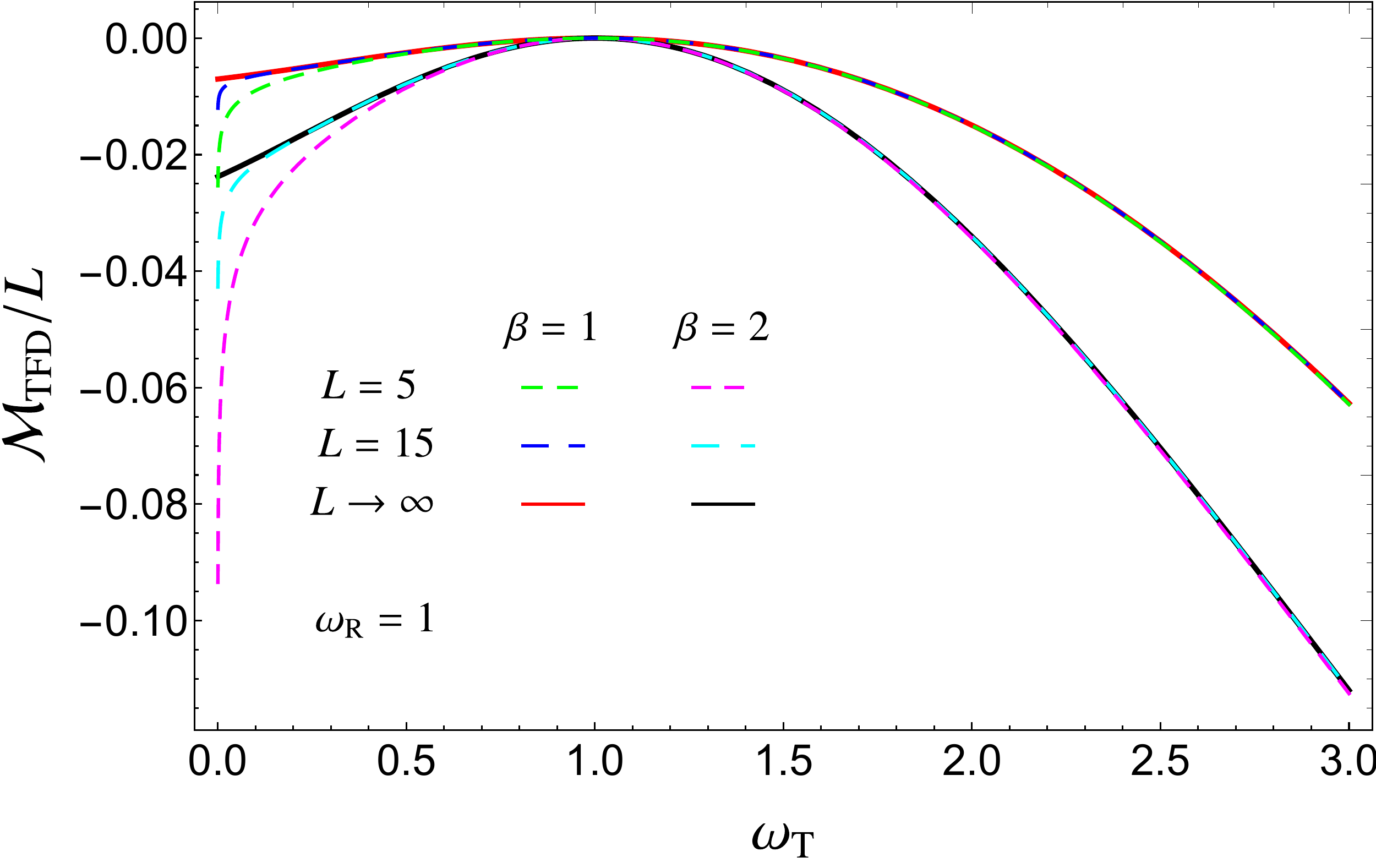}}
\subfigure
{
\hspace{.1cm}\includegraphics[width=.51\textwidth]{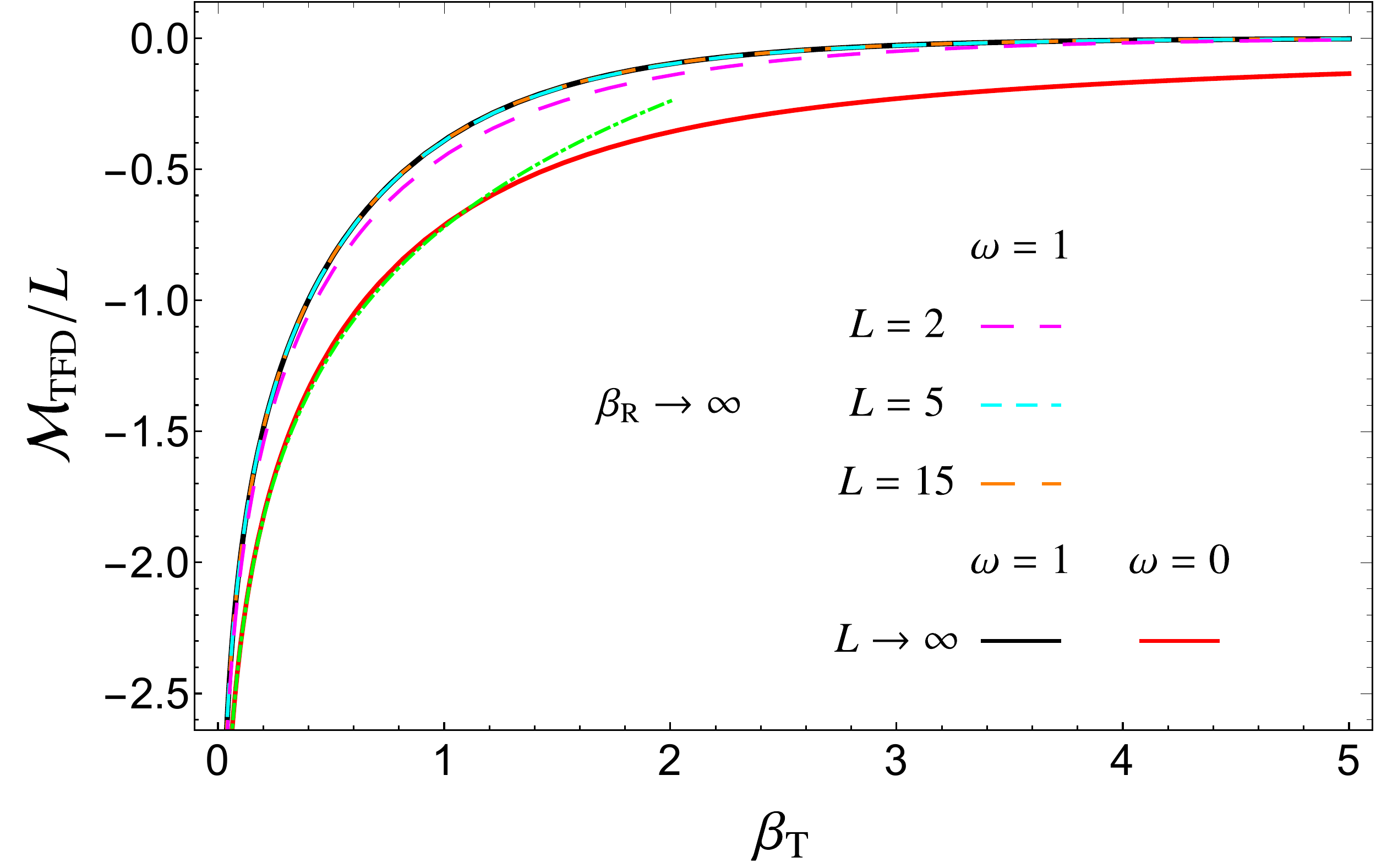}}

\vspace{-0.2cm}
\caption{Comparison between the mutual complexity for the TFD state (\ref{mutualcompTFDv2}) and its thermodynamic limit in the periodic harmonic chain (here $\kappa=m=1$). In the left panel we show $\mathcal{M}_{\textrm{\tiny TFD}}/L$ as function of $\omega_{\textrm{\tiny T}}$ setting $\omega_{\textrm{\tiny R}}=1$ and considering two values of $\beta_{\textrm{\tiny R}}=\beta_{\textrm{\tiny T}}\equiv \beta$ and two values of $L$ (dashed curves). We also report $a_{\textrm{\tiny TFD}}$ in (\ref{a-funct_TD_mutualTFDv0}) for the same values of the parameters (solid curves). In the right panel the dependence on $\beta_{\textrm{\tiny T}}$ is investigated by plotting (\ref{mutualcompTFDv2}) for three different values of $L$ (dashed curves) and (\ref{a-funct_TD_mutualTFDv0}) (black solid curve) both for $\beta_{\textrm{\tiny R}}\to\infty$ and $\omega_{\textrm{\tiny R}}=\omega_{\textrm{\tiny T}}\equiv\omega=1$. The massless limit in (\ref{a-funct_TD_mutualTFD}) is also reported (red solid curve) and its small $\beta_{\textrm{\tiny T}}$ behaviour in (\ref{a-funct-massless-TD-smallbeta}) is checked (green dot-dashed curve).
}
\vspace{.2 cm}
\label{fig:MutualComplexityTFD}
\end{figure}

In Fig.\,\ref{fig:MutualComplexityTFD} we compare the mutual complexity for the TFD in  (\ref{mutualcompTFDv2}) with its thermodynamic limit in (\ref{mutualTFD_TD}) for various values of the parameters. 
In the left panel we show $\mathcal{M}_{\textrm{\tiny TFD}}/L$ (dashed lines) as function of $\omega_{\textrm{\tiny T}}$ for fixed $\omega_{\textrm{\tiny R}}=1$
and for two values of $\beta_{\textrm{\tiny R}}=\beta_{\textrm{\tiny T}}\equiv \beta$.
As $L$ increases, the dashed curves approach the solid curves representing $a_{\textrm{\tiny TFD}}$ given in (\ref{a-funct_TD_mutualTFDv0}).
When $\omega_{\textrm{\tiny T}}\to 0$, $\mathcal{M}_{\textrm{\tiny TFD}}/L$ evalueted for finite $L$ diverges, while its thermodynamic limit is finite,
as observed above. 
In the right panel we show $\mathcal{M}_{\textrm{\tiny TFD}}/L$ as function of $\beta_{\textrm{\tiny T}}$ 
when $\beta_{\textrm{\tiny R}}\to\infty$ and $\omega_{\textrm{\tiny R}}=\omega_{\textrm{\tiny T}}\equiv\omega=1$. 
Remarkably, the curves obtained for finite number of sites coincide with their thermodynamic limit already for $L=5$. 
In the same panel we plot $a_{\textrm{\tiny TFD}}\big(\omega_{\textrm{\tiny R}}=0,\omega_{\textrm{\tiny T}}=0,
\beta_{\textrm{\tiny R}}\to	\infty, \beta_{\textrm{\tiny T}}\big)$ in (\ref{a-funct_TD_mutualTFD}) (red solid curve), 
checking also  that its behaviour for $\beta_{\textrm{\tiny T}}\ll 1$ is well reproduced by (\ref{a-funct-massless-TD-smallbeta}) (green dot-dashed curve). 

We find it worth considering the mutual complexity for the thermofield double state
in the continuum limit procedure where $\epsilon=\sqrt{m/\kappa}\to 0$ and $L\to \infty$,
while $\tilde{L}\equiv\epsilon L $ is kept fixed, as done also in \cite{Caceres:2019pgf} for the purification complexity. 
\\
In this limiting regime, the sum over dimensionless numbers $k$ becomes an integral over the positive momenta $p$.
Moreover, from (\ref{dispersion relation}), (\ref{omega_tilde}) and (\ref{T_tilde}), we find that
$\Omega_{\textrm{\tiny S},k}\to\epsilon\sqrt{\omega_{\textrm{\tiny S}}^2+p^2}\equiv\epsilon E_{\textrm{\tiny S},p}$ and $\tilde{\beta}_\textrm{\tiny S}\Omega_{\textrm{\tiny S},k}\to \beta_\textrm{\tiny S}E_{\textrm{\tiny S},p}$.
Thus, (\ref{mutualcompTFDv2}) becomes
\bea
\label{mutualcompTFDv2 continuum}
\mathcal{M}_{\textrm{\tiny TFD}}\big(\omega_{\textrm{\tiny R}},\omega_{\textrm{\tiny T}},
\beta_{\textrm{\tiny R}}, \beta_{\textrm{\tiny T}}\big) 
&=&
\\
&&
\hspace{-3.4cm}
=\,
\frac{\tilde{L}}{4\pi}\int_0^{\infty}
\Bigg\{\bigg[\log\bigg(\frac{E_{\textrm{\tiny R},p}\coth\big(\beta_{\textrm{\tiny R}}E_{\textrm{\tiny R},p}/2\big)}{E_{\textrm{\tiny T},p} \coth\big(\beta_{\textrm{\tiny T}}E_{\textrm{\tiny T},p}/2\big)}\bigg) \bigg]^2-\bigg[\log\bigg(\frac{E_{\textrm{\tiny R},p}\coth\big(\beta_{\textrm{\tiny R}}E_{\textrm{\tiny R},p}/4\big)}{E_{\textrm{\tiny T},p} \coth\big(\beta_{\textrm{\tiny T}}E_{\textrm{\tiny T},p}/4\big)}\bigg) \bigg]^2
\nonumber
\\
&&\hspace{-1cm}
+\bigg[\log\bigg(\frac{E_{\textrm{\tiny R},p}\coth\big(\beta_{\textrm{\tiny T}}E_{\textrm{\tiny T},p}/2\big)}{E_{\textrm{\tiny T},p}\coth\big(\beta_{\textrm{\tiny R}}E_{\textrm{\tiny R},p}/2\big)}\bigg)\bigg]^2-
\bigg[\log\bigg(\frac{E_{\textrm{\tiny R},p}\coth\big(\beta_{\textrm{\tiny T}}E_{\textrm{\tiny T},p}/4\big)}{E_{\textrm{\tiny T},p}\coth\big(\beta_{\textrm{\tiny R}}E_{\textrm{\tiny R},p}/4\big)}\bigg)\bigg]^2
\Bigg
\}\,dp
\nonumber
\eea
Notice that, given $\textrm{S}=\{\textrm{R},\textrm{T}\}$, since $E_{\textrm{\tiny S},p}\sim p$ when $p\gg\omega_\textrm{\tiny S}$, 
all the four terms of the integrand vanish when $p\to \infty$ and the four resulting integrals in (\ref{mutualcompTFDv2 continuum}) are separately UV finite.
\\
Instead, if we consider the mutual complexity of the TFD when the reference state is the unentangled product state given in (\ref{mutualcompTFDv2 refprodstate})
in this limiting regime, the UV finiteness is due to a non trivial cancellation among divergent contributions. 
\\
In this case, from (\ref{symp-spectrum-phys}), we have $\sigma_{\textrm{\tiny phys,S},k}\to\sqrt{\omega_{\textrm{\tiny S}}^2+p^2}\equiv E_{\textrm{\tiny S},p}$ with $\textrm{S}=\{\textrm{R},\textrm{T}\}$;
hence (\ref{mutualcompTFDv2 refprodstate}) becomes
\bea
\label{mutualcompTFDv2 refprodstate cont}
\mathcal{M}_{\textrm{\tiny TFD}}\big(\omega_{\textrm{\tiny R}},\omega_{\textrm{\tiny T}},
\beta_{\textrm{\tiny R}}\to\infty, \beta_{\textrm{\tiny T}}\big)
&=&
\\
&&
\hspace{-4cm}
=\;
\frac{\tilde{L}}{4\pi}\int_0^{\infty}
\Bigg\{\bigg[\log\bigg(\frac{\omega_{\textrm{\tiny R}}}{E_{\textrm{\tiny T},p} \coth\big(\beta_{\textrm{\tiny T}}E_{\textrm{\tiny T},p}/2\big)}\bigg) \bigg]^2-\bigg[\log\bigg(\frac{\omega_{\textrm{\tiny R}}}{E_{\textrm{\tiny T},p} \coth\big(\beta_{\textrm{\tiny T}}E_{\textrm{\tiny T},p}/4\big)}\bigg) \bigg]^2
\nonumber
\\
&&\hspace{-1.5cm}
+
\bigg[\log\bigg(\frac{\omega_{\textrm{\tiny R}}\coth\big(\beta_{\textrm{\tiny T}}E_{\textrm{\tiny T},p}/2\big)}{E_{\textrm{\tiny T},p}}\bigg)\bigg]^2-
\bigg[\log\bigg(\frac{\omega_{\textrm{\tiny R}}\coth\big(\beta_{\textrm{\tiny T}}E_{\textrm{\tiny T},p}/4\big)}{E_{\textrm{\tiny T},p}}\bigg)\bigg]^2
\Bigg
\}\,dp
\nonumber
\eea
Up to the global factor $\tfrac{\tilde{L}}{4\pi}$,
both the terms of the integrand coming from $2\mathcal{C}^2_{\textrm{\tiny th}} $ in (\ref{mutualcompTFD}), 
i.e. the first one and the third one, diverge as $2[\log(p/\omega\textrm{\tiny R})]^2$ when $p\to\infty$ 
while both the second term and the fourth term, which originate from $\mathcal{C}^2_{\textrm{\tiny TFD}} $ in (\ref{mutualcompTFD}), 
diverge as $-2[\log(p/\omega\textrm{\tiny R})]^2$. 
Because of the relative factor 2 between $\mathcal{C}^2_{\textrm{\tiny th}} $ 
and $\mathcal{C}^2_{\textrm{\tiny TFD}} $ in the definition (\ref{mutualcompTFD}),
these UV divergences cancel in (\ref{mutualcompTFDv2 refprodstate cont}).
This feature has been first observed in \cite{Caceres:2019pgf} for the mutual complexity of the thermofield double state evaluated through the thermal purification complexity.

\subsection{Reduced density matrices}
\label{sec-subregion-examples}

Important mixed states to explore are the reduced density matrices of a subsystem $A$.

Consider the density matrix $\hat{\rho}_{\textrm{\tiny R}}$ and $\hat{\rho}_{\textrm{\tiny T}}$  
of the reference and of the target states respectively and introduce a spatial bipartition $A \cup B$
of the system that induces a factorisation of the Hilbert space, as already discussed in Sec.\,\ref{sec:comp-eh}.
For the Gaussian states that we are interested in, 
let us denote by $\gamma_{\textrm{\tiny R},A} $ and $ \gamma_{\textrm{\tiny T},A}$
the reduced covariance matrices corresponding to the subsystem $A$,
that characterise the reduced density matrices 
$\hat{\rho}_{\textrm{\tiny R},A}\equiv \textrm{Tr}_B \hat{\rho}_{\textrm{\tiny R}}$ 
and $\hat{\rho}_{\textrm{\tiny T},A}\equiv \textrm{Tr}_B  \hat{\rho}_{\textrm{\tiny T}}$
respectively.
We remark that, whenever  $B\neq \emptyset$, 
the reduced density matrices $\hat{\rho}_{\textrm{\tiny R},A}$ and $\hat{\rho}_{\textrm{\tiny T},A}$ 
are mixed states, even when $\hat{\rho}_{\textrm{\tiny R}}$ and $\hat{\rho}_{\textrm{\tiny T}}$ are pure states.
The reduced covariance matrix $\gamma_A$ is obtained by 
just restricting the indices of the covariance matrix of the entire system to the ranges identifying the subsystem $A$.

By applying (\ref{c2 complexity}) to these mixed states, one obtains the subregion complexity
\be
\label{c2-complexity-rdm}
\mathcal{C}_2
\,=\,
\frac{1}{2\sqrt{2}}\;
\sqrt{
\textrm{Tr} \,\Big\{ \big[
\log \!\big( \gamma_{\textrm{\tiny T},A} \, \gamma_{\textrm{\tiny R},A} ^{-1}\, \big)
\big]^2  \Big\}}
\,.
\ee
In the context of the gauge/gravity correspondence, 
the subregion complexity has been studied e.g. in 
\cite{Alishahiha:2015rta,Agon:2018zso,Alishahiha:2018lfv,Auzzi:2019vyh}.

In the following we provide numerical results of this complexity 
only for the simplest case where $A$ is an interval in an infinite harmonic chain
and for some convenient reference and target states. 
In order to construct the reduced covariance matrices, in this case we need the 
two-point correlators of the harmonic chain in the thermodynamic limit.
For a thermal state, they can be found by taking the limit $L \to \infty$ of (\ref{corrs-periodic-thermal}) and (\ref{corrs-periodic-thermalv2}).
The results can be written in terms of $\Omega_\theta =\sqrt{\tilde{\omega}^2 +4 \, (\sin\theta)^2}$ (see (\ref{dispersion relation TD}))
as follows
\bea
\label{qq-corr-th-TD}
\langle \hat{q}_i \hat{q}_j \rangle_\beta
&=&
\frac{1}{2 \pi} \int_0^{\pi} 
\frac{\coth[\Omega_\theta /(2 \widetilde{T})]}{\Omega_\theta} \, \cos[2\theta\, (i-j)] \,d\theta
\\
\rule{0pt}{.9cm}
\label{pp-corr-th-TD}
\langle \hat{p}_i \hat{p}_j \rangle_\beta
&=&
\frac{1}{2 \pi} \int_0^{\pi}\!
\Omega_\theta\coth[\Omega_\theta /(2 \widetilde{T})]\cos[2\theta\, (i-j)]\, d\theta\,.
\eea

\begin{figure}[t!]
\subfigure
{\hspace{-.8cm}
\includegraphics[width=.5\textwidth]{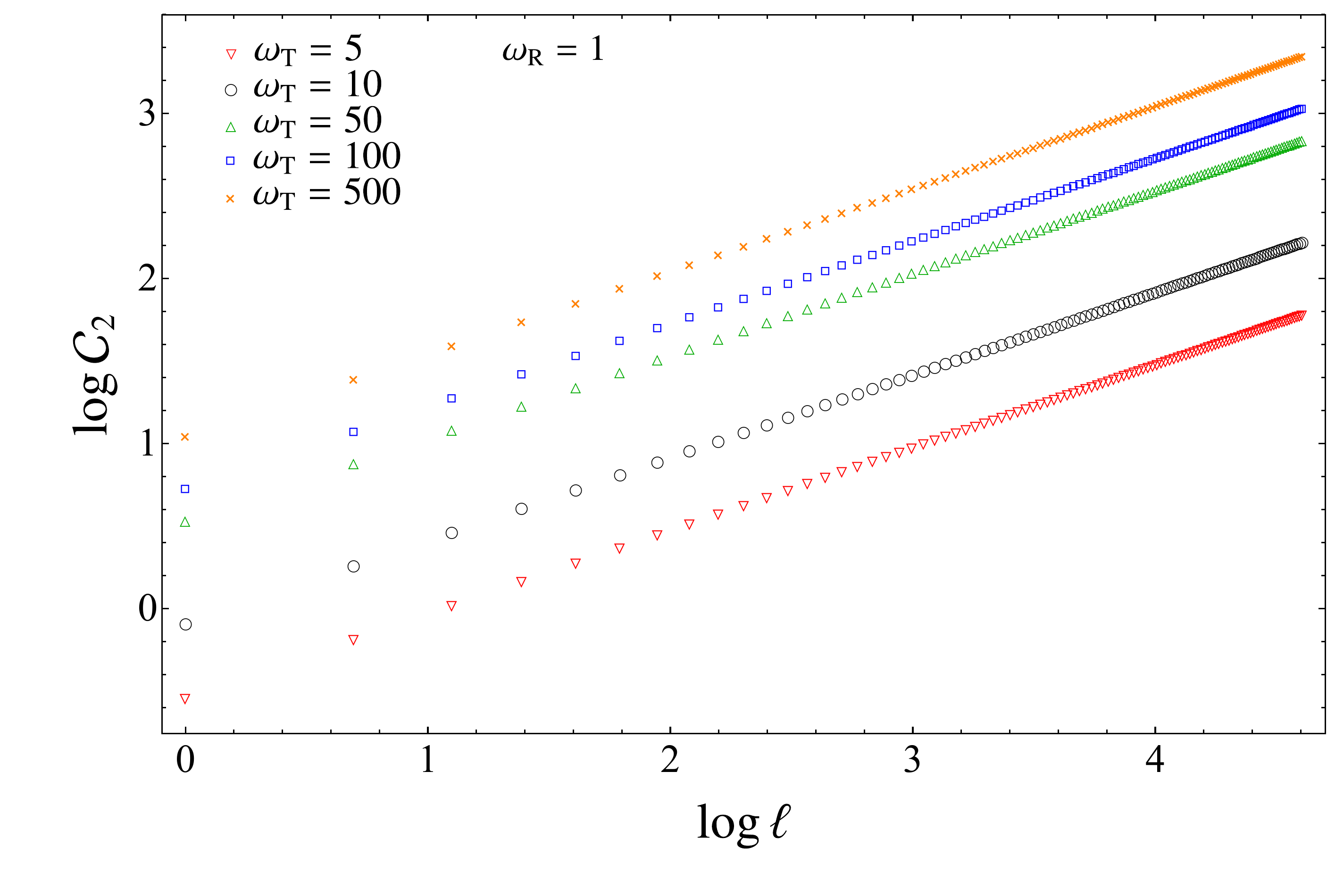}}
\subfigure
{
\hspace{.1cm}\includegraphics[width=.5\textwidth]{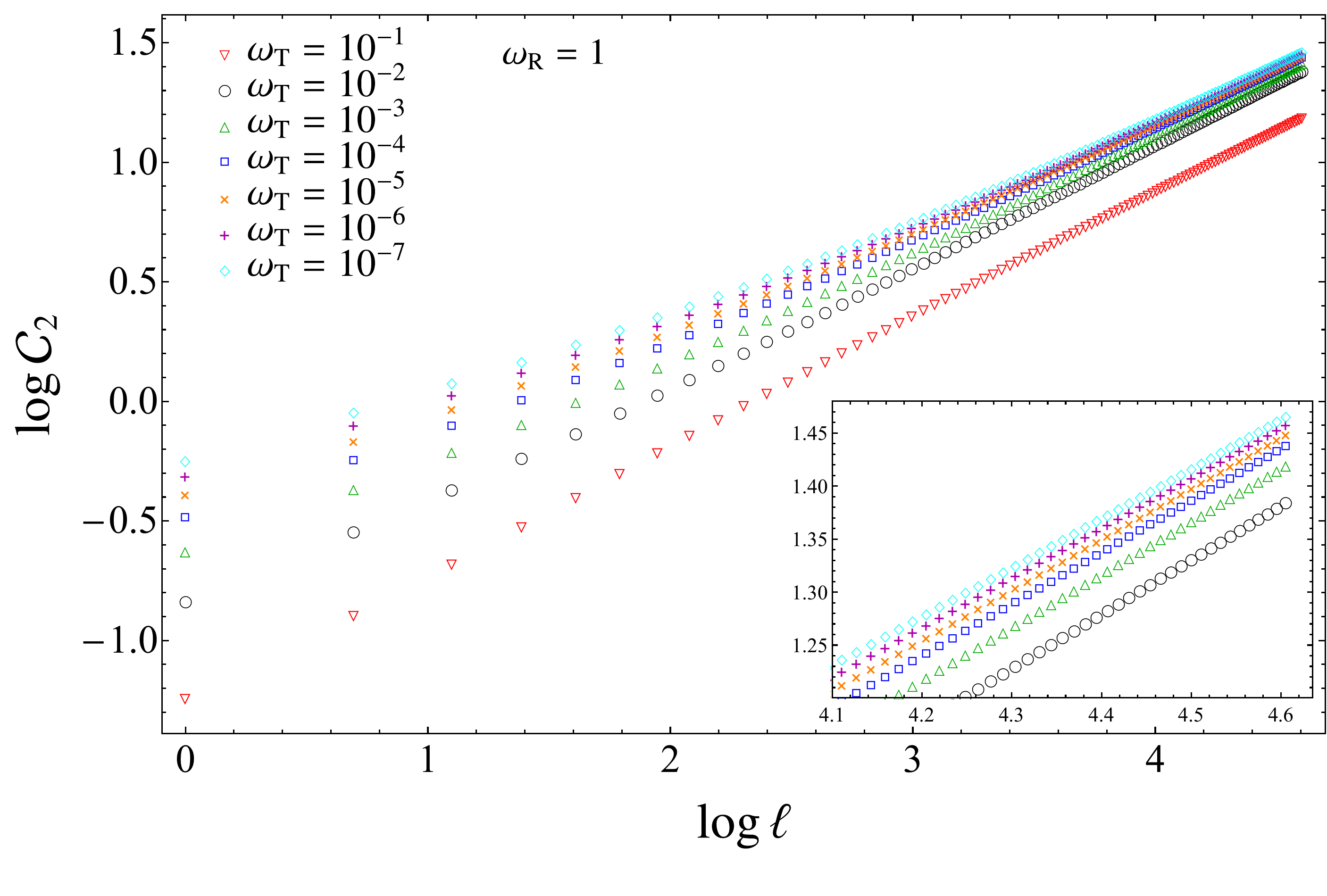}}
\subfigure
{
\hspace{-.8cm}\includegraphics[width=.5\textwidth]{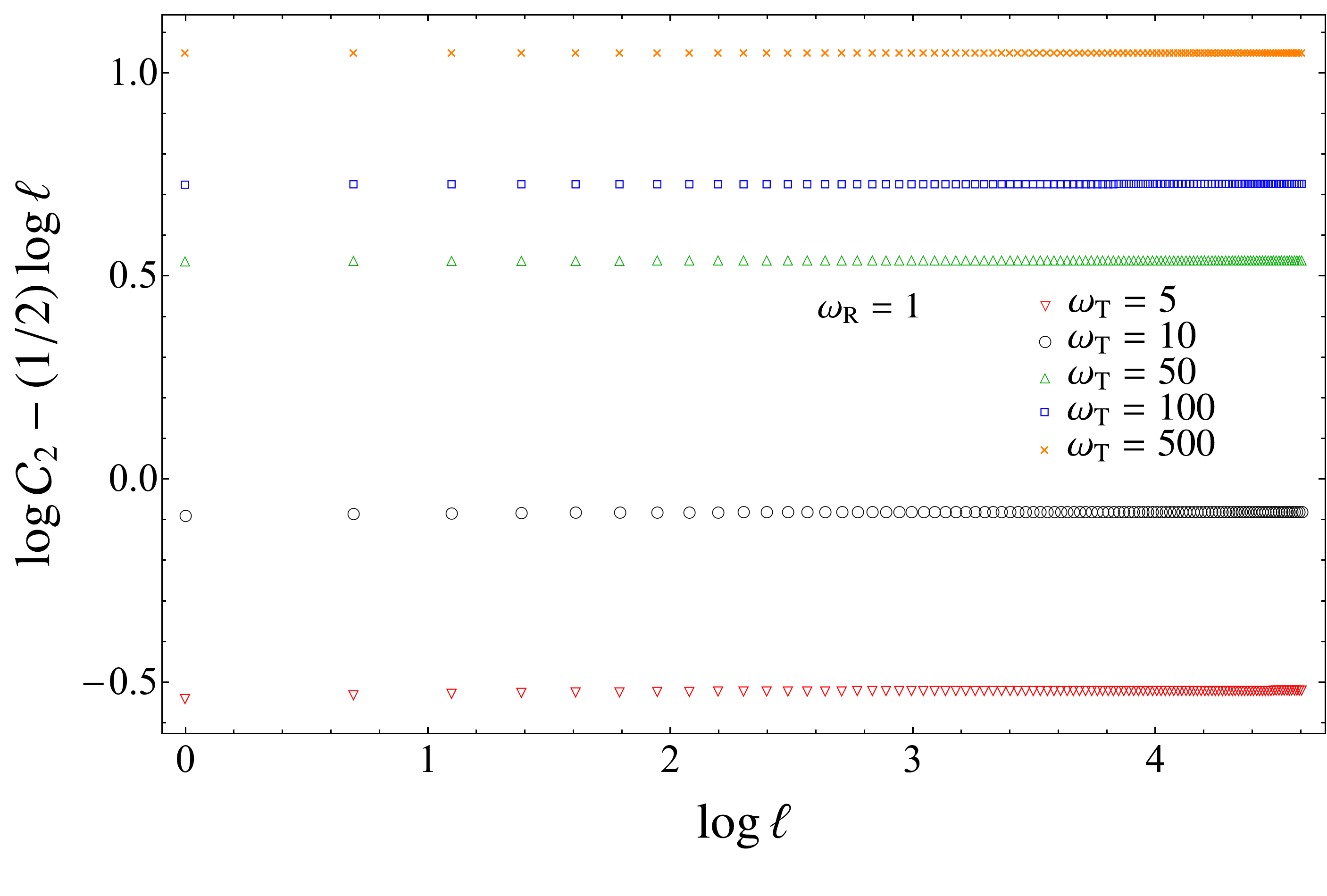}}
\subfigure
{\hspace{.45cm}
\includegraphics[width=.5\textwidth]{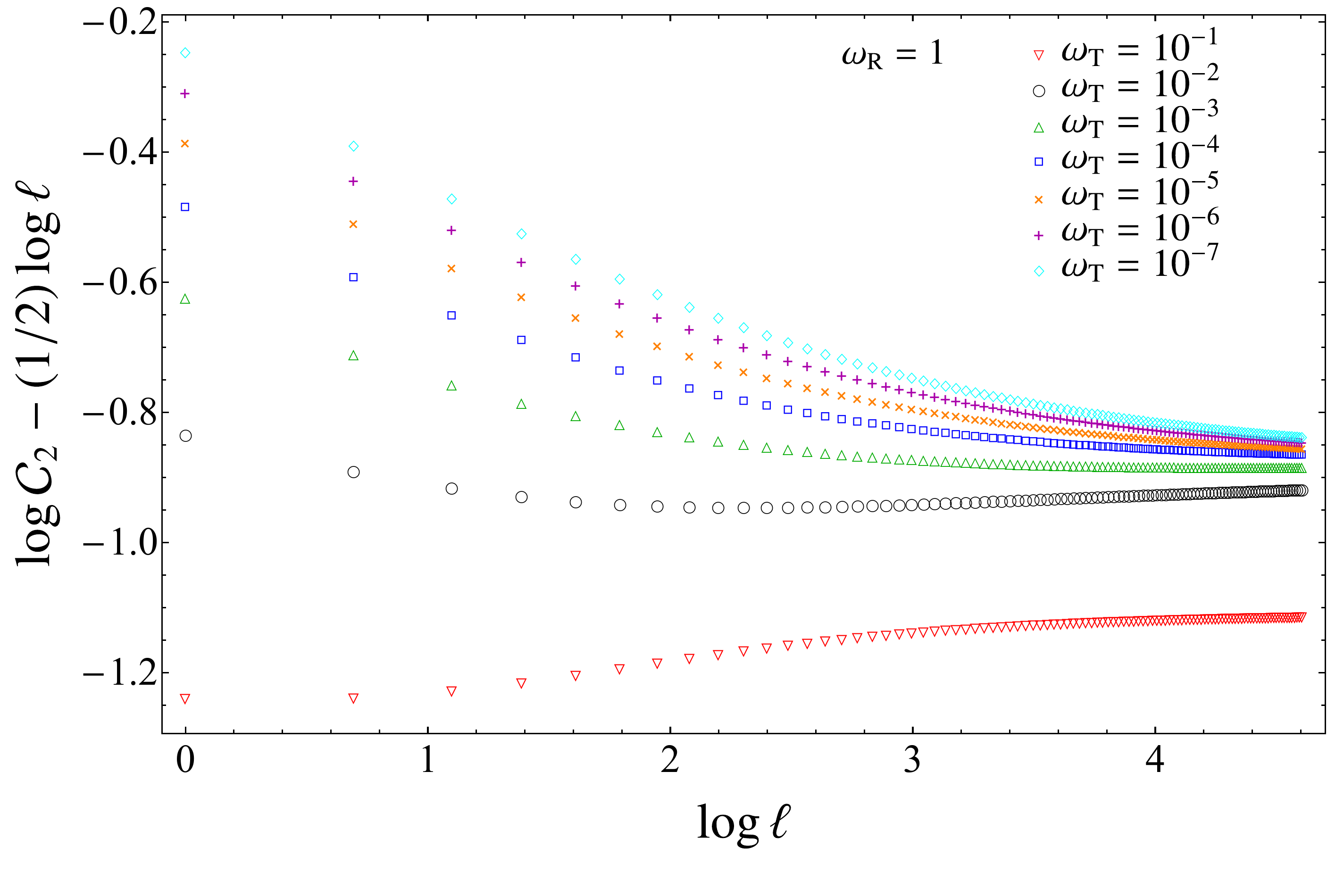}}
\caption{Subregion complexity $\mathcal{C}_2$ for an interval made by $\ell$ sites in an infinite harmonic chain as function of $\ell$. 
The chain is in its ground state and 
$\omega_\textrm{\tiny R} \neq \omega_\textrm{\tiny T}$.
Here $\kappa=m=1$; hence $\tilde{\omega}_\textrm{\tiny R}=\omega_\textrm{\tiny R}$ and 
$\tilde{\omega}_\textrm{\tiny T}=\omega_\textrm{\tiny T}$.
We fix $\omega_\textrm{\tiny R}=1$, considering various values for $\omega_\textrm{\tiny T}$:
$\omega_\textrm{\tiny T} > \omega_\textrm{\tiny R}$ in the left panels and
$\omega_\textrm{\tiny T} < \omega_\textrm{\tiny R}$ in the right panels.
The subregion complexity $\mathcal{C}_2$ is reported in the top panels,
while its subleading term is studied in the bottom panels. 
} 
\vspace{0.4cm}
\label{fig:FRCompSubRegionfunctionofL}
\end{figure}

\begin{figure}[t!]
\subfigure
{\hspace{-.8cm}
\includegraphics[width=.5\textwidth]{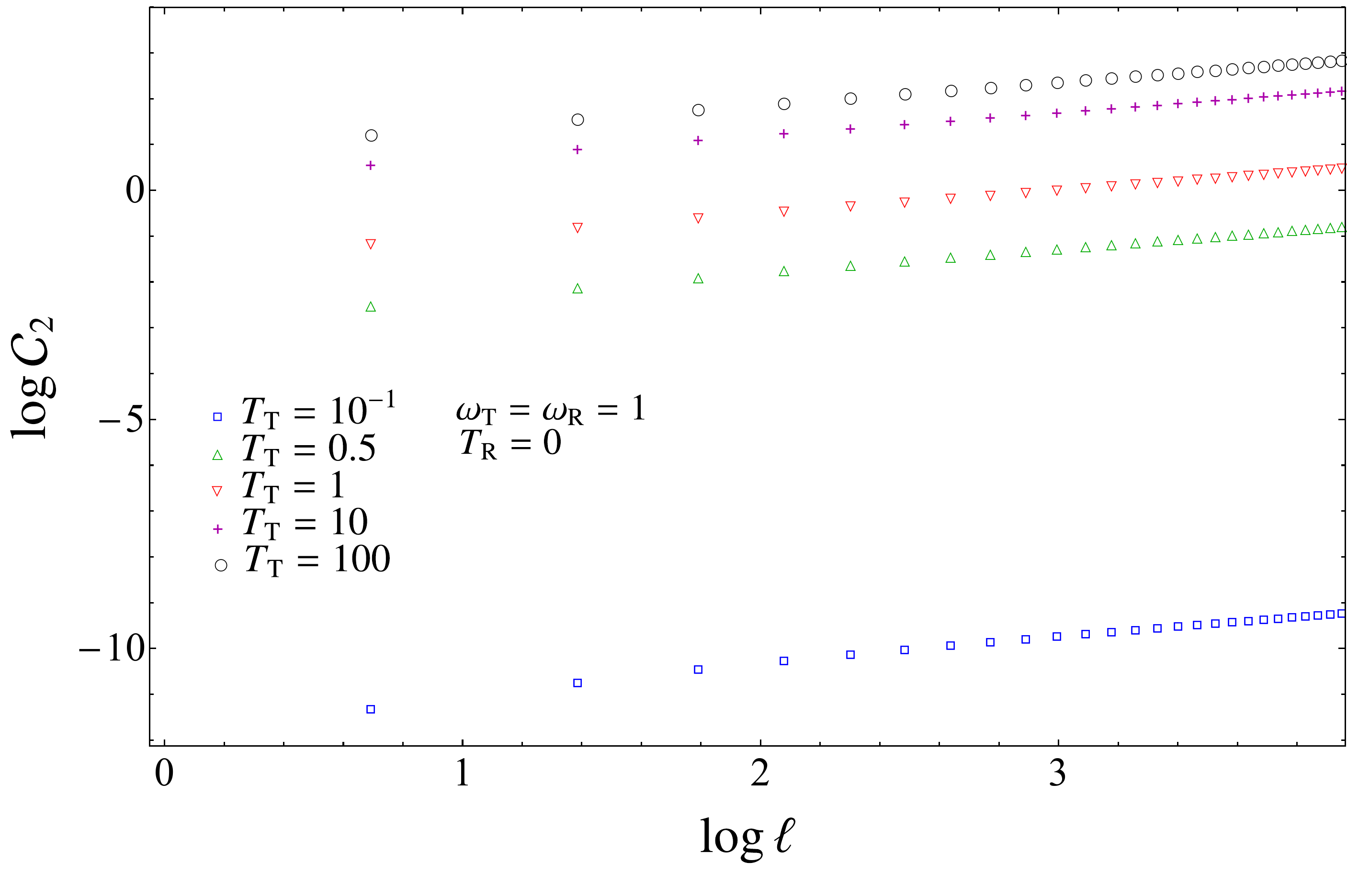}}
\subfigure
{
\hspace{.1cm}\includegraphics[width=.5\textwidth]{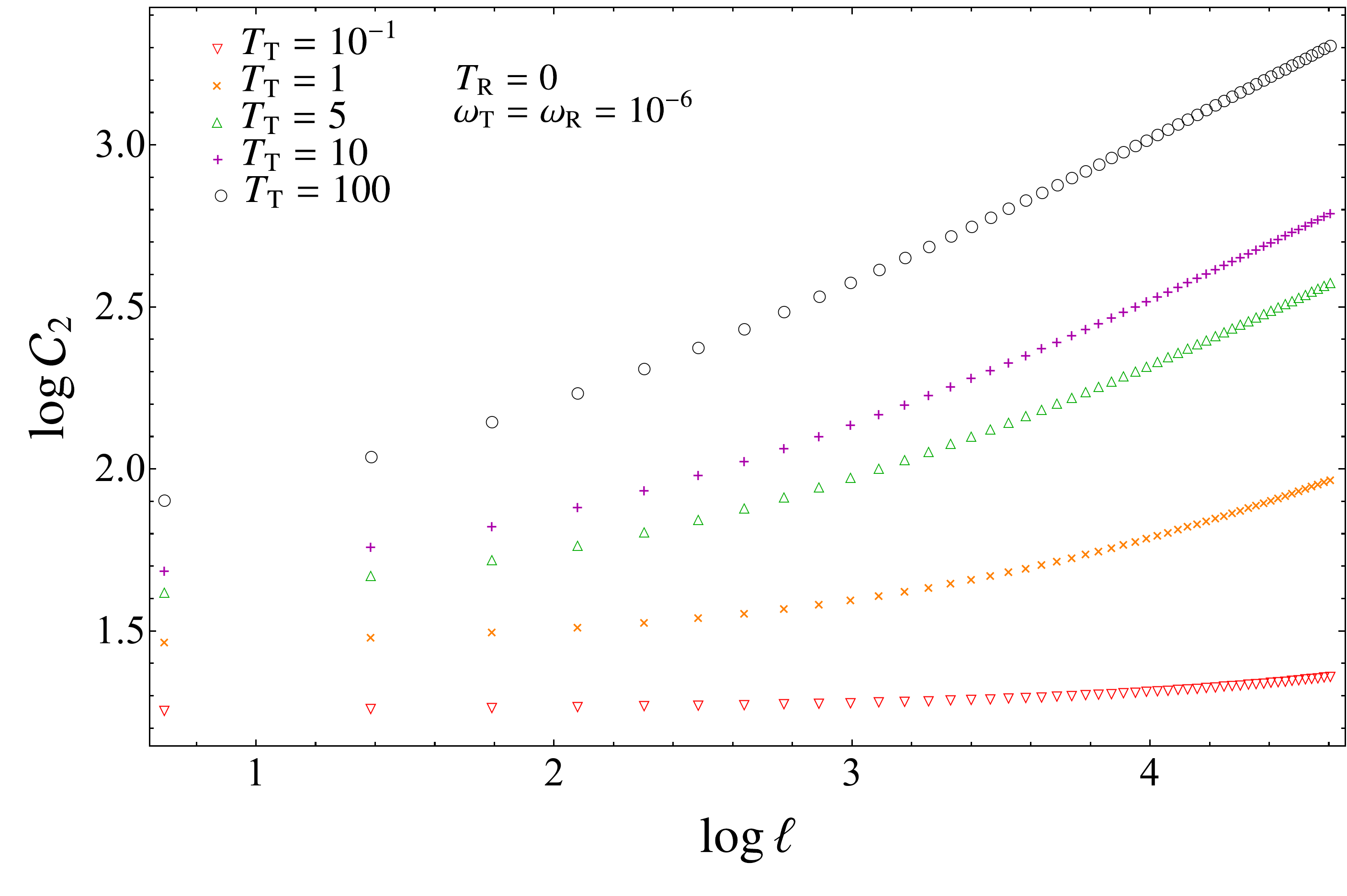}}
\subfigure
{
\hspace{-.8cm}\includegraphics[width=.5\textwidth]{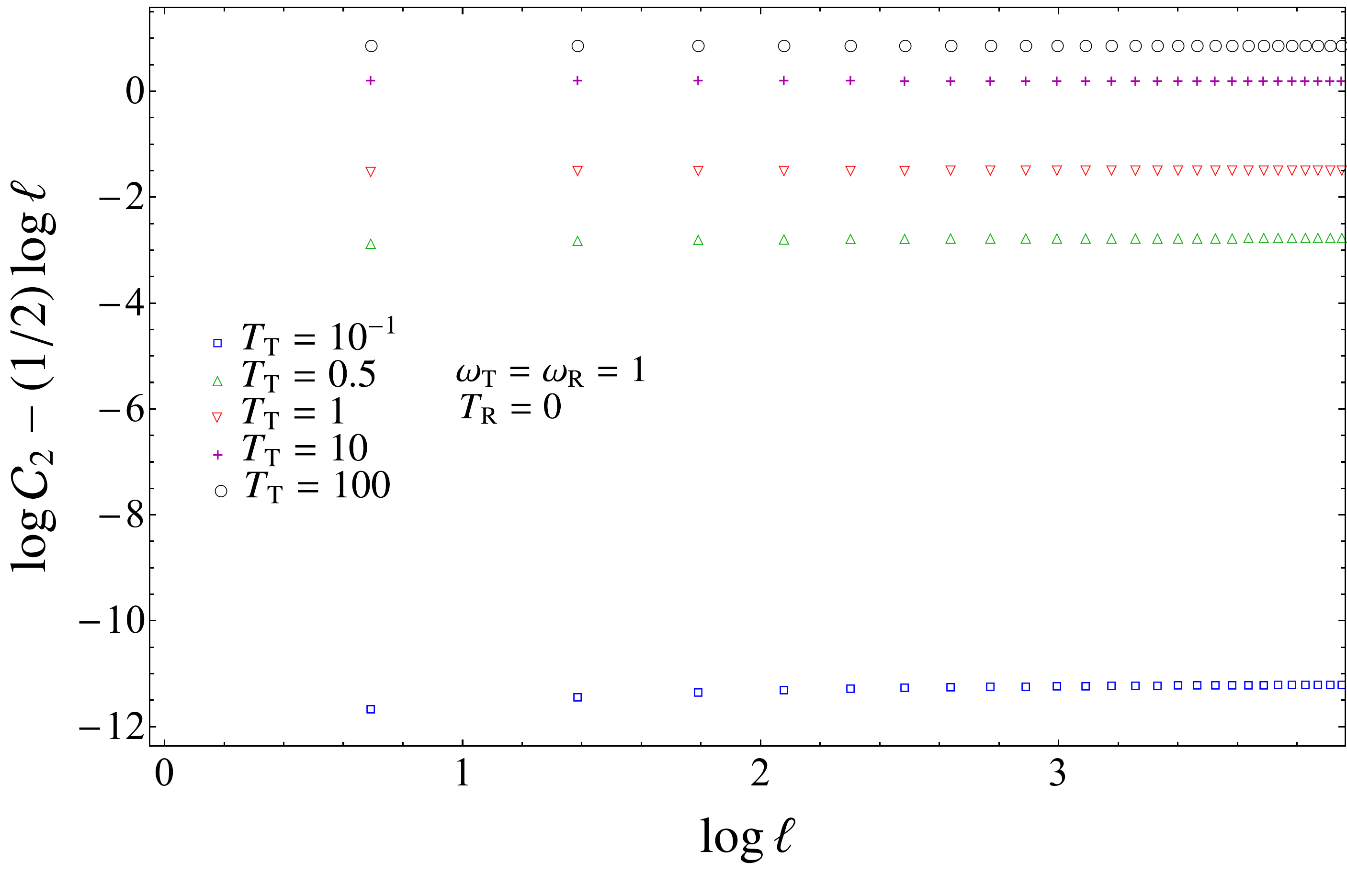}}
\subfigure
{\hspace{.45cm}
\includegraphics[width=.5\textwidth]{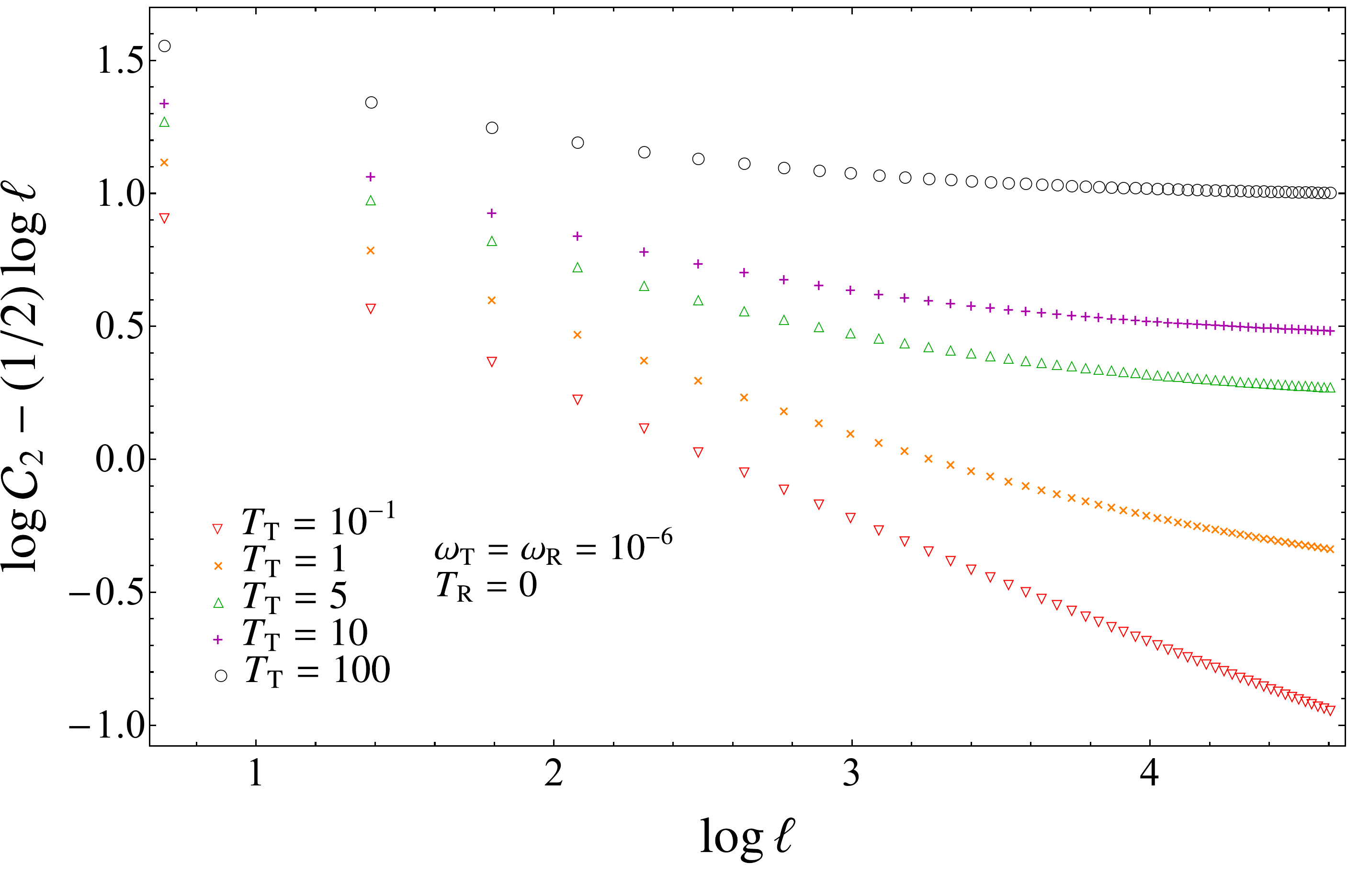}}
\caption{Subregion complexity $\mathcal{C}_2$ for an interval made by $\ell$ sites in an infinite harmonic chain as function of $\ell$. The infinite harmonic chain is in a thermal state and $T_\textrm{\tiny R} \neq T_\textrm{\tiny T}$.
Here $\kappa=m=1$; hence $\tilde{\omega}_\textrm{\tiny R}=\omega_\textrm{\tiny R}$, 
$\tilde{\omega}_\textrm{\tiny T}=\omega_\textrm{\tiny T}$, $\tilde{T}_\textrm{\tiny R}=T_\textrm{\tiny R}$ and 
$\tilde{T}_\textrm{\tiny T}=T_\textrm{\tiny T}$.
We fix $T_\textrm{\tiny R}=0$, considering various values for $T_\textrm{\tiny T}$.
We set $\omega_\textrm{\tiny T} = \omega_\textrm{\tiny R}=1$ in the left panels and
$\omega_\textrm{\tiny T} = \omega_\textrm{\tiny R}=10^{-6}$ in the right panels.
The subregion complexity $\mathcal{C}_2$ is reported in the top panels,
while its subleading term is studied in the bottom panels. 
} 
\vspace{0.4cm}
\label{fig:FRCompThSubRegionfunctionofL}
\end{figure}

The limit $\widetilde{T}\to 0$ of these expressions provides the two-point correlators in the ground state, namely
\bea
\label{qq-corr-TD}
\langle \hat{q}_i \hat{q}_j \rangle
&=&
\frac{1}{2 \pi} \int_0^{\pi} \!
\frac{1}{\Omega_\theta}  \cos[2\theta\, (i-j)] d\theta
\\
\rule{0pt}{.9cm}
\label{pp-corr-TD}
\langle \hat{p}_i \hat{p}_j \rangle
&=&
\frac{1}{2 \pi} \int_0^{\pi}\!
\Omega_\theta\cos[2\theta\, (i-j)]\, d\theta
\eea
whose analytic expressions read \cite{Botero04}
\bea
\label{qq-corr-pbc-int}
& & \hspace{-1.5cm}
\langle \hat{q}_i \hat{q}_j \rangle
\,=\,
\frac{\mu^{i-j+1/2}}{2}\,
\binom{i-j-1/2}{i-j}\,
_2F_1 \big(  1/2\, , i-j+ 1/2\, , i-j+1 \,,  \mu^2 \,\big)
\\
\rule{0pt}{.9cm}
\label{pp-corr-pbc-int}
& & \hspace{-1.5cm}
\langle \hat{p}_i \hat{p}_j \rangle
\,=\,
\frac{\mu^{i-j-1/2}}{2}\,
\binom{i-j-3/2}{i-j}\;
_2F_1 \big( - 1/2\, , i-j- 1/2\, , i-j+1 \,,  \mu^2 \,\big)
\eea
where the parameter $\mu$ depends only on $\tilde{\omega}$ as follows
\be
\label{z-def}
\mu \equiv \frac{1}{4}\Big(\tilde{\omega}-\sqrt{\tilde{\omega}^2+4} \,\Big)^2\,.
\ee

In Fig.\,\ref{fig:FRCompSubRegionfunctionofL}
we consider the subregion complexity for a block made by $\ell$ consecutive sites in an infinite harmonic chain 
when the entire system is in its ground state and $\omega_\textrm{\tiny R} \neq \omega_\textrm{\tiny T}$.
The data reported in  Fig.\,\ref{fig:FRCompSubRegionfunctionofL},
where we have fixed $\omega_\textrm{\tiny R}=1$, allow to conclude that
\be
\label{subregionCompnumerics}
\log\mathcal{C}_2=\frac{1}{2}\log\ell + O(1)
\;\;\;\;\qquad\;\;\;\;
\ell\to \infty
\ee
where the additive constant depends on $\omega_\textrm{\tiny T}$.
Comparing the left panels and the right panels, we observe that larger values for $\ell$ are needed to 
reach the behaviour (\ref{subregionCompnumerics}) for these choices of $\omega_\textrm{\tiny R} > \omega_\textrm{\tiny T}$.
We checked numerically that, when $\omega_\textrm{\tiny R} \neq1$,
(\ref{subregionCompnumerics}) holds with a subleading term that depends also on $\omega_\textrm{\tiny R}$.

In Fig.\,\ref{fig:FRCompThSubRegionfunctionofL} we have reported
the subregion complexity for a block made by $\ell$ consecutive sites in an infinite harmonic chain 
when the entire system is in a thermal state and $\omega_\textrm{\tiny R} = \omega_\textrm{\tiny T}$.
In particular, we have chosen $T_\textrm{\tiny R}=0$ and various values of $T_\textrm{\tiny T}>0$.
In the left panels we have considered $\omega_\textrm{\tiny T}=\omega_\textrm{\tiny R}=1$,
finding a reasonable agreement with (\ref{subregionCompnumerics}), 
where the subleading constant term depends on $T_\textrm{\tiny T}$.
In the right panels we have fixed $\omega_\textrm{\tiny T}=\omega_\textrm{\tiny R}=10^{-6}$,
finding that the behaviour (\ref{subregionCompnumerics}) is more difficult to observe 
as $T_\textrm{\tiny T} \to 0$ because larger values for $\ell$ are needed.

\subsection{Mutual complexity of reduced density matrices}
\label{subsec:mutualsubregion}

The complexity of the ground states and of the thermal states, 
considered in Sec.\,\ref{sec-pure-states-examples} and Sec.\,\ref{sec-thermal-states-examples} respectively, 
grow like $\sqrt{L}$ as $L \to \infty$, where $L$ is the number of sites composing the entire periodic chain.
Furthermore, considering an interval made by $\ell$ sites in an infinite harmonic chain,
the numerical results reported in Sec.\,\ref{sec-subregion-examples} 
tell us that the subregion complexity for  this interval grows like $\sqrt{\ell}$ as $\ell \to \infty$.

Given a spatial subregion $A$ and 
the density matrices $\hat{\rho}_{\textrm{\tiny R}}$ and $\hat{\rho}_{\textrm{\tiny T}}$,
which can correspond to pure or mixed states,
let us denote by $\mathcal{C}_{\textrm{\tiny R,T}}(A)$ the subregion complexity between the reduced
density matrices $\hat{\rho}_{\textrm{\tiny R},A}$ and $\hat{\rho}_{\textrm{\tiny T},A}$
introduced in Sec.\,\ref{sec-subregion-examples}.

In this subsection we consider the cases where the spatial subregion $A$ is bipartite
into two complementary spatial subregions $A_1$ and $A_2$ such that 
$A=A_1\cup A_2$.
For this spatial configuration,
various entanglement quantifiers like the entanglement entropies
\cite{Calabrese:2009ez,Calabrese:2010he,Cardy:2013nua,Coser:2013qda,DeNobili:2015dla,Coser:2015dvp}
(see e.g. \cite{Hubeny:2007re,Headrick:2010zt,Tonni:2010pv,Fonda:2014cca} 
for related calculations in the gauge/gravity correspondence)
and the entanglement negativity
\cite{Calabrese:2012ew,Calabrese:2012nk, Calabrese:2013mi,Eisler_2015,Coser:2015mta,Eisler_2016,DeNobili:2016nmj}
have been studied.

\begin{figure}[t!]
\subfigure
{\hspace{-.8cm}
\includegraphics[width=.51\textwidth]{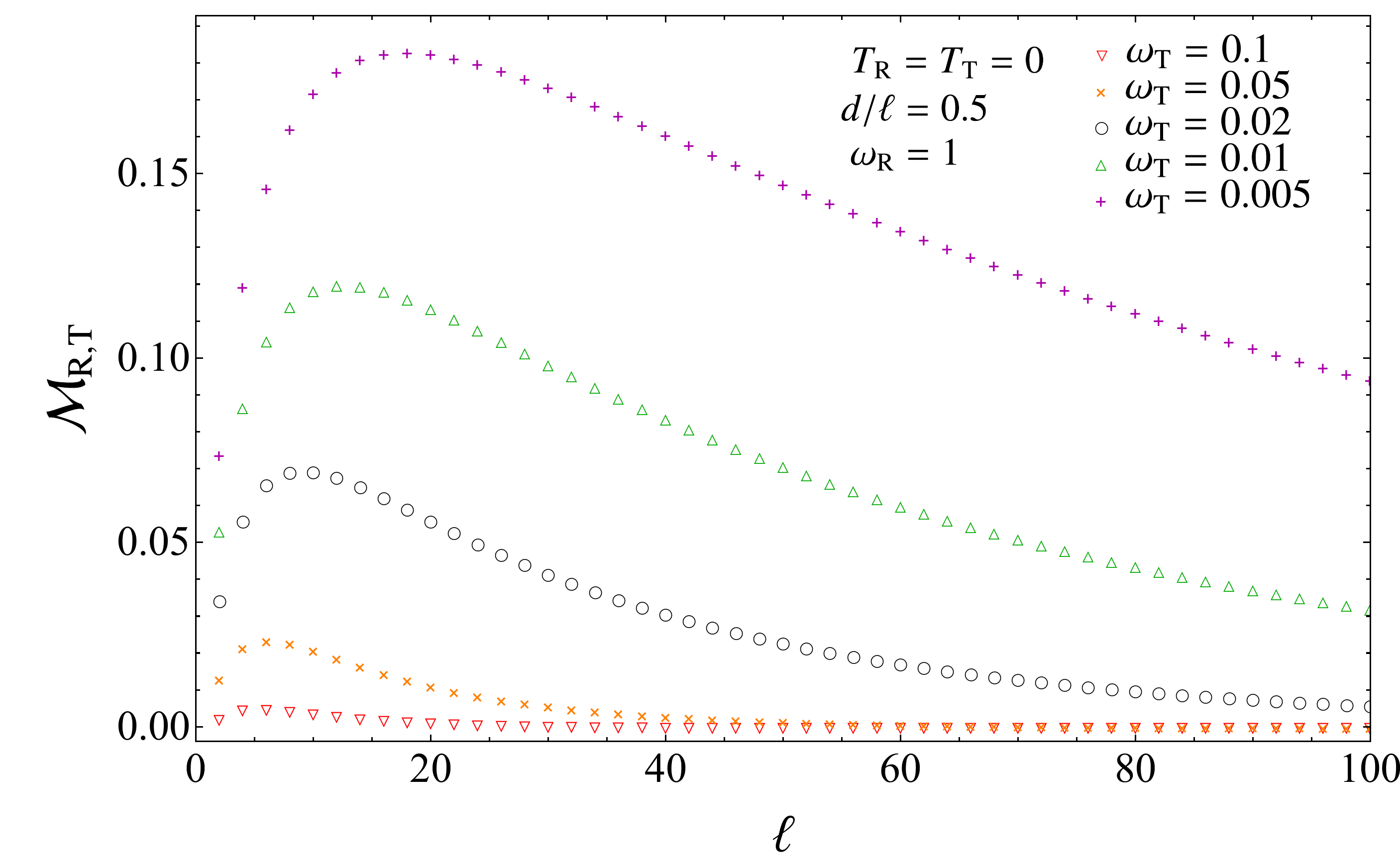}}
\subfigure
{
\hspace{.1cm}\includegraphics[width=.51\textwidth]{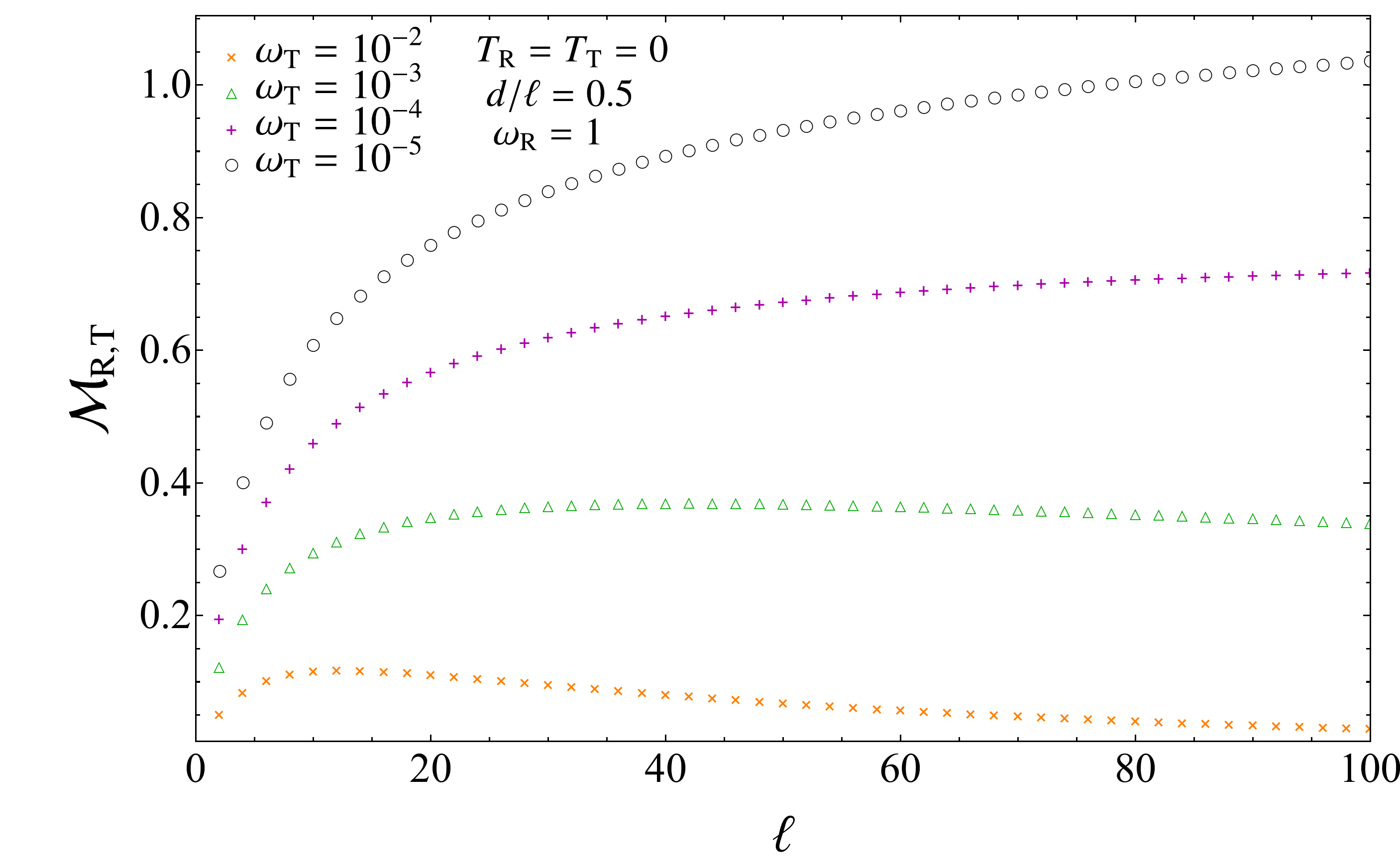}}
\subfigure
{
\hspace{-.8cm}\includegraphics[width=.51\textwidth]{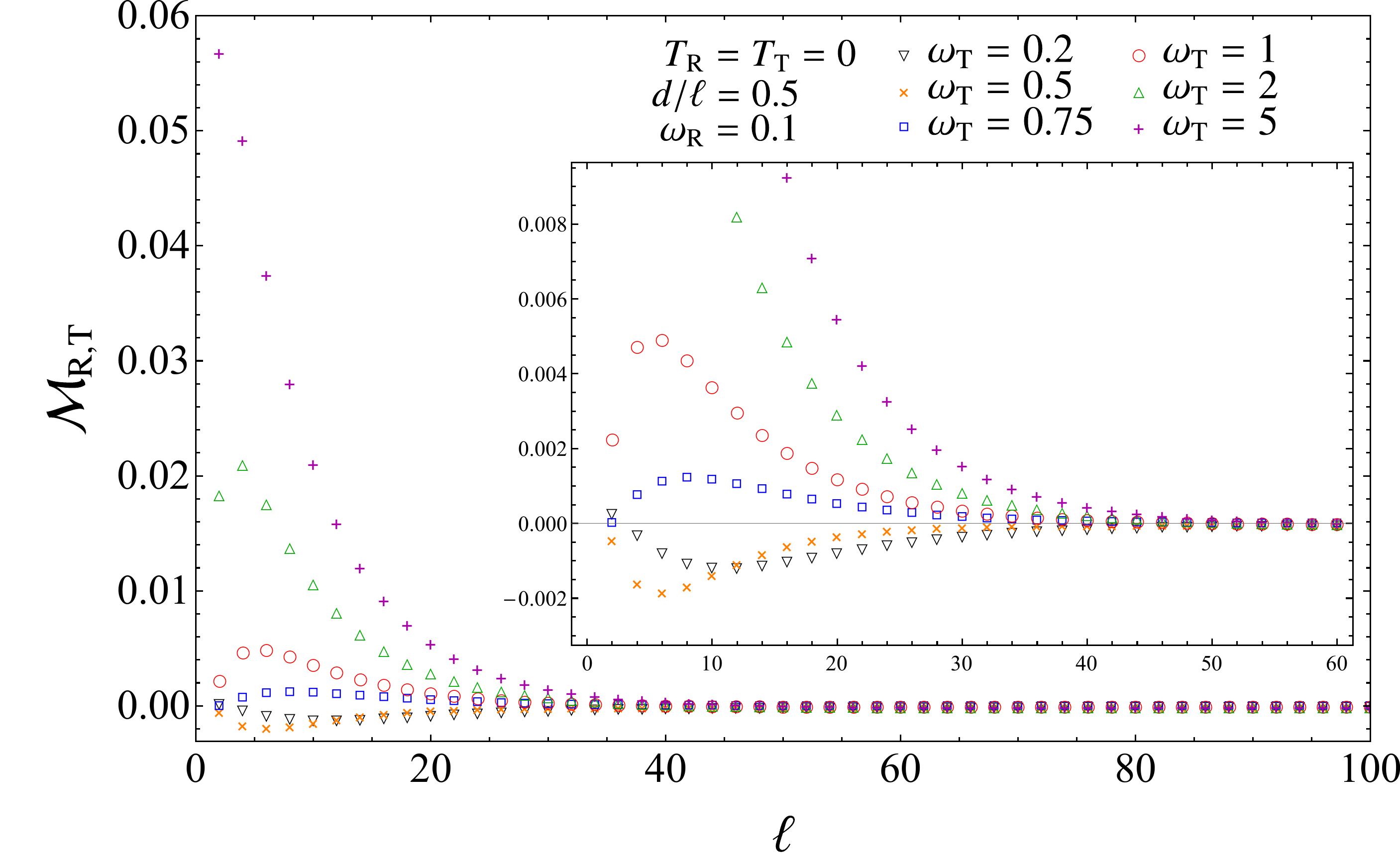}}
\subfigure
{
\hspace{-.3cm}\includegraphics[width=.51\textwidth]{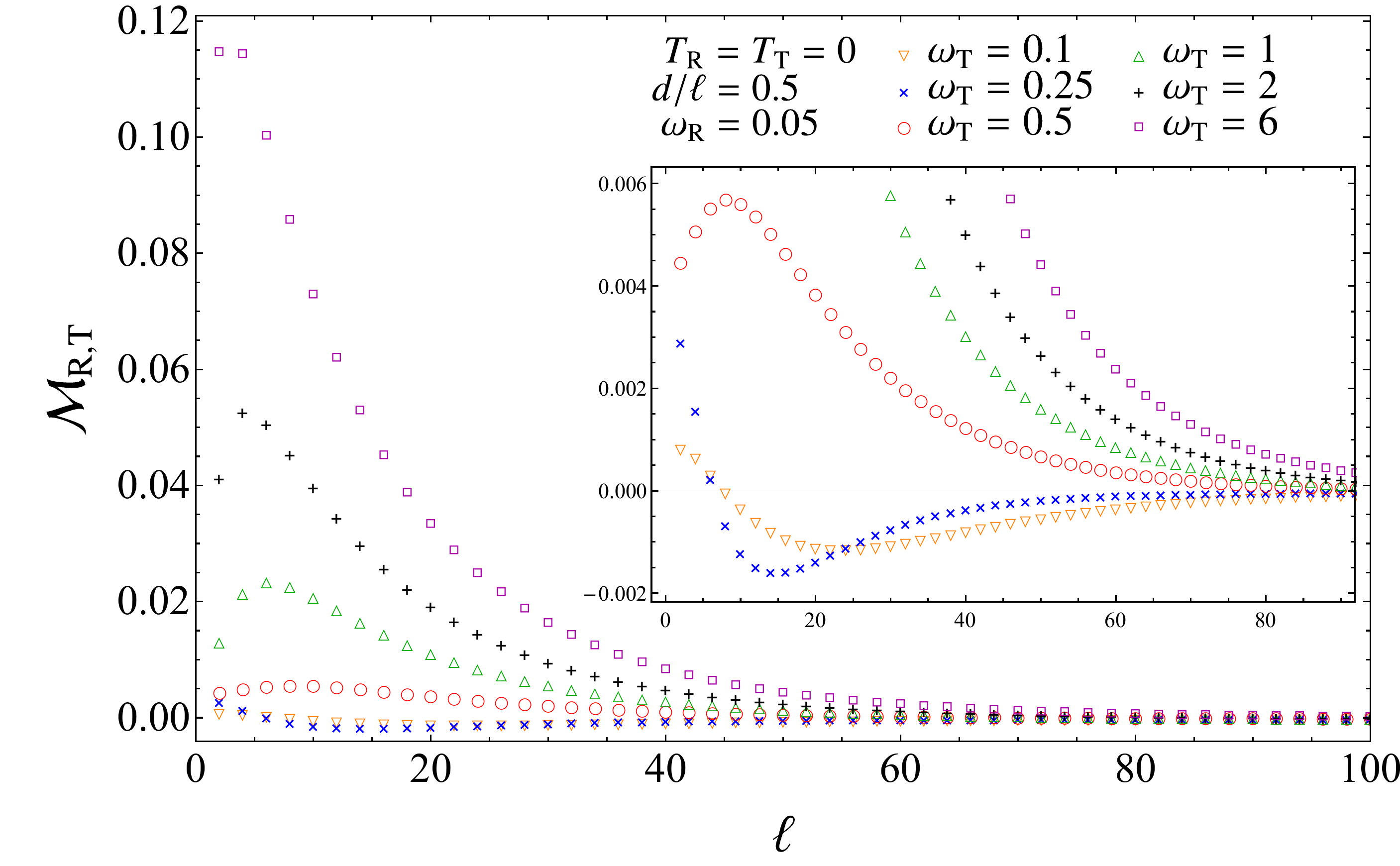}}
\subfigure
{
\hspace{-.8cm}\includegraphics[width=.51\textwidth]{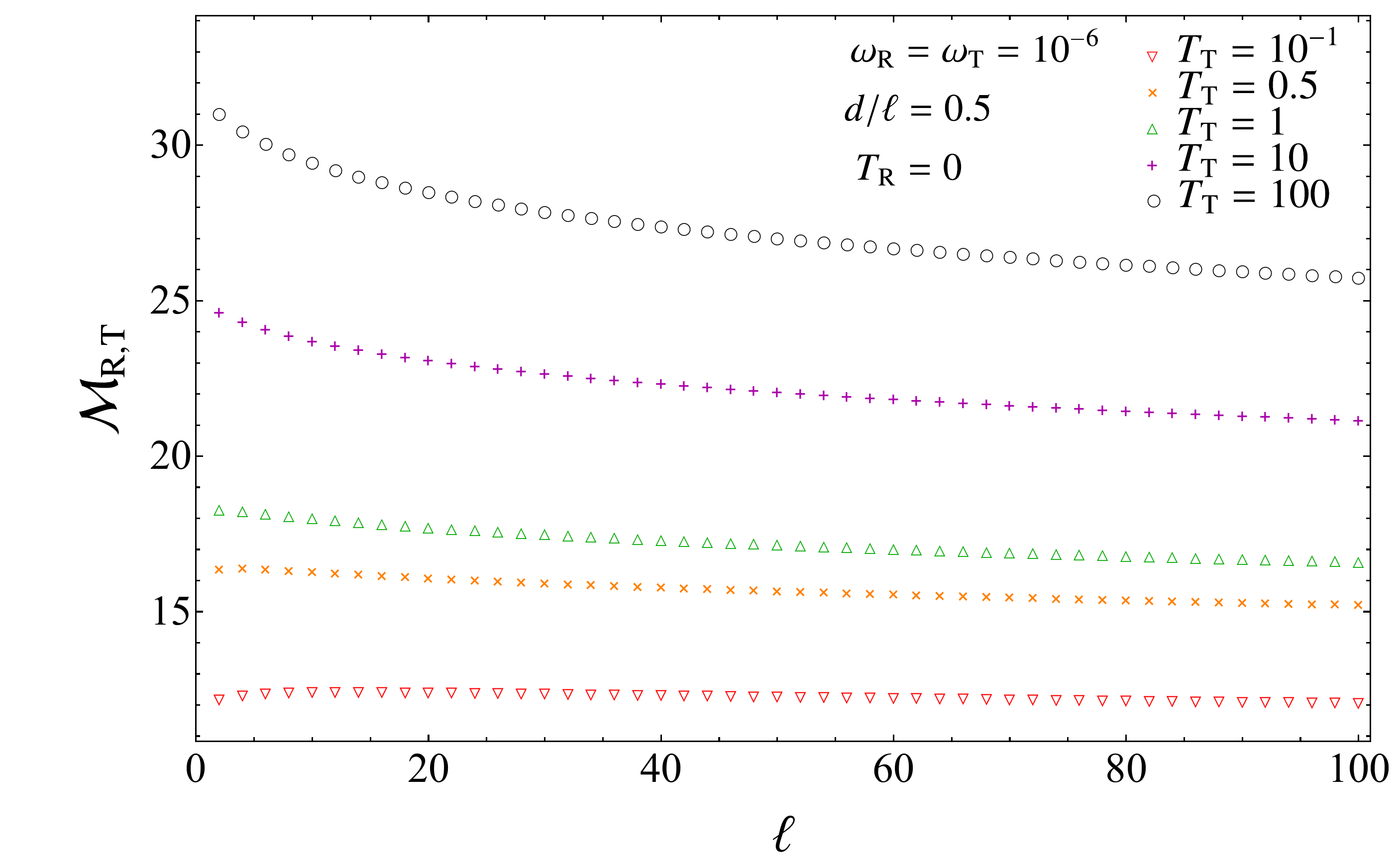}}
\subfigure
{
\hspace{.1cm}\includegraphics[width=.51\textwidth]{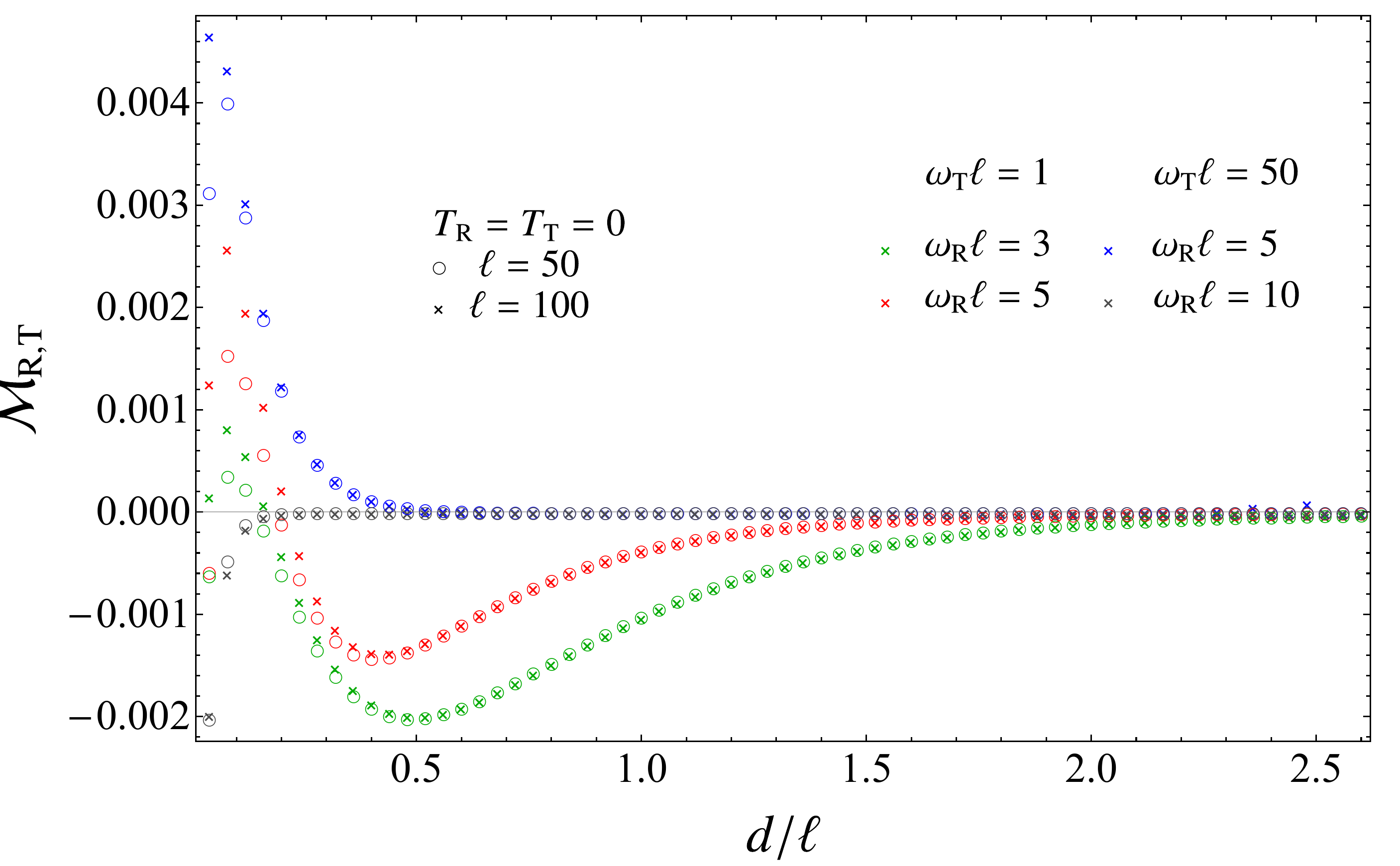}}
\caption{
The mutual complexity $\mathcal{M}_{\textrm{\tiny R,T}}$ for two disjoint equal intervals made by $\ell$ sites and separated by  $d$ sites in an infinite harmonic chain. 
Here $\kappa=m=1$, hence $\tilde{\omega}_\textrm{\tiny R}=\omega_\textrm{\tiny R}$, $\tilde{\omega}_\textrm{\tiny T}=\omega_\textrm{\tiny T}$, $\widetilde{T}_\textrm{\tiny R}=T_\textrm{\tiny R}$ and $\widetilde{T}_\textrm{\tiny T}=T_\textrm{\tiny T}$.
In the top panels, in the middle panels and in the bottom right panel 
the chain is in its ground state and $\omega_\textrm{\tiny R}\neq \omega_\textrm{\tiny T}$:
both $\omega_\textrm{\tiny T}<\omega_\textrm{\tiny R}$ (top panels)
and $\omega_\textrm{\tiny T}>\omega_\textrm{\tiny R}$ (middle panels) are considered. 
In the bottom left panel, a case involving thermal states with  
$T_\textrm{\tiny R}=0$ and various $T_\textrm{\tiny T}>0$ is explored.
The ratio $d/\ell$ is kept fixed, except in the bottom right panel, where $\mathcal{M}_{\textrm{\tiny R,T}}$ is shown as function of $d/\ell$.
}
\vspace{0cm}
\label{fig:MutualCompTnon0omegasmallfuncelldoverl0d5}
\end{figure}

The subregions $A_1$ and $A_2$ can be either disjoint or have a non vanishing intersection.
Since $\mathcal{C}_{\textrm{\tiny R,T}}(A)^2$ grows with the volume of $A$
as the number of sites in $A$ diverges, we are naturally led to introduce 
the mutual complexity for subregions as follows
\cite{Alishahiha:2018lfv,Caceres:2019pgf}
\be
\label{mutualcomp_def}
\mathcal{M}_{\textrm{\tiny R,T}}(A_1, A_2)
\equiv\,
\mathcal{C}_{\textrm{\tiny R,T}}(A_1)^2 + \mathcal{C}_{\textrm{\tiny R,T}}(A_2)^2
- \mathcal{C}_{\textrm{\tiny R,T}}(A_1 \cup A_2)^2
- \mathcal{C}_{\textrm{\tiny R,T}}(A_1 \cap A_2)^2
\ee
which is finite as the number of sites in $A_1$ and $A_2$ diverges.

In an infinite chain, let us consider the mutual complexity when $A_1$ and $A_2$ are 
two equal and disjoint intervals made by $\ell$ sites and separated by $d$ sites. 
In Fig.\,\ref{fig:MutualCompTnon0omegasmallfuncelldoverl0d5} we report the numerical results 
of the mutual complexity for this configuration 
as function of $\ell$, while $d/\ell$ is kept fixed ($d/\ell= 1/2$ for the data in the figure).

In the top and middle panels of Fig.\,\ref{fig:MutualCompTnon0omegasmallfuncelldoverl0d5},
the reference state and the target state are the ground states of the chains characterised by 
$\omega_\textrm{\tiny R}$ and $\omega_\textrm{\tiny T}$  respectively.
The mutual complexity is shown as function of $\ell$: 
for the data shown in each panel $\omega_\textrm{\tiny R}$ is fixed and the different curves
are associated to different values of $\omega_\textrm{\tiny T}$.
When $\omega_\textrm{\tiny T} < \omega_\textrm{\tiny R}$ (top panels)
the numerical curves for small values of $\ell$ are increasing until they reach a maximum
at a value of $\ell$ that depends on $\omega_\textrm{\tiny T}$ (top left panel). 
After the maximum, the mutual complexity decreases with $\ell$, 
but for many values of $\omega_\textrm{\tiny T}$ we cannot appreciate the finite asymptotic 
limit as $\ell \to \infty$ because larger values of $\ell$ are needed. 
In the top right panel, for small enough values of $\omega_\textrm{\tiny T}$, 
the values of  $\ell$ that we consider are too small to appreciate the occurrence of a maximum.

When $\omega_\textrm{\tiny T} > \omega_\textrm{\tiny R}$ (middle panels) a similar behaviour is observed:
also in these cases the position of the maximum of the curve depends on $\omega_\textrm{\tiny T}$. 
In these cases we observe that, as $\ell \to \infty$, 
the mutual complexity decreases until the zero value is reached. 
By comparing the two panels in the middle, one observes that
the value of $\ell$ where the data vanish increases when $\omega_\textrm{\tiny R}$ decreases.
Furthermore, from the middle panels we can appreciate also the fact 
that the sign of $\mathcal{M}_{\textrm{\tiny R,T}}$ is not definite:
it is mainly positive, but for some values of the parameters 
($\omega_\textrm{\tiny T}$ close to $\omega_\textrm{\tiny R}$ and $\ell$ sufficiently small) 
the curve is negative.

In the bottom left panel of Fig.\,\ref{fig:MutualCompTnon0omegasmallfuncelldoverl0d5},
the reference state is a ground state again,
while $\hat{\rho}_{\textrm{\tiny T}}$ is a thermal state at temperature $T_\textrm{\tiny T}>0$.
The curves corresponding to different values of $T_\textrm{\tiny T}>0$ decrease with $\ell$ 
and the asymptotic value depends on $T_\textrm{\tiny T}$.

In the bottom right panel of Fig.\,\ref{fig:MutualCompTnon0omegasmallfuncelldoverl0d5},
the dependence of the mutual complexity on the ratio $d/\ell$ is considered,
when the chain is in its ground state and $\omega_\textrm{\tiny R}\neq  \omega_\textrm{\tiny T}$.
We observe an interesting collapse for data corresponding to fixed values of $\omega_\textrm{\tiny R}\ell$ and  $\omega_\textrm{\tiny T}\ell$, while $\ell$ changes. 
Furthermore, the resulting curve vanishes after a critical value of $d/\ell$.
This critical ratio increases as either $\omega_\textrm{\tiny R}\ell$ or  $\omega_\textrm{\tiny T}\ell$ decreases.

\subsection{A comparison with the approach based on the purification complexity}
\label{subsec:purif_numerics}

\begin{figure}[t!]
\subfigure
{\hspace{-.8cm}
\includegraphics[width=.51\textwidth]{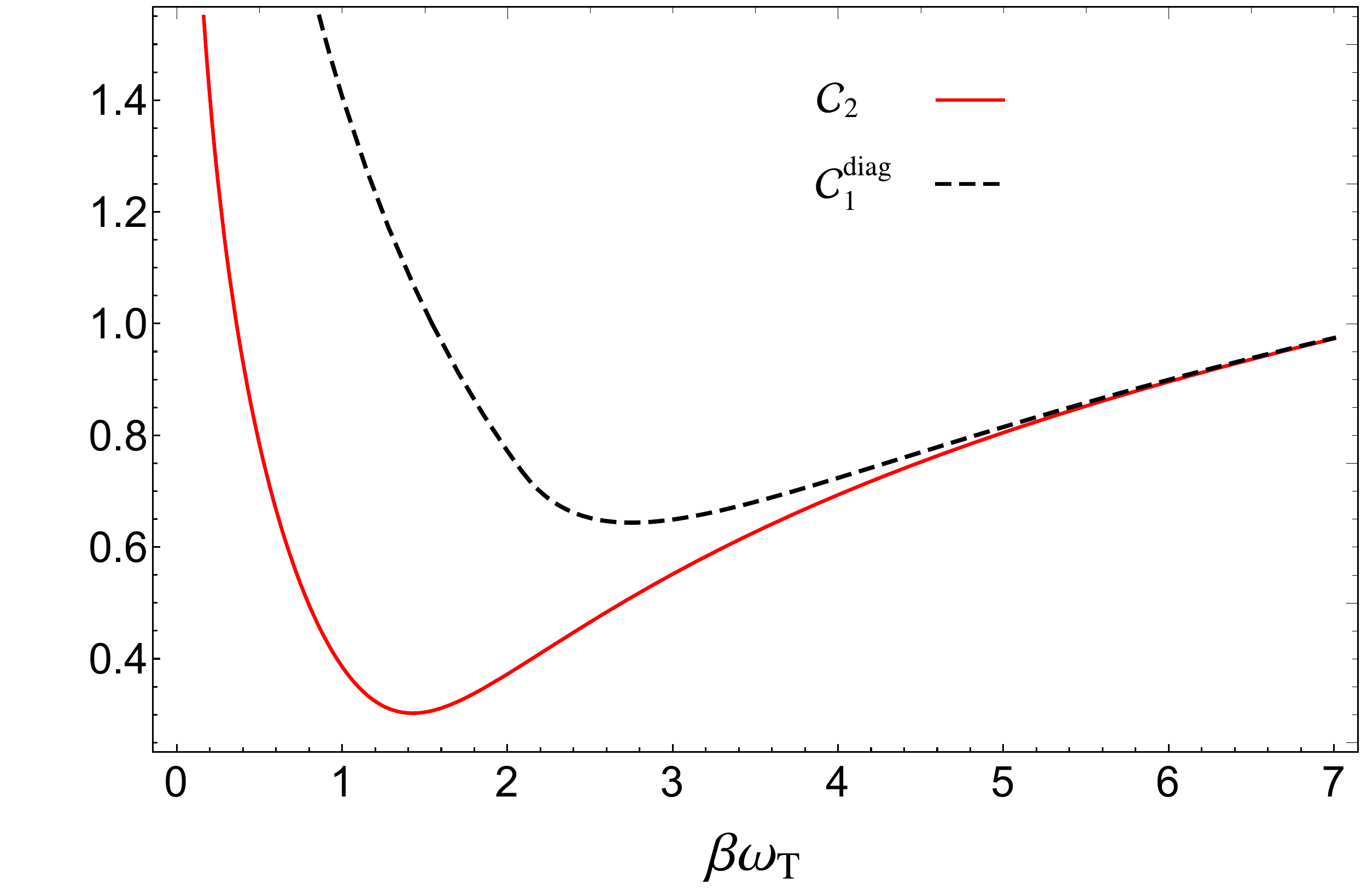}}
\subfigure
{
\hspace{.1cm}\includegraphics[width=.51\textwidth]{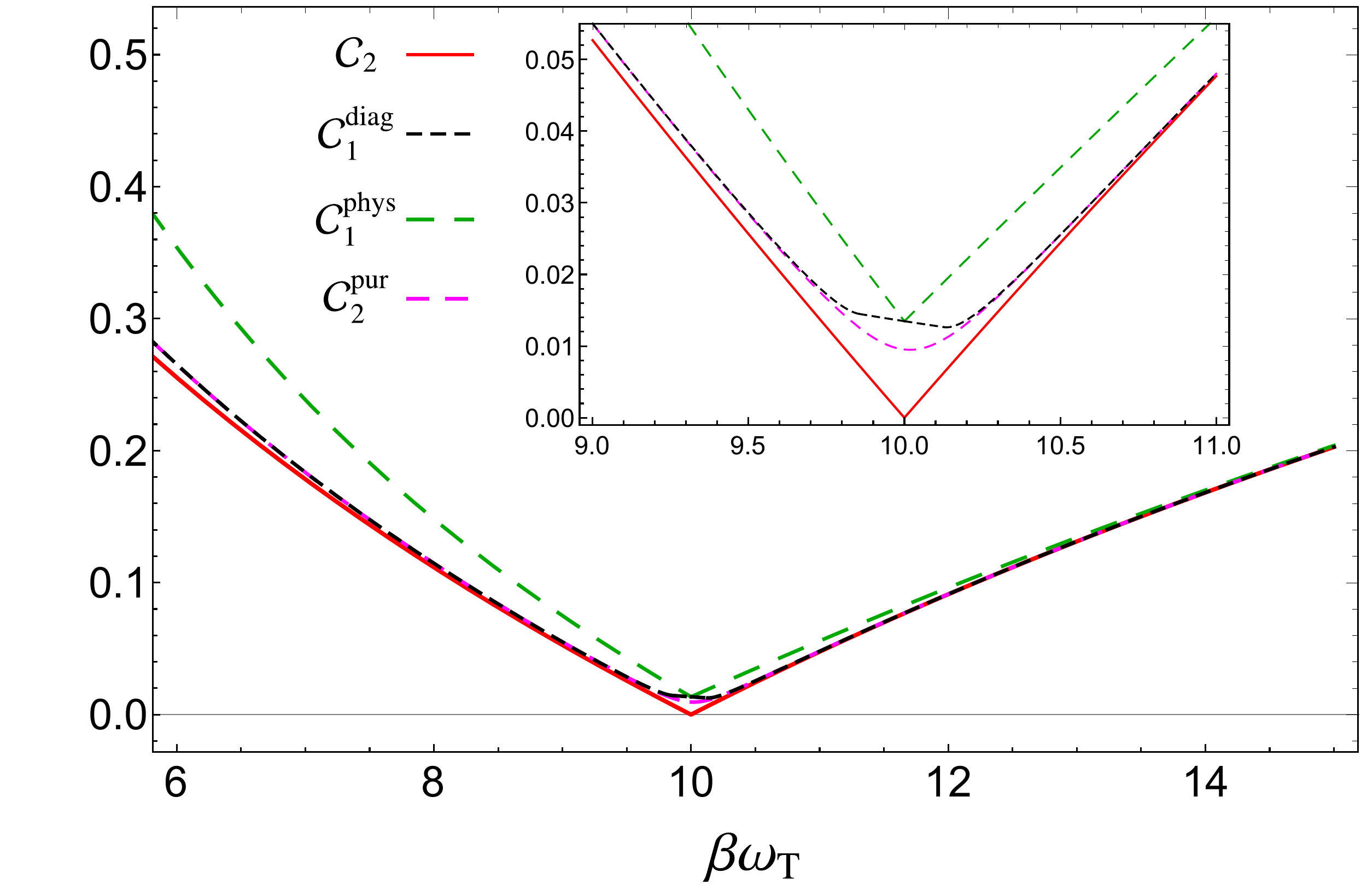}}
\vspace{-0.2cm}
\caption{
One-mode mixed states:
The complexity $\mathcal{C}_2$ from (\ref{FRComp1modefrompuretothermal}) (red solid line)
and the complexities $\mathcal{C}_1$ and $\mathcal{C}_2^{\textrm{\tiny pur}}$ based on the purification complexity,
as functions of $\beta\omega_{\textrm{\tiny T}}$. The complexity $\mathcal{C}_1$ is evaluated
for the diagonal basis from (\ref{purification complexity}) (black dashed line) 
and for  the physical basis from (\ref{purcompl_physicalbasis}) (green dashed line). 
The complexity $\mathcal{C}_2^{\textrm{\tiny pur}}$ from (\ref{C2_purificationcomp}) is independent on the basis (magenta dashed line). 
In the left panel $\beta\omega_{\textrm{\tiny R}}=1$, 
while in the right panel $\beta\omega_{\textrm{\tiny R}}=10$, 
hence also $\mathcal{C}^{\textrm{\tiny phys}}_1$ and $\mathcal{C}_2^{\textrm{\tiny pur}}$ are shown.
}
\vspace{.2 cm}
\label{fig:ComparisonFRvsPurComp}
\end{figure}

We find it worth comparing our results in Sec.\,\ref{sec-thermal-states-examples} for the complexity of thermal states 
with the corresponding ones obtained in \cite{Caceres:2019pgf} 
through the approach based on the purification complexity, that has been discussed in Sec.\,\ref{subsec:selectioncriterion}.

The results from \cite{Caceres:2019pgf} that we consider have been obtained using the $F_1$ cost function (see Appendix \ref{app:c1} for details), 
while the complexity (\ref{c2 complexity}) is based on the $F_2$ cost function.
These different cost functions lead to a different scaling with the total size $L$ of the chain. 
In particular, the $F_1$ cost function provides a complexity that diverges with $L$, 
while the $F_2$ cost function leads to the milder divergence given by $\sqrt{L}$.
This feature, which has been observed already in \cite{Jefferson:2017sdb} for pure states,
holds also for thermal states, 
as remarked in \cite{Caceres:2019pgf} for the $F_1$ cost function 
and in  (\ref{Thermal complexity TD}) for our approach, that is based on the
$F_2$ cost function.
Because of this different scaling, a meaningful comparison between these two approaches 
can be done only for one-mode mixed states, where $L=1$.
When both the reference and the target states are pure and $L=1$, 
both the $F_1$ cost function and the $F_2$ cost function provide the same complexity \cite{Jefferson:2017sdb}.

Let us consider the circuit made by one-mode mixed states where
the reference state is the ground state with frequency $\omega_{\textrm{\tiny R}}$ 
and the target state is a thermal state at inverse temperature $\beta$ with frequency $\omega_{\textrm{\tiny T}}$.

Specialising the complexity in (\ref{thermal complexity from pure to th}) to this case 
and for $m = \kappa =1$ we obtain
\be
\label{FRComp1modefrompuretothermal}
\mathcal{C}_2\,
=\,
\frac{1}{2\sqrt{2}}\,
\sqrt{
 \left[\log\left( \frac{ \omega_{{\textrm{\tiny R}}}  }
 { \omega_{{\textrm{\tiny T}}}} \,\coth\frac{\beta \omega_{{\textrm{\tiny T}}}}{2 }
 \right)
  \right]^2
\! +
\left[\log\left( \frac{ \omega_{{\textrm{\tiny T}}} }
 { \omega_{{\textrm{\tiny R}}} }\,\coth\frac{\beta \omega_{{\textrm{\tiny T}}}}{2 }
 \right)
  \right]^2}\,.
\ee

In this simple case, analytical results have been found in  \cite{Caceres:2019pgf}  
also through the approach based on the purification complexity. 
The results for the $\mathcal{C}_1$ complexity, 
which is defined through the $F_1$ cost function,
are basis dependent. 
In \cite{Caceres:2019pgf} two particular bases have been considered,
which have been called physical basis and diagonal basis 
(see Appendix\;\ref{app:c1} for their definitions).
In  the diagonal basis, the analytic result found in  \cite{Caceres:2019pgf}  for this case reads
\be
\label{purification complexity}
\mathcal{C}^{\textrm{\tiny diag}}_{1}
=
\begin{cases}
\displaystyle
\,\frac{1}{2} \log\! \left(  \frac{\omega_\textrm{\tiny R}}{\omega_\textrm{\tiny T}} \right)
+ 
\frac{1}{2} \log \! \left(\frac{\frac{\omega_\textrm{\tiny R}}{\omega_\textrm{\tiny T}} \coth{\left(\beta\omega_\textrm{\tiny T}/2\right)}-1}{\frac{\omega_\textrm{\tiny R}}{\omega_\textrm{\tiny T}}-\coth{(\beta\omega_\textrm{\tiny T}/2)}}\right)
\qquad 
&
\displaystyle
\coth \!\bigg(\! \frac{\beta\omega_\textrm{\tiny T}}{4}\! \bigg)
\leqslant 
\frac{\omega_{{\textrm{\tiny R}}}}{\omega_\textrm{\tiny T}}
\\
\rule{0pt}{.7cm}
\, \log\! \big[\coth{\left(\beta\omega_\textrm{\tiny T} /4 \right)}\big]
\qquad 
&
\displaystyle
\tanh \!\bigg(\! \frac{\beta\omega_\textrm{\tiny T}}{4}\! \bigg)
\leqslant 
\frac{\omega_{{\textrm{\tiny R}}}}{\omega_\textrm{\tiny T}}
\leqslant 
\coth \!\bigg(\! \frac{\beta\omega_\textrm{\tiny T}}{4}\! \bigg)
\\
\rule{0pt}{.8cm}
\displaystyle\,
\frac{1}{2} \log \left(\frac{\omega_\textrm{\tiny T}}{\omega_\textrm{\tiny R}}\right)
+ \frac{1}{2} \log \left(\frac{\frac{\omega_\textrm{\tiny T}}{\omega_\textrm{\tiny R}} \coth{\left(\beta\omega_\textrm{\tiny T}/2\right)}-1}{\frac{\omega_\textrm{\tiny T}}{\omega_\textrm{\tiny R}}-\coth{\left(\beta\omega_\textrm{\tiny T}/2\right)}}\right)
\qquad 
&
\displaystyle
\tanh \!\bigg(\! \frac{\beta\omega_\textrm{\tiny T}}{4}\! \bigg)
\geqslant 
\frac{\omega_{{\textrm{\tiny R}}}}{\omega_\textrm{\tiny T}}\,.
\end{cases}
\ee
For the physical basis analytic results are not available.
In the regime $\beta\omega_{\textrm{\tiny T}}\gg1$,
the following perturbative expansion has been found \cite{Caceres:2019pgf}
\be
\label{purcompl_physicalbasis}
\mathcal{C}^{\textrm{\tiny phys}}_{1}
=
\frac{1}{2}\,
\big| \log(\omega_{\textrm{\tiny R}}/\omega_{\textrm{\tiny T}}) \big|
+
\frac{\log\big[\coth\big(\beta\omega_{\textrm{\tiny T}}/4\big)\big] \, \log(\omega_{\textrm{\tiny R}} / \omega_{\textrm{\tiny T}})}{
\sqrt{\omega_{\textrm{\tiny R}}/\omega_{\textrm{\tiny T}}}-\sqrt{\omega_{\textrm{\tiny T}}/\omega_{\textrm{\tiny R}}}}+O(e^{-\beta\omega_{\textrm{\tiny T}}})\,.
\ee
As for the $F_2$ purification complexity, an analytic result valid in the entire range of the parameter is not available in the literature. 
However, in the regime where $\beta\omega_{\textrm{\tiny T}}\gg 1$, it has been found \cite{Caceres:2019pgf}
\be
\label{C2_purificationcomp}
\mathcal{C}_2^{\textrm{\tiny pur}}
=
\frac{1}{2}\,\sqrt{
\log\big(\omega_{\textrm{\tiny R}}/\omega_{\textrm{\tiny T}})
\left[
\log\big(\omega_{\textrm{\tiny R}}/\omega_{\textrm{\tiny T}})
+
\big[\log\! \big(\coth{\left(\beta\omega_\textrm{\tiny T} /4 \right)}\big)\big]^2
\bigg(\frac{\omega_{\textrm{\tiny R}}/\omega_{\textrm{\tiny T}}+1}{\omega_{\textrm{\tiny R}}/\omega_{\textrm{\tiny T}}-1}\bigg)
\right]}+
O\big(e^{-\beta\omega_{\textrm{\tiny T}}}\big)
\ee

The expressions for the complexity in (\ref{FRComp1modefrompuretothermal}), (\ref{purification complexity}), (\ref{purcompl_physicalbasis}) and (\ref{C2_purificationcomp})
depend only on $\beta\omega_{\textrm{\tiny T}} $ and on the ratio $\omega_{\textrm{\tiny R}}/\omega_{\textrm{\tiny T}}$. 
As consistency check, we notice that in the limit $\beta\omega_{\textrm{\tiny T}}\to\infty$,
where the circuit is made by pure states, 
all the expressions in (\ref{FRComp1modefrompuretothermal}), (\ref{purification complexity}), (\ref{purcompl_physicalbasis}) and (\ref{C2_purificationcomp})
become $\tfrac{1}{2} |\log( \omega_{\textrm{\tiny R}} / \omega_{\textrm{\tiny T}}) |$,
as expected from \cite{Jefferson:2017sdb}.

In Fig.\,\ref{fig:ComparisonFRvsPurComp} we show 
the expressions for the complexity in (\ref{FRComp1modefrompuretothermal}), (\ref{purification complexity}), (\ref{purcompl_physicalbasis}) and (\ref{C2_purificationcomp})
in terms of $\beta\omega_{\textrm{\tiny T}} $ for a fixed value of $\beta\omega_{\textrm{\tiny R}} $
(we choose $\beta\omega_{\textrm{\tiny R}}=1$ in the left panel and $\beta\omega_{\textrm{\tiny R}}=10$ in the right panel).
The curves for $\mathcal{C}^{\textrm{\tiny phys}}_{1}$ ad $\mathcal{C}^{\textrm{\tiny pur}}_{2}$ occur only in the right panel because
they exist only in the regime of $\beta\omega_{\textrm{\tiny T}}\gg1$.
We find it worth remarking that curves for $\mathcal{C}_2$ always lie below the curves corresponding to the
complexity evaluated through the purification complexity. 
Furthermore, as $ \beta \omega_{\textrm{\tiny T}}$ growths, all the curves collapse on the same curve,
as expected from the above observation, 
since $ \beta \omega_{\textrm{\tiny T}} \to \infty$ corresponds to the limit where the circuit is made by pure states. 
In the right panel one also notices that $\mathcal{C}^{\textrm{\tiny pur}}_{2}$ is smaller than the complexity $\mathcal{C}_{1}$ in the diagonal basis, which in turn is smaller than $\mathcal{C}_{1}$ in the physical basis,
as already remarked in \cite{Caceres:2019pgf}.

\subsection{A comparison with holography}

In the context of the gauge/gravity correspondence, 
the procedures introduced for the gravitational dual of the circuit complexity are known as
"complexity $=$ volume" proposal (CV) \cite{Susskind:2014rva,Stanford:2014jda}, 
"complexity $=$ action" proposal (CA) \cite{Brown:2015lvg,Brown:2015bva} 
and "complexity $=$ spacetime volume" proposal (CV2.0) \cite{Couch:2016exn}.

The subregion holographic complexity corresponding to these three proposals have been also explored \cite{Alishahiha:2015rta,Carmi:2016wjl, Caceres:2019pgf}.
Considering a subregion $A$ in the constant time slice of the $d$-dimensional Conformal Field Theory (CFT) on the boundary of $\textrm{AdS}_{d+1}$,
it has been found that the leading divergence of the holographic complexity is proportional to 
$V(A)/\epsilon^{d-1}$, where $V(A)$ is the volume of $A$ and $\epsilon$ is the UV cutoff. 
This divergence suggests that the holographic results should be compared with the square of the complexity $\mathcal{C}_2$
mainly explored in this manuscript. 

Denoting by $\mathcal{C}_{\textrm{\tiny  AdS}}$ the holographic complexity evaluated through one of the proposals mentioned above
and by the $\mathcal{C}_{\textrm{\tiny  AdS}}(A)$ the corresponding subregion holographic complexity, 
$\mathcal{C}_{\textrm{\tiny  AdS}}$ is superadditive when,
given two disjoint subregions $A$ and $B$ on the boundary of the AdS space, the following inequality holds
\cite{Agon:2018zso}
\be 
\label{superadditivity}
\mathcal{C}_{\textrm{\tiny AdS}}(A)+\mathcal{C}_{\textrm{\tiny AdS}}(B)\leqslant\mathcal{C}_{\textrm{\tiny AdS}}(A\cup B)
\ee
while $\mathcal{C}_{\textrm{\tiny  AdS}}$ is subadditive when the opposite inequality holds.
Equivalently, in terms of the holographic mutual complexity \cite{Alishahiha:2018lfv}
\be
\label{holomutualcom}
\mathcal{M}_{\textrm{\tiny  AdS}}(A,B)
=
\mathcal{C}_{\textrm{\tiny  AdS}}(A)+\mathcal{C}_{\textrm{\tiny  AdS}}(B)-\mathcal{C}_{\textrm{\tiny  AdS}}(A\cup B)
\ee
which should be the gravitational dual of (\ref{mutualcomp_def}), 
the holographic complexity is superadditive when $\mathcal{M}_{\textrm{\tiny  AdS}}(A,B) \leqslant 0$ for any choice of the regions $A$ and $B$.

When the gravitational background is the eternal black hole, which is the gravitational dual of the TFD state \cite{Maldacena:2001kr},
and the subregions L and R are constant time-slices of the two disconnected boundaries where the two copies of the same CFT are defined,
the definition of the holographic mutual complexity becomes \cite{Caceres:2019pgf}
\be
\label{holomutualcomTFD}
\mathcal{M}_{\textrm{\tiny  AdS}}(\textrm{TFD})=\mathcal{C}_{\textrm{\tiny  AdS}}(\textrm{L})+\mathcal{C}_{\textrm{\tiny  AdS}}(\textrm{R})-\mathcal{C}_{\textrm{\tiny  AdS}}(\textrm{TFD})
\ee

It has been found that, 
while the CV and the CV2.0 proposals for the holographic complexity are superadditive, 
this property is not always satisfied for the CA proposal \cite{Agon:2018zso,Caceres:2018blh}.

It has been shown that the mutual complexity (\ref{holomutualcomTFD}) is negative in any number of spacetime dimensions
for all the three proposals for the holographic complexity \cite{Agon:2018zso,Caceres:2019pgf}. 
This qualitatively agrees with the results that we have obtained in Sec.\;\ref{subsec:mutualcomp-TFD}
for the mutual complexity given in (\ref{mutualcompTFD}) and (\ref{mutualcompTFDv2}).

When the dual CFT is in its ground state (hence its gravitational dual is the empty AdS spacetime) and a spatial subregion is considered,
for $\textrm{AdS}_3$ it has been shown that
the holographic mutual complexity (\ref{holomutualcom}) is negative for all the three proposals \cite{Alishahiha:2015rta,Carmi:2016wjl,Caceres:2019pgf}.
This qualitatively disagrees with the results reported in Sec.\,\ref{subsec:mutualsubregion} 
because we do not observe a definite sign for the mutual complexity (\ref{mutualcomp_def}) 
(see e.g. Fig.\,\ref{fig:MutualCompTnon0omegasmallfuncelldoverl0d5}).
\\

\section{Conclusions}
\label{sec:discussion}

In this manuscript we have studied the circuit complexity of the mixed bosonic Gaussian states
occurring in the Hilbert space of harmonic lattices in any number of dimensions
by employing  the Fisher-Rao distance between Gaussian Wigner functions. 

Considering mixed states with vanishing first moments, 
by applying  a known result for the symmetric and positive definite matrices 
\cite{PETZ199681,Bhatia07book,Bathia15,Forstner2003,Lawson01}
to the covariance matrices of the model, 
we have provided the optimal circuit (\ref{optimal circuit}),
which holds when the set of the allowed gates provides circuits made only by Gaussian states.
The length (\ref{deltaAB}) of this optimal circuit
in the geometry determined by the Fisher information matrix
is identified with the circuit complexity (\ref{c2 complexity})
to obtain a target state from a given reference state
(the tolerance is zero for these circuits).
In the special case of pure states,
the known results of \cite{Jefferson:2017sdb,Chapman:2018hou}
for the $\mathcal{C}_2$ complexity have been recovered. 
For thermal states originating from the same hamiltonian,
the expression (\ref{C-2-complexity-th}) has been obtained.

The Williamson's decomposition of the covariance matrix
(see Sec.\,\ref{sec:williamson}) is the main tool
employed throughout our analysis. 
This decomposition identifies the symplectic spectrum, 
that is invariant under changes of basis
that preserve the canonical commutation relations. 
The role of the symplectic spectra and of the basis 
in the computation of the $\mathcal{C}_2$ complexity 
is made manifest in the expression (\ref{C-2-complexity}).
Furthermore, 
the Williamson's decomposition leads to natural ways
to introduce the spectrum complexity and the basis complexity 
for mixed bosonic Gaussian states
(see Sec.\,\ref{subsec:spectrum-comp} and Sec.\,\ref{subsec:basis-comp}).
This provides an explicit realisation of the proposal
made in \cite{Agon:2018zso}.

The optimal circuits described in this manuscript allow us to 
study the purification of mixed states without ancillae.
Motivated by the first law of complexity, 
in Sec.\,\ref{sec:purification} we have mainly considered the 
purification of a given mixed state through 
the $W$ path (\ref{optimal purification path}).
Further future analyses could lead to find the optimal purification path.

The Gaussian mixed states that are not pure in harmonic lattices
can be characterised also through their entanglement hamiltonian matrices.
The optimal circuit and the corresponding complexity 
in terms of the entanglement hamiltonian matrices
have been investigated in Sec.\,\ref{sec:comp-eh}.

It is important to understand how to construct the optimal circuits.
A preliminary analysis has been carried out in Sec.\,\ref{sec:gauss-channel},
where the possibility to express the optimal circuit 
in terms of Gaussian channels has been explored.
We have not been able to find 
a general solution to this interesting problem,
hence further future investigations are needed.

It is instructive to compare alternative 
quantitative approaches to the complexity of mixed states. 
The approach described in this manuscript 
holds only for the bosonic Gaussian states occurring in harmonic lattices
and it provides computable expressions for a generic number of 
degrees of freedom. 
The method discussed in \cite{Caceres:2019pgf} 
(see Sec.\,\ref{sec:purification-approaches}),
that is based on the introduction of ancillary degrees of freedom, 
can be formulated for every model
but it leads to expressions that are more difficult to evaluate.

A detailed analysis has been carried out 
in the simplest case of the harmonic chain 
either on the circle or on the infinite line (see Sec.\,\ref{sec:examples}).
Analytic or numerical results have been reported. 
For the thermal states we have explored the 
optimal path, the corresponding circuit complexity (see e.g. (\ref{ThermalComplexityHC}))
and the purification. 
Analytic and numerical results have been found 
for the mutual complexity of thermofield double states
(see Sec.\,\ref{subsec:mutualcomp-TFD}).
Finally, for the mixed states given by reduced density matrices, 
we have studied the circuit complexity for an interval in the infinite line
and the mutual complexity of two disjoint intervals
(see Sec.\,\ref{subsec:mutualsubregion}).
Interestingly, in Fig.\,\ref{fig:MutualCompTnon0omegasmallfuncelldoverl0d5}
we observe that, for two disjoint and equal intervals of length $\ell$ separated by $d$ sites, 
the mutual complexity vanishes as $\ell$ increases, while the ratio $d/\ell$ is kept fixed.
Furthermore, considering this quantity as function of $d/\ell$,
we observe that data corresponding to given value of $\omega_\textrm{\tiny T}\ell$ and $\omega_\textrm{\tiny R}\ell$ collapse 
and the resulting curves vanish after a critical value of the ratio $d/\ell$.

Our analysis mainly focusses on bosonic Gaussian states with vanishing first moments.
It is very interesting to explore also the complexity of mixed Gaussian states
whose first moments are non vanishing.
The expression (\ref{c2 complexity}) for the circuit complexity 
holds also when 
the reference state and the target state have the same first moments, 
that can be non vanishing \cite{Atkinson81,PineleCosta19,PineleCosta20}.
In Sec.\,\ref{sec-coherent-states} we have provided results for the coherent states
(pure states with non vanishing first moments) and 
the complexity (\ref{c2 complexity cs}) has been discussed \cite{Strapasson16,PineleCosta19},
finding agreement with \cite{Guo:2018kzl}.
We emphasise that an explicit expression of the Fisher-Rao distance 
in the most general case of mixed states with non vanishing first moments
is not available in the literature \cite{PineleCosta19,PineleCosta20}.
Upper and lower bounds for the complexity have been discussed in Sec.\,\ref{sec:bounds}.


The circuit complexity of mixed states is a challenging task 
deserving many future studies.

The analysis reported in this manuscript in the simple setup of bosonic Gaussian states
can be extended in various directions. 
For instance, it is a straightforward application to study 
the $\mathcal{C}_2$ circuit complexity in harmonic lattices 
in the presence of boundary, defects
 \cite{ Sato:2019kik,Braccia:2019xxi, Chapman:2018bqj}
 or in time dependent scenarios \cite{Camargo:2018eof,Chapman:2018dem,Chapman:2018lsv,Auzzi:2019mah}.

One of the main motivations of our work is to provide some tools to study complexity in quantum field theories.
Evaluating complexity of mixed states in quantum field theories remains an important challenge. 
The complexity of pure states in quantum field theories has been explored in various studies
\cite{Bhattacharyya:2018bbv,Caputa:2017urj,Czech:2017ryf,Caputa:2017yrh,
Chapman:2017rqy,
Bhattacharyya:2018wym,Caputa:2018kdj,Camargo:2018eof,Chapman:2018bqj,Camargo:2019isp,Ge:2019mjt,Bueno:2019ajd,Sato:2019kik,Erdmenger:2020sup,Flory:2020eot} 
and it would be instructive to extend these analyses to mixed states. 
The tools of Information Geometry, 
that we have largely employed in our analysis,
could provide further tools to handle this interesting problem \cite{Erdmenger:2020vmo}.

 
Let us remark that our analysis has been performed by assuming
that all the states of the quantum circuits are Gaussian.
It is important to go beyond this limitation by exploring the complexity
of circuits involving mixed states that are not Gaussian. 

Finally, we remind that 
the results reported in this manuscript have been obtained 
in the ideal situation where the maximal freedom is allowed in the choice of the gates.
Typically, only a limited number of gates 
can be employed in the  construction of  quantum circuits. 
It is worth trying to adapt our analysis to 
more realistic cases by introducing
a tolerance and various kinds of restrictions in the set of the allowed gates 
\cite{Watrous2008quantum,Aaronson:2016vto}.

\newpage
\vskip 30pt 
\centerline{\bf Acknowledgments} 
\vskip 10pt

It is a pleasure to thank Juan Hernandez, Rob Myers and Shan-Ming Ruan
for collaboration and important discussions throughout the development of this project. 
We are grateful to 
Leonardo Banchi, Fabio Benatti, Vittorio Giovannetti, Ro Jefferson and Domenico Seminara
for useful discussions. 
We thank Rob Myers also for his comments on the draft. 
ET acknowledges
the organisers of the following meetings and workshops, 
where parts of this work have been done,
for the kind hospitality and the financial support:
the program {\it The Dynamics of Quantum Information} at KITP in Santa Barbara (October 2018),
the meeting {\it Qubits on the Horizon} in Aruba (January 2019),
the long term workshop {\it Quantum Information in String Theory and Many-body Systems} 
at YITP in Kyoto (June 2019)
and the IV annual meeting of the It from Qubit Simons collaboration in New York (December 2019). 
ET's research has been conducted within the framework of the 
Trieste Institute for Theoretical Quantum Technologies (TQT).

\vskip 10pt 
\appendix


\section{Schr\"odinger representation}
\label{app:wwm}

In this Appendix we briefly discuss two aspects of the Gaussian mixed states described
in Sec.\,\ref{subsec:phase-space} in the Schr\"odinger representation.
In Sec.\,\ref{app:ww-transform} we report the kernel $\rho(\boldsymbol{q},  \tilde{\boldsymbol{q}}) 
= \langle \boldsymbol{q} |\,\hat{\rho}\,| \tilde{\boldsymbol{q}} \rangle$ of the density matrix 
corresponding to the Gaussian Wigner function (\ref{WignerGaussian-zero-moments}).
In Sec.\,\ref{app:rdm-schrod} we consider the spatial bipartition $A\cup B$
of a system in a pure state, writing the kernel $\rho_A(\boldsymbol{q}_A,  \tilde{\boldsymbol{q}}_A) $ 
for the reduced density matrix of the spatial subsystem $A$ 
in terms of the parameters occurring in the wave function of the pure state of the entire system.

\subsection{Wigner-Weyl transform}
\label{app:ww-transform}

The density matrix $\hat{\rho}$, that fully characterises a mixed state,
is hermitean, positive definite and its trace is equal to one.
Being a linear operator on the Hilbert space, 
$\hat{\rho}$ is completely determined by its kernel 
$\rho(\boldsymbol{q} , \tilde{\boldsymbol{q}})  =
\langle \boldsymbol{q} |\,\hat{\rho}\,| \tilde{\boldsymbol{q}} \rangle$
in the Schr\"odinger representation.

In this manuscript we are interested in the states whose kernel $\rho(\boldsymbol{q} , \tilde{\boldsymbol{q}}) $
is Gaussian \cite{Simon87}. This means that 
\be
\label{WWtransf}
\rho(\boldsymbol{q} , \tilde{\boldsymbol{q}}) 
=
\mathcal{N}_\rho^2 \,
\exp\bigg\{\!
-\frac{1}{2} \big(  \boldsymbol{q}^{\textrm t} ,  \tilde{\boldsymbol{q}}^{\textrm t} \big) \,
\bigg(\begin{array}{cc}
\Theta \,& - \,\Phi
\\
- \,\Phi^\ast \,& \Theta^\ast
\end{array}\bigg)
\bigg(\begin{array}{cc}
\boldsymbol{q} 
\\
\tilde{\boldsymbol{q}}
\end{array}\bigg)
\bigg\}
\ee
where $\Theta$ are $\Phi$ are $N \times N$ complex matrices.
Since the argument of the exponential in (\ref{WWtransf}) must be invariant 
under transposition, we have $\Theta=\Theta^{\textrm t}$ and $\Phi=\Phi^{\dagger}$.
This implies that (\ref{WWtransf}) is fixed by choosing $N(2N+1)$ real parameters:
$N(N+1)$ real parameters in $\Theta$ 
and $N^2$ real parameters in $\Phi$.
The normalisation condition $\int_{\mathbb{R}^N}\rho(\boldsymbol{q} , \boldsymbol{q})  \,d\boldsymbol{q}  =1$
for (\ref{WWtransf}) gives
\be
\label{norm-rho}
\mathcal{N}_\rho^2 = \frac{1}{\pi^{N/2}}\,
\sqrt{\textrm{det} [ \textrm{Re} (\Theta) - \textrm{Re} (\Phi) ]}
\ee
which is well defined when $\textrm{Re} (\Theta) - \textrm{Re} (\Phi)$ is strictly positive.
We remark that $\hat{\rho}= \hat{\rho}^\dagger$ is equivalent to 
$\rho(\boldsymbol{q} , \tilde{\boldsymbol{q}})^\ast  = \rho( \tilde{\boldsymbol{q}}, \boldsymbol{q}) $ 
 \cite{Simon87}. This condition is satisfied by (\ref{WWtransf}).

The Wigner-Weyl transform (also called Moyal transform) of 
$\rho(\boldsymbol{q} , \tilde{\boldsymbol{q}}) $ is defined as
\be
\label{wwm-transform}
w(\boldsymbol{q} , \boldsymbol{p}) =
\frac{1}{(2\pi)^N}
\int_{\mathbb{R}^N}
\rho\bigg(
\boldsymbol{q} - \frac{1}{2}\, \tilde{\boldsymbol{q}}\,  , 
\boldsymbol{q} + \frac{1}{2}\, \tilde{\boldsymbol{q}}
\bigg) 
\;e^{\textrm{i} \,\tilde{\boldsymbol{q}}^{\textrm t} \boldsymbol{p} }
\,d \tilde{\boldsymbol{q}}\,.
\ee

By using (\ref{WWtransf}) and (\ref{norm-rho}) into (\ref{wwm-transform})
and performing a Gaussian integral\footnote{The following Gaussian integral
\be
\label{Gaussian-integral}
\int_{\mathbb{R}^n} 
e^{- \,\boldsymbol{x}^{\textrm t} A\, \boldsymbol{x} + \boldsymbol{b}^{\textrm t}  \boldsymbol{x}}
\,d\boldsymbol{x}
\,=
\sqrt{ \frac{\pi^n}{\textrm{det} (A)}}\; \,
e^{\frac{1}{4} \boldsymbol{b}^{\textrm t} A^{-1} \,\boldsymbol{b}}
\ee
has been employed, where $d\boldsymbol{x} \equiv \prod_{i=1}^n dx_i$},
one finds that the Wigner-Weyl transform of (\ref{WWtransf}) is Gaussian as well.
In particular, it reads 
\be
\label{WignerfromWW}
w_{\textrm{\tiny G}}(\boldsymbol{q} , \boldsymbol{p}) 
=
\frac{1}{\pi^N} \,\sqrt{\frac{\textrm{det}(T)}{\textrm{det}(T+C)}}\;
\exp\bigg\{\!
-\frac{1}{2} 
\Big[\,
4\,\boldsymbol{q}^{\textrm t} \, T\, \boldsymbol{q}
+ \big(\boldsymbol{p} - I \boldsymbol{q}\big)^{\textrm t}  \, 
(T+C)^{-1} 
\big(\boldsymbol{p} - I \boldsymbol{q}\big)
\Big]
\bigg\}
\ee
where 
\be
\label{TIC-Re-Im}
T+C =  \frac{1}{2} \,\big[\, \textrm{Re} (\Theta) + \textrm{Re} (\Phi) \,\big]
\;\qquad\;
 T = \frac{1}{2} \,\big[\, \textrm{Re} (\Theta) - \textrm{Re} (\Phi) \,\big]
\;\qquad\;
I \,= \,\textrm{Im} (\Phi) - \textrm{Im} (\Theta)\,.
\ee
Since $\textrm{Re} (\Theta)$ and $\textrm{Re} (\Phi)$ are symmetric, 
$T$ and $C$ are $N \times N$ real and symmetric matrices,
while $I$ is a generic $N \times N$ real matrix,
given that $\textrm{Im} (\Theta)$ is symmetric and $\textrm{Im} (\Phi)$ is antisymmetric;
hence (\ref{WignerfromWW}) is determined by $N (2N+1)$ real parameters, as expected. 
The complex matrices $\Theta$ and $\Phi$ can be written in terms of the real matrices
$T$, $C$ and $I$ by inverting (\ref{TIC-Re-Im}). The result is
\be
\label{ThetaPhi-TIC}
\Theta = 2\,T+C  - \textrm{i}\,\frac{I + I^{\textrm t}}{2}
\;\;\qquad\;\;
\Phi = C  + \textrm{i}\,\frac{I - I^{\textrm t}}{2}\,.
\ee

The expression (\ref{WignerfromWW})
can be written in the form (\ref{WignerGaussian-zero-moments})
with $\gamma$ given by (\ref{gamma-block-QPM}) with
\be
\label{blocksfromWW}
Q = \frac{1}{4}\, T^{-1}
\;\;\qquad\;\;
P = T + C +\frac{1}{4}\, I^{\textrm t}\, T^{-1} I^{\textrm t}
\;\;\qquad\;\;
M =  \frac{1}{4}\, T^{-1} I^{\textrm t}\,.
\ee
This is obtained by noticing that
\be
\textrm{det}(\gamma) 
= \textrm{det}(Q)\; \textrm{det}\big(P-M^{\textrm t} \,Q^{-1}M\big)
= \frac{\textrm{det}(T+C) }{4^N\, \textrm{det}(T) }\,.
\ee
The matrices $T$, $C$ and $I$ can be expressed in terms of the blocks of $\gamma$ 
in (\ref{gamma-block-QPM}) by inverting (\ref{blocksfromWW}).
The result is
\be
T = \frac{1}{4}\, Q^{-1}
\;\;\qquad\;\;
C = P - \frac{1}{4}\, Q^{-1} - M^{\textrm t}\, Q^{-1}\, M
\;\;\qquad\;\;
I  =  M^{\textrm t}\, Q^{-1}\,.
\ee
Thus, in terms of the blocks of $\gamma$ in (\ref{gamma-block-QPM}),
the complex matrices in (\ref{WWtransf}) and (\ref{ThetaPhi-TIC}) read
\bea
\label{ThetaQPM}
& &
\Theta 
= 
P  - M^{\textrm t}\, Q^{-1}\, M + \frac{1}{4}\, Q^{-1}
- \textrm{i} \,M^{\textrm t}\, Q^{-1}
\\
\label{PhiQPM}
\rule{0pt}{.7cm}
& &
\Phi 
= 
P - M^{\textrm t}\, Q^{-1}\, M  - \frac{1}{4}\, Q^{-1}
+ \textrm{i} \,M^{\textrm t}\, Q^{-1}
\eea
These matrices are real when $\gamma = Q \oplus P$ 
in (\ref{gamma-block-QPM}) is block diagonal.

We remark that the Wigner characteristic function
in (\ref{CharFuncDef}) is related to the kernel $\rho(\boldsymbol{q} , \tilde{\boldsymbol{q}}) $ through the following relation
\be
\label{characWignervsWWMtrans} 
\rho\bigg(\boldsymbol{q}-\frac{\tilde{\boldsymbol{q}}}{2} , \boldsymbol{q}+\frac{\tilde{\boldsymbol{q}}}{2}\bigg) 
=
\int_{\mathbb{R}^N} \!
\chi(\boldsymbol{\xi})\,
e^{- \textrm{i}\, \boldsymbol{q}^{\textrm{t}}\tilde{\boldsymbol{p}}}
\,
d\tilde{\boldsymbol{p}}
\;\;\qquad\;\;
\boldsymbol{\xi} = 
\bigg( \hspace{-.0cm} 
\begin{array}{c}
\tilde{\boldsymbol{q}} \\  \tilde{\boldsymbol{p}}
\end{array} \hspace{-.0cm}  \bigg)\,.
\ee
Indeed, the Wigner function (\ref{WignerFuncDef}) is recovered 
by plugging (\ref{characWignervsWWMtrans}) into (\ref{wwm-transform}).

\subsection{Reduced density matrix}
\label{app:rdm-schrod}

In the Schr\"odinger representation, the kernel 
$\rho_A(\boldsymbol{q}_A , \tilde{\boldsymbol{q}}_A) $
corresponding to the reduced density matrix $\hat{\rho}_A$ of the subsystem $A$
of a bipartite harmonic lattice in a pure state can be computed as follows.

Considering the wavefunction (\ref{wf-ps}) for the pure state of the entire system,
the spatial bipartition $A\cup B$ of the harmonic lattice naturally leads to 
write the real and symmetric matrices $E$ and $F$ in (\ref{wf-ps}) as 
the following block matrices
\be
\label{EF-block-mat}
E
\equiv
\bigg(\begin{array}{cc}
E_A & E_{AB}
\\
E_{AB}^{\textrm t} & E_B
\end{array}\bigg)
\;\;\qquad\;\;
F
\equiv
\bigg(\begin{array}{cc}
F_A & F_{AB}
\\
F_{AB}^{\textrm t} & F_B
\end{array}\bigg)\,.
\ee
In terms of the blocks introduced in (\ref{EF-block-mat}), 
the wave function (\ref{wf-ps})  becomes
\be
\label{wf-ps-bipart}
\psi (\boldsymbol{q}_A , \boldsymbol{q}_B) \,
=
\mathcal{N}_\psi \,
\exp\bigg\{\!
-\frac{1}{2} 
\Big[\,
\boldsymbol{q}^{\textrm t}_A \, \Omega_A\, \boldsymbol{q}_A
+ \boldsymbol{q}^{\textrm t}_B \, \Omega_B\, \boldsymbol{q}_B
+ 2\,\boldsymbol{q}^{\textrm t}_A \, \Omega_{AB}\, \boldsymbol{q}_B
\,\Big]
\bigg\}
\ee
where
\be
\label{omega-matrices}
\Omega_A \equiv E_A +\textrm{i}\, F_A
\;\;\qquad\;\;
\Omega_B \equiv E_B +\textrm{i}\, F_B
\;\;\qquad\;\;
\Omega_{AB} \equiv E_{AB} +\textrm{i}\, F_{AB}
\ee
and
\be
\label{N-psi-0}
\mathcal{N}_\psi 
= 
\left( \frac{\textrm{det}(E)}{\pi^N}\right)^{1/4}
=
\frac{1}{\pi^{N/4}}\, \sqrt[4]{\textrm{det}(E_B)\; \textrm{det}\big(E_A - E_{AB}\, E_B^{-1}\, E_{AB}^{\textrm t} \big)}\,.
\ee

The kernel $\rho_A(\boldsymbol{q}_A , \tilde{\boldsymbol{q}}_A) $
corresponding to the reduced density matrix $\hat{\rho}_A$
is obtained by tracing out the degrees of freedom corresponding to the part $B$ of the harmonic lattice.
By employing (\ref{wf-ps-bipart}) and the Gaussian integral (\ref{Gaussian-integral}), 
one obtains
\bea
\label{rdm-schrod}
\rho_A(\boldsymbol{q}_A , \tilde{\boldsymbol{q}}_A) 
&=&
\int_{\mathbb{R}^{N_B}}\!
 \psi (\boldsymbol{q}_A , \boldsymbol{q}_B) \, 
\psi (\tilde{\boldsymbol{q}}_A , \boldsymbol{q}_B)^\ast\, d \boldsymbol{q}_B 
\\
\label{rdm-schrod-gauss}
\rule{0pt}{.8cm}
&=&
\mathcal{N}_\rho^2 \,
\exp\bigg\{\!
-\frac{1}{2} \big(  \boldsymbol{q}^{\textrm t}_A\, , \, \tilde{\boldsymbol{q}}^{\textrm t}_A \big) \,
\bigg(\begin{array}{cc}
\Theta_A \,& - \,\Phi_A
\\
- \,\Phi_A^\ast \,& \Theta_A^\ast
\end{array}\bigg)
\bigg(\begin{array}{cc}
\boldsymbol{q}_A 
\\
\tilde{\boldsymbol{q}}_A
\end{array}\bigg)
\bigg\}
\nonumber
\eea
where
\be
\label{ThetaPhi-A}
\Theta_A 
\,\equiv\, 
\Omega_A -\frac{1}{2}\; \Omega_{AB}\, E^{-1}_B\, \Omega_{AB}^{\textrm t}
\;\;\qquad\;\;
\Phi_A 
\,\equiv\,
\frac{1}{2}\; \Omega_{AB}\, E^{-1}_B\, \Omega_{AB}^\dagger\,.
\ee
Notice that $\Theta_A $ is symmetric and $\Phi_A $ is hermitean, as expected from the general expression in (\ref{WWtransf}).
We find it worth remarking that $F_B$ does not occur in (\ref{ThetaPhi-A}).
The real and the imaginary parts of $\Theta_A $ read respectively
\bea
\label{retheta_A}
\textrm{Re} (\Theta_A )
&=&
E_A -\frac{1}{2} \Big( E_{AB}\, E^{-1}_B  E_{AB}^{\textrm t} - F_{AB}\, E^{-1}_B  F_{AB}^{\textrm t} \Big)
\\
\label{imtheta_A}
\rule{0pt}{.7cm}
\textrm{Im} (\Theta_A )
&=&
F_A -\frac{1}{2} \Big( F_{AB}\, E^{-1}_B  E_{AB}^{\textrm t} + E_{AB}\, E^{-1}_B  F_{AB}^{\textrm t} \Big)
\eea
and for $\Phi_A $ we have respectively
\bea
\label{rephi_A}
\textrm{Re} (\Phi_A )
&=&
\frac{1}{2} \Big( E_{AB}\, E^{-1}_B  E_{AB}^{\textrm t} + F_{AB}\, E^{-1}_B  F_{AB}^{\textrm t} \Big)
\\
\label{inphi_A}
\rule{0pt}{.7cm}
\textrm{Im} (\Phi_A )
&=&
\frac{1}{2} \Big( F_{AB}\, E^{-1}_B  E_{AB}^{\textrm t} - E_{AB}\, E^{-1}_B  F_{AB}^{\textrm t} \Big)\,.
\eea
Imposing that the trace of (\ref{rdm-schrod-gauss}) is one leads to
\be
\mathcal{N}_\rho^2 = \frac{1}{\pi^{N_A/2}}\,
\sqrt{\textrm{det}\big(E_A - E_{AB}\, E_B^{-1}\, E_{AB}^{\textrm t} \big)}
\ee
which is consistent with (\ref{norm-rho}), once (\ref{retheta_A}) and (\ref{rephi_A}) have been employed.

If $\Omega_{AB}$ is left invertible
i.e. the $N_B \times N_A$ matrix $\Omega_{AB}^{-1}$ exists such that $\Omega_{AB}^{-1} \,\Omega_{AB} = \boldsymbol{1}$, 
we have that $\Omega_{AB}^\dagger$ is right invertible with 
$(\Omega_{AB}^\dagger)^{-1} = (\Omega_{AB}^{-1})^\dagger$.
Assuming also that $\Phi_A$ is invertible, 
we can isolate $\Omega_A$ and $E_B$ in (\ref{ThetaPhi-A}), finding
\be
\Omega_A \,=\,
\Theta_A + \Omega_{AB} \,\Omega_{AB}^{-1}\, \Phi_A \, \big( \Omega_{AB}^{-1}\big)^\dagger \,\Omega_{AB}^{\textrm t}
\;\;\qquad\;\;
E_B \,=\,
\frac{1}{2}\; \Omega_{AB}^\dagger\, \Phi_A^{-1}\, \Omega_{AB}\,.
\ee
Thus, for given $\Theta_A$ and $\Phi_A$, 
we have the freedom to choose the $N_B \times N_B$ real symmetric matrix $F_B$ 
and  the $N_A\times N_B$ complex matrix $\Omega_{AB}$,
namely $\tfrac{N_B(N_B+1)}{2} + 2N_A N_B$ real parameters.

In the special case $F= \boldsymbol{0}$, that has been considered e.g. in  
\cite{Caceres:2019pgf, Bhattacharyya:2018sbw}, 
the matrices in (\ref{omega-matrices}) are real.
Furthermore, from (\ref{imtheta_A}) and (\ref{inphi_A}),
we have that also $\Theta_A$ and $\Phi_A$ are real.

In Sec.\,\ref{sec:purification-approaches} we have discussed
the purification of a mixed state with $N$ sites through the introduction 
of an auxiliary lattice with $N_{\textrm{\tiny anc}}$ sites.
The results reported in this appendix can be employed 
in Sec.\,\ref{sec:purification-approaches} by setting
$N=N_A$ and $N_{\textrm{\tiny anc}}=N_B$. 
In particular, in the simplest case, which is given by $N_A=N_B=1$, 
the above counting tells us that we have three parameters to choose. 
This result has been found also in 
Sec.\,\ref{sec-purification-ext-1mode} 
by using (\ref{onemodepure_solexpli_GammaMnon0}).

The above results provide a lower bound for the number $N_B$ of ancillary degrees of freedom 
that must be introduced to purify a mixed state. 
A theorem of matrix algebra \cite{Lang04book}  guarantees that, given two matrices $M$ and $N$,
the rank of their product is bounded by $\textrm{rank}(M N)\leqslant \min[\textrm{rank}(M),\textrm{rank}(N)]$.
Applying this result to the second equation in (\ref{ThetaPhi-A}), we have that
$\textrm{rank}(\Phi_A)\leqslant\min[\textrm{rank}(E_B^{-1}),\textrm{rank}(\Omega_{AB})]$, 
where the fact that $\textrm{rank}(\Omega_{AB})=\textrm{rank}(\Omega_{AB}^\dagger)$
has been used. 
Then, since $\textrm{rank}(E_B^{-1}) = N_B$
(given that $E_B^{-1}$ is invertible)
and the rank of the $N_A\times N_B$ rectangular matrix $\Omega_{AB}$ is bounded by 
$\textrm{rank}(\Omega_{AB})\leqslant \min[N_A,N_B]\leqslant N_B$, 
we can conclude that $N_B \geqslant \textrm{rank}(\Phi_A ) $. 
In \cite{Caceres:2019pgf} this argument has been applied for real matrices.


\section{On the Fisher-Rao distance between Gaussian PDF's}
\label{app:first-moments}

In this appendix we report some known results about 
the Fisher-Rao distance between Gaussian probability distribution functions 
\cite{COSTA201559,Mahalanobis36,Bhattacharyya43,Hotelling30,
Rao45,Fisher21,Amari16book, PineleCosta19,Atkinson81,PineleCosta20}
in order to apply them to the analysis of the complexity of mixed bosonic Gaussian states.

A Gaussian probability distribution function (PDF) 
in one variable (also called univariate PDF) reads
\be
\label{univar-pdf}
p(x; \boldsymbol{\theta} )
\equiv
\frac{e^{-\frac{1}{2} (x-\mu)^2/v^2}}{\sqrt{2\pi}\; v}
\;\;\qquad\;\;
\boldsymbol{\theta} \equiv (\mu, v )
\ee
where $\mu\in\mathbb{R}$ is the mean and $v >0$ is the  standard deviation.

These Gaussian PDF's provide a manifold $\mathcal{M}_1$
once the metric is introduced through the {\it Fisher information matrix} \cite{Fisher21,Rao45,Amari16book}
\be
\label{FR-metric-2}
g_{i,j}( \boldsymbol{\theta} ) \equiv
\int_{\mathbb{R}}
\frac{\partial \log[p(x; \boldsymbol{\theta} )]}{\partial \theta_i}
\;\frac{\partial \log[p(x; \boldsymbol{\theta} )]}{\partial \theta_j}
\;p(x; \boldsymbol{\theta} )\,
dx
\ee
where $i,j \in \{1,2\}$.
Plugging (\ref{univar-pdf}) into (\ref{FR-metric-2}), one obtains the diagonal matrix
$\textrm{diag}(1/v^2, 2/v^2)$, that provides the following infinitesimal distance \cite{Rao45,Atkinson81}
\be
\label{H2metric}
ds^2=\frac{d\mu^2+2\, d v^2 }{v^2}
\ee
which characterises the hyperbolic upper half plane $\mathbb{H}_2$ after the rescaling $\mu\to\sqrt{2}\,\mu$.
Thus, by equipping the space of the univariate Gaussian PDF's parameterised by the pair $(\mu, v)$
with the metric characterised by the Fisher information matrix (\ref{FR-metric-2}),
the geodesics are either the lines with constant $\mu$ 
or the half-ellipses with eccentricity $1/\sqrt{2}$ ending on the axis $v=0$.
By evaluating the length of these geodesics, one finds that
the Fisher-Rao distance between two univariate Gaussian PDF's
associated to the parameters $\boldsymbol{\theta}_1=(\mu_1, v_1)$ and $\boldsymbol{\theta}_2=(\mu_2, v_2)$ is
\cite{Rao45, Atkinson81}
\be
\label{univariateGaussian_FR}
d^{(1)}_{\textrm{\tiny FR}}(\boldsymbol{\theta}_1,\boldsymbol{\theta}_2)
\equiv
2\,\textrm{arccosh}\bigg(1+\frac{1}{2 v_1 v_2}\bigg[\,\frac{(\mu_1-\mu_2)^2}{2}+(v_1-v_2)^2\,\bigg]\bigg)\,.
\ee
When $ \mu_1=\mu_2=\mu$, by using the relation $\textrm{arccosh}(x)=\log(x+\sqrt{x^2-1}\,)$, 
one finds that (\ref{univariateGaussian_FR}) becomes $2\log(|v_2/v_1|)$, 
which is the distance (\ref{deltaAB}) specialised to Gaussian PDF's in one variable\footnote{
The normalisation of (\ref{univariateGaussian_FR}) is different from the one used in \cite{PineleCosta19} 
}.

We are interested in the manifold $\mathcal{M}_n$ made by the Gaussian PDF's in $n$ variables $\boldsymbol{x}^{\textrm{t}} \in \mathbb{R}^n$, 
which are (see (\ref{WignerGaussian}) with $n=2N$)
\be
\label{pdf-multi}
p(\boldsymbol{x}; \boldsymbol{\theta})
\equiv
\frac{e^{-\frac{1}{2} (\boldsymbol{x} - \boldsymbol{\mu})^{\textrm{t}}  \,\Sigma^{-1}  (\boldsymbol{x} - \boldsymbol{\mu})}}{(2\pi)^{n/2} \, \sqrt{\textrm{det}(\Sigma)}}
\;\;\qquad\;\;
\boldsymbol{\theta} \equiv (\boldsymbol{\mu} , \Sigma)
\ee
where $\boldsymbol{\mu}^{\textrm{t}} \in \mathbb{R}^n$ is the mean vector
and $\Sigma$ is a $n \times n$ positive definite symmetric matrix called covariance matrix.
The parameter space for $\boldsymbol{\theta}$ has $n + n(n+1)/2$ real dimensions:
$n$ parameters for $\boldsymbol{\mu}$ and $n(n+1)/2$ for $\Sigma$.
In this space, it would be interesting to have 
a closed form for the Fisher-Rao distance that generalises (\ref{univariateGaussian_FR})  to $n\geqslant 1$.
Nonetheless, important explicit results have been obtained for some interesting submanifolds of $\mathcal{M}_n$.

In 1976, S.\,T.\,Jensen \citep{Atkinson81}  found that the $n(n+1)/2$ dimensional submanifold 
$\mathcal{M}_{\boldsymbol{\mu}_0}$ defined by the Gaussian PDF's with the same $\boldsymbol{\mu}= \boldsymbol{\mu}_0$
is totally geodesic\footnote{A submanifold $\widetilde{\mathcal{M}} \subset \mathcal{M}$ is totally geodesics if 
any geodesics in $\widetilde{\mathcal{M}}$ is also geodesics in $\mathcal{M}$ \cite{Helgason78book}.}
and that the Fisher-Rao distance in this case becomes
\be
\label{FR-distance-mu}
d_{\boldsymbol{\mu}_0} (\boldsymbol{\theta}_1, \boldsymbol{\theta}_2) 
\equiv 
\Bigg[\sum_{i=1}^n \big( \log(\lambda_i)\big)^2\,\Bigg]^{1/2}
\;\;\qquad\;\;
\boldsymbol{\theta}_j \equiv (\boldsymbol{\mu}_0 , \Sigma_j)
\ee
where $\lambda_i$ are the eigenvalues of $\Sigma_1^{-1/2} \,\Sigma_2\, \Sigma_1^{-1/2}$. 
The distance (\ref{FR-distance-mu}) is employed throughout this manuscript
to evaluate the complexity of mixed bosonic Gaussian states 
(see (\ref{deltaAB})).

Another interesting submanifold $\mathcal{M}_{\Sigma_0}$ to consider
is given by the Gaussian PDF's with the same covariance matrix $\Sigma = \Sigma_0$.
The Fisher-Rao distance on this submanifold becomes the 
Mahalanobis distance \cite{Mahalanobis36,PineleCosta19} 
\be
\label{Mahalanobisdistance}
d_{\Sigma_0}(\boldsymbol{\theta}_1, \boldsymbol{\theta}_2) 
\equiv 
\sqrt{2\,}\big[ (\boldsymbol{\mu}_1 - \boldsymbol{\mu}_2)^{\textrm{t}}  \,\Sigma_0^{-1}  (\boldsymbol{\mu}_1 - \boldsymbol{\mu}_2) \big]^{1/2}
\;\;\qquad\;\;
\boldsymbol{\theta}_j \equiv (\boldsymbol{\mu}_j , \Sigma_0)\,.
\ee
We remark that $\mathcal{M}_{\Sigma_0}$ is not a totally geodesic submanifold of $\mathcal{M}_n$ \cite{COSTA201559,PineleCosta19}.

It is worth considering also the submanifold $\mathcal{M}_{\textrm{\tiny diag}}$ made by the Gaussian PDF's
whose covariance matrix is diagonal, 
namely $\Sigma = \textrm{diag}(v_1^2, \dots ,v_n^2)$, with $v_i>0$. 
In this submanifold the infinitesimal distance becomes 
\cite{Skovgaard84,COSTA201559}
\be
\label{productFRmetric}
ds^2=\sum_{i=1}^{n} \frac{d\mu_i^2+2 \,d v_i^2 }{v_i^2}
\ee
which suggests that it is convenient to arrange the parameters
as $\boldsymbol{\theta} = (\boldsymbol{\theta}^{(1)}_1, \dots , \boldsymbol{\theta}^{(1)}_n)$,
with  $\boldsymbol{\theta}^{(1)}_i \equiv (\mu_i, v_i)$  in this case.
The infinitesimal distance (\ref{productFRmetric}) leads to write
the distance between two PDF's in $\mathcal{M}_{\textrm{\tiny diag}}$
in terms of (\ref{univariateGaussian_FR}) as follows
\be
\label{d_diag-def}
d_{\textrm{\tiny diag}} (\boldsymbol{\theta}_1, \boldsymbol{\theta}_2) 
\equiv 
\bigg[\, \sum_{i=1}^n d^{(1)}_{\textrm{\tiny FR}} \big(\boldsymbol{\theta}^{(1)}_{i,1}, \boldsymbol{\theta}^{(1)}_{i,2}\big) ^2\,\bigg]^{1/2}.
\ee
From (\ref{productFRmetric}) one concludes that the geodesics in $\mathcal{M}_{\textrm{\tiny diag}}$
are the curves $\boldsymbol{\theta}(s)$ such that $\boldsymbol{\theta}^{(1)}_i(s)$ is a geodesic in hyperbolic upper half plane equipped with the metric (\ref{H2metric}), 
for all $1\leqslant i \leqslant n$.
Notice that we are not guaranteed that a geodesic in $\mathcal{M}_{\textrm{\tiny diag}}$ is also a geodesic in $\mathcal{M}_n$
because $\mathcal{M}_{\textrm{\tiny diag}}$ is not a totally geodesic submanifold of $\mathcal{M}_n$.
Instead, 
a totally geodesics submanifold of $\mathcal{M}_n$ is 
$\widetilde{\mathcal{M}}_{\textrm{\tiny diag}} \subset \mathcal{M}_{\textrm{\tiny diag}}$, 
which  is made by the elements of $\mathcal{M}_{\textrm{\tiny diag}}$  such that
$\boldsymbol{\mu}$ is a given eigenvector of $\Sigma$ 
(see e.g. Proposition II.1 in \cite{Strapasson16}\footnote{In \cite{PineleCosta19} the submanifold $\widetilde{\mathcal{M}}_{\textrm{\tiny diag}}$ is denoted by $\mathcal{M}_{D \mu}$.}).
For instance, the Gaussian PDF's whose covariance matrices are proportional to the identity
are contained in $\widetilde{\mathcal{M}}_{\textrm{\tiny diag}}$ and in this case
$\boldsymbol{\mu}$ is the generic element of $\mathbb{R}^n$.

Consider a diagonal $\Sigma$ and the eigenvector 
$\boldsymbol{\mu}^{\textrm{t}}=(\mu,0,\dots,0)$ \cite{Strapasson16}. 
In this case the metric (\ref{productFRmetric}) becomes
\be
\label{productFRmetric_v2}
ds^2=\frac{d\mu^2+2 \,d v_1^2 }{v_1^2}+2\sum_{i=2}^{n} \frac{ d v_i^2 }{v_i^2}
\ee
and the corresponding geodesics can be found as discussed above \cite{COSTA201559,Strapasson16}. 
By specialising (\ref{d_diag-def}) to this case and employing (\ref{univariateGaussian_FR}),
one obtains
\be
\label{FR-mu-diag}
d_{\textrm{\tiny diag}} (\boldsymbol{\theta}_1, \boldsymbol{\theta}_2) 
=
\sqrt{ \bigg[\,2\,\textrm{arccosh}\bigg(1+\frac{(\mu_{1}-\mu_{2})^2/2+(v_{1,1}-v_{2,1})^2}{2 \,v_{1,1} v_{2,1}}\bigg)\bigg]^2
+\sum_{i=2}^n \bigg[\,2\log\!\bigg(\frac{v_{2,i}}{v_{1,i}}\bigg) \bigg]^2}\,.
\ee
Notice that, when $\Sigma$ has a degenerate spectrum, 
its eigenvectors can have more than one non vanishing components.

The Mahalanobis distance (\ref{Mahalanobisdistance}) can be applied on the submanifold $\mathcal{M}_{\Sigma_0}$,
which is not totally geodesic.
Very recently, a closed form for the distance 
$d_{\textrm{\tiny FR}}(\boldsymbol{\theta}_1 ,\boldsymbol{\theta}_2 )$
between PDF's in $\mathcal{M}_n$ 
having the same covariance matrix $\Sigma_0$  has been found \cite{PineleCosta20}.
Since $\mathcal{M}_{\Sigma_0}$ is not a totally geodesic submanifold of $\mathcal{M}_n$, 
the Mahalanobis  distance (\ref{Mahalanobisdistance}) does not necessarily correspond to
the length of a geodesic connecting two PDF's with the same covariance matrix in $\mathcal{M}_n$. 
Instead, the distance $d_{\textrm{\tiny FR}}(\boldsymbol{\theta}_1 ,\boldsymbol{\theta}_2 )$ provides the length of the 
shortest path in $\mathcal{M}_n$ between two PDF's with the same $\Sigma_0$.
Since we are not restricting to a submanifold of $\mathcal{M}_n$,
this is the proper Fisher-Rao distance in $\mathcal{M}_n$ between two PDF's with the same covariance matrix.
Thus, given two Gaussians PDF's with the same covariance matrix $\Sigma_0$,
we have that
$d_{\textrm{\tiny FR}}(\boldsymbol{\theta}_1 ,\boldsymbol{\theta}_2 )
\leqslant 
d_{\Sigma_0}(\boldsymbol{\theta}_1, \boldsymbol{\theta}_2) $.

Given two Gaussian PDF's in $\mathcal{M}_n$ parametrised by 
$\boldsymbol{\theta}_1 \equiv (\boldsymbol{\mu}_1 , \Sigma_0)$ 
and $\boldsymbol{\theta}_2 \equiv (\boldsymbol{\mu}_2 , \Sigma_0)$, 
let us consider the orthogonal matrix $\Pi$ such that 
$\Pi(\boldsymbol{\mu}_2-\boldsymbol{\mu}_1)
=(|\boldsymbol{\mu}_2-\boldsymbol{\mu}_1|,0,\dots,0)
\equiv|\boldsymbol{\mu}_2-\boldsymbol{\mu}_1|\boldsymbol{e}_1$.  
Since $\Sigma_0$ is symmetric and positive definite and $\Pi$ is orthogonal,
also the matrix $\Pi \,\Sigma_0 \,\Pi^{\textrm{t}}$ is symmetric and positive definite,
hence it can be decomposed as \cite{PineleCosta20}
\be
\label{LDL decomposition}
\Pi \,\Sigma_0 \,\Pi^{\textrm{t}}=U \,\mathcal{S}_{\Sigma_0} \,U^{\textrm{t}}
\ee
where $U$ is an upper triangular matrix with all the diagonal entries equal to one and $\mathcal{S}_{\Sigma_0}$ is a diagonal matrix with positive entries.
The Fisher-Rao distance between $\boldsymbol{\theta}_1 = (\boldsymbol{\mu}_1 , \Sigma_0)$ 
and $\boldsymbol{\theta}_2 = (\boldsymbol{\mu}_2 , \Sigma_0)$ in $\mathcal{M}_n$ found in \cite{PineleCosta20} reads
\be
\label{FR-distance-same-sigma}
d_{\textrm{\tiny FR}}(\boldsymbol{\theta}_1 ,\boldsymbol{\theta}_2 )
=
d_{\textrm{\tiny diag}} (\boldsymbol{\theta}_0, \boldsymbol{\theta}_\mu)
\,\,\qquad\,\,
\boldsymbol{\theta}_0\equiv(\boldsymbol{0},\mathcal{S}_{\Sigma_0})
\,\,\qquad\,\,
\boldsymbol{\theta}_\mu\equiv(|\boldsymbol{\mu}_2-\boldsymbol{\mu}_1|\boldsymbol{e}_1,\mathcal{S}_{\Sigma_0})
\ee
where $d_{\textrm{\tiny diag}} $ is defined in (\ref{FR-mu-diag}).

In order to construct the matrices $\Pi$ and $\mathcal{S}_{\Sigma_0}$,
let us introduce the unit vector
$\boldsymbol{m}\equiv(\boldsymbol{\mu}_2-\boldsymbol{\mu}_1)/|\boldsymbol{\mu}_2-\boldsymbol{\mu}_1|$,
observing that the orthogonal matrix $\Pi$ satsfies
$\Pi\,\boldsymbol{m}=\boldsymbol{e}_1$.
This matrix can be constructed by considering 
the basis of $\mathbb{R}^n$ given by 
$\mathfrak{B}=\big\{\boldsymbol{m},\boldsymbol{e}_1,\dots,
\boldsymbol{e}_{k-1},\boldsymbol{e}_{k+1},\dots,\boldsymbol{e}_n \big\}$, 
where $m_k \neq 0$ is a non vanishing component of $\boldsymbol{m}$
and $\boldsymbol{e}_i$ is the unit vector having only the $i$-th component equal to one. 
The standard Gram-Schmidt procedure \cite{Lang04book} allows to construct an orthonormal basis
$\bar{\mathfrak{B}}=\big\{\boldsymbol{m},\boldsymbol{u}_1,\dots,\boldsymbol{u}_{n-1} \big\}$
from $\mathfrak{B}$. Then,
the orthogonal matrix $\Pi$ in (\ref{LDL decomposition}) is the matrix whose columns are the vectors of $\bar{\mathfrak{B}}$.

The Cholesky decomposition \cite{Watkins02book} allows to write
a symmetric and positive definite matrix $M$ in a unique way as 
$M=L_c \,L_c^{\textrm{t}}$,
where $L_c$ is a lower triangular matrix.
This result can be related to (\ref{LDL decomposition})
by considering the matrix $\mathcal{I}$ having 1 on the antidiagonal and 0 elsewhere,
which satisfies $\mathcal{I}=\mathcal{I}^{\textrm{t}}=\mathcal{I}^{-1}$.
The matrix $\mathcal{I} \,\Pi \,\Sigma_0 \,\Pi^{\textrm{t}} \,\mathcal{I}$ is symmetric and positive definite,
hence its Cholesky decomposition tells us that it can be written as
$\mathcal{I} \,\Pi \,\Sigma_0 \,\Pi^{\textrm{t}}\, \mathcal{I}=L_c \,L_c^{\textrm{t}}$
in term of a lower triangular matrix $L_c$.
This gives $ \Pi \,\Sigma_0 \,\Pi^{\textrm{t}}=\mathcal{I} \,L_c \,L_c^{\textrm{t}} \,\mathcal{I}
=\mathcal{I}\, L_c \,\mathcal{I} \,(\mathcal{I}\, L_c \,\mathcal{I})^{\textrm{t}}$. 
Since $L_c$ is a lower triangular matrix, we have that $U_c \equiv \mathcal{I}\, L_c \,\mathcal{I}$
is an upper triangular matrix and it satisfies
\be
\label{Cholesky}
\Pi \,\Sigma_0\, \Pi^{\textrm{t}}=U_c \,U_c^{\textrm{t}}\,.
\ee
This decomposition agrees with (\ref{LDL decomposition}), provided that $U_c=U \mathcal{S}_{\Sigma_0}^{1/2}$.

For any upper triangle matrix $U$, we have that\footnote{Writing (\ref{upper triangle propr}) in components we have $U_{j,k}=\sum_l\widetilde{U}_{j,l}\delta_{l,k}U_{k,k}=\widetilde{U}_{j,k}U_{k,k}$. 
When $j=k$, the identity is verified because $\widetilde{U}_{j,j}=1$. 
When $j>k$, we have that $U_{j,k}=0 $ implies $\widetilde{U}_{j,k}=0 $, given that $U_{k,k}> 0$ (which comes from the Cholesky decomposition).}
\be
\label{upper triangle propr}
U = \widetilde{U} \, \textrm{diag}(U)
\ee
where $\widetilde{U} $ has $1$ along the diagonal.
Applying this to $U_c$ gives $\mathcal{S}_{\Sigma_0}^{1/2} =  \textrm{diag}(U_c)$ and $U = \widetilde{U}_c$.

When $n=1$, the distance $d_{\textrm{\tiny FR}}(\boldsymbol{\theta}_1 ,\boldsymbol{\theta}_2 )$ in (\ref{FR-distance-same-sigma})
becomes $d^{(1)}_{\textrm{\tiny FR}}(\boldsymbol{\theta}_1 ,\boldsymbol{\theta}_2 )$ in (\ref{univariateGaussian_FR}).

The above discussion can be employed to define the complexity for coherent states,
which are pure states described by Gaussian Wigner functions 
with non vanishing first moments (see Sec.\,\ref{subsec:phase-space} and Sec.\,\ref{sec-coherent-states})
\cite{Serafini17book}. 
Let us restrict to the coherent states with diagonal covariance matrices and first moments with a single non vanishing component. 
Since the coherent states are pure states, their covariance matrices are constrained by (\ref{pure-state-condition}) \cite{HolevoWerner01,Serafini17book}. 
Applying the constraints to (\ref{productFRmetric_v2}), one obtains the metric (\ref{metricFR_pure_mu}). 
This metric and the distance (\ref{FR-mu-diag}) lead to the expression (\ref{c2 complexity cs}) for the complexity for coherent states.
This is consistent with the results found in \cite{Guo:2018kzl}, as discussed in Sec.\,\ref{sec-coherent-states} in a more detailed way.
In Sec.\,\ref{sec-coherent-states} we have also exploited 
the distance (\ref{FR-distance-same-sigma}) to compute 
the complexity (\ref{c2 complexity cs same gamma0}) 
between two coherent states defined by (\ref{coherentstate def}) from the same ground state.
These states have the same covariance matrix, but different first moments.


\section{Bures distance and Hilbert-Schmidt distance}
\label{app:other_dist}

In the literature of quantum information, different distances have been constructed for mixed states,
even in the simple case of the bosonic Gaussian states.
In this appendix we discuss the Bures distance and the Hilbert-Schmidt distance \cite{GeomQuantumStates_book},
that have been introduced in Sec.\,\ref{subsec:FRdist}.
In particular, we report their expressions in terms of the covariance matrices and then
consider the special case of thermal states. 
An application of the Bures metric in the context of the complexity is discussed in \cite{Ruan:2020vze}.

The Bures distance between quantum states
(defined in (\ref{Bures distance Hilbert spacev1}) from the fidelity) 
is Riemannian and contractive
\cite{spehner2016geometric} (see Sec.\,\ref{subsec:FRdist}). 
An explicit expression for the fidelity between two bosonic Gaussian states 
in terms of the corresponding covariance matrices $\gamma_1$ and $\gamma_2$
has been found in \cite{BanchiPirandola-15}.
For vanishing first moments, it reads\footnote{In \cite{BanchiPirandola-15} 
the fidelity (\ref{Bures distance phase space}) 
between two Gaussian states with non vanishing first moments is also provided.}
\be
\label{Bures distance phase space}
\mathcal{F}(\gamma_1 , \gamma_2)
=
\frac{F_{\textrm{\tiny tot}}}{\sqrt[4]{\textrm{det}\left(\gamma_1 +\gamma_2\right)}}
\ee
where $F_{\textrm{\tiny tot}}$ is defined as
\be
\label{def_Ffunc_Bures}
F_{\textrm{\tiny tot}}^4
\,=\,
\textrm{det}\! 
\left[
2 \left(
\sqrt{\boldsymbol{1}+\frac{ (\gamma_{\textrm{\tiny aux}}\,J)^{-2}}{4}}
+\boldsymbol{1}
 \right) 
 \gamma_{\textrm{\tiny aux}} \right]
 \,\,\qquad\,\,
 \gamma_{\textrm{\tiny aux}}
 =
 J^{\textrm{t}}(\gamma_1+\gamma_2)^{-1}
 \left(
 \frac{J}{4}
 +
 \gamma_2 \,J \,\gamma_1
 \right)\,.
\ee
The Bures distance in terms of the covariance matrices can be easily obtained 
by plugging (\ref{Bures distance phase space}) into (\ref{Bures distance Hilbert spacev1}).
A canonical transformation characterised by the symplectic matrix $S$ induces the change
$ \gamma_i  \to  S \gamma_i S^{\textrm{t}}$ for the covariance matrices $\gamma_i$, with $i=1,2$.
Simple matrix algebra based on the property of the symplectic matrices
leads to conclude that the auxiliary covariance matrix $\gamma_{\textrm{\tiny aux}}$ in (\ref{def_Ffunc_Bures}) changes as $ \gamma_{\textrm{\tiny aux}}  \to  S \gamma_{\textrm{\tiny aux}} S^{\textrm{t}}$
and also that $\gamma_{\textrm{\tiny aux}} J  \to  S (\gamma_{\textrm{\tiny aux}} J)S^{-1}$.
Thus, both $F_{\textrm{\tiny tot}}^4$ in (\ref{Bures distance phase space}) 
and $\mathcal{F}(\gamma_1 , \gamma_2)$ in (\ref{def_Ffunc_Bures}) 
are left invariant by a canonical transformation.
We refer to \cite{Monras_2010,Monras13,BanchiPirandola-15} 
for the Bures distance between two density matrices that are infinitesimally close.

Let us focus on $\gamma_1$ and $\gamma_2$ corresponding to thermal states
having temperatures $T_i$ in harmonic chains with frequencies $\omega_i$, elastic constants $\kappa_i$ 
and masses $m_i$, where $i=1,2$.

By using (\ref{QP-diag-V-th}) and exploiting the fact that $V$ depends only on the size of the chain, we can easily diagonalise $\gamma_1+\gamma_2$ as follows
\be
\label{diagon_gammaplusgamma}
\gamma_1+\gamma_2 = V \big[\big( \mathcal{Q}_1+\mathcal{Q}_2 \big)\oplus\big( \mathcal{P}_1+\mathcal{P}_2 \big)\big] \,V^{-1} 
\ee
where the elements of $\mathcal{Q}_i$ and $\mathcal{P}_i$ with $i=1,2$ are defined in (\ref{diag-corr-FT-th}) and the matrix $V$ 
has been introduced in Sec.\,\ref{subsec-examples-hc}.
By employing (\ref{QP-diag-V-gs}) and the fact that $V$ is orthogonal and symplectic, we observe that
for $\gamma_{\mathrm{\tiny aux}}$ in (\ref{def_Ffunc_Bures}) we have
\be
\label{diagon_gammaaux-J}
\gamma_{\mathrm{\tiny aux}}
\,=\,
V \,\mathcal{M}_{1,2}  \,V^{-1}
\;\;\qquad\;\;
\gamma_{\mathrm{\tiny aux}} \,J
\,=\,
V \,\mathcal{M}_{1,2}  \,J\,V^{-1}
\ee
where
\be
\mathcal{M}_{1,2} 
\equiv
 \Big[\big( \mathcal{Q}_1+\mathcal{Q}_2 \big)^{-1}\oplus\big( \mathcal{P}_1+\mathcal{P}_2 \big)^{-1}\Big] \, 
\bigg(
 \frac{J}{4}
 +
 \big( \mathcal{Q}_2 \oplus\mathcal{P}_2 \big) \,J \,\big( \mathcal{Q}_1 \oplus \mathcal{P}_1 \big)
 \bigg)\,.
\ee
Notice that $\mathcal{M}_{1,2} $ is not diagonal, while $\mathcal{M}_{1,2} \, J$ is diagonal. 
From (\ref{diagon_gammaplusgamma}) and (\ref{diagon_gammaaux-J}) 
one realises that $V$ cancels in  (\ref{Bures distance phase space}) and (\ref{def_Ffunc_Bures}),
leaving the diagonal matrices $\mathcal{Q}_i$ and $\mathcal{P}_i$.
After some algebra,  we find that the fidelity
(\ref{Bures distance phase space}) for the thermal states $\gamma_i$ with $i=1,2$ becomes
\be
\label{FidelityThermalStates}
\mathcal{F}(\gamma_1 , \gamma_2)
\,=\,
\frac{1}{2^{L/2}}
\prod_{k=1}^L
\left[\,
\frac{\Omega_{1,k}\, \Omega_{2,k}\, (1+\sqrt{1-4 B_k})^2}{(\sigma_{1,k}\Omega_{1,k}+\sigma_{2,k}\Omega_{2,k})(\sigma_{2,k}\Omega_{1,k}+\sigma_{1,k}\Omega_{2,k})\,B_k}
\,\right]^{1/4}
\ee
where
\be
\label{DefinitionBThermal}
B_k
\equiv
\frac{(\sigma_{1,k}\Omega_{1,k}+\sigma_{2,k}\Omega_{2,k})(\sigma_{2,k}\Omega_{1,k}+\sigma_{1,k}\Omega_{2,k})}{(4\sigma_{1,k}\sigma_{2,k}\Omega_{1,k}+\Omega_{2,k})(4\sigma_{1,k}\sigma_{2,k}\Omega_{2,k}+\Omega_{1,k})}
\;\;\qquad\;\;
\sigma_{i,k}=
\frac{1}{2}
\coth\! \big(
\Omega_{i,k}/(2 \widetilde{T}_i)\big)\,.
\ee

The Bures distance is easily obtained by 
substituting (\ref{FidelityThermalStates}) into (\ref{Bures distance Hilbert spacev1}). The result reads
\be
\label{BuresDistThermalStates}
d_{\textrm{\tiny B}}
\,=\,
\sqrt{2}\;
\sqrt{\,1-
\frac{1}{2^{L/2}}\prod_{k=1}^L
\left[\,
\frac{\Omega_{1,k}\,\Omega_{2,k}\, (1+\sqrt{1-4 B_k})^2}{(\sigma_{1,k}\Omega_{1,k}+\sigma_{2,k}\Omega_{2,k})(\sigma_{2,k}\Omega_{1,k}+\sigma_{1,k}\Omega_{2,k}) \, B_k}
\,\right]^{1/4}
}\,.
\ee

As consistency check of this expression, we can consider the limit $\widetilde{T}_i\to 0$,
which provides the Bures distance between pure states.
In this limit all the symplectic eigenvalues are  $\tfrac{1}{2}$;
hence, from (\ref{DefinitionBThermal}) we get $B_k=\frac{1}{4}$.
Then, the fidelity (\ref{FidelityThermalStates}) simplifies to 
\be
\mathcal{F}(\gamma_1 , \gamma_2)
\,=\,
\prod_{k=1}^L
\left[\,
\frac{(\Omega_{\textrm{\tiny 1},k}+\Omega_{\textrm{\tiny 2},k})^2}{4\,\Omega_{\textrm{\tiny 1},k}\Omega_{\textrm{\tiny 2},k}}
\,\right]^{-1/4}
\!\! =\,
\prod_{k=1}^L
\left[\cosh\left(\frac{1}{2}\log\left(\frac{\Omega_{\textrm{\tiny 2},k}}{\Omega_{\textrm{\tiny 1},k}}\right) \right)\right]^{-1/2}
\ee
and the Bures distance (\ref{BuresDistThermalStates})  becomes
\bea
\label{Bures distance HC pure}
d_{\textrm{\tiny B}}(\gamma_1 , \gamma_2)
&\,=\,&
\sqrt{2}\;
\sqrt{\,
1-
\prod_{k=1}^L
\left[
\frac{1}{4}\frac{(\Omega_{\textrm{\tiny 1},k}+\Omega_{\textrm{\tiny 2},k})^2}{\Omega_{\textrm{\tiny 1},k}\Omega_{\textrm{\tiny 2},k}}
\right]^{-1/4}
}
\nonumber
\\
\rule{0pt}{1.1cm}
&=&
\sqrt{2}\;
\sqrt{
\,
1-
\prod_{k=1}^L
\left[\cosh\left(\frac{1}{2}\log\left(\frac{\Omega_{\textrm{\tiny 2},k}}{\Omega_{\textrm{\tiny 1},k}}\right) \right)\right]^{-1/2}
}
\eea
which is equal to the Fubini-Study distance between the two states, as expected.

The other distance that we consider is the Hilbert-Schmidt distance,
which has been defined in (\ref{HS distance hilbert v1})
for two generic density operators. 
When the two density matrices are infinitesimally close to each other
(i.e. $\hat{\rho}' = \hat{\rho}+ d\hat{\rho}$), this definition gives
\be
\label{HS distance hilbert infinitesimal v1}
ds^2_{\textrm{\tiny HS}}
=
\textrm{Tr} (d\rho)^2\,.
\ee

Focussing on Gaussian states, the Hilbert-Schmidt distance (\ref{HS distance hilbert v1}) 
between two mixed states can be written in terms of their covariance matrices as follows \cite{LinkStrunz15}
\be
\label{HS distance phase sp v1}
d_{\textrm{\tiny HS}}(\gamma_1 , \gamma_2) 
=
\sqrt{
\frac{1}{\sqrt{\textrm{det}(2\gamma_1)}} + \frac{1}{\sqrt{\textrm{det}(2\gamma_2)}}
- 
\frac{2}{\sqrt{\textrm{det}[\gamma_1 +\gamma_2])}}
}\,.
\ee
Since a canonical transformation characterised by the symplectic matrix $S$ induces the transformation 
$\gamma_i \to \gamma'_i = S\, \gamma_i \,S^{\textrm t}$ on the covariance matrices
and $\textrm{det}(S)=1$, it is straightforward to check that $d_{\textrm{\tiny HS}}$ 
is invariant under canonical transformations. 
The infinitesimal distance for (\ref{HS distance phase sp v1}) reads \cite{LinkStrunz15}
\be
\label{HS distance phase sp infinitesimal v1}
ds^2_{\textrm{\tiny HS}}
=
\frac{1}{16 \,\sqrt{\textrm{det} (2\gamma)}}\;
\Big\{
\big[ \textrm{Tr} (\gamma^{-1} d\gamma) \big]^2
+ 2\,
\textrm{Tr}\big[ (\gamma^{-1} d\gamma)^2 \big]
\Big\}\,.
\ee

The Hilbert-Schmidt distance (\ref{HS distance phase sp v1}) between 
the thermal states introduced in the text above (\ref{diagon_gammaplusgamma})
can be evaluated by employing (\ref{QP-diag-V-th})
and (\ref{diagon_gammaplusgamma}), where the diagonal matrices are given by (\ref{diag-corr-FT-th}).
Thus, for the determinants involved in (\ref{HS distance phase sp v1}), one finds
\be
\label{det-corr}
\textrm{det}(2\gamma_i)=
\prod_{k=1}^L
\coth ^2\!\big(
\Omega_{i,k}/(2 \widetilde{T}_i)
\big)
\ee
where $i=1,2$ and
\bea
\label{det-sum-corr}
& &\hspace{-.3cm}
\textrm{det}\!
\left[ (\gamma_1 +\gamma_2) \right]
=
\\
\rule{0pt}{.8cm}
& & \hspace{.0cm}
=\,
\prod_{k=1}^L
\left\{
\frac{1}{4}
\left[
\coth ^2\!\left(
\frac{\Omega_{1,k}}{2 \widetilde{T}_1}
\right)+
\coth ^2\!\left(
\frac{\Omega_{2,k}}{2 \widetilde{T}_2}
\right)+
\left(
\frac{\Omega_{2,k}}{\Omega_{1,k}}
+
\frac{\Omega_{1,k}}{\Omega_{2,k}}
\right)
\coth\!\left(
\frac{\Omega_{1,k}}{2 \widetilde{T}_1}
\right)
\coth\!\left(
\frac{\Omega_{2,k}}{2 \widetilde{T}_2}
\right)
\right]
\right\}\,.
\nonumber
\eea
Plugging (\ref{det-corr}) and (\ref{det-sum-corr}) into (\ref{HS distance phase sp v1}), 
in terms of the notation in (\ref{DefinitionBThermal})
we get \footnote{Our definition of covariance matrix differs from the one adopted in 
\cite{LinkStrunz15} by a factor of $2$.}
\bea
\label{HS distance th states}
d_{\textrm{\tiny HS}}(\gamma_1 , \gamma_2) 
= & &
\\
& & \hspace{-2cm}
=\,
\frac{1}{2^{L/2}}\,
\sqrt{\;
\prod_{k=1}^L \sigma_{1,k}^{-1} + 
\prod_{k=1}^L \sigma_{2,k}^{-1} -
2^{L+1}\prod_{k=1}^L
\left\lbrace
\left[
\sigma_{1,k}^2 
+ \sigma_{2,k}^2
+ \left(\frac{\Omega_{2,k}}{\Omega_{1,k}}+\frac{\Omega_{1,k}}{\Omega_{2,k}} \right)  \sigma_{1,k}  \sigma_{2,k}
\right]^{-1/2}
\right\rbrace
}\,.
\nonumber
\eea

In the special case of pure states, all the symplectic eigenvalues are equal to $\tfrac{1}{2}$;
hence (\ref{HS distance th states}) simplifies to
\bea
\label{eq:HS distance HC pure}
d_{\textrm{\tiny HS}}(\gamma_1 , \gamma_2) 
&\,=\,& 
\sqrt{
2\left\lbrace 
1-
\prod_{k=1}^L
\left[
\frac{(\Omega_{\textrm{\tiny 1},k}+\Omega_{\textrm{\tiny 2},k})^2}{4\,\Omega_{\textrm{\tiny 1},k}\,\Omega_{\textrm{\tiny 2},k}}
\right]^{-1/2}
\right\rbrace}
\nonumber
\\
\rule{0pt}{1.1cm}
&=&
 \sqrt{
2\left\lbrace 
1-
\prod_{k=1}^L
\left[\cosh\left(\frac{1}{2}\log\left(\frac{\Omega_{\textrm{\tiny 2},k}}{\Omega_{\textrm{\tiny 1},k}}\right) \right)\right]^{-1}
\right\rbrace
}\,.
\eea

It is worth comparing the Bures and the Hilbert-Schmidt distances in the case of pure states.
From (\ref{Bures distance HC pure}) and (\ref{eq:HS distance HC pure}),
one obtains
\be
d_{\textrm{\tiny HS}}
\,=\,
\sqrt{2}\,d_{\textrm{\tiny B}} \,\sqrt{1-\left(d_{\textrm{\tiny B}}/2\right)^2}\,.
\ee
The occurrence of this relation 
should be related to the fact that
the Fubini-Study distance is the natural distance between pure states
\cite{GeomQuantumStates_book,spehner2016geometric}.


\section{Comments on some matrix identities}
\label{app:matrix_id}

In this appendix we discuss some matrix identities employed throughout this manuscript.

In many matrix computations  we have used the following property
\be
\label{f(MN) identity app}
f(MN) \,=\,   N^{-1} f(NM)\, N = M\,f(NM)\, M^{-1} \,.
\ee
It is straightforward to prove
these matrix identities when $f(x) = x^n$ and $n$ is an integer number.
Nonetheless, (\ref{f(MN) identity app}) has been often employed 
for $f(x)=\log x$ or for $f(x)= x^s$ with $0\leqslant s\leqslant 1$;
hence in the following we show that (\ref{f(MN) identity app}) holds also for these functions.

The logarithm of a matrix $M$ is defined through the series expansion \cite{Hall15book}
\be
\label{log-def}
\log M
=
\sum_{k=1}^\infty \frac{(-1)^{k+1}}{k} \left( M-\boldsymbol{1} \right)^k\,.
\ee
This definition gives
\be
\label{appB-step-1}
N^{-1} \big[ \log (NM) \big] N
=
\sum_{k=1}^\infty \frac{(-1)^{k+1}}{k} \; N^{-1}\big( NM-\boldsymbol{1} \big)^k N\,.
\ee
Since for the $k$-th term of this series we have $ N^{-1}\left( NM-\boldsymbol{1} \right)^k N = \left( MN-\boldsymbol{1} \right)^k$,
(\ref{appB-step-1}) becomes
\be
\label{f(MN) identity for log}
N^{-1}\left[\log\left(NM\right)\right]N
=
\sum_{k=1}^\infty \frac{(-1)^{k+1}}{k}\left( MN-\boldsymbol{1} \right)^k
=
\log\left(M N\right)
\ee
which provides the first equality in (\ref{f(MN) identity app}) for $f(x)=\log x$.
The second equality in (\ref{f(MN) identity app}) can be obtained by 
repeating the steps in (\ref{appB-step-1}) and (\ref{f(MN) identity for log}) for
$M\big[\log(N M)\big]M^{-1}$ in (\ref{appB-step-1}).

In order to check that (\ref{f(MN) identity app}) holds also when $f(x)= x^s$, let us observe that
\be
\label{expansionExp}
\left(
NM
\right)^s
=e^{s\log(NM)}
=\sum_ {k=0}^\infty \frac{s^k}{k!}\big[\log\left( NM\right)\big]^k\,.
\ee
This leads to
\be 
\label{appB-step7}
N^{-1}
\left(
N M
\right)^s
N
=\sum_ {k=0}^\infty \frac{s^k}{k!} \, N^{-1}\big[\log\left( N M\right)\big]^k N
\ee
whose $k$-th term can be written as $N^{-1}\left[\,\log\left( N M\right)\,\right]^k N = \left[\,\log\left(M N\right)\,\right]^k$,
once (\ref{f(MN) identity for log}) has been employed. 
Thus, (\ref{appB-step7}) becomes
\be
\label{appB-step8}
N^{-1}
\left(
N M
\right)^s
N
=
\sum_ {k=0}^\infty \frac{s^k}{k!}\, \big[ \log\left( MN\right)\big]^k 
=
\left(
M N 
\right)^s
\ee
which corresponds to the first equality in (\ref{f(MN) identity app}) for $f(x)=x^s$.
The second equality in (\ref{f(MN) identity app}) for $f(x)=x^s$ can be found 
by repeating the steps in (\ref{appB-step7}) and (\ref{appB-step8}) for $M(N M)^sM^{-1}$.

Another remark deserving more detailed comments concerns (\ref{deltaAB}), where we introduced the Fisher-Rao distance
$d(\gamma_{\textrm{\tiny R}}, \gamma_{\textrm{\tiny T}})$ as 
$|\!|  \log (\gamma_{\textrm{\tiny R}}^{-1/2}\,\gamma_{\textrm{\tiny T}} \,\gamma_{\textrm{\tiny R}}^{-1/2} )  |\!|_2$, 
pointing out that this expression is not equal to $ |\!|  \log (\gamma_{\textrm{\tiny T}} \,\gamma_{\textrm{\tiny R}}^{-1} )  |\!|_2 $. 
Indeed, since  $|\!|  M  |\!|_2 \equiv \sqrt{\textrm{Tr}\left(M^{\dagger} M\right)}$ \cite{Bhatia07book},
we can exploit that $\log (\gamma_{\textrm{\tiny R}}^{-1/2}\,\gamma_{\textrm{\tiny T}} \,\gamma_{\textrm{\tiny R}}^{-1/2} )$ 
is real and symmetric to write
\be
|\!|  \log (\gamma_{\textrm{\tiny R}}^{-1/2}\,\gamma_{\textrm{\tiny T}} \,\gamma_{\textrm{\tiny R}}^{-1/2} )  |\!|_2
=
\sqrt{\textrm{Tr} \big[\log \!\big(\gamma_{\textrm{\tiny R}}^{-1/2}\,\gamma_{\textrm{\tiny T}} \,\gamma_{\textrm{\tiny R}}^{-1/2} \,\big)\big]^2}
=
\sqrt{\textrm{Tr}  \big[\log (\gamma_{\textrm{\tiny T}} \,\gamma_{\textrm{\tiny R}}^{-1} )\big]^2}
\ee
where in the last equality the cyclic property of the trace has been used. 
On the other hand, since 
$[\log (\gamma_{\textrm{\tiny T}} \,\gamma_{\textrm{\tiny R}}^{-1} )]^{\mathrm{t}}
= \log (\gamma_{\textrm{\tiny R}}^{-1} \,\gamma_{\textrm{\tiny T}} ) $, 
the matrix $ \log (\gamma_{\textrm{\tiny T}} \,\gamma_{\textrm{\tiny R}}^{-1} )$  is not symmetric, we have
\be
|\!|  \log (\gamma_{\textrm{\tiny T}} \,\gamma_{\textrm{\tiny R}}^{-1} )  |\!|_2
=
\sqrt{\textrm{Tr}\left[\log (\gamma_{\textrm{\tiny R}}^{-1}\,\gamma_{\textrm{\tiny T}} )\log (\gamma_{\textrm{\tiny T}}\,\gamma_{\textrm{\tiny R}}^{-1} )\right]}
\neq
\sqrt{\textrm{Tr}\left[\log (\gamma_{\textrm{\tiny T}} \,\gamma_{\textrm{\tiny R}}^{-1} )\right]^2}
\ee
which tells us that $d(\gamma_{\textrm{\tiny R}}, \gamma_{\textrm{\tiny T}})$ cannot be written as  
$|\!|  \log (\gamma_{\textrm{\tiny T}} \,\gamma_{\textrm{\tiny R}}^{-1} )  |\!|_2$.

We find it worth providing further details about the construction of the symplectic matrices 
occurring in the Williamson's decompositions of $\widehat{H}^{\textrm{\tiny phys}}$ 
and of some covariance matrices in Sec.\,\ref{sec:examples}.

Let us consider two symmetric and positive definite $N \times N$ real matrices $A$ and $B$
that are diagonalised by the same orthogonal real matrix $\widetilde{O}$.
It is straightforward to write $A\oplus B = O^{\textrm{t}} (\mathcal{A} \oplus \mathcal{B}) O$,
where $O\equiv \widetilde{O} \oplus \widetilde{O}$ is orthogonal and symplectic,
while the diagonal matrices
$\mathcal{A} =\textrm{diag}(\alpha_1, \dots, \alpha_N)$ and 
$\mathcal{B} =\textrm{diag}(\beta_1, \dots, \beta_N)$ collect the eigenvalues of $A$ and $B$
respectively.
The Williamson's decomposition of $A\oplus B$ reads 
$A\oplus B = W^{\textrm{t}} (\mathcal{D} \oplus \mathcal{D}) W$, where
$\mathcal{D} = \textrm{diag}(\sqrt{\alpha_1 \beta_1}, \dots, \sqrt{\alpha_N \beta_N}\,)$
and the symplectic matrix $W$ is given by
\be
\label{W-chi-O}
W \equiv \chi\, O
\;\;\qquad\;\;
\chi \equiv
\textrm{diag}\big((\alpha_1/ \beta_1)^{1/4} , \dots, (\alpha_N /\beta_N)^{1/4},  
(\alpha_1 / \beta_1)^{-1/4} , \dots, (\alpha_N / \beta_N)^{-1/4}\big)
\ee
where $\chi$ is a diagonal symplectic matrix. 
We remark that (\ref{W-chi-O}) provides the Euler decomposition (\ref{EulerDec}) with 
$\mathcal{X} = \chi$, $R=O$ and $L =\boldsymbol{1}$.


\section{Details on the first law of complexity}
\label{app:sec_first_law}

In this appendix we report some technical details concerning the first law of complexity
and the derivations of the results reported in Sec.\,\ref{sec:spectrum-basis}.

The variation of the square of the distance (\ref{deltaAB}) 
under the independent variations $ \gamma_{\textrm{\tiny T}} \to  \gamma_{\textrm{\tiny T}}+ \delta  \gamma_{\textrm{\tiny T}}$ 
and $ \gamma_{\textrm{\tiny R}} \to  \gamma_{\textrm{\tiny R}}+ \delta  \gamma_{\textrm{\tiny R}}$
of the covariance matrices of the reference and of the target state reads
\be
\label{delta dFR step1}
\delta d^2
=\,
 2\,\textrm{Tr} \Big\{ \!
\log \Delta_{\textrm{\tiny TR}}
\, \delta \big[ \log \Delta_{\textrm{\tiny TR}} \big]
\!\Big\}
=
2\,
\textrm{Tr} \Big\{ 
\big( \log \Delta_{\textrm{\tiny TR}} \big)\,
\Delta_{\textrm{\tiny TR}}^{-1} \, \delta \Delta_{\textrm{\tiny TR}}
\Big\}\,.
\ee
The last expression can be found by 
first observing that, since $M$ and $\delta M$ do not commute in general,
we can exploit the following formula \cite{AdlerNotes}
\be
\label{formula delta log}
\delta \log M 
\,=
\int_0^1 \big[ (1-b) M + b \,\boldsymbol{1}  \big]^{-1} \delta M\, \big[ (1-b) M + b \,\boldsymbol{1}  \big]^{-1}  db
\ee
When $M$ and $\delta M$ commute, by employing the matrix that diagonalises them simultaneously, 
one can easily check that 
(\ref{formula delta log}) becomes $M^{-1} \delta M = \delta M \, M^{-1}$.
In the general situation where $M$ and $\delta M$ do not necessarily commute,
from the cyclic property of the trace and the fact that different functions of the same matrix commute
we find that the first expression in (\ref{delta dFR step1}) becomes
\bea
\label{var-ds-app-step0}
\delta d^2
&=&
2\,\int_0^1
 \textrm{Tr} \bigg\{ 
  \log \Delta_{\textrm{\tiny TR}}\,
 \big[ (1-b) \Delta_{\textrm{\tiny TR}} + b \,\boldsymbol{1}  \big]^{-2}  \,
 \delta \Delta_{\textrm{\tiny TR}}
\bigg\}\, db
\nonumber
\\
&=&
2\,
 \textrm{Tr} \bigg\{ 
  \log \Delta_{\textrm{\tiny TR}}
\bigg(
 \int_0^1 \big[ (1-b) \Delta_{\textrm{\tiny TR}} + b \,\boldsymbol{1}  \big]^{-2} db 
 \bigg)
\,\delta \Delta_{\textrm{\tiny TR}}
\bigg\}\,.
\eea
Notice that $\log \Delta_{\textrm{\tiny TR}}$ and $\delta \Delta_{\textrm{\tiny TR}}$ do not commute in general. 
The last expression in (\ref{delta dFR step1}) is obtained 
from (\ref{var-ds-app-step0}) and 
\be
 \int_0^1 \big[ (1-b) \Delta_{\textrm{\tiny TR}} + b \,\boldsymbol{1}  \big]^{-2} db 
 \,=\,
 \Delta_{\textrm{\tiny TR}}^{-1}\,.
\ee

Straightforward matrix manipulations and the identity $\delta M^{-1} = -  M^{-1}\, \delta M\, M^{-1}$
lead to write (\ref{delta dFR step1}) as 
\bea
\label{variation dFR gen1}
 \delta d^2
&=&
2\,
\textrm{Tr} \Big\{ 
\big( \log \Delta_{\textrm{\tiny TR}} \big)\,
 \gamma_{\textrm{\tiny R}} \,\gamma_{\textrm{\tiny T}}^{-1}
\big( \delta \gamma_{\textrm{\tiny T}} \, \gamma_{\textrm{\tiny R}}^{-1} 
- \gamma_{\textrm{\tiny T}} \, \gamma_{\textrm{\tiny R}}^{-1} \,\delta\gamma_{\textrm{\tiny R}} \gamma_{\textrm{\tiny R}}^{-1} \big)
\Big\}
\\
\label{variation dFR gen1 step2}
&=&
2\,
\textrm{Tr} \Big\{ 
\big( \log \Delta_{\textrm{\tiny TR}} \big)\,
\Big(  \Delta_{\textrm{\tiny TR}}^{-1} \,
 \delta \gamma_{\textrm{\tiny T}} 
-  \delta\gamma_{\textrm{\tiny R}} 
\Big)
  \gamma_{\textrm{\tiny R}}^{-1} 
\Big\}\,.
\eea
Finally, $\delta d^2=2d\,\delta d$ and (\ref{f(MN) identity app}) with $f(x)=\log x$
provide the expression (\ref{first variation d-squared v1}).


In the following we compute separately the two sides of (\ref{first law complexity}).
Considering the r.h.s. of (\ref{first law complexity}) first, from (\ref{FR cost function}) one obtains
\be
\label{first law step 1 app}
\sum_{i j} 
\frac{\partial F}{\partial \dot{\gamma}_{i j}}\, \delta \gamma_{i j}\,
=\,
\frac{ \mathrm{Tr}\big[\gamma^{-1}\, \dot{\gamma}\, \gamma^{-1}\, \delta \gamma \big] }{\sqrt{\mathrm{Tr}
\big[
\left(
\gamma^{-1} \dot{\gamma}
\right)^2
\big]}}
\,.
\ee
From the expression (\ref{optimal circuit}) for the geodesic, it is not difficult to find that
\bea
\label{inverse G}
G_s^{-1}
&=&
\gamma_{\textrm{\tiny R}}^{-1/2} \,
\Big(  \gamma_{\textrm{\tiny R}}^{- 1/2}  \,\gamma_{\textrm{\tiny T}} \,\gamma_{\textrm{\tiny R}}^{-1/2}  \Big)^{-s}
\gamma_{\textrm{\tiny R}}^{-1/2} 
\\
\label{der G}
\partial_s G_s\,
&=&
\gamma_{\textrm{\tiny R}}^{1/2} \,
\Big[\log \Big(  \gamma_{\textrm{\tiny R}}^{- 1/2}  \,\gamma_{\textrm{\tiny T}} \,\gamma_{\textrm{\tiny R}}^{-1/2}  \Big)\Big]\,
\Big(  \gamma_{\textrm{\tiny R}}^{- 1/2}  \,\gamma_{\textrm{\tiny T}} \,\gamma_{\textrm{\tiny R}}^{-1/2}  \Big)^{s}
\gamma_{\textrm{\tiny R}}^{1/2} 
\\
\label{der inverse G}
\partial_s\,G^{-1}_s\,
&=&
-\,\gamma_{\textrm{\tiny R}}^{-1/2} \,
\Big[\log \Big(  \gamma_{\textrm{\tiny R}}^{- 1/2}  \,\gamma_{\textrm{\tiny T}} \,\gamma_{\textrm{\tiny R}}^{-1/2}  \Big)\Big]\,
\Big(  \gamma_{\textrm{\tiny R}}^{- 1/2}  \,\gamma_{\textrm{\tiny T}} \,\gamma_{\textrm{\tiny R}}^{-1/2}  \Big)^{-s}
\gamma_{\textrm{\tiny R}}^{-1/2}\,.
\eea
For the subsequent discussion, let us remark that, 
by specifying (\ref{der inverse G}) to $s=0$ and $s=1$, and using (\ref{f(MN) identity app}), we find
\bea
\label{dG1}
& &
\partial_s G_s^{-1} \big|_{s=1}
=
- \,\gamma^{-1}_{\textrm{\tiny T}} \log(\Delta_{\textrm{\tiny TR}})
=
\log\big(\gamma^{-1}_{\textrm{\tiny T}}\gamma_{\textrm{\tiny R}}\big)\,\gamma^{-1}_{\textrm{\tiny T}}
\\
\label{dG0}
\rule{0pt}{.5cm}
& &
\partial_s G_s^{-1} \big|_{s=0}
=
- \,\gamma^{-1}_{\textrm{\tiny R}} \log(\Delta_{\textrm{\tiny TR}})
=
\log\big(\gamma^{-1}_{\textrm{\tiny T}}\gamma_{\textrm{\tiny R}}\big)\,\gamma^{-1}_{\textrm{\tiny R}}\,.
\eea

The denominator in the r.h.s of (\ref{first law step 1 app})
along the geodesic (\ref{optimal circuit}) reads
\be
\label{den-tr-geod}
\sqrt{\mathrm{Tr}
\left[
\left(
G_s^{-1} \,\partial_s G_s
\right)^2
\right]}
\ee 
Combining (\ref{inverse G}) and (\ref{der G}), we observe that, for any $0\leqslant s \leqslant 1$,
this expression is equal to $d$ defined in (\ref{deltaAB}).
Furthermore, the numerator in the r.h.s. of (\ref{first law step 1 app}) at the endpoints of the geodesics
can be expressed by using (\ref{inverse G}), (\ref{der G}) and (\ref{der inverse G}).
Thus, from (\ref{first law step 1 app}) we get
\be
\label{first law step 2}
\sum_{ij}\,\frac{\partial F}{\partial \dot{\gamma}_{i j}}\, \delta \gamma_{i j}\,\bigg |_{0}^1\,
=\,
 \frac{1}{d}
 \bigg(
 \textrm{Tr} \Big\{ 
\partial_s G_s^{-1} \big|_{s=0} \, \delta \gamma_{\textrm{\tiny R}}
\Big\}
 -
\textrm{Tr} \Big\{ 
\partial_s G_s^{-1} \big|_{s=1} \, \delta \gamma_{\textrm{\tiny T}}
\Big\}
\bigg)\,.
\ee

As for the l.h.s. of (\ref{first law complexity}), let us consider $d^2$ from (\ref{deltaAB}).
First, one notices that (\ref{f(MN) identity app}) gives
\be
\label{auxiliary relation}
\gamma_{\textrm{\tiny R}}^{-1}
f \big( \gamma_{\textrm{\tiny T}}  \gamma_{\textrm{\tiny R}}^{- 1}  \big) 
=
\gamma_{\textrm{\tiny R}}^{-1/2} \,
f\big(  \gamma_{\textrm{\tiny R}}^{- 1/2}  \,\gamma_{\textrm{\tiny T}} \,\gamma_{\textrm{\tiny R}}^{-1/2} \big)
\, \gamma_{\textrm{\tiny R}}^{-1/2}\,.
\ee
The expressions obtained by specialising this result to $f(x)=\log x$ and $\left(\log x\right) x^{-1}$
allow to write (\ref{variation dFR gen1 step2}) as follows
\bea
\label{delta d squared}
\delta d^2
&=&
2\,
\textrm{Tr} \Big\{ 
  \,\gamma_{\textrm{\tiny R}}^{-1/2} \,
\Big[\log \Big(  \gamma_{\textrm{\tiny R}}^{- 1/2}  \,\gamma_{\textrm{\tiny T}} \,\gamma_{\textrm{\tiny R}}^{-1/2}  \Big)\Big]\,
\Big(  \gamma_{\textrm{\tiny R}}^{- 1/2}  \,\gamma_{\textrm{\tiny T}} \,\gamma_{\textrm{\tiny R}}^{-1/2}  \Big)^{-1}
\gamma_{\textrm{\tiny R}}^{-1/2} 
\, \delta\gamma_{\textrm{\tiny T}}
\Big\}\,
\nonumber
\\
\rule{0pt}{.6cm}
& &
-\,2\,\textrm{Tr} \Big\{ 
  \,\gamma_{\textrm{\tiny R}}^{-1/2} \,
\Big[\log \Big(  \gamma_{\textrm{\tiny R}}^{- 1/2}  \,\gamma_{\textrm{\tiny T}} \,\gamma_{\textrm{\tiny R}}^{-1/2}  \Big)\Big]\,
\gamma_{\textrm{\tiny R}}^{-1/2}
\, \delta\gamma_{\textrm{\tiny R}} 
\Big\}
\,.
\eea
Then, by using (\ref{der inverse G}) in  (\ref{delta d squared}), 
we obtain that $\delta d = (2d)^{-1}\delta d^2$ can be written as 
\be
\label{1st law final}
\delta d
=
 \frac{1}{d}
 \bigg(
 \textrm{Tr} \Big\{ 
\partial_s G_s^{-1} \big|_{s=0} \, \delta \gamma_{\textrm{\tiny R}}
\Big\}
 -
\textrm{Tr} \Big\{ 
\partial_s G_s^{-1} \big|_{s=1} \, \delta \gamma_{\textrm{\tiny T}}
\Big\}
\bigg)
\ee
whose r.h.s. coincides with the r.h.s. of (\ref{first law step 2}); hence (\ref{first law complexity}) is satisfied.
Furthermore,
by plugging (\ref{dG1}) and (\ref{dG0}) into (\ref{1st law final}),
the expression (\ref{first variation d-squared v1}) is obtained.


It is worth verifying that the Fisher-Rao cost function $F$ defined in (\ref{FR cost function}) evaluated along the path (\ref{optimal circuit}) satisfies the Euler-Lagrange equations, namely
\be
\label{EL eq}
\bigg[\,
\frac{\partial F}{\partial \gamma_{ij}}-\frac{d}{ds}\left(\frac{\partial F}{\partial \dot{\gamma}_{ij}}\right)
\bigg]\bigg|_{\gamma\,=\, G_s}
\!\!=\,0\,.
\ee
This consistency check can be performed first by observing that from (\ref{FR cost function}) we have 
\be
\label{appA-step2}
\frac{\partial F}{\partial \gamma}
=
-\frac{\gamma^{-1}\, \dot{\gamma}\, \gamma^{-1}\, \dot{\gamma}\, \gamma^{-1}}{
\sqrt{\textrm{Tr}\big[(\gamma^{-1}\dot{\gamma})^2\big]}}
\ee
while $\frac{\partial F}{\partial \dot{\gamma}}$ can be read from (\ref{first law step 1 app}). 
The expression  (\ref{appA-step2}) and $\frac{\partial F}{\partial \dot{\gamma}}$
along the geodesics (\ref{optimal circuit}) can be found 
by using (\ref{inverse G}), (\ref{der G}) and (\ref{der inverse G}).
Then, some algebra leads to (\ref{EL eq}).

We find it instructive specialising the above results to pure states.
Considering the geodesic given by (\ref{geodesic_pure}), 
whose initial and final covariance matrices are given in (\ref{gamma mat pure simple form}),
we have that both $\Delta_{\textrm{\tiny TR}}$ and $\Delta_{\textrm{\tiny TR}} + \delta \Delta_{\textrm{\tiny TR}} $ are diagonal,
hence they commute. 
This implies that (\ref{variation dFR gen1 step2}) can be obtained directly from (\ref{delta dFR step1}).
Indeed, since in this basis $\Delta_{\textrm{\tiny TR}} = \mathcal{X}^2_{\textrm{\tiny TR}}$, we have that (\ref{delta dFR step1}) becomes
\be
\label{variation dFR squared pure}
 \delta d^2
=\,
2\,
\textrm{Tr}\big[
\big( \log \mathcal{X}^2_{\textrm{\tiny TR}} \big)\,
\mathcal{X}^{-2}_{\textrm{\tiny TR}} \; \delta \mathcal{X}^2_{\textrm{\tiny TR}}
\,\big]\,.
\ee 
From (\ref{gamma mat pure simple form}) we find $\delta\gamma'_{\textrm{\tiny R}}=0$ 
and $\delta \gamma'_{\textrm{\tiny T}}=\frac{1}{2}\delta  \mathcal{X}^2_{\textrm{\tiny TR}}$. 
Thus, (\ref{variation dFR squared pure}) and (\ref{variation dFR gen1 step2})
are equivalent in the case of pure states.

In order to write (\ref{variation dFR squared pure}) in terms of the geodesics (\ref{optimal circuit}),
let us consider the $a$-th power (with $a\neq 0$)
of the geodesic (\ref{geodesic_pure})
and compute the derivative of the resulting matrix w.r.t. to the parameter $s$ along the geodesic.
The result reads
\be
\label{partial G_s a generic}
\big( \log \mathcal{X}^2_{\textrm{\tiny TR}} \big)\, \left(\mathcal{X}^2_{\textrm{\tiny TR}}\right)^{s\, a}
=
\frac{2^a}{a}\,
\partial_s \big(G_s\big)^a\,.
\ee
Setting $sa =-1$ in this expression, one finds
\be
\big( \log \mathcal{X}^2_{\textrm{\tiny TR}} \big)\, \left(\mathcal{X}^{2}_{\textrm{\tiny TR}}\right)^{-1}
=
-\frac{1}{2}\,\partial_s G_s^{-1}\big|_{s=1}\,.
\ee
This leads to write (\ref{variation dFR gen1 step2}) for pure states as follows
\be
\label{1st law final purestate}
 \delta d^2
 \,=\,
 -\,
 \textrm{Tr} \Big\{ 
 \,\partial_s G_s^{-1} \big|_{s=1} \, \delta \gamma_{\textrm{\tiny T}}
\Big\}
\ee
where we have also used that $ \delta\gamma'_{\textrm{\tiny T}}=\frac{1}{2}\delta  \mathcal{X}^2_{\textrm{\tiny TR}} $ 
in the basis that we are considering.
Since $\delta\gamma'_{\textrm{\tiny R}}=0$, one immediately realises 
that (\ref{1st law final purestate}) corresponds to (\ref{1st law final}) for pure states.

Another natural value for $s$ to choose in (\ref{partial G_s a generic}) 
is $s=1/2$, that corresponds to the middle point of the geodesic.
Comparing (\ref{partial G_s a generic}) with (\ref{variation dFR squared pure}),
it is natural to consider $s\,a =-1$, i.e. $a=-2$,
finding that (\ref{variation dFR squared pure}) can be written as
\be
 \delta d^2
=
-\frac{1}{4}\,\textrm{Tr} \Big\{ 
\partial_s G_s^{-2} \big|_{s=1/2}
\; \delta \big( \mathcal{X}^2_{\textrm{\tiny TR}} \big)
\Big\}\,.
\ee
We remark that $\mathcal{X}_{\textrm{\tiny TR}}$
is the diagonal matrix providing the complexity in the case of pure states.

Another useful expression for $\delta d$ comes from
the Williamson's decompositions (\ref{gamma R and T williamson}).
Considering variations $\delta \gamma$ such that $\gamma+\delta \gamma$ is also a covariance matrix
(in particular, $\delta \gamma$ is symmetric).
Given the Williamson's decomposition (\ref{williamson th gammaA}) for $\gamma$,
let us express $\delta \gamma$ in terms of the variations $\delta\mathcal{D}$ and $\delta W$
of the symplectic spectrum and of the symplectic matrix $W$ respectively. 
By using $\delta W^{\textrm{t}}= (\delta W)^{\textrm{t}}$, 
$\textrm{Tr}(MN^{\textrm{t}} ) = \textrm{Tr}(M^{\textrm{t}} N) $ 
and the fact that $\partial_s G_s^{-1} $ is symmetric\footnote{This property can be proved by first transposing (\ref{der inverse G}) and then using $f(M)^{\textrm{t}}=f(M^{\textrm{t}})$ and $[f(M),g(M)]=0$, that hold for generic functions of the same matrix.},
we can write (\ref{1st law final})  as 
\bea
\label{deltad_williamson_v1}
\rule{0pt}{.7cm}
\delta d
&=&
 \frac{1}{d}
 \bigg(
 \textrm{Tr} \Big\{ 
\partial_s G_s^{-1} \big|_{s=0} \, 
\big[\,
2\,W^{\textrm{t}}_{\textrm{\tiny R}}\, \mathcal{D}_{\textrm{\tiny R}} \,\delta W_{\textrm{\tiny R}}  
+ W^{\textrm{t}}_{\textrm{\tiny R}} \, \delta\mathcal{D}_{\textrm{\tiny R}} \,W_{\textrm{\tiny R}} 
\,\big] 
\Big\}
\\
& & \hspace{1.2cm}
 -\,
\textrm{Tr} \Big\{ 
\partial_s G_s^{-1} \big|_{s=1} \, 
\big[\,
2\,W^{\textrm{t}}_{\textrm{\tiny T}}\, \mathcal{D}_{\textrm{\tiny T}} \,\delta W_{\textrm{\tiny T}}  
+ W^{\textrm{t}}_{\textrm{\tiny T}} \, \delta\mathcal{D}_{\textrm{\tiny T}} \,W_{\textrm{\tiny T}} 
\,\big] 
\Big\}
\bigg)
\nonumber
\eea
in terms of the four contributions coming from the basis variations 
$\delta W_{\textrm{\tiny R}}  $ and $\delta W_{\textrm{\tiny T}}  $
and from the spectra variations $\delta\mathcal{D}_{\textrm{\tiny R}}$ and $\delta\mathcal{D}_{\textrm{\tiny T}}$.
The expression  (\ref{first-law-williamson}) for $\delta d$
can be easily obtained by
plugging (\ref{dG1}) and (\ref{dG0}) into (\ref{deltad_williamson_v1}).


\section{Thermofield double states}
\label{app-tfd}

In this appendix we consider the thermofield double states (TFD's) for the harmonic lattices \cite{Khanna09book},
whose circuit complexity has been explored in various studies over the last few years
\cite{Chapman:2018hou,Chapman:2019clq,Doroudiani:2019llj}.

The TFD's are pure states constructed by entangling two equal copies of the harmonic lattice 
in such a way that a thermal state of the original system is obtained by tracing out one of the two copies.
In Sec.\,\ref{app-TFD-cv} we provide the Williamson's decomposition of these pure states,
showing also that they are special cases of the analysis reported in Sec.\,\ref{sec:purification-approaches}.
In Sec.\,\ref{app-tfd-c2} the circuit complexity for TFD's is discussed.

Consider two harmonic lattices 
(that will be denoted as ``left'' and ``right''  in the following)
made by the same number $N$ of sites.
These two systems can be combined creating a system made by
$2N$ sites (denoted as ``doubled'' system)
whose hamiltonian reads
\be
\label{Hphys_TFD}
\widehat{H}_\textrm{\tiny d}
\equiv\frac{1}{2}\, \hat{\boldsymbol{r}}_\textrm{\tiny d}^{\textrm t} \,
H^{\textrm{\tiny phys}}_\textrm{\tiny d}  \,
\hat{\boldsymbol{r}}_\textrm{\tiny d}
=
\frac{1}{2}\, \hat{\boldsymbol{r}}_\textrm{\tiny d}^{\textrm t} \bigg(Q^{\textrm{\tiny phys}}_\textrm{\tiny d}
\oplus\, 
\frac{1}{m}\,\boldsymbol{1}\bigg) \, \hat{\boldsymbol{r}}_\textrm{\tiny d}
\;\qquad\;
Q^{\textrm{\tiny phys}}_{\textrm{\tiny d}}=Q^{\textrm{\tiny phys}}\oplus Q^{\textrm{\tiny phys}}
\ee
where
$\hat{\boldsymbol{r}}_\textrm{\tiny d}^{\textrm t} 
\equiv
(\hat{\boldsymbol{q}}_\textrm{\tiny l}^{\textrm t}, \hat{\boldsymbol{q}}_\textrm{\tiny r}^{\textrm t},
\hat{\boldsymbol{p}}_\textrm{\tiny l}^{\textrm t}, \hat{\boldsymbol{p}}_\textrm{\tiny r}^{\textrm t} )$,
where the subiindices refer to the left and right part of the doubled system,
and $Q^{\textrm{\tiny phys}}$ has been introduced in (\ref{Hphys_blocks}).
For the periodic chain the matrix $Q^{\textrm{\tiny phys}}$
has been written explicitly in (\ref{Hphys-mat-form}).

It is not difficult to adapt the diagonalisation procedure 
described in Sec.\,\ref{subsec:thermal} to (\ref{Hphys_TFD}).
This leads to construct the $4N\times 4N$ matrix 
$H^{\textrm{\tiny phys}}_\textrm{\tiny d}$ 
defined in (\ref{Hphys_TFD}) as follows (see (\ref{Hphys-Wdec}))
\be 
\label{Hphys_TFD_diag}
H^{\textrm{\tiny phys}}_\textrm{\tiny d}
=
V_\textrm{\tiny d}\,\mathcal{X}_\textrm{\tiny d}\,
\mathcal{D}_{\textrm{\tiny d}} \, \mathcal{X}_\textrm{\tiny d}\,V^{\textrm{t}}_{\textrm{\tiny d}}
\ee
where we have introduced the diagonal matrix 
$\mathcal{D}_\textrm{\tiny d}\equiv \mathcal{D}_{\textrm{\tiny phys}}\oplus \mathcal{D}_{\textrm{\tiny phys}}$, the symplectic and orthogonal matrix $V_\textrm{\tiny d}\equiv V\oplus V$ and the symplectic diagonal matrix
\be
\label{chi_phys_TFD}
\mathcal{X}_\textrm{\tiny d}=S\oplus S^{-1}
\,\,\qquad\,\,
S\equiv 
\mathcal{S} \oplus \mathcal{S}
\ee
where 
\be
\label{S-mat-def}
\mathcal{S}\equiv
\sqrt{m} \; \textrm{diag}\Big(\sqrt{\Omega_1},\dots,\sqrt{\Omega_N}\,\Big)
\ee
with $\Omega_k$ dispersion relation introduced through (\ref{Hphysgen_diag}).
These matrices are defined in terms of the $2N \times 2N$ matrices 
$V$ and $\mathcal{D}_{\textrm{\tiny phys}}$
in (\ref{Hphysgen_diag}) and (\ref{williamson-Hphys-gen}) respectively.

Since $V_\textrm{\tiny d}$ and $\mathcal{X}_\textrm{\tiny d}$ are symplectic matrices, 
the expression (\ref{Hphys_TFD_diag}) leads us to write the
Williamson's decomposition of $H^{\textrm{\tiny phys}}_\textrm{\tiny d}$ as follows
\be
\label{HphysTFD_Williamson}
H^{\textrm{\tiny phys}}_{\textrm{\tiny d}}= W^{\textrm{t}}_{\textrm{\tiny d}}\,\mathcal{D}_{\textrm{\tiny d}}\, W_{\textrm{\tiny d}}
\,\,\qquad\,\,
W_{\textrm{\tiny d}}= \mathcal{X}_\textrm{\tiny d}\,V^{\textrm{t}}_{\textrm{\tiny d}}\,.
\ee

This decomposition suggests to introduce the following set of canonical conjugated variables 
(see Sec.\,\ref{subsec:thermal})
\be
\label{s-var-TFD}
\hat{\boldsymbol{s}}_\textrm{\tiny d}\equiv W_\textrm{\tiny d} \,\hat{\boldsymbol{r}}_\textrm{\tiny d}
\;\;\qquad\;\;
\hat{\boldsymbol{s}}_\textrm{\tiny d}^{\textrm{t}}
\equiv
(\hat{\mathfrak{q}}_{\textrm{l},1} , \dots , \hat{\mathfrak{q}}_{\textrm{l},N}, \hat{\mathfrak{q}}_{\textrm{r},1} , \dots , \hat{\mathfrak{q}}_{\textrm{r},N},\hat{\mathfrak{p}}_{\textrm{l},1}, \dots, \hat{\mathfrak{p}}_{\textrm{l},N}, \hat{\mathfrak{p}}_{\textrm{r},1}, \dots, \hat{\mathfrak{p}}_{\textrm{r},N})\,.
\ee 
Defining the annihilation operators and the creation operators  for the two parts of the system as in 
(\ref{b_operators def}), one obtains a vector
$\hat{\boldsymbol{b}}_\textrm{\tiny d}^{\textrm t}
\equiv \big(
\hat{\boldsymbol{b}}_\textrm{\tiny l}^{\textrm t},
\hat{\boldsymbol{b}}_\textrm{\tiny r}^{\textrm t},
(\hat{\boldsymbol{b}}_\textrm{\tiny l}^\dagger)^{\textrm t},
(\hat{\boldsymbol{b}}_\textrm{\tiny r}^\dagger)^{\textrm t}
\big)$,
where $ \hat{\mathfrak{b}}_{l,k}$ and $ \hat{\mathfrak{b}}_{r,k}$ are
the $k$-th element of 
$\hat{\boldsymbol{b}}_\textrm{\tiny l}$ and $\hat{\boldsymbol{b}}_\textrm{\tiny r}$ respectively.
In terms of the components of $\hat{\boldsymbol{b}}_\textrm{\tiny d}$, 
the hamiltonian (\ref{Hphys_TFD}) becomes
\be
\widehat{H}_\textrm{\tiny d}
=
\sum_{k=1}^N 
\Omega_{k} \! \left( 
\hat{\mathfrak{b}}_{l,k}^\dagger\, \hat{\mathfrak{b}}_{l,k}
+\hat{\mathfrak{b}}_{r,k}^\dagger\, \hat{\mathfrak{b}}_{r,k} 
+1 \right) .
\ee
The standard quantisation procedure leads to introduce the eigenstates 
$|\boldsymbol{n}_{\textrm{\tiny r}} , \boldsymbol{n}_\textrm{\tiny l}\rangle 
\equiv
|\boldsymbol{n}_\textrm{\tiny l} \rangle_{\textrm{\tiny l}} 
\, |\boldsymbol{n}_\textrm{\tiny r} \rangle_{\textrm{\tiny r}} $
of the number operator, 
that can be factorised through the eigenstates 
$|\boldsymbol{n}_\textrm{\tiny l} \rangle_\textrm{\tiny l}$ 
and $|\boldsymbol{n}_\textrm{\tiny r}\rangle_\textrm{\tiny r} $
of the number operators corresponding to the two parts.
The eigenstates with $\boldsymbol{n}_{\textrm{\tiny r}}= \boldsymbol{n}_{\textrm{\tiny l}}
\equiv \boldsymbol{n}$
allow to define the thermofield double state (TFD) as follows \cite{Khanna09book}
\be
\label{TFD_Hlattice}
|\textrm{TFD}\rangle
=
\prod_{k=1}^{N}\sqrt{1-e^{-\beta\Omega_{k}}}\;
\sum_{\boldsymbol{n}} 
e^{-\frac{\beta}{2}\sum_{k=1}^N \Omega_{k} n_k } 
|\boldsymbol{n}\rangle_\textrm{\tiny l} \, |\boldsymbol{n}\rangle_\textrm{\tiny r} \,.
\ee
When $\beta \to \infty$, the TFD
becomes the product state of the two ground states
$ |\boldsymbol{0}\rangle_\textrm{\tiny l} \, |\boldsymbol{0}\rangle_\textrm{\tiny r}$.

Tracing out the degrees of freedom corresponding to one of the two parts, e.g. the right part, in (\ref{TFD_Hlattice})
one obtains
\be
\label{TFD-traced}
\textrm{Tr}_{\mathcal{H}_\textrm{r}}\big(|\textrm{TFD}\rangle\langle \textrm{TFD}|\big)
=
\prod_{k=1}^{N} \big(1-e^{-\beta\Omega_{k}} \big)
\sum_{\boldsymbol{n}}
e^{-\beta\sum_{k=1}^N \Omega_{k} n_k }
\,|\boldsymbol{n}\rangle_\textrm{\tiny l} \;  {}_\textrm{\tiny l}\langle\boldsymbol{n}|
\ee
which is the thermal density matrix for the left half system at temperature $1/\beta$.

\subsection{Covariance matrix}
\label{app-TFD-cv}

The covariance matrix of the TFD can be found 
through a slight generalisation of the procedure described in Sec.\,\ref{subsec:thermal} for the thermal states. 
From (\ref{def_CM_zero_mom}) and (\ref{s-var-TFD}), the covariance matrix of this pure state can be written as 
\be
\label{CM_TFD}
\gamma_{\textrm{\tiny TFD}}
=
\textrm{Re}\langle\textrm{TFD}|\,
\hat{\boldsymbol{r}}_\textrm{\tiny d} \, \hat{\boldsymbol{r}}_\textrm{\tiny d}^{\textrm t}\,
|\textrm{TFD}\rangle
=
W^{-1}_{\textrm{\tiny d}}\,
\textrm{Re}\big[ \langle\textrm{TFD}|\,
\hat{\boldsymbol{s}}_\textrm{\tiny d}\hat{\boldsymbol{s}}_\textrm{\tiny d}^{\textrm t}\,
|\textrm{TFD}\rangle\big] 
\,W^{-\textrm{t}}_{\textrm{\tiny d}}\,.
\ee
In order to compute the matrix 
$\textrm{Re}\langle\textrm{TFD}|\hat{\boldsymbol{s}}_\textrm{\tiny d}\hat{\boldsymbol{s}}_\textrm{\tiny d}^{\textrm t}|\textrm{TFD}\rangle$, 
one first expresses the operators in $\hat{\boldsymbol{s}}_\textrm{\tiny d}$ in terms of the creation and annihilation operators in $\hat{\boldsymbol{b}}_\textrm{\tiny d}$ 
and then  exploits their action on (\ref{TFD_Hlattice}). 
The non vanishing elements of 
$\textrm{Re}\langle\textrm{TFD}|\hat{\boldsymbol{s}}_\textrm{\tiny d}\hat{\boldsymbol{s}}_\textrm{\tiny d}^{\textrm t}|\textrm{TFD}\rangle$, 
are
\be
\label{qq-pp-tfd}
\begin{split}
\textrm{Re}\langle \hat{\mathfrak{q}}_{l,k} \hat{\mathfrak{q}}_{l,k} \rangle
&=
\textrm{Re}\langle \hat{\mathfrak{q}}_{r,k} \hat{\mathfrak{q}}_{r,k} \rangle=
\textrm{Re}\langle \hat{\mathfrak{p}}_{l,k} \hat{\mathfrak{p}}_{l,k} \rangle=
\textrm{Re}\langle \hat{\mathfrak{p}}_{r,k} \hat{\mathfrak{p}}_{r,k} \rangle=
 \frac{1}{2}
\coth(\beta \Omega_{k}/2)
\\
\rule{0pt}{.6cm}
\textrm{Re}\langle \hat{\mathfrak{q}}_{r,k} \hat{\mathfrak{q}}_{l,k} \rangle
&=
\textrm{Re}\langle \hat{\mathfrak{q}}_{l,k} \hat{\mathfrak{q}}_{r,k} \rangle=
-\textrm{Re}\langle \hat{\mathfrak{p}}_{r,k} \hat{\mathfrak{p}}_{l,k} \rangle=
-\textrm{Re}\langle \hat{\mathfrak{p}}_{l,k} \hat{\mathfrak{p}}_{r,k} \rangle=
 \frac{1}{2}\;
\frac{1}{\sinh(\beta \Omega_{k}/2)}
\end{split}
\ee
where the notation $\langle \widehat{O} \rangle \equiv \langle\textrm{TFD}|\, \widehat{O}\,|\textrm{TFD} \rangle$
has been adopted.
By using (\ref{qq-pp-tfd}), the covariance matrix 
$\textrm{Re}\langle\textrm{TFD}|\hat{\boldsymbol{s}}_\textrm{\tiny d}\hat{\boldsymbol{s}}_\textrm{\tiny d}^{\textrm t}|\textrm{TFD}\rangle$
in (\ref{CM_TFD}) can be written as
\be
\label{UpsilonTFD}
\textrm{Re}
\langle \hat{\boldsymbol{s}}_\textrm{\tiny d}\,
\hat{\boldsymbol{s}}_\textrm{\tiny d}^{\textrm t}\,\rangle
=
\Upsilon_\textrm{\tiny TFD}^{(+)}
\oplus \Upsilon_\textrm{\tiny TFD}^{(-)}
\,\,\qquad\,\,
\Upsilon_\textrm{\tiny TFD}^{(\pm)}\equiv 
\bigg(\,\begin{array}{cc}
\Lambda_{\textrm{\tiny TFD}} & \pm\tilde{\Lambda}_{\textrm{\tiny TFD}}
\\
 \pm\tilde{\Lambda}_{\textrm{\tiny TFD}} & \Lambda_{\textrm{\tiny TFD}}
\end{array}\bigg)
\ee
where we have introduced the following $N \times N$ diagonal matrices 
\be
\label{LambdaTFD}
\begin{split}
\Lambda_{\textrm{\tiny TFD}}
&
\equiv
\frac{1}{2}\,
\textrm{diag}\Big( 
\coth(\beta \Omega_{1}/2),\dots, 
\coth(\beta \Omega_{N}/2)\Big)
\\
\rule{0pt}{.8cm}
\tilde{\Lambda}_{\textrm{\tiny TFD}}
&
\equiv
\frac{1}{2}\,
\textrm{diag}\bigg( 
\frac{1}{\sinh(\beta \Omega_{1}/2)}\,,\dots, 
\frac{1}{\sinh(\beta \Omega_{N}/2)}\bigg)
\end{split}
\ee
which satisfy 
\be
\label{lambasq-lambdasq}
\Lambda_{\textrm{\tiny TFD}}^2-\tilde{\Lambda}^2_{\textrm{\tiny TFD}}=\frac{1}{4}\,\boldsymbol{1}
\ee
and
\be
\label{lambasqploverlambadamin}
\big(\Lambda_{\textrm{\tiny TFD}}+\tilde{\Lambda}_{\textrm{\tiny TFD}}\big)\big(\Lambda_{\textrm{\tiny TFD}}-\tilde{\Lambda}_{\textrm{\tiny TFD}}\big)^{-1}=\textrm{diag}\big(\coth^2\big(\beta\Omega_1/4\big),\dots,\coth^2\big(\beta\Omega_N/4\big)\big)\,.
\ee

Plugging (\ref{UpsilonTFD}) into (\ref{CM_TFD})
and employing the definition of $W_{\textrm{\tiny d}} $ in (\ref{HphysTFD_Williamson}),
for the covariance matrix of the TFD we find 
\be
\label{gammaTFDv2}
\gamma_{\textrm{\tiny TFD}}
=
V_{\textrm{\tiny d}} \, \mathcal{X}^{-1}_{\textrm{\tiny d}}
\Big(\Upsilon_\textrm{\tiny TFD}^{(+)}
\oplus \Upsilon_\textrm{\tiny TFD}^{(-)}
\Big)
\mathcal{X}^{-1}_{\textrm{\tiny d}} \,V_{\textrm{\tiny d}}^{\textrm{t}}
=
Q_{\textrm{\tiny TFD}}\oplus P_{\textrm{\tiny TFD}}
\ee
where (\ref{chi_phys_TFD}) and  $V_\textrm{\tiny d}\equiv V\oplus V$
have been employed to write the last expression,
which is given in terms of the following $2N\times 2N$ symmetric matrices
\be
\label{QandPTFDv2}
Q_{\textrm{\tiny TFD}}
=
V S^{-1} \,\Upsilon_\textrm{\tiny TFD}^{(+)} \,S^{-1} \,V^{\textrm{t}}
\,\,\qquad\,\,
P_{\textrm{\tiny TFD}}=V S\,\Upsilon_\textrm{\tiny TFD}^{(-)} \,S \, V^{\textrm{t}}\,.
\ee

By using that $V$ is an orthogonal matrix and that
\be 
\label{Upsilonpmcondition}
\Upsilon_\textrm{\tiny TFD}^{(+)} \, \Upsilon_\textrm{\tiny TFD}^{(-)}=\frac{1}{4}\,\boldsymbol{1}
\ee
which can be obtained from (\ref{UpsilonTFD}) and (\ref{lambasq-lambdasq}),
one finds
\be
\label{Puritycond_TFD}
Q_{\textrm{\tiny TFD}}\,P_{\textrm{\tiny TFD}}
=
\frac{1}{4}\,\boldsymbol{1}
\ee
as expected,
since the TFD is a pure state whose covariance matrix has non vanishing blocks only along the diagonal.

In order to write the Williamson's decompositions of $\gamma_{\textrm{\tiny TFD}}$,
let us observe that the matrices $\Upsilon_\textrm{\tiny TFD}^{(\pm)}
$ can be diagonalised by the $2N\times2N$ symplectic and orthogonal matrix $O$ as follows
\be
\label{Upsilonplminv2}
\Upsilon_\textrm{\tiny TFD}^{(\pm)}
=O^{\textrm{t}}\bigg(\big(\Lambda_{\textrm{\tiny TFD}}\pm\tilde{\Lambda}_{\textrm{\tiny TFD}}\big)\oplus\big(\Lambda_{\textrm{\tiny TFD}}\mp\tilde{\Lambda}_{\textrm{\tiny TFD}}\big)\bigg)O
\,\,\qquad\,\,
O=\frac{1}{\sqrt{2}}
\bigg(
\begin{array}{cc}
\boldsymbol{1}&\boldsymbol{1}
\\
-\boldsymbol{1}\;&\boldsymbol{1}
\end{array}
\bigg)\,.
\ee
By using (\ref{W-chi-O}), (\ref{lambasq-lambdasq}) and (\ref{lambasqploverlambadamin}), 
these matrices can be written as
\be
\label{Upsilonplmin}
\Upsilon_\textrm{\tiny TFD}^{(\pm)}
=
\frac{1}{2}\,O^{\textrm{t}}\mathcal{X}_{\textrm{\tiny TFD}} ^{\pm 2} \,O 
\ee
where
\be
\label{chimat_TFD}
\mathcal{X}_{\textrm{\tiny TFD}}
=\textrm{diag}\bigg(\sqrt{\coth\big(\beta\Omega_1/4\big)}\,,\dots,\sqrt{\coth\big(\beta\Omega_N/4\big)}\,,
\sqrt{\tanh\big(\beta\Omega_1/4\big)}\,,
\dots, \sqrt{\tanh\big(\beta\Omega_N/4\big)}
\,\bigg).
\ee
Plugging (\ref{chi_phys_TFD}) and (\ref{Upsilonplmin}) into (\ref{gammaTFDv2}),
one gets the Williamson's decompositions of $\gamma_{\textrm{\tiny TFD}}$ as
\be
\label{gammaTFD_Willi}
\gamma_{\textrm{\tiny TFD}}=\frac{1}{2}\,W^{\textrm{t}}_{\textrm{\tiny TFD}}\,W_{\textrm{\tiny TFD}}
\ee
where the  $4N\times 4N$ symplectic matrix $W_{\textrm{\tiny TFD}}$ is 
\be
\label{W_TFD-explicit}
W_{\textrm{\tiny TFD}}
=
\big(\mathcal{X}_{\textrm{\tiny TFD}} \,O\, S^{-1}V^{\textrm{t}}\big)
\oplus
\big(\mathcal{X}_{\textrm{\tiny TFD}}^{-1} \,O \,S\,V^{\textrm{t}}\big)\,.
\ee

It is instructive to express the fact that 
the TFD is a particular purification of a thermal state (see (\ref{TFD-traced}))
by identifying it within the analysis reported in Sec.\,\ref{sec:purification-approaches}.
This can be done by setting $N_{\textrm{\tiny anc}} = N$
and by rewriting the covariance matrix of the TFD in terms of the 
matrices occurring in (\ref{gamma_extended blocks-dec}).

Comparing (\ref{gammaTFDv2}) with (\ref{gamma-block-QPM-ext}), we easily conclude that
in this case  $Q_{\textrm{\tiny TFD}}$ and $P_{\textrm{\tiny TFD}}$ 
correspond to $Q_{\textrm{\tiny ext}}$ and $P_{\textrm{\tiny ext}}$ respectively,
while $M_{\textrm{\tiny ext}}=\boldsymbol{0}$.
Then, by employing the block diagonal matrices (\ref{chi_phys_TFD}), (\ref{UpsilonTFD}) and 
$V=\widetilde{V}\oplus\widetilde{V}$,
where $\widetilde{V}$ is the $N \times N$ orthogonal matrix 
(see (\ref{Hphysgen_diag}) and (\ref{Vtilde-def-even})-(\ref{Vtilde-def-odd}) 
for the periodic harmonic chain),
we can write $Q_{\textrm{\tiny TFD}}$ and $P_{\textrm{\tiny TFD}}$ 
as the partitioned matrices in (\ref{gamma_extended blocks-dec}) with 
\be
\label{QQPP_TFD}
Q=Q_{\textrm{\tiny anc}}
= \widetilde{V}\, \mathcal{S}^{-1}\Lambda_{\textrm{\tiny TFD}}\,\mathcal{S}^{-1}\,\widetilde{V}^{\textrm{t}}
\;\;\qquad\;\;
P=P_{\textrm{\tiny anc}}
= \widetilde{V}\, \mathcal{S}\,\Lambda_{\textrm{\tiny TFD}}\,\mathcal{S}\,\widetilde{V}^{\textrm{t}}
\ee
and
\be
\label{GammaPGammaQ_TFD}
\Gamma_Q
\,=\, 
\widetilde{V}\,\mathcal{S}^{-1}\tilde{\Lambda}_{\textrm{\tiny TFD}}\,\mathcal{S}^{-1}\,\widetilde{V}^{\textrm{t}}
\;\;\qquad\;\;
\Gamma_P\,=\,
 -\,\widetilde{V}\, \mathcal{S}\,\tilde{\Lambda}_{\textrm{\tiny TFD}}\,\mathcal{S}\,\widetilde{V}^{\textrm{t}}
 \,.
\ee

We remark that (\ref{QQPP_TFD}) 
and (\ref{GammaPGammaQ_TFD})
satisfy the conditions
 in (\ref{pure-state-exteded-conds-M0-bis}).
Furthermore, $Q \oplus P = Q_{\textrm{\tiny anc}} \oplus P_{\textrm{\tiny anc}}$ 
constructed from (\ref{QQPP_TFD}) 
provides the covariance matrix of a thermal state given in (\ref{Williamson_CM_th}), as expected. 
Thus,  the TFD is a purification of the thermal state 
and its covariance matrix satisfies (\ref{pure-state-exteded-conds-M0}).

\subsection{Complexity}
\label{app-tfd-c2}

The TFD are pure states; hence the complexity 
of a target TFD with respect to a reference TFD 
can be computed by employing (\ref{C-2-complexity-pure}).
In the most general case where 
the target TFD and the  reference TFD originate  from different hamiltonians, 
complicated expressions occur
because $W_{\textrm{\tiny TFD}}$ depends on the physical hamiltonian
through $S$ and $V$ in a non trivial way.

For the sake of simplicity, let us  focus on the special case where 
the same hamiltonian underlies both  the target TFD and the  reference TFD,
which are only distinguished by their inverse temperatures
$\beta_{\textrm{\tiny R}}$ and $\beta_{\textrm{\tiny T}} $.
In this case both the reference state and the target state have the same $S$ and $V$.
Moreover, since (\ref{Upsilonplminv2}) tells us that $O$ does not contain parameters,
the reference and target states that we are considering 
can be distinguished only through their matrices $\mathcal{X}_{\textrm{\tiny TFD,R}}$
and $\mathcal{X}_{\textrm{\tiny TFD,T}}$.
In this case, by employing (\ref{W_TFD-explicit}) we find that
the matrix defined in (\ref{W_TR def}) crucially simplifies to the following diagonal matrix
\be
\label{WTR_TFD}
W_{\textrm{\tiny TFD,TR}}= 
\big( \mathcal{X}_{\textrm{\tiny TFD,T}} \, \mathcal{X}_{\textrm{\tiny TFD,R}}^{-1} \big)
\oplus
\big( \mathcal{X}_{\textrm{\tiny TFD,T}}^{-1}\,\mathcal{X}_{\textrm{\tiny TFD,R}} \big)\,.
\ee

The circuit complexity corresponding to this choice of TFD's
can be obtained by  plugging (\ref{WTR_TFD}) into (\ref{C-2-complexity-pure}).
The result reads
\be
\label{C2_TFD_v1}
\mathcal{C}_2=\frac{1}{2\sqrt{2}}\sqrt{
\textrm{Tr} \,\Big\{\big[\log\! \big( \mathcal{X}_{\textrm{\tiny TFD,T}}^2\mathcal{X}_{\textrm{\tiny TFD,R}}^{-2} \big)\big]^2 \oplus \big[\log\! \big( \mathcal{X}_{\textrm{\tiny TFD,T}}^{-2}\mathcal{X}_{\textrm{\tiny TFD,R}}^{2} \big)\big]^2 \Big\}} 
\ee
which can be written more explicitly by employing (\ref{chimat_TFD}) for these TFD's, finding 
\be
\label{ComplexityTFDsamephysHam}
\mathcal{C}_2
=
\frac{1}{\sqrt{2}}\;
\sqrt{
\sum_{k=1}^N  \bigg[\log\! \bigg( \frac{\coth\big(\beta_{\textrm{\tiny T}}\Omega_{k}/4\big)}{\coth\big(\beta_{\textrm{\tiny R}}\Omega_{k}/4\big)} \bigg)\bigg]^2  }
\,.
\ee

An interesting regime to consider corresponds to $\beta_{\textrm{\tiny R}}\Omega_{k} \gg 1$.
In this limit the reference state is the product of the ground states of the two parts
because only $\boldsymbol{n} = \boldsymbol{0} $ contributes in (\ref{TFD_Hlattice}).
In this regime the complexity (\ref{ComplexityTFDsamephysHam}) simplifies to 
\be
\label{ComplexityTFDvsGSsamephysHam}
\mathcal{C}_2
=
\frac{1}{\sqrt{2}}\;\sqrt{
\sum_{k=1}^N \,\big[\log\! \big( \!\coth\big(\beta_{\textrm{\tiny T}}\Omega_{k}/4\big) \big)\big]^2  }
\ee
which is consistent with the results reported in \cite{Chapman:2018hou}.

We find it worth generalising (\ref{ComplexityTFDsamephysHam}) by considering 
a circuit where the reference state and the target state correspond 
to different hamiltonians that have 
the same matrix $V_\textrm{\tiny d}$ in their decompositions (\ref{Hphys_TFD_diag}). 
This is the case e.g. for the periodic harmonic chain explored in Sec.\,\ref{subsec-examples-hc},
where $\widetilde{V}$ defined in (\ref{Vtilde-def-even}) and (\ref{Vtilde-def-odd})
depends only on the number of sites of the chain,
hence it is independent of the parameters occurring in the hamiltonian of the chain. 
From (\ref{gammaTFDv2}) and (\ref{Puritycond_TFD}),
we have that $\gamma_{\textrm{\tiny TFD}}=Q_{\textrm{\tiny TFD}}\oplus 4 Q^{-1}_{\textrm{\tiny TFD}}$, which implies
\be
\gamma_{\textrm{\tiny TFD,T}} \, \gamma_{\textrm{\tiny TFD,R}}^{-1}
=
Q_{\textrm{\tiny TFD,T}} \,Q^{-1}_{\textrm{\tiny TFD,R}}
\oplus  
Q^{-1}_{\textrm{\tiny TFD,T}} \, Q_{\textrm{\tiny TFD,R}}\,.
\ee
This allows to write the complexity (\ref{c2 complexity}) as follows
\be
\label{C2simplified}
\mathcal{C}_2
=
\frac{1}{2\sqrt{2}} \,
\sqrt{\textrm{Tr} \Big\{ \big[\log (Q_{\textrm{\tiny TFD,T}} \,Q^{-1}_{\textrm{\tiny TFD,R}}
\oplus  Q^{-1}_{\textrm{\tiny TFD,T}} \,Q_{\textrm{\tiny TFD,R}}) \big]^2 \Big\}}
=
\frac{1}{2}\,
\sqrt{\textrm{Tr} \Big\{ \big[\log (Q_{\textrm{\tiny TFD,T}} \,Q^{-1}_{\textrm{\tiny TFD,R}})\big]^2 \Big\}}\,.
\ee

By applying (\ref{QandPTFDv2}) to this case, where the reference and target states have the same matrix $V$,
the argument of the logarithm in (\ref{C2simplified}) becomes 
\be
\label{QTQR_TFD}
Q_{\textrm{\tiny TFD,T}} \,Q^{-1}_{\textrm{\tiny TFD,R}}
\,=\,
V \,S_\textrm{\tiny T}^{-1} \,\Upsilon_\textrm{\tiny TFD,T}^{(+)} \,S_\textrm{\tiny T}^{-1} 
\,S_\textrm{\tiny R} \,\big(4 \Upsilon_\textrm{\tiny TFD,R}^{(-)}\big) \,S_\textrm{\tiny R} V^{\textrm{t}}
\ee
where we have used that $\big(\Upsilon_\textrm{\tiny TFD,R}^{(+)}\big)^{-1}=4 \Upsilon_\textrm{\tiny TFD,R}^{(-)}$
(see (\ref{Upsilonpmcondition})).
The relations in  (\ref{chi_phys_TFD}), (\ref{S-mat-def}), (\ref{UpsilonTFD}) and (\ref{LambdaTFD})
lead to write (\ref{QTQR_TFD}) as follows
\be
\label{QTQR_TFD_v2}
Q_{\textrm{\tiny TFD,T}} \, Q^{-1}_{\textrm{\tiny TFD,R}}
=
V \bigg(\,\begin{array}{cc}
\Lambda_{\textrm{\tiny TR}} & \tilde{\Lambda}_{\textrm{\tiny TR}}
\\
 \tilde{\Lambda}_{\textrm{\tiny TR}} & \Lambda_{\textrm{\tiny TR}}
\end{array}\bigg)V^{\textrm{t}}
\ee
where $\Lambda_{\textrm{\tiny TR}}$ and $\tilde{\Lambda}_{\textrm{\tiny TR}}$ 
are $N\times N$ diagonal matrices whose entries read
\bea
\label{LambdaTR}
\big(\Lambda_{\textrm{\tiny TR}}\big)_{k,k}
&=&
\frac{\Omega_{\textrm{\tiny R},k}}{\Omega_{\textrm{\tiny T},k}}\bigg[ \coth\big(\beta_{\textrm{\tiny R}}\Omega_{\textrm{\tiny R},k}/2\big)\coth\big(\beta_{\textrm{\tiny T}}\Omega_{\textrm{\tiny T},k}/2\big)-\frac{1}{\sinh\big(\beta_{\textrm{\tiny R}}\Omega_{\textrm{\tiny R},k}/2\big)\sinh\big(\beta_{\textrm{\tiny T}}\Omega_{\textrm{\tiny T},k}/2\big)}\bigg]
\nonumber
\\
\rule{0pt}{.8cm}
\big(\tilde{\Lambda}_{\textrm{\tiny TR}}\big)_{k,k}
&=&
\frac{\Omega_{\textrm{\tiny R},k}}{\Omega_{\textrm{\tiny T},k}}
\bigg[\,
\frac{\coth\big(\beta_{\textrm{\tiny R}}\Omega_{\textrm{\tiny R},k}/2\big)}{\sinh\big(\beta_{\textrm{\tiny T}}\Omega_{\textrm{\tiny T},k}/2\big)}-\frac{\coth\big(\beta_{\textrm{\tiny T}}\Omega_{\textrm{\tiny T},k}/2\big)}{\sinh\big(\beta_{\textrm{\tiny R}}\Omega_{\textrm{\tiny R},k}/2\big)} \,\bigg]
\hspace{1.5cm}
1\leqslant k \leqslant N\,.
\eea

By adapting the result given in (\ref{Upsilonplminv2}) for $\Upsilon_\textrm{\tiny TFD}^{(+)}$, 
we can diagonalise the matrix containing $\Lambda_{\textrm{\tiny TR}}$ and $\tilde{\Lambda}_{\textrm{\tiny TR}}$  
in the r.h.s. of (\ref{QTQR_TFD_v2}) through the orthogonal matrix $O$.
This leads to write (\ref{QTQR_TFD_v2}) as follows
\be
\label{QTQR_TFD_v3}
Q_{\textrm{\tiny TFD,T}} Q^{-1}_{\textrm{\tiny TFD,R}}
=
V O^{\textrm{t}}\bigg[\big(\Lambda_{\textrm{\tiny TR}}+\tilde{\Lambda}_{\textrm{\tiny TR}}\big)\oplus\big(\Lambda_{\textrm{\tiny TR}}-\tilde{\Lambda}_{\textrm{\tiny TR}}\big)\bigg]OV^{\textrm{t}}
\ee
where the entries of the diagonal matrices within the square brackets are given by 
\be
\label{LambdaTRcombintation}
\big(\Lambda_{\textrm{\tiny TR}}+\tilde{\Lambda}_{\textrm{\tiny TR}}\big)_{k,k}
=\frac{\Omega_{\textrm{\tiny R},k}\coth\big(\beta_{\textrm{\tiny T}}\Omega_{\textrm{\tiny T},k}/4\big)}{\Omega_{\textrm{\tiny T},k} \coth\big(\beta_{\textrm{\tiny R}}\Omega_{\textrm{\tiny R},k}/4\big)} 
\,\,\qquad\,\,
\big(\Lambda_{\textrm{\tiny TR}}-\tilde{\Lambda}_{\textrm{\tiny TR}}\big)_{k,k}
=\frac{\Omega_{\textrm{\tiny R},k}\coth\big(\beta_{\textrm{\tiny R}}\Omega_{\textrm{\tiny R},k}/4\big)}{\Omega_{\textrm{\tiny T},k}\coth\big(\beta_{\textrm{\tiny T}}\Omega_{\textrm{\tiny T},k}/4\big)}\,.
\ee

Plugging (\ref{QTQR_TFD_v3}) into (\ref{C2simplified}),
we find that the orthogonal matrices $V$ and $O$ do not contribute to the complexity.
By employing also (\ref{LambdaTRcombintation}), one obtains
\be
\label{ComplexityTFDHC}
\mathcal{C}_2
=
\frac{1}{2}\,
\sqrt{\sum_{k=1}^N
\Bigg\{\bigg[\log\bigg(\frac{\Omega_{\textrm{\tiny R},k}\coth\big(\beta_{\textrm{\tiny R}}\Omega_{\textrm{\tiny R},k}/4\big)}{\Omega_{\textrm{\tiny T},k} \coth\big(\beta_{\textrm{\tiny T}}\Omega_{\textrm{\tiny T},k}/4\big)}\bigg) \bigg]^2+\bigg[\log\bigg(\frac{\Omega_{\textrm{\tiny R},k}\coth\big(\beta_{\textrm{\tiny T}}\Omega_{\textrm{\tiny T},k}/4\big)}{\Omega_{\textrm{\tiny T},k}\coth\big(\beta_{\textrm{\tiny R}}\Omega_{\textrm{\tiny R},k}/4\big)}\bigg)\bigg]^2
\Bigg\}}\,.
\ee

In the regime $\beta_{\textrm{\tiny R}}\Omega_{\textrm{\tiny R},k}\gg 1$ this expression simplifies to
\be
\mathcal{C}_2
=
\frac{1}{2}\,\sqrt{
\sum_{k=1}^N
\Bigg\{\bigg[\log\bigg(\frac{\Omega_{\textrm{\tiny T},k} \coth\big(\beta_{\textrm{\tiny T}}\Omega_{\textrm{\tiny T},k}/4\big)}{\Omega_{\textrm{\tiny R},k} }\bigg) \bigg]^2+\bigg[\log\bigg(\frac{\Omega_{\textrm{\tiny R},k}\coth\big(\beta_{\textrm{\tiny T}}\Omega_{\textrm{\tiny T},k}/4\big)}{\Omega_{\textrm{\tiny T},k}}\bigg)\bigg]^2\Bigg\}}
\ee
which is consistent with the results reported in 
\cite{Chapman:2018hou}\footnote{See Eq.\;(192) of  \cite{Chapman:2018hou} at $t=0$.}.

\section{Diagonal and physical bases for the $\mathcal{C}_1$  complexity}
\label{app:c1}

In this appendix we briefly discuss the definition of the $\mathcal{C}_1$ complexity,
which is based on the $F_1$ cost function, 
hence it is a base dependent quantity. 
We also introduce the diagonal basis and the physical basis, 
slightly extending the definition given in  \cite{Caceres:2019pgf}.
Some results reported in Sec.\,\ref{subsec:purif_numerics}
have been obtained by employing these bases.

In the Nielsen's geometric approach to complexity between pure states
\cite{Nielsen06, NielsenDowling06, DowlingNielsen08},  
the circuit connecting the reference and the target states
is made by the unitary matrices $\widehat{U}_{\textrm{\tiny N}}(s)$, with $s\in [0,1]$, 
which are written as follows
\be
\label{circuitunitary}
\widehat{U}_{\textrm{\tiny N}}(s) = \overleftarrow{\mathcal{P}}\,e^{-\textrm{i}\int_0^{s} H_{\textrm{\tiny N}}(\sigma)\, d\sigma}
\,\,\qquad\,\,
H_{\textrm{\tiny N}}(\sigma) = \sum_{I} Y^{I}(\sigma)\,\widehat{K}_I
\ee
where $\overleftarrow{\mathcal{P}}$ is the path-ordered exponential indicating that the 
circuit is constructed from right to left as $s$ increases,
$ \widehat{K}_I$ are the hermitian generators of the unitary transformation
and the functions $Y^{I}(\sigma)$, that are called control functions, 
characterise the gates at a given point of the circuit.
The circuit depth is defined through cost function $F$ as follows
\be
\label{circuitdepth}
D\big(\widehat{U}_{\textrm{\tiny N}}\big)=\int_0^1 \! F\big(\widehat{U}_{\textrm{\tiny N}}(s),Y^{I}(s)\big)\, ds\,.
\ee
The complexity corresponds to the minimal circuit depth,
obtained by comparing all the possible unitary circuits connecting the reference state to the target state.
The allowed cost functions must satisfy some properties that have been discussed e.g. in 
\cite{Jefferson:2017sdb}.

In this manuscript we consider only the $F_1$ cost function and the $F_2$ cost function.
These cost functions are defined respectively as
\be
\label{costfunction}
F_1=\sum_I \big|Y^{I}\big|
\,\,\qquad\,\,
F_2=\Big[\, \sum_I \big(Y^{I} \big)^2 \,\Big]^{1/2}
\ee
and, through (\ref{circuitdepth}),
they provide the $\mathcal{C}_1$ complexity and the $\mathcal{C}_2$  complexity respectively.

Consider the harmonic lattice and the corresponding covariance matrix introduced in Sec.\,\ref{sec:FR-distance}.

In \cite{Chapman:2018hou}, the complexity of pure states has been studied
by employing  the fact that, since a unitary circuit can be represented as a circuit in $\textrm{Sp}(2N, \mathbb{R})$,
instead of (\ref{circuitunitary}),  for this model we can equivalently consider
\be
\label{circuitCMMyersChapman}
U(s)=\mathcal{P}\,e^{\int_0^s  K(\sigma)\, d\sigma}
\,\,\qquad\,\,
K(\sigma)=\sum_{I} \mathcal{Y}^{I}(\sigma) \,K_I
\ee
where $K_I$ are the generators of $\textrm{Sp}(2N, \mathbb{R})$,
hence the index $1\leqslant I\leqslant N(2N+1)$.

The symplectic matrix $U(s)$ in (\ref{circuitCMMyersChapman}) has been discussed also in Sec.\,\ref{subsec:pure}.
In particular, from (\ref{Ws-circuit-pure-states}) and Eq.\,(57) in \cite{Chapman:2018hou}, we have 
\be
G_s
=
U(s)\, \gamma_{\textrm{\tiny R}} \,U(s)^{\textrm{t}}
=
\frac{1}{2}\, W_s^{\textrm{t}} \, W_s
\,\,\qquad\,\,
G_{1}=\gamma_{\textrm{\tiny T}}
\ee
where $\gamma_{\textrm{\tiny R}}$ and $\gamma_{\textrm{\tiny T}}$ are the covariance matrices of the reference pure state and the target pure state respectively.
In \cite{Chapman:2018hou} the symplectic matrix $U(s)$ has been written 
in terms of  the matrix $\Delta_{\textrm{\tiny TR}}$ defined in (\ref{def relative CM})
as 
\be
\label{optimalpathMyersChapman}
U(s)=e^{s K}=e^{\frac{s}{2}\log\Delta_{\textrm{\tiny TR}}}
\;\;\qquad\;\;
K =\frac{1}{2}\log\Delta_{\textrm{\tiny TR}}\,.
\ee
By setting $U(s)=U_s$ defined in (\ref{newoptimalcircuitderivation}), one observes that $U(s)$
in (\ref{optimalpathMyersChapman}) coincides with (\ref{Ws-circuit-pure-states}).
Comparing (\ref{optimalpathMyersChapman}) with (\ref{circuitCMMyersChapman}), 
one observes that both $K$ and $\mathcal{Y}^I$
(obtained by expanding $K$ as in (\ref{circuitCMMyersChapman}))
are independent of $\sigma$.
Because of this feature, the integral in (\ref{circuitdepth}) is trivial to perform.
For the cost functions in (\ref{costfunction}), 
the results read respectively
\be
\label{optimal control func}
\mathcal{C}_1=\sum_I \big|\mathcal{Y}^{I}\big|
\,\,\qquad\,\,
\mathcal{C}_2=\Big[ \,\sum_I \big(\mathcal{Y}^{I}\big)^2 \,\Big]^{1/2}\,.
\ee

After a change of basis, the generators $K_I$ and the control functions in (\ref{circuitCMMyersChapman}) change respectively as follows
\cite{Jefferson:2017sdb,Chapman:2018hou} 
\be
\label{change basis generators}
 K_I=\sum_J O_{IJ} \,\widetilde{K}_J
 \;\;\qquad\;\;
 \mathcal{Y}^{I}=\sum_J O_{JI}\, \widetilde{\mathcal{Y}}^{J}
\ee
where $O_{IJ}$ are the entries of an orthogonal $N(2N+1)\times N(2N+1)$ real matrix $O$.
Given that $O$ is orthogonal,
in (\ref{optimal control func}) the $\mathcal{C}_2$ complexity is invariant 
while the $\mathcal{C}_1$ complexity is not.

In \cite{Caceres:2019pgf} the complexity of mixed states based on the purification complexity
(see Sec.\,\ref{sec:purification-approaches})
is mainly studied by employing the $F_1$ cost function. 
In particular, this $\mathcal{C}_1$ complexity is investigated in two different bases: 
the diagonal basis and the physical basis.

The diagonal basis in the extended system, which is in a pure state, 
is defined by the change of basis corresponding to 
the symplectic and orthogonal matrix $R$ introduced in (\ref{gamma-ps-generic}).

In order to introduce the physical basis, let us consider the 
wave function (\ref{wf-ps}) of the pure state characterising the extended system.  
This wave function is completely described by 
$N_{\textrm{\tiny ext}}\times N_{\textrm{\tiny ext}}$ 
complex symmetric matrix $E_{\textrm{\tiny ext}} +\textrm{i} F_{\textrm{\tiny ext}}$,
that can be written as follows
\be
\label{EextFext_blockdec}
E_{\textrm{\tiny ext}}+\textrm{i} F_{\textrm{\tiny ext}} 
\equiv
\bigg(\,\begin{array}{cc}
E+\textrm{i}F & \Gamma_E+\textrm{i}\Gamma_F
\\
\Gamma_E^{\textrm t}+\textrm{i}\Gamma_F^{\textrm t} \; & E_{\textrm{\tiny anc}}+\textrm{i}F_{\textrm{\tiny anc}}
\end{array}\bigg)
\ee
where $E$ and $F$ are $N\times N$ real symmetric matrices, 
$E_{\textrm{\tiny anc}}$ and $F_{\textrm{\tiny anc}}$ are $N_{\textrm{\tiny anc}}\times N_{\textrm{\tiny anc}}$ real symmetric matrices,
while $\Gamma_E$ and $\Gamma_F$ are $N \times N_{\textrm{\tiny anc}}$ real matrices.

In the physical basis both $E+\textrm{i}F$ and 
$E_{\textrm{\tiny anc}}+\textrm{i}F_{\textrm{\tiny anc}}$ are diagonal matrices. 
By employing a result of matrix algebra (see Corollary 4.4.4 of \cite{Horn13book}),
the complex and symmetric matrices 
$E+\textrm{i}F$ 
and $E_{\textrm{\tiny anc}}+\textrm{i}F_{\textrm{\tiny anc}}$ can be diagonalised as follows
\be
D=X (E+\textrm{i}F) X^{\textrm{t}}
\,\,\qquad\,\,
D_{\textrm{\tiny anc}}=X_{\textrm{\tiny anc}}(E_{\textrm{\tiny anc}}+\textrm{i}F_{\textrm{\tiny anc}})X_{\textrm{\tiny anc}}^{\textrm{t}}
\ee
where $D$ and $D_{\textrm{\tiny anc}}$ are real diagonal matrices with non negative entries
and the matrices $X$ and $X_{\textrm{\tiny anc}}$ are unitary. 
The physical basis is defined through the change of basis characterised by the matrix 
$X^{\textrm{\tiny phys}}\equiv X \oplus X_{\textrm{\tiny anc}}$, 
that brings the blocks on the diagonal of
$E_{\textrm{\tiny ext}}+\textrm{i} F_{\textrm{\tiny ext}}$ in (\ref{EextFext_blockdec})
in their diagonal forms. 
In the special case of $F_{\textrm{\tiny ext}}=\boldsymbol{0}$,
the definition of physical basis given in  \cite{Caceres:2019pgf} is recovered.


\bibliographystyle{nb}

\bibliography{refsMSC}

\end{document}